\documentclass[fleqn,usenatbib]{mnras}
\usepackage{newtxtext,newtxmath}
\usepackage{float}

\usepackage[T1]{fontenc}

\usepackage{amsmath}	
\usepackage{graphicx, kantlipsum}
\usepackage{caption}
\usepackage{subcaption}
\usepackage{chngpage}
\usepackage{natbib}
\usepackage{pgffor}
\usepackage{rotating}
\usepackage{geometry}
\usepackage{lscape}
\usepackage{longtable}
\usepackage{wrapfig}
\usepackage[noabbrev]{cleveref}
\usepackage[section]{placeins}
\usepackage{multirow}
\usepackage{array}
\usepackage{tikz}
\usepackage{tikz-3dplot}
\usetikzlibrary{quotes,angles}
\newcommand{\PreserveBackslash}[1]{\let\temp=\\#1\let\\=\temp}
\newcolumntype{C}[1]{>{\PreserveBackslash\centering}p{#1}}
\newcolumntype{R}[1]{>{\PreserveBackslash\raggedleft}p{#1}}
\newcolumntype{L}[1]{>{\PreserveBackslash\raggedright}p{#1}}
\usepackage{tablefootnote}

\title[Multiple components in the outflow of NML~Cyg]{Multiple components in the molecular outflow of the red supergiant NML~Cyg}

\author[H. Andrews et al.]{
H. Andrews\thanks{E-mail: holly.andrews@chalmers.se},
E. De Beck,
P. Hirvonen
\\
Division of Astronomy and Plasma Physics, Department of Space, Earth and Environment, Chalmers University of Technology, Gothenburg, Sweden
}

\date{Accepted 2021 November 04. Received 2021 November 04; in original form 2021 September 14}

\pubyear{2021}

\begin{document}
\label{firstpage}
\pagerange{\pageref{firstpage}--\pageref{lastpage}}
\maketitle

\begin{abstract}
Despite their large impact on stellar and galactic evolution, the properties of outflows from red supergiants are not well characterised. We used the Onsala 20m telescope to perform a spectral survey at 3\,mm and 4\,mm (68\,–\,116 \,GHz) of the red supergiant NML~Cyg, alongside the yellow hypergiant IRC~+10420. Our observations of NML~Cyg were combined with complementary archival data to enable a search for signatures of morphological complexity in the circumstellar environment, using emission lines from 15 molecular species. The recovered parameters imply the presence of three distinct, coherent and persistent components, comprised of blue-shifted and red-shifted components, in addition to an underlying outflow centred at the stellar systemic velocity. Furthermore, to reproduce $^{12}$CO emission with three-dimensional radiative transfer models required a spherical outflow with three superposed conical outflows, one towards and one away from the observer, and one in the plane of the sky. These components are higher in density than the spherical outflow by up to an order of magnitude. We hence propose that NML Cyg's circumstellar environment consists of a small number of high-density large-scale coherent outflows embedded in a spherical wind. This would make the mass-loss history similar to that of  VY~CMa, and distinct from $\mu$~Cep, where the outflow contains many randomly distributed smaller clumps. A possible correlation between stellar properties, outflow structures and content is critical in understanding the evolution of massive stars and their environmental impact.
\end{abstract}




\defcitealias{QL2016}{QL16}

\section{Introduction}
Massive stars, with initial masses $> 8 M_{\odot}$, are important astrophysical objects that act as powerful sources of chemical and mechanical enrichment through feedback processes from their stellar winds. Key stages for feedback from massive stars are the short-lived post-main sequence phases of red supergiants (RSGs) and yellow hypergiants (YHGs), which experience strong mass loss, at rates reaching up to $10^{-3} M_{\odot}$\,yr$^{-1}$ via episodic outflows \citep{deJager1998, levesque2010_physicsofRSGs}. These stellar stages are characterised by high luminosities ($10^{5} L_{\odot}$) and relatively cool temperatures ($T_{\textrm{eff}} \sim 3000 - 8000$\,K), allowing for the formation of molecules and dust close to the star.

Observations of molecular emission allow us to probe the circumstellar environment and provide constraints on the chemistry and dynamics. Ultimately, this provides information on the mass-loss process(es) at work. This is essential to improve our understanding of the evolution of massive stars as RSGs and YHGs, the resulting supernovae (SNe), and the final stellar remnants, neutron stars (NS) and black holes (BHs). Molecular emission has been observed around a small number of YHGs, including IRC~+10420, AFGL~2343 and IRAS 17163–3907 \citep{Castro-Carrizo2007,Wallstrom2015, QL2016}, as well as around several RSGs, including Betelgeuse, $\mu$ Cep, PZ Cas, and VY CMa \citep[e.g.,][]{muller2007_vycma,ziurys2007_vycma_complexity, debeck2010_comdot, tenenbaum2010_vycma_irc10216_ARO, teyssier2012, Kaminski2013TiO,  alcolea2013_vycma_hifistars, adande2013_vycma_ARO_so_so2, matsuura2014_vycma_spire_pacs,  Montarges2019}. All of these previous studies show evidence of a rich chemistry present in the environments surrounding massive evolved stars, for which a full understanding has not been reached.

Spectral surveys with wide bandwidths offer the ability to systematically detect a large number of transitions across multiple molecular species, and provide an unbiased manner in which to probe the chemical content of the stellar outflows. Spectral studies have only been carried out for a small number of massive evolved stars, and most have focused on lines in the sub-mm and IR regimes, rather than the transitions present at longer wavelengths. This paper presents the results of a spectral survey carried out at millimetre wavelengths for a RSG that has previous evidence for the presence of a large number of  molecules in its circumstellar environment, NML~Cyg. 

NML~Cyg is a M6 RSG with a luminosity of $6 \times 10^{5}L_{\odot}$ and high initial mass of $40\,M_{\odot}$, located 1.6\,kpc away in the Cygnus X region near the Cyg~OB2 massive stellar association  \citep{schuster2009, Zhang2012}. NML~Cyg has an estimated mass-loss rate of $4-5\times10^{-4}\,M_{\odot}$\,yr$^{-1}$, which has led to the production of a dense circumstellar environment \citep{gordon2018_nmlcyg_dust}. Previous observations of NML~Cyg have included dust-scattered light and thermal dust emission \citep[e.g.][]{schuster2009, gordon2018_nmlcyg_dust}, and maser and thermal emission from the circumstellar gas \citep[e.g.][]{etoka_diamond_2004, ziurys2009_hcnco, pulliam2011, teyssier2012, Zhang2012}. Asymmetries in the dust emission at a few 100~AU north-west (NW) from the star have been attributed to irradiation caused by the Cyg~OB2 association \citep{schuster2009}. The initial mass, bolometric luminosity, and dust mass-loss rate of NML~Cyg are similar to those of  the more extensively studied RSG VY CMa \citep{Zhang2012}. However, relatively little is known about the extent, morphology, physical conditions, and chemical composition of the outflow of NML~Cyg, making it an excellent candidate for detailed studies.

In this paper we report on millimetre observations of molecular emission from the circumstellar environment of NML~Cyg. Section~\ref{sect:obs} discusses the  Onsala Space Observatory (OSO) 20m radio telescope observations. Section~\ref{sect:results} presents the molecular content observed around NML~Cyg and constraints on the multi-component nature of the circumstellar environment, based on line decomposition and three-dimensional radiative transfer modelling. Section~\ref{sect:discussion} discusses our results, comparing the results from different modelling techniques, and placing the results for NML~Cyg in context with studies of other RSGs. Section~\ref{sect:conclusions} provides the main conclusions of the paper. 

\section{Observations}\label{sect:obs}

Observations were gathered using the Onsala Space Observatory 20m radio telescope, over 2016 Feb $11 - 15$ and  $21 - 24$, of the RSG NML~Cyg and the YHG IRC~+10420. The data were initially gathered as science verification data for the 3\,mm and 4\,mm receivers installed in 2015 and 2016 on the OSO 20m telescope \citep{oso_3mmrec_instrumentpaper, 4mm_oso_rec_instrument2016}. The use of these data as science verification for these receivers is given in more detail in Appendix \ref{sect:verification} (available online in the supplementary material). Frequencies, typical beam efficiencies and system temperatures for each tuning of the observations are listed in Table~\ref{tab:obssummary}. Each tuning had a width of 4$\,$GHz, allowing for a continuous frequency coverage across the total observed spectral range.

The data were primarily reduced with the use of CLASS in GILDAS \citep{CLASS}, with the additional use of XS \citep{OSOHandbook}, a bespoke data package designed for the reduction and analysis of spectra from the OSO telescopes.  In CLASS, spectra were checked for quality and expected noise levels. Scaling was applied to the individual spectra to convert from antenna temperature to main-beam temperature scales, with the use of specific main-beam efficiencies for each spectrum which varied with elevation (Table~\ref{tab:obssummary}). The typical aperture efficiency for the OSO 20m telescope is 22$\,$Jy/K at 86$\,$GHz \citep{OSOHandbook}. Spikes in the central observed channels were removed and 20$\,$MHz of edge channels were clipped due to the presence of sharp discontinuities. An additional scaling was applied to the spectral axis to account for possible errors from diurnal motion with final uncertainties on the order of a few hundred kHz, well below the final spectral resolution of the smoothed dataset which was used for analysis.

Reflections of waves from the inner surface of the telescope's radome caused standing waves in the spectra with a frequency of 250\,MHz, visible as spikes in the Fourier transform of the data, for frequency tunings centered at 70, 73, 76 and 79\,GHz. These standing waves were removed via manual inspection with the use of the data processing software XS \citep{OSOHandbook}, improving the measured root-mean squared noise levels (rms) of the final spectra.

Baseline corrections were applied to each individual spectrum, after masking regions with molecular emission. The baselines were adequately reproduced by first-degree polynomials. The individual spectra were then added together to create one spectrum per frequency tuning, with an additional iteration of baseline corrections.  

With a final spectrum for each frequency tuning, the spectra could then be combined to generate a final spectrum covering the total frequency range  $68\,-\,116$\,GHz. The final spectra had rms noise levels of $5 - 10$\,mK at a spectral resolution of 4\,km\,s$^{-1}$.

\begin{table}
\centering
		\caption{Summary of observation properties. The observations are split specifying the use of the 4\,mm or the 3\,mm receiver. Listed in each row is the central frequency of the frequency tuning (in GHz), the system temperature (in Kelvin), the total integration time on source (in hours), the size of the beam, $\theta$, (in arcseconds), the beam efficiencies, and the elevation of the observations (in degrees).}
		\label{tab:obssummary}
		\begin{tabular}{p{0.9cm}p{0.6cm}p{1.3cm}p{0.5cm}p{1.5cm}p{1.5cm}}
			\hline
			\noalign{\smallskip}
			$\nu_{\rm LO}$ & $T_{\rm sys}$ & Int Time & $\theta$ & $B_{\rm eff}$ & Elevation \\
			(GHz) & (K) & (hr)& ($^{\prime\prime}$)  & & (degrees) \\
			\noalign{\smallskip}
			\hline
			\noalign{\smallskip}
			4\,mm &&&&& \\
			\hline
			 70 & 362 & 5.7$^{\mathrm{a}}$,\,6.8$^{\mathrm{b}}$ & 54 & $0.53-0.63$ & $24.0-59.2$  \\
			 73 & 295 & 3.3 & 52 & $0.55 - 0.63$ & $45.0 - 68.0$ \\ 
		     76 & 240 & 1.6  & 50 & $0.51 - 0.55$ & $33.7 - 46.5$   \\
			 79 & 197 & 1.6 & 48& $0.54 - 0.59$ & $47.3 -  59.9$ \\
			 82 & 180 & 1.6 & 46 & $0.47 -  0.60$ & $25.8 - 67.5$ \\
			 85 & 191 & 3.1$^{\mathrm{a}}$,\,1.6$^{\mathrm{b}}$ & 44 & $0.46 - 0.55$ & $30.7 - 57.2$ \\
			\hline
			\noalign{\smallskip}
			3\,mm &&&&& \\
			\hline
			 87  &  169 & 1.0 & 43 & $0.45 - 0.54$ & $30.9 - 55.3$ \\
			 90  &  211 & 1.2$^{\mathrm{a}}$,\,1.0$^{\mathrm{b}}$ & 42 & $0.42 - 0.44$ & $25.9 - 32.2$ \\
			 93  &  163 & 1.2 & 41 & $0.41 - 0.42$ & $23.4 - 31.3$  \\
			 96  &  164 & 1.0$^{\mathrm{a}}$,\,1.4$^{\mathrm{b}}$ & 40 & $0.39 -  0.41$ & $25.9 - 33.8$ \\
			 99  &  143 & 0.8$^{\mathrm{a}}$,\,1.7$^{\mathrm{b}}$ & 38  &$0.40 - 0.43$ & $35.3 - 46.1$ \\
			 102 &  166 & 4.0$^{\mathrm{a}}$,\,3.6$^{\mathrm{b}}$ & 37 & $0.37 - 0.50$ & $30.5 - 65.9$ \\
			 105 &  169 & 2.2 & 36 & $0.36 - 0.41$ & $30.8 -  47.7$\\
		     108 &  190 & 2.5 & 35 & $0.35 - 0.52$ & $33.5 - 52.4$\\
			 111 & 229  & 3.0 & 34 & $0.34 - 0.44$ & $35.0 - 58.9$ \\
			 114 &  324 & 3.5 & 33 & $0.33 - 0.44$ & $35.7 - 62.7$ \\
			
			\noalign{\smallskip}
			\hline
		\end{tabular}
		\footnotesize{1.$^{\mathrm{a}}$ lists the integration times for a frequency tuning specific to IRC~+10420 and $^{\mathrm{b}}$ lists integration times specific to NML~Cyg.}
\end{table}

\onecolumn
\begin{landscape}
	\begin{longtable}
	{l crr rrr rrr rrr c}
		\caption{Gaussian line fits for NML~Cyg. For each emission line, the molecular species is listed, followed by the transition, the upper energy level of the transition, and the rest frequency. This is followed by the different fit parameters; the peak temperature $T_{\mathrm{p}}$, the mean velocity, $\varv_{\mathrm{m}}$, and the velocity width, $\Delta \varv$, where the additional subscripts ($b,\,c,\,r$) indicate whether the parameter value is for the blue-shifted, central, or red-shifted component, respectively. The last column indicates the observational data set where the line was detected; $1$ indicates from OSO, $2$ from JCMT \citep{debeck2010_comdot} and $3$ from HIFI \citep{teyssier2012}. Errors taken from the \textsc{curvefit} procedure for each parameter are given in parentheses.}
		\label{tab:gausslinefits}
		\\ \hline\\[-2ex]
		Species & Transition & $E_{\mathrm{upp}}/k$& Rest Freq.\footnotemark &  $T_{\mathrm{p,\,b}}$  &  $\varv_{\mathrm{m,b}}$ & $\Delta \varv_{\mathrm{b}}$ & $T_{\mathrm{p,\,c}}$ & $\varv_{\mathrm{m,\,c}}$ & $\Delta \varv_{\mathrm{c}}$ & $T_{\mathrm{p,\,r}}$ & $\varv_{\mathrm{m,r}}$ & $\Delta \varv_{\mathrm{r}}$ & D  \\
		&&(K)&(MHz)&(mK)&(km\,s$^{-1}$)&(km\,s$^{-1}$)&(mK)&(km\,s$^{-1}$)&(km\,s$^{-1}$)&(mK)&(km\,s$^{-1}$)&(km\,s$^{-1}$)&\\
		\hline\\[-2ex]
		\endfirsthead
		\\ \hline\\[-2ex]
		Species & Transition & $E_{\mathrm{upp}}/k$& Rest Freq.\footnotemark &  $T_{\mathrm{p,\,b}}$  &  $\varv_{\mathrm{m,b}}$ & $\Delta \varv_{\mathrm{b}}$ & $T_{\mathrm{p,\,c}}$ & $\varv_{\mathrm{m,\,c}}$ & $\Delta \varv_{\mathrm{c}}$ & $T_{\mathrm{p,\,r}}$ & $\varv_{\mathrm{m,r}}$ & $\Delta \varv_{\mathrm{r}}$ & D  \\
		&&(K)&(MHz)&(mK)&(km\,s$^{-1}$)&(km\,s$^{-1}$)&(mK)&(km\,s$^{-1}$)&(km\,s$^{-1}$)&(mK)&(km\,s$^{-1}$)&(km\,s$^{-1}$)&\\
		\hline
		\endhead
		\hline\\[-2ex]
        \multicolumn{3}{l}{Continued on next page\ldots} \\[0.5ex]
        \endfoot
        \\[-1.8ex]
        \endlastfoot
        \hline \\[-2ex]
		$^{12}$CO & $1 - 0$ & 5.5 & 115271.202         & 287 (12) & -20.0 (0.2) & 16.3 (0.2) & 361 (31) & 0.1 (0.3) & 22.6 (1.2) & 319 (30) & 16.3 (0.2) & 10.9 (0.4) & 1\\
		$^{12}$CO & $2 - 1$ & 16.6 &  230538.000       & 1368 (1) & -20.7 (0.1) & 16.1 (0.1) & 2220 (1) & 0.0 (0.1) & 28.4 (0.1) & 1269 (1) & 17.1 (0.1) & 8.8 (0.1) & 2 \\
	    $^{12}$CO & $3 - 2$ & 33.2 & 345795.990        & 3421 (23) & -19.7 (0.1) & 17.9 (0.1) & 4144 (45) & 0.0 (0.1) & 22.6 (0.2) & 3228 (30) & 16.5 (0.1) & 13.3 (0.1) & 2 \\
		$^{12}$CO & $4 - 3$ & 55.3 & 461040.768        & 1854 (18) & -20.5 (0.1) & 16.6 (0.1) & 2637 (30) & 0.0 (0.1) & 26.7 (0.2) & 1800 (30) & 15.8 (0.1) & 13.8 (0.1) & 2 \\
		$^{12}$CO & $6 - 5$ & 116.2 & 691473.076       & 1867 (90) & -19.4 (0.3) & 16.2 (0.4) & 2106 (272) & 0.0 (0.5) & 19.9 (1.5) & 1909 (198) & 16.0 (0.5) & 17.0 (0.5) & 2 \\
    	$^{12}$CO & $6 - 5$ & 116.2 & 691473.076       & 307 (1) & -18.7 (0.1) & 17.9 (0.1) & 437 (3) & 0.0 (0.1) & 21.5 (0.1) & 312 (2) & 15.3 (0.1) & 15.7 (0.1) & 3\\
		$^{12}$CO & $10 - 9$ & 304.2 & 1151985.452     & 308 (10) & -17.4 (0.3) & 23.7 (0.2) & 448 (20) & 0.0 (0.1) & 21.0 (0.4) & 262 (13) & 14.1 (0.2) & 17.2 (0.1) & 3  \\
		$^{12}$CO & $16 - 15$	& 751.7  & 1841346.506  & 280 (15) & -16.9 (0.5) & 21.6 (0.5) & 528 (40) & 0.0 (0.2) & 17.8 (0.4) & 176 (20) & 15.5 (1.0) & 21.0 (1.0) & 3 \\
		\hline \\[-2ex]
		$^{13}$CO&	$1 - 0$ & 5.3 & 110201.354  & 36 (6) & -13.7 (1.6) & 26.9 (2.5) & 38 (23) & 0.0 (3.0) & 10.5 (4.0) & 55 (7) & 14.7 (0.5) &  9.7 (0.8) & 1 \\
		$^{13}$CO&	$2 - 1$ & 15.9 & 220398.684 & 224 (1) & -23.4 (0.1) & 18.2 (0.1) & 190 (1) & 1.5 (0.1) & 36.8 (0.3) & 105 (1) & 25.6 (0.1) & 17.0 (0.1)& 2 \\
		$^{13}$CO& $6 - 5$	& 111.1 & 661067.277& 69 (1) & -22.5 (0.1) & 14.0 (0.1) & 112 (2) & 0.0 (0.1) & 26.6 (0.1) & 52 (2) & 16.4 (0.1) & 12.6 (0.1) & 3 \\
		$^{13}$CO& $10 - 9$ & 290.8 &1101349.597& 44 (30) & -16.1 (2.7) & 13.5 (3.0) & 161 (50) & 0.0 (0.6) & 17.1 (3.5) & 43 (30) & 14.6 (2.2) & 11.8 (3.5) & 3 \\
		$^{13}$CO & $16 - 15$& 718.7&1760485.983& 74 (10) & -9.5 (0.1)  & 3.5 (0.4)  & 198 (8) & 2.1 (0.1) & 13.2 (0.4) & 112 (7) & 17.6 (0.2) & 10.8 (0.2) & 3 \\
		\hline \\[-2ex]
		$^{28}$SiO & $2 - 1$  & 6.3  & 86846.985       & 138 (1) & -18.4 (0.1) & 17.0 (0.1) & 262 (1) & 0.0 (0.1) & 27.1 (0.1) & 103 (1) & 18.6 (0.1) & 15.0 (0.1) & 1 \\
		$^{28}$SiO & $16 - 15$  & 283.3 & 694294.114   & 40 (1) & -16.4 (0.1) & 11.8 (0.1) & 135 (1) & 0.0 (0.1) & 18.8 (0.1) & 33 (1) & 15.0 (0.1) & 11.8 (0.1) & 3 \\
		$^{28}$SiO, $v$=1 & $2 - 1$ & 1775.3 &86243.428& 663 (1) & -14.8 (0.1) & 14.2 (0.1) & 923 (1) & 0.0 (0.1) & 10.2 (0.1) & 35 (1) & 10.0 (0.1) & 15.7 (0.1) & 1 \\
		$^{28}$SiO, $v$=1& $13 - 12$ &1957.4&560325.870& 4 (1) & -19.1 (0.1) & 8.8 (0.2)  & 15 (1) & 0.9 (0.1) & 18.5 (0.2) & 4 (1) & 7.0 (0.1)  & 7.8 (0.5)& 3 \\
        $^{28}$SiO, $v$=1& $15 - 14$&2017.4&646429.438 & 15 (1) & -17.6 (0.1) & 5.9 (0.1)  & 32 (1) & 0.0 (0.1) & 15.4 (0.1) & 11 (1) & 12.7 (0.1) & 3.7 (0.1)& 3 \\
		\hline \\[-2ex]
		$^{29}$SiO & $2 - 1$  & 6.2 & 85759.194 & 22 (3) & -20.8 (0.4) & 11.3 (1.0) & 33 (3) & 1.7 (0.4) & 22.2 (1.7) & 21 (4) & 22.1 (0.4) & 10.5 (1.6) & 1 \\
		$^{29}$SiO & $13 - 12$ &187.2&557184.861& 22 (1) & -12.1 (0.1) & 17.5 (0.1) & 29 (1) & 1.6 (0.1) & 11.5 (0.1) & 20 (1) & 13.1 (0.1) & 14.9 (0.1)& 3 \\
		$^{29}$SiO & $26 - 25$&721.6&1112834.020& 46 (55) & -6.7 (4.1)  & 9.4 (4.1)  & 65 (77) & 1.1 (2.1) & 8.8 (5.0)  & 31 (22) & 11.3 (2.8) & 10.3 (4.0) & 3 \\
		\hline \\[-2ex]
		$^{30}$SiO & $2 - 1$  & 6.1 & 84746.166 & 19 (1) & -20.7 (0.1) & 19.7 (0.1) & 17 (1) & 1.5 (0.1) & 32.2 (0.9) & 8 (1) & 25.0 (0.1) & 19.4 (0.2) & 1 \\
		$^{30}$SiO & $26 - 25$&713.1&1099712.552& 29 (1) & -12.0 (0.1) & 8.0 (0.1)  & 37 (1) & 0.0 (0.1) & 12.1 (0.2) & 54 (1) & 5.7 (0.1) & 4.7 (0.1) & 3 \\
		\hline \\[-2ex]
		H$^{12}$CN & $1 - 0$  & 4.3 & 88631.602        & 44 (1) & -21.2 (0.1) & 18.9 (0.1) & 60 (1) & 1.5 (0.1) & 37.7 (0.1) & 28 (1) & 19.2 (0.1) & 15.1 (0.1) & 1 \\
		H$^{12}$CN & $13 - 12$ & 386.9 & 1151449.088   & 17 (4) & -18.0 (0.8) & 11.8 (1.6) & 91 (4) & 0.0 (0.2) & 18.7 (0.6) & 66 (5) & 17.0 (0.1) & 3.9 (0.2)& 3 \\
		\hline \\[-2ex]
		H$^{13}$CN & $1 - 0$& 4.1& 86339.921    & 8 (1) & -14.5 (0.1) & 15.4 (0.1) & 7 (1) & 4.5 (0.1) & 24.0 (0.1) & 7 (1) & 22.6 (0.1) & 7.5 (0.1) & 1 \\
		\hline \\[-2ex]
		o-H$_{2}^{16}$O, $v = 0$&$1_{1,0} - 1_{0,1}$& 61.0 &556935.988 & 318 (1) & -9.7 (0.1)  & 17.8 (0.1) & 171 (1) & 1.5 (0.1) & 11.0 (0.1) & 225 (1) & 16.1 (0.1) & 17.0 (0.1) & 3 \\
		o-H$_{2}^{16}$O, $v = 0$&$3_{1,2} - 2_{2,1}$ &249.4&1153126.821& 605 (32) & -12.8 (0.2) & 17.4 (0.2) & 741 (92) & 1.5 (0.3) & 19.9 (0.8) & 407 (58) & 15.0 (0.9) & 20.9 (0.6) & 3 \\
		o-H$_{2}^{16}$O, $v = 0$&$ 3_{2,1} - 3_{1,2}$&305.2&1162911.602& 636 (17) & -9.9 (0.3)  & 25.0 (0.2) & 536 (46) & 3.0 (0.2) & 17.5 (0.4) & 475 (24) & 15.4 (0.4) & 20.0 (0.3) & 3 \\
		p-H$_{2}^{16}$O, $v = 0$&$ 1_{1,1} - 0_{0,0}$&53.4&1113343.007 & 705 (1) & -11.4 (0.1) & 13.3 (0.1) & 797 (1) & 0.0 (0.1) & 11.8 (0.1) & 510 (1) & 15.2 (0.1) & 18.4 (0.1) & 3 \\
		p-H$_{2}^{16}$O, $v = 0$&$ 4_{2,2} - 4_{1,3}$&454.3&1207638.730& 395 (42) & -13.0 (1.5) & 35.3 (2.0) & 513 (116) & 2.9 (0.3) & 22.7 (1.0) & 189 (77) & 18.0 (6.0) & 35.3 (5.0) & 3 \\
	    p-H$_{2}^{16}$O, $v = 0$ &$ 6_{3,3} - 6_{2,4}$&951.8&1762042.791& 368 (47) & -8.0 (0.6)  & 15.5 (0.6) & 470 (49) & 5.0 (0.5) & 15.9 (0.6) & 73 (19) & 15.0 (4.0) & 35.3 (5.0) & 3 \\
		o-H$_{2}^{16}$O, $v_{2}$=1&$ 1_{1,0} - 1_{0,1}$&2360.3&658006.251& 451 (1) & -5.5 (0.1) & 10.4 (0.1) & 658 (1) & 0.0 (0.1) & 7.2 (0.1) & 85 (1) & 5.0 (0.1) & 7.5 (0.1) & 3 \\
		p-H$_{2}^{16}$O, $v_{2}$=1&$ 1_{1,1} - 0_{0,0}$&2352.4&1205789.095 & 134 (14) & -9.4 (0.5) & 11.8 (0.8) & 176 (39) & 1.6 (0.4) & 9.5 (1.0) & 21 (24) & 11.0 (5.2) & 11.8 (6.7) & 3 \\
		\hline \\[-2ex]
		p-H$^{18}_{2}$O, $v_{2}$=1 & $ 1_{1,1} - 0_{0,0}$ & 52.9 & 1101698.256    & 66 (20) & -6.3 (0.9) & 16.2 (0.5) & 132 (24) & 3.3 (0.7) & 18.2 (1.0) & 13 (5) & 18.0 (3.4) & 23.9 (3.0) & 3 \\
		\hline \\[-2ex]
		SO & $2_{2} - 1_{1}$ &19.3&86093.950        & 7 (1) & -20.5 (0.1) & 17.9 (0.1) & 11 (1) & 1.0 (0.1) & 18.1 (0.1) & 6 (1) & 13.6 (0.1) & 7.9 (0.1) & 1 \\
		SO & $2_{3} - 1_{2}$ &9.2& 99299.870        & 49 (1) & -21.3 (0.1) & 13.4 (0.1) & 43 (1) & 1.0 (0.1) & 27.4 (0.1) & 54 (1) & 19.0 (0.1) & 6.3 (0.1) & 1 \\
		SO & $3_{2} - 2_{1}$ &21.1&109252.220       & 22 (1) & -18.6 (0.1) & 24.1 (0.1) & 16 (1) & 5.0 (0.1) & 29.6 (0.1) & 20 (1) & 19.8 (0.1) & 5.7 (0.1) & 1 \\
		SO&$13_{14} - 12_{13}$&192.7&560178.650     & 12 (1) & -15.0 (0.1) & 10.4 (0.1) & 20 (1) & 0.0 (0.1) & 11.4 (0.1) & 13 (1) & 13.6 (0.1) & 7.9 (0.1) & 3 \\
		SO&$12_{12} - 11_{11}$&194.4&558087.642     & 11 (1) & -20.6 (0.1) & 6.4 (0.1)  & 16 (1) & 0.2 (0.1) & 19.0 (0.1) & 4 (1) & 13.0 (0.1) & 14.1 (0.1) & 3 \\
		SO&$13_{13} - 12_{12}$&201.2 &559319.721    & 15 (1) & -18.2 (0.1) & 9.0 (0.1)  & 14 (1) & 0.6 (0.1) & 20.0 (0.1) & 3 (1) & 15.4 (0.1) & 14.1 (0.2) & 3  \\
		SO&$15_{16} - 14_{15}$&252.6&645875.924     & 13 (1) & -9.7 (0.2)  & 21.5 (0.3) & 18 (1) & 0.0 (0.1) & 13.1 (0.1) & 15 (1) & 12.9 (0.1) & 10.5 (0.1) & 3 \\
		\hline \\[-2ex]
		SO$_{2}$&$8_{1,7} - 8_{0,8}$&36.7&83688.093 & 12 (1) & -20.6 (0.1) & 18.2 (0.1) & 11 (1) & 0.0 (0.1) & 23.5 (0.4) & 5 (1) & 16.7 (0.1) & 13.1 (0.1) & 1 \\
		SO$_{2}$&$16_{2,14} - 15_{3,13}$&137.5&104033.582& 12 (1) & -13.7 (0.1) & 12.4 (0.1) & 16 (1) & 0.0 (0.1) & 14.2 (0.1) & 1 (1) & 16.0 (1.1) & 28.3 (1.6) & 1 \\
		SO$_{2}$&$10_{1,9} - 10_{0,10}$&54.7&104239.295  & 23 (1) & -24.1 (0.3) & 7.9 (0.1)  & 15 (1) & 0.0 (0.1) & 23.0 (0.1) & 12 (1) & 17.6 (0.1) & 7.1 (0.1) & 1 \\
		\hline \\[-2ex]
		SiS &  $6 - 5$ & 18.3 & 108924.301             & 23 (1) & -15.8 (0.1) & 16.5 (0.1) & 23 (4) & 4.0 (1.0) & 23.5 (2.0) & 11 (4) & 18.3 (1.0) & 18.4 (0.7) & 1 \\
		\hline \\[-2ex]
		NH$_{3}$&$1_{0} - 0_{0}$ &27.5&572498.160      & 72 (60) & -18.4 (1.8) & 17.1 (3.7) & 70 (100) & 0.0 (5.0) & 29.8 (6.8) & 32 (160) & 14.7 (5.0) & 20.7 (7.6) & 3 \\
		\hline \\[-2ex]
		H$_{2}$S & $3_{1,2} - 2_{2,1}$&136.8&1196012.118 & 70 (34) & -10.0 (4.7) & 23.5 (4.6) & 112 (55) & 0.0 (0.4) & 13.9 (2.0) & 82 (12) & 11.3 (0.4) & 9.4 (0.6) & 3 \\
		\hline \\[-2ex]
		OH & $^{2}\Pi_{1/2}, 3/2 - 1/2$ & 269.8 & 1834746.855 & 570 (15) & -11.7 (0.2) & 16.5 (0.2) & 660 (32) & 2.2 (0.1) & 14.2 (0.3) & 609 (17) & 15.6 (0.2) & 15.4 (0.2)& 3 \\
		\hline
	\end{longtable}
	\footnotesize{2. Rest frequencies for all molecular transitions are taken from the CDMS database \citep{Muller2005, Muller2001}, with the exception of o- and p-H$_{2}$O and NH$_{3}$ where rest frequencies are taken from the JPL database \citep{JPL}. Quantum numbers listed are the following: $J$ for CO, HCN, SiO, SiS, $N_J$ for SO, $J_{K_{\mathrm{a}},K_{\mathrm{c}}}$ for SO$_2$, H$_2$O and H$_2$S, and $N_J$ for OH.}
\end{landscape}
\twocolumn

\section{Results}\label{sect:results}

We detected 15 emission lines from 10 molecular species and isotopologues towards NML~Cyg with the OSO 20m radio telescope (Table~\ref{tab:gausslinefits}). The majority of detected species were S-bearing and Si-bearing. We also detected $^{12}$CO and H$^{12}$CN, and their isotopologues $^{13}$CO and H$^{13}$CN. 
\subsection{Multi-component fits to emission profiles} 
Previous observations of thermal and maser molecular emission towards NML~Cyg have shown clear evidence of multiple velocity components  \citep[][see also H$_2$O emission in Fig.~\ref{fig:h2o_lines}]{teyssier2012, Zhang2012}. The OSO 20m observations show multiple peaks, rather than flat-topped, two-horned, or close-to-parabolic line profiles, implying deviation from a simple spherical circumstellar geometry. Furthermore, whereas high optical depths can cause self-absorption in the blue wings of emission lines from spherical outflows  \citep[see e.g. SiO towards R Dor and CO towards IRC +10216;][]{debeck2012,debeck_2018_rdor}, the observed profiles for NML Cyg often exhibit blue-shifted components which are as strong or stronger than the rest of the profile (e.g. Figs.~\ref{fig:so_lines} and~\ref{fig:so2_lines}, available online in the supplementary material). This finding motivated an investigation into whether we could quantify the presence of multiple thermal line velocity components for the first time for this source. Multiple outflow components have previously been reported for the RSG VY~CMa \citep{ziurys2007_vycma_complexity, adande2013_vycma_ARO_so_so2}. Spatially resolved observations of its circumstellar environment show clear evidence that thermal molecular emission from different species traces different regions within the stellar winds, as seen for, e.g., TiO, TiO$_{2}$, and NaCl \citep{Kaminski2013TiO, debeck_2015_tio2_vycma, decin_2016_vycmanacl}.

In order to determine whether any trends were present across molecular species and excitation conditions, we included archival data from Herschel/HIFI \citep{teyssier2012} and JCMT \citep{debeck2010_comdot}. 
In this first attempt at characterising the multi-component nature of the outflow from NML~Cyg, we assumed that the components could be represented in the line profile by
(a) a central component centered at the systemic velocity, (b) a red-shifted directed outflow, and (c) a blue-shifted directed outflow. This approximation was applied as the superposition of three Gaussian fits to the line profiles, with each Gaussian representing one of these components. Each Gaussian component was determined by implementing a non-linear least squares fit. We acknowledge that Gaussian line profiles are not fully representative of a spherical wind or directional outflows, but use these as a first approximation to derive initial values for the properties of the components, as was carried out in similar procedures for VY~CMa \citep{ adande2013_vycma_ARO_so_so2,alcolea2013_vycma_hifistars}. 

The output fit parameters determined for each Gaussian component were the peak amplitude, the full width at half maximum (FWHM), and the position of the peak. This corresponded to nine final output parameters per spectral line: the peak temperatures $T_{\mathrm{p,\,b}}$, $T_{\mathrm{p,\,c}}$, $T_{\mathrm{p,\,r}}$, the central velocities (relative to the systemic velocity v$_{\mathrm{LSR}}$) $\varv_{\mathrm{m,b}}$, $\varv_{\mathrm{m,\,c}}$, $\varv_{\mathrm{m,r}}$, and the velocity widths $\Delta \varv_{\mathrm{b}}$, $\Delta \varv_{\mathrm{c}}$, and $\Delta \varv_{\mathrm{r}}$, where the subscripts b, c, and r refer to the blue-shifted, central, and red-shifted components, respectively. The derived parameters for each molecular line transition are presented in Table~\ref{tab:gausslinefits}.

The fits were determined with a consideration of appropriate boundary conditions, as summarised in Table~\ref{tab:boundary_conditions}. The central component was assumed to fit within 5\,km\,s$^{-1}$ of the  systemic velocity ($\varv_{\mathrm{LSR}}$) of the source, taken to be 2.5\,km\,s$^{-1}$. This value is based on the central peak of the 658$\,$GHz $J_{K_{\mathrm{a}},K_{\mathrm{c}}} = 1_{1,0} - 1_{0,1} (v_{2} = 1)$ o-H$_{2}$O maser line\footnote{Previous measurements in \citet{teyssier2012} from this line quoted $\varv_{\mathrm{LSR}}$ =  0\,km\,s$^{-1}$.}, and is also found to fit well to the central component of many of the thermal emission lines across multiple molecular species (see Table \ref{tab:gausslinefits}). The range of possible values within which $\varv_{\mathrm{m,r}}$ and $\varv_{\mathrm{m,b}}$ could vary were not allowed to overlap with this central region on either side. The values for $\varv_{\mathrm{m,r}}$ and $\varv_{\mathrm{m,b}}$ were also given outer boundary conditions of $\pm30\,$km\,s$^{-1}$ from the $\varv_{\mathrm{LSR}}$, as outflow components around the star were expected to fall within the expected linewidth of NML~Cyg. This was taken as $\sim30\,$km\,s$^{-1}$, in line with previous observations of NML Cyg that have indicated an expansion velocity of $\sim33\,$km\,s$^{-1}$ \citep{debeck2010_comdot}, and is supported by the measurement of the central components for the $^{13}$CO line transitions for NML Cyg.
The FWHM of each Gaussian ($\Delta\varv$) was given an upper boundary constraint of $50\,$km\,s$^{-1}$. 

\begin{table}
\caption{List of boundary conditions for multi-component fits. The first column lists the parameter and the second column lists the boundary conditions constraining the parameters for each component.
    }
    \label{tab:boundary_conditions}
    \centering
   \begin{tabular}{c|C{4cm}}
        \hline
        Parameter & Conditions \\
        \hline
        Peak Temperature (K) &  $T_{\mathrm{p,\,b}} > 0$  \\
        & $T_{\mathrm{p,\,c}} > 0$  \\
        & $T_{\mathrm{p,\,r}} > 0$ \\
       Mean Velocity (km\,s$^{-1}$) 
        & $-30.0 < \varv_{\mathrm{m,\,b}} < -5.0$ \\
        & $-5.0 < \varv_{\mathrm{m,\,c}} < 5.0$ \\
        & $5.0 < \varv_{\mathrm{m,\,r}} < 30.0$  \\
        Velocity Width (km\,s$^{-1}$) & $\Delta \varv_{\mathrm{b}} < 50.0$ \\
        & $\Delta \varv_{\mathrm{c}} < 50.0$ \\
        & $\Delta \varv_{\mathrm{r}} < 50.0$ \\
        \hline
   \end{tabular}
    
\end{table}

Our approximation did not assume bipolar outflows or symmetric directional outflows, allowing $T_{\mathrm{p}}$, $\varv_{\mathrm{m}}$, and $\Delta \varv$ to vary independently for the three components. Uncertainties on the fit parameters were obtained from the co-variance matrices returned by the $\textsc{scipy curve\_fit}$ procedure, extrapolated from the rms noise in the observations \citep{scipy_reference}. These uncertainties are lower limits to the true level of uncertainty, as we do not consider optical depth effects, which will be more significant for the spectral lines of highly abundant species such as CO, SiO, SO, SO$_{2}$, and H$_{2}$O.

We applied multi-component fits to 49 lines from 15 molecular species and isotopologues. Central, red-shifted, and blue-shifted components were fit to all the lines, as shown in Table~\ref{tab:gausslinefits}. A demonstration of the multi-component fits is provided in the main body of text with the example of $^{12}$CO (Fig. \ref{fig:co_nml_multifits}), with fits to all other species/isotopologues provided in the appendix, as part of the online supplementary material. We investigated the fit parameters for different transitions of the same molecular species to explore possible trends with excitation temperature. For each molecular species where multiple transitions were observed we created three plots: (1) a plot comparing $\varv_{\mathrm{m,b}}$, $\varv_{\mathrm{m,\,c}}$, and $\varv_{\mathrm{m,r}}$, (2) a plot of $\Delta \varv_{\mathrm{b}}$, $\Delta \varv_{\mathrm{c}}$, and $\Delta \varv_{\mathrm{r}}$ against excitation temperature of the transition's upper level, $E_{\mathrm{upp}}/k$, and (3) a plot comparing the line strengths (integrated intensities) of the red- and blue-shifted emission components normalised to the central component, to the excitation temperature, $E_{\mathrm{upp}}/k$.

We find a good fit of multiple components to the majority of the lines investigated. There are a small number of $^{28}$SiO and H$_{2}$O lines for which this is not the case, but these are found to be affected by masering. The results for each molecular species are discussed separately in the following  subsections.

\begin{figure*}
	\begin{subfigure}{0.32\linewidth}
		\centering
		\includegraphics[height=1.85in]{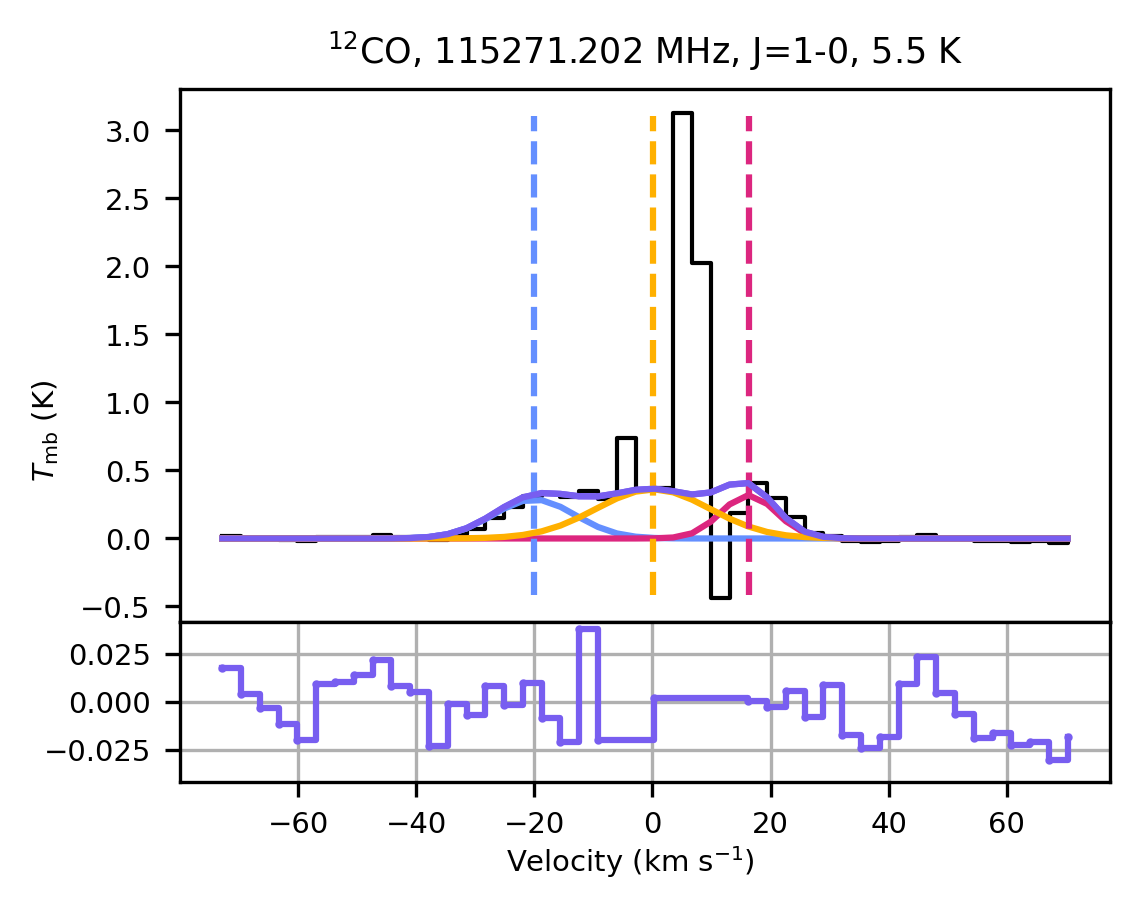}
	\end{subfigure}\hfill
	\begin{subfigure}{0.32\linewidth}
		\centering
		\includegraphics[height=1.85in]{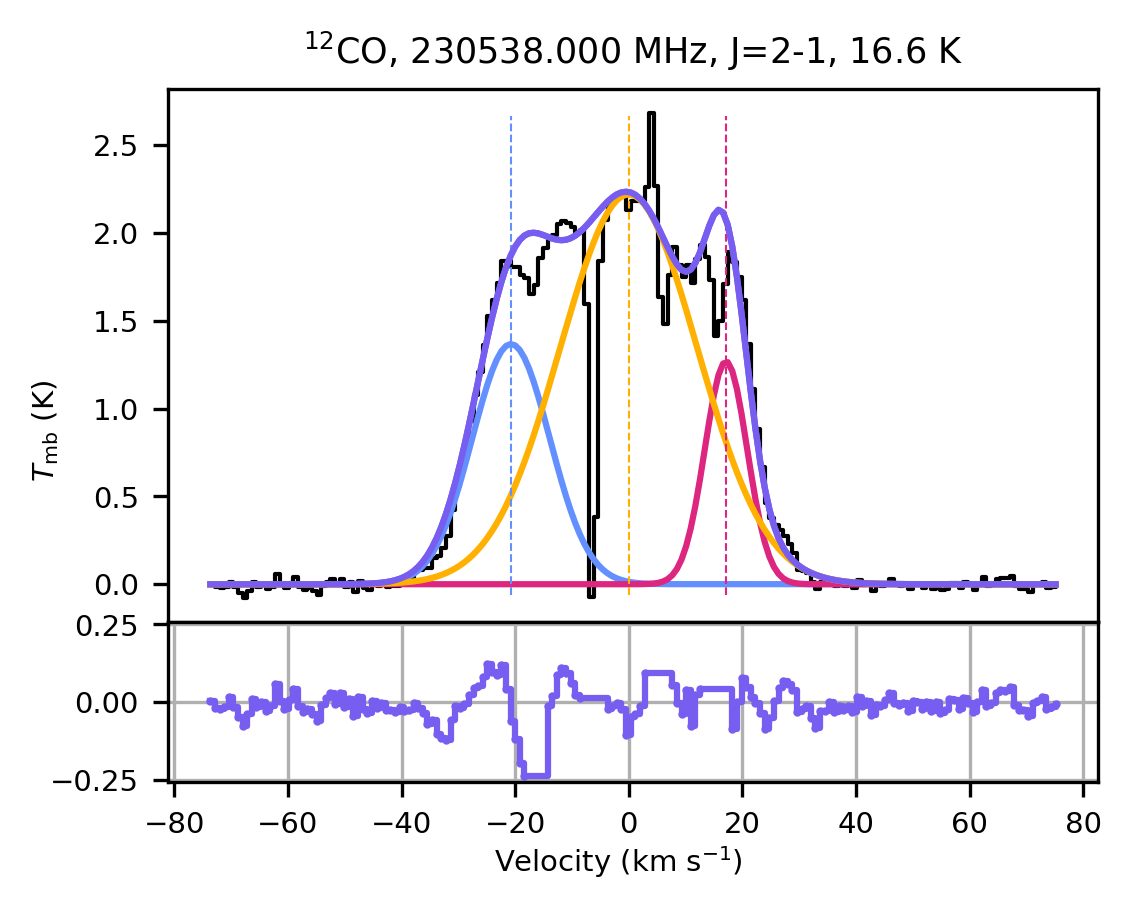}
	\end{subfigure}\hfill
	\begin{subfigure}{0.32\linewidth}
		\centering
		\includegraphics[height=1.85in]{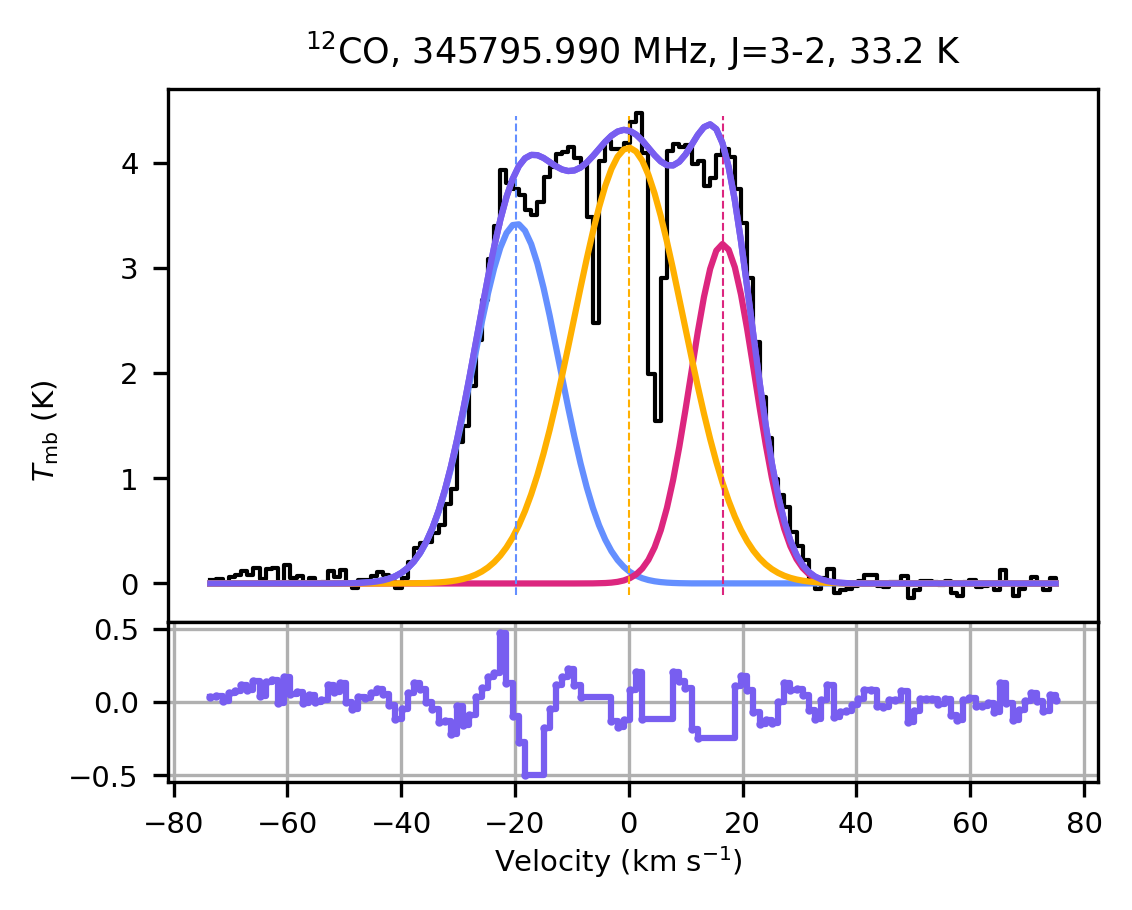}
	\end{subfigure}\hfill
	\begin{subfigure}{0.32\textwidth}
		\includegraphics[height=1.85in]{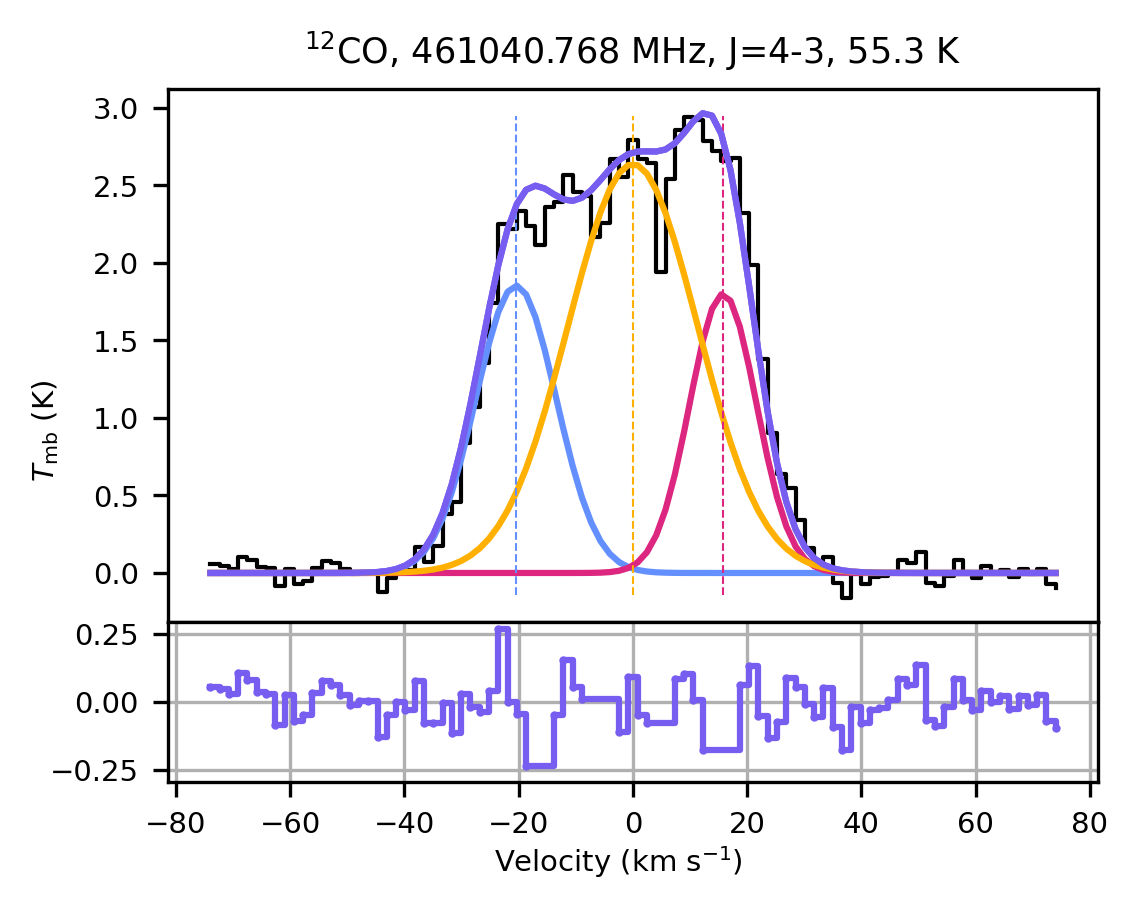}
	\end{subfigure}\hfill
	\begin{subfigure}{0.32\textwidth}
		\includegraphics[height=1.85in]{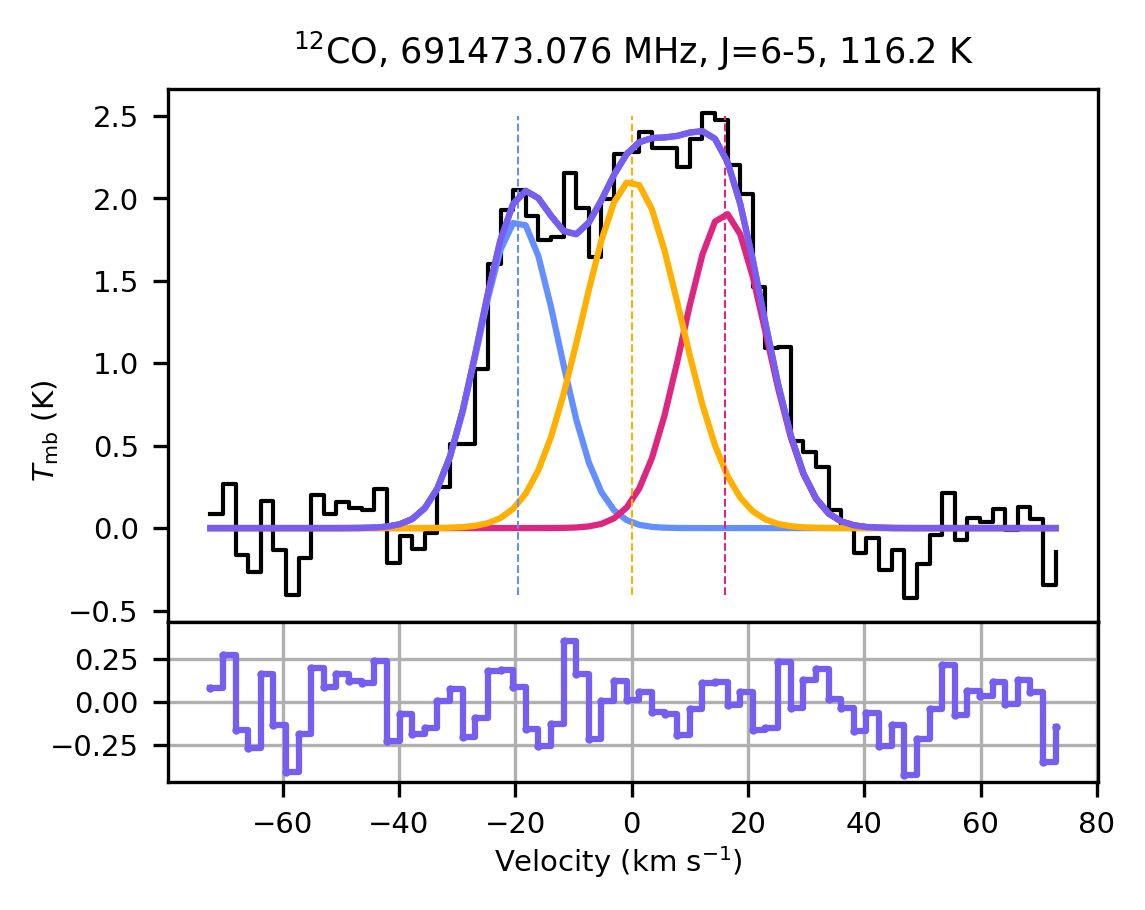}
	\end{subfigure}\hfill
	\begin{subfigure}{0.32\textwidth}
		\includegraphics[height=1.85in]{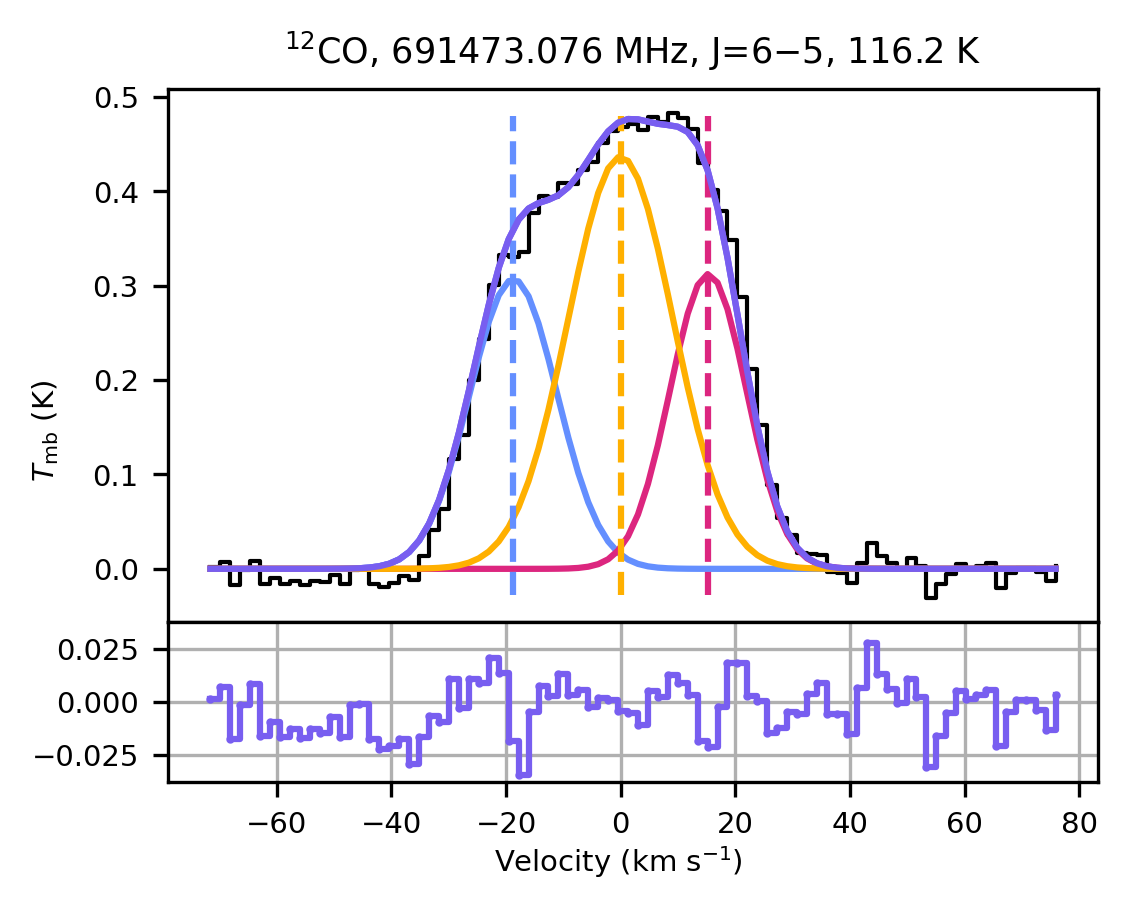}
	\end{subfigure}\hfill
	\begin{subfigure}{0.32\textwidth}
		\includegraphics[height=1.85in]{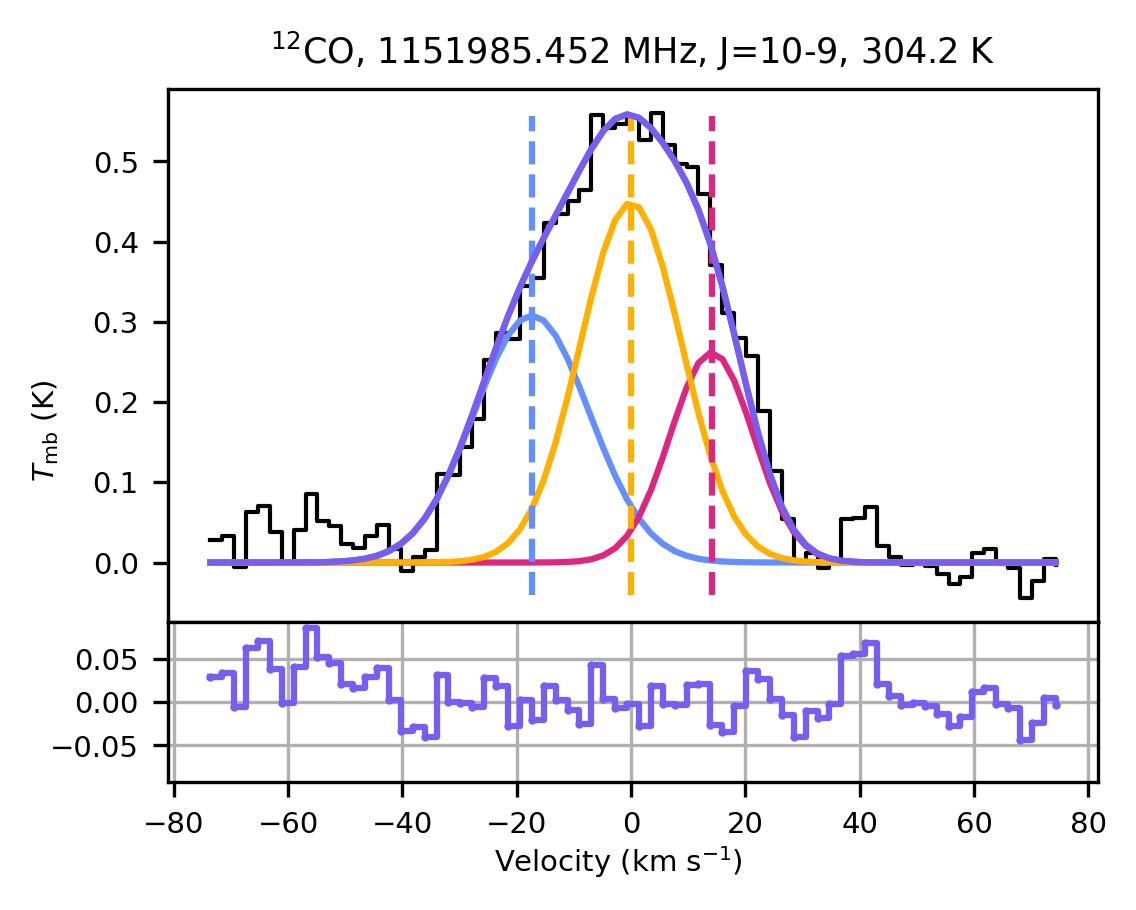}
	\end{subfigure}\hfill
	\begin{subfigure}{0.32\textwidth}
		\includegraphics[height=1.85in]{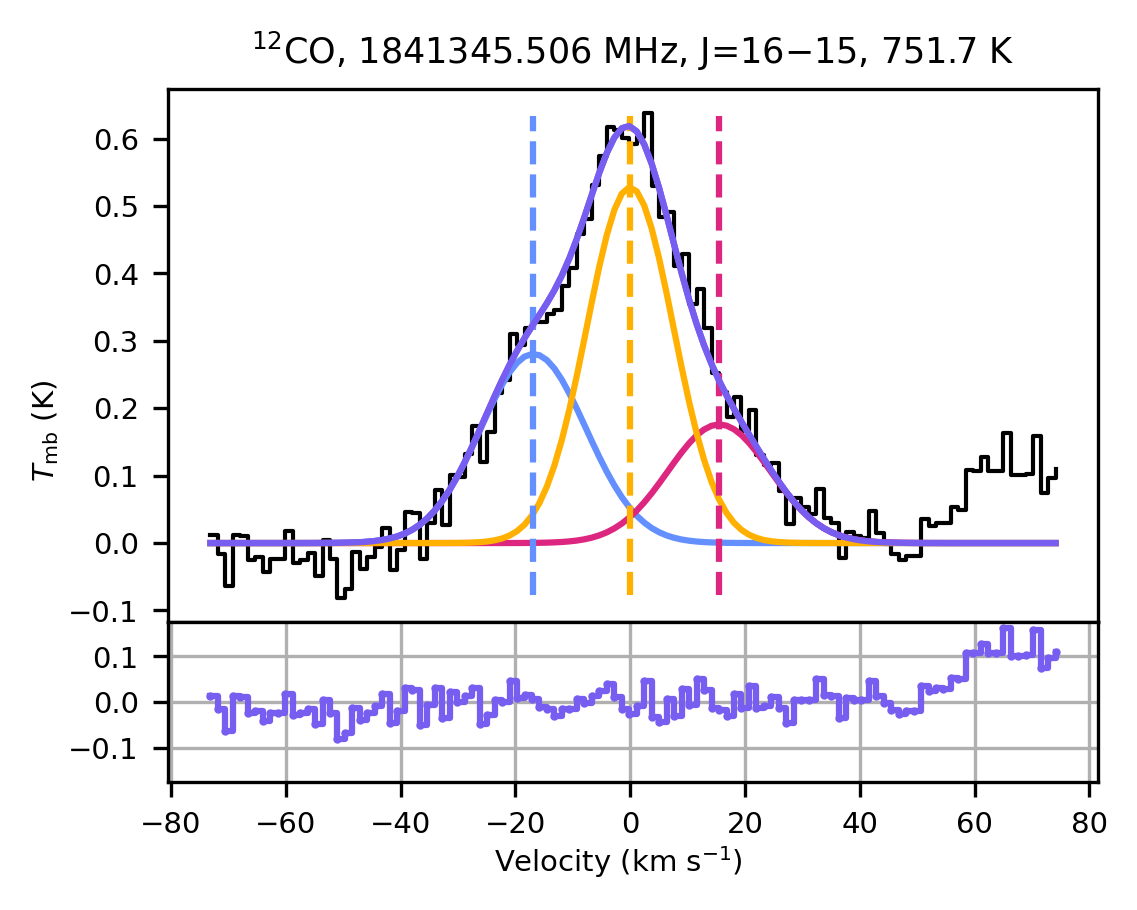}
	\end{subfigure}\hfill
	\begin{minipage}[c]{0.32\textwidth}
	\hspace{\fill}
	\end{minipage}\hfill
	\caption{Multi-component fits to $^{12}$CO emission around NML~Cyg, observed with OSO, HIFI and JCMT. For $^{12}$CO transitions $J = 1 - 0, 2 - 1$ and $J = 4 - 3$, where Galactic absorption and emission from the ISM is present, the contaminated channels are masked for the fitting procedure. The \emph{blue} line indicates the blue-shifted component fit, the \emph{red} line indicates the red-shifted component fit and the \emph{yellow} line indicates the central emission component. Vertical dashed lines indicate the position of the central velocity for each component in the respective associated colours. The combined fit of the three components is indicated by a \emph{purple} line, and the residuals between the combined fit and the observed line profile is given in the grid below the main plot in \emph{purple}.}
	\label{fig:co_nml_multifits}
\end{figure*}

\begin{figure*}
	\begin{subfigure}[c]{0.32\textwidth}
		\centering
		\includegraphics[height=1.85in]{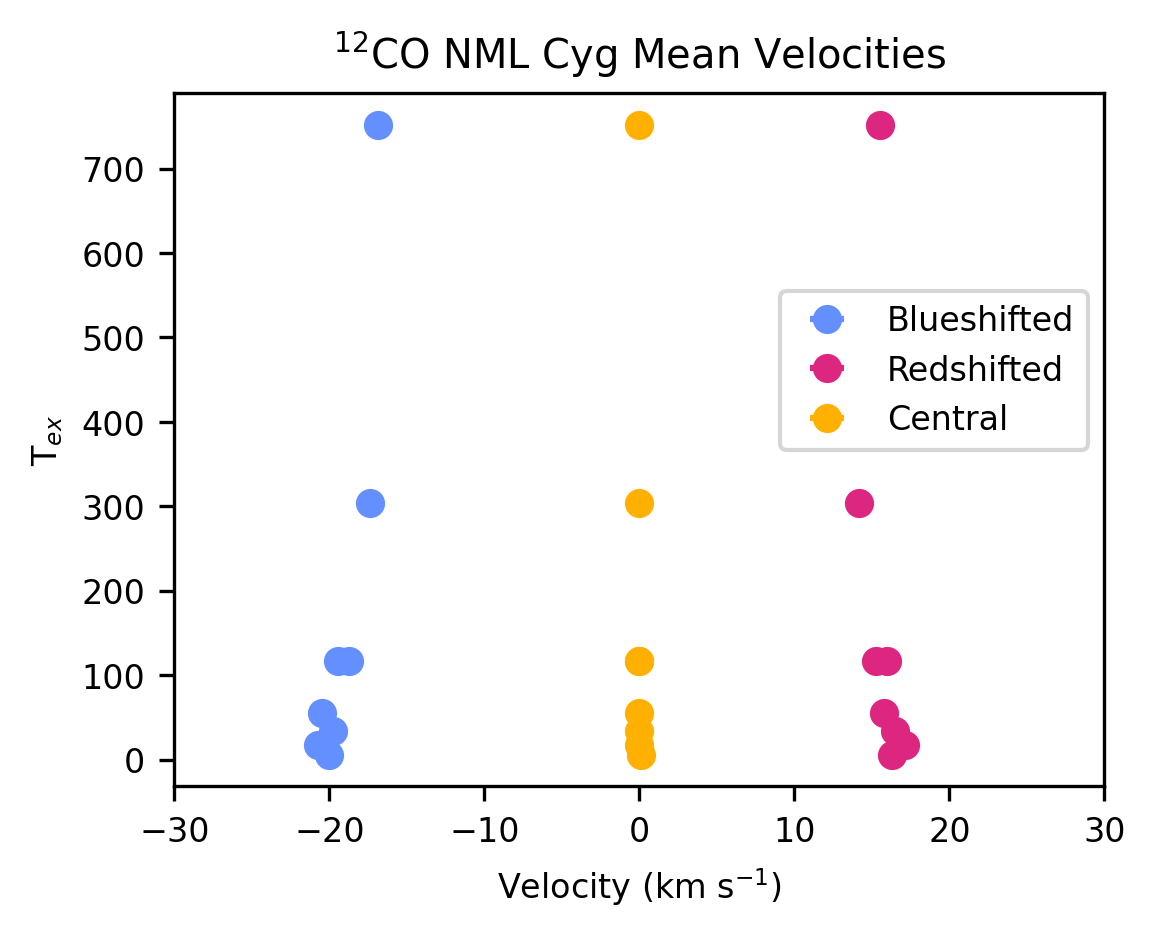}
	\end{subfigure}\hfill
	\begin{subfigure}[c]{0.32\textwidth}
	\centering
	\includegraphics[height=1.85in]{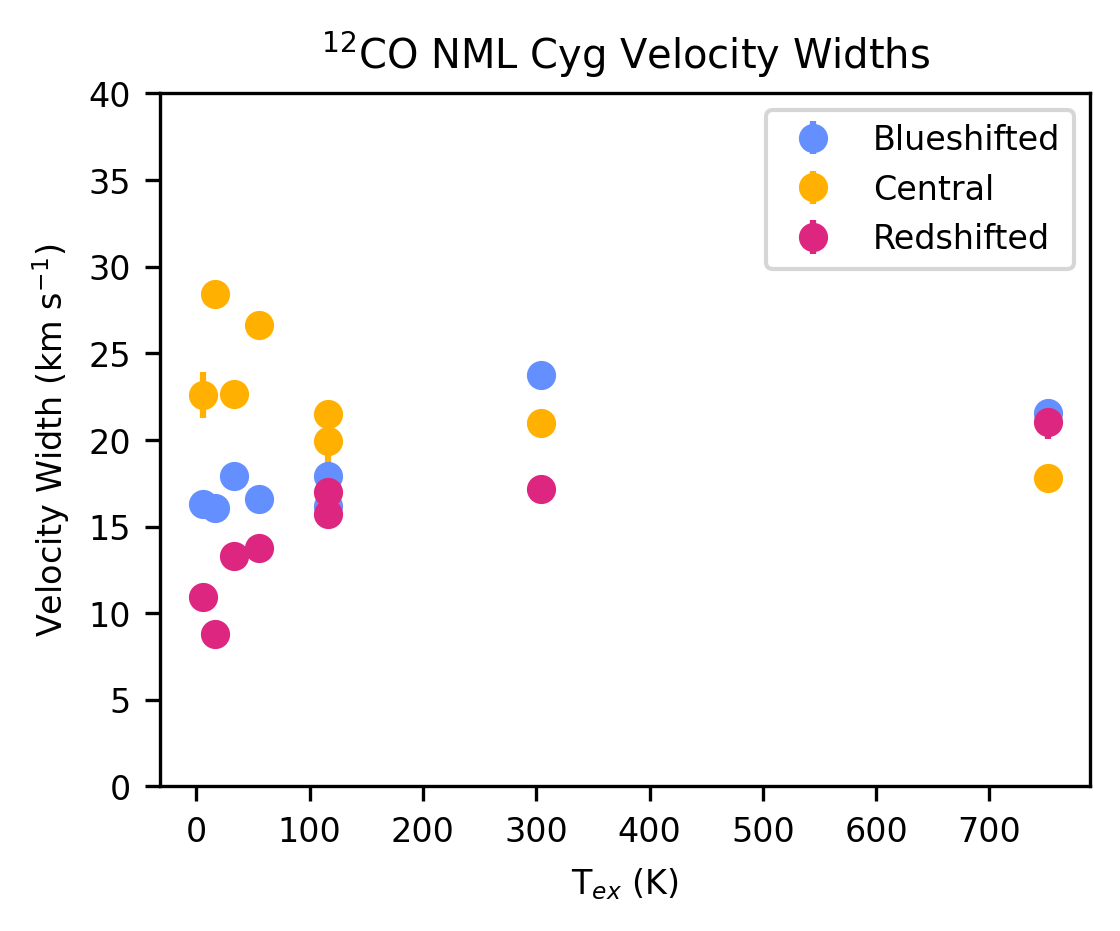}
	\end{subfigure}\hfill
	\begin{subfigure}[c]{0.32\textwidth}
		\centering
		\includegraphics[height=1.85in]{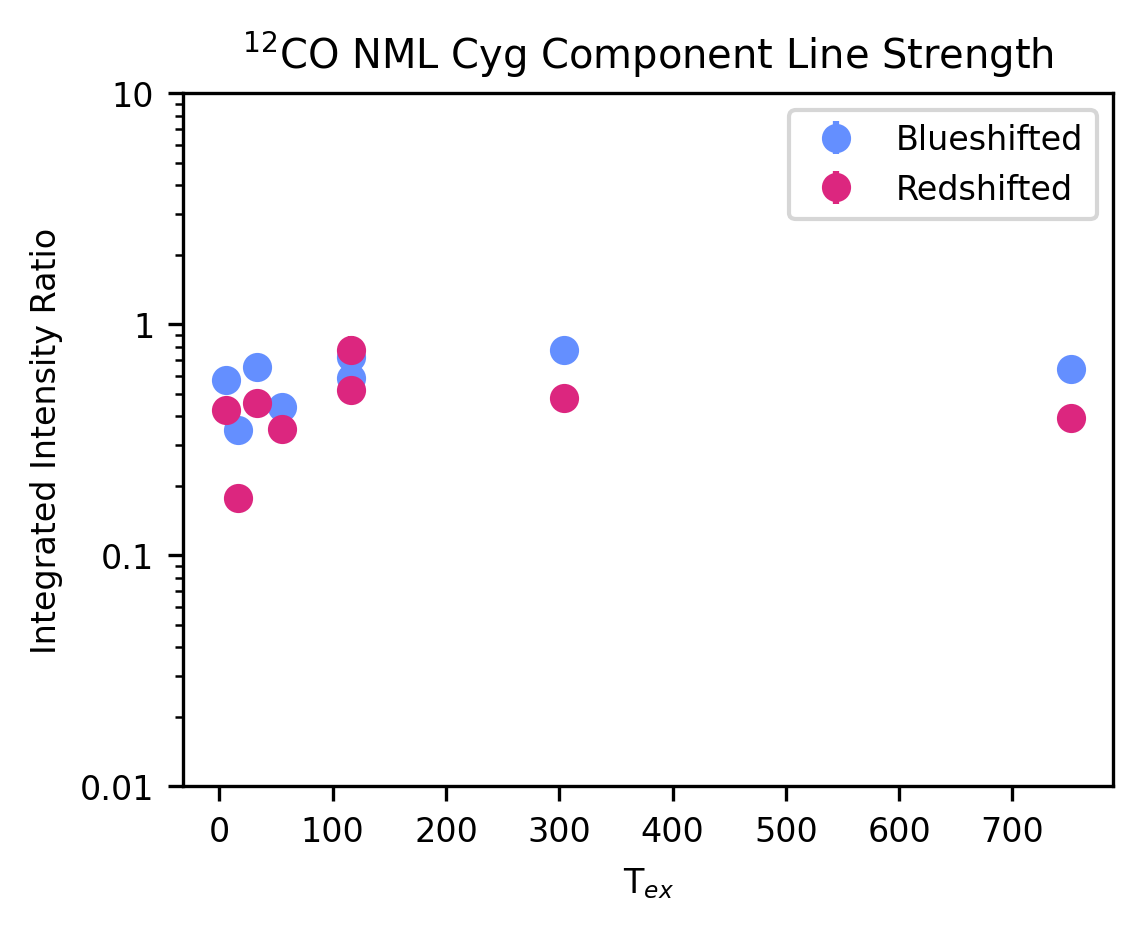}
	\end{subfigure}\hfill
	\caption{Parameter fits for different transitions of $^{12}$CO detected around NML~Cyg. \emph{Left}: Excitation temperature ($E_{\mathrm{upp}}/k$) against the velocity offset from $\varv_{\mathrm{LSR}}$ of different components.
	\emph{Centre}: Velocity widths of the different central, red-shifted and blue-shifted components.
	\emph{Right}: Integrated intensity ratio of the blue and red components, normalised to the central component of the three Gaussian fits, against excitation temperature ($E_{\mathrm{upp}}/k$).}
	\label{fig:co_fits}
\end{figure*}

\subsubsection{$^{12}$CO and $^{13}$CO}\label{sect:co}

Emission lines from both $^{12}$CO ($J = 1 - 0$ up to $J = 6 - 5$) and $^{13}$CO ($J = 1 - 0$) were found to be contaminated by interstellar emission and absorption across the range $-18$ to $18$\,km\,s$^{-1}$ (Figs.~\ref{fig:co_nml_multifits} and \ref{fig:13co_lines}). We masked these features for our multi-component fitting, but they cause an additional limitation to the conclusions that can be drawn from the final parameter fits. $^{12}$CO $J = 6 - 5$ was present in both the JCMT and HIFI datasets. Initial parameter fits gave consistent values for the blue-shifted and central component on both observations, but led to a discrepancy between the red-shifted components. We fixed the fit parameters for  the red-shifted component on the JCMT $J = 6 - 5$ transition to the red-shifted component fit measured for the HIFI $J = 6 - 5$ transition, which was taken as more accurate. 

We found good fits with multiple components for all $^{12}$CO and $^{13}$CO lines (Figs.~\ref{fig:co_nml_multifits} and \ref{fig:13co_lines}). For $^{12}$CO, we see a tentative indication that the magnitudes of the mean velocities for the blue-shifted and red-shifted components $\varv_{\mathrm{m,b}}$ and $\varv_{\mathrm{m,r}}$ appear smaller as the excitation temperature, $E_{\mathrm{upp}}/k$, increases (Fig.~\ref{fig:co_fits}). For $^{13}$CO we found there is stronger support for smaller magnitudes for the mean velocities for the blue-shifted component, and tentative support for smaller magnitudes for $\varv_{\mathrm{m,r}}$, as $E_{\mathrm{upp}}/k$ increases (Fig.~\ref{fig:13co_fits}). 

The component widths, $\Delta \varv$, were found to decrease for all $^{13}$CO components with increasing excitation temperature, as was also found to be the case for $\Delta \varv_{\mathrm{c}}$ for $^{12}$CO. However, the component widths $\Delta \varv_{\mathrm{b}}$ and $\Delta \varv_{\mathrm{r}}$ of $^{12}$CO were found to increase with increasing $E_{\mathrm{upp}}/k$. 
This could possibly in part be explained by an increase in self-absorption in the blue part of the spectrum for higher-$J$ transitions, which would artificially lower $\Delta\varv_{\mathrm{c}}$ and cause more emission to be fit by the blue- and red-shifted components, making those broader, increasing $\Delta \varv_{\mathrm{b}}$ and $\Delta \varv_{\mathrm{r}}$.

\subsubsection{SiO and isotopologues}\label{sect:sio}

In the combined OSO and HIFI datasets, we investigated five $^{28}$SiO emission lines\footnote{We applied a 5\,mK offset to the HIFI spectrum of the $^{28}$SiO $J  = 15 - 14\,(v = 1)$ transition, compared to the data presented by \citet{teyssier2012} to account for an uneven baseline.}. The lines are well fitted by three components, with the exception of the $^{28}$SiO $J = 2 - 1\,(v = 1)$ maser line at 86.243\,GHz,  where only a very weak red-shifted component was found (Fig.~\ref{fig:sio_lines}). Previous observations of $^{28}$SiO masers towards NML~Cyg revealed double-peaked profiles with the mean velocities of the blue-shifted and red-shifted components varying significantly from $-18$ to $-5$\,km\,s$^{-1}$ and $4$ to $11$\,km\,s$^{-1}$, respectively \citep{boboltz_2000_siomasers_nmlcyg, Zhang2012}. The $^{28}$SiO $J = 2 - 1\,(v = 1)$ maser detected by OSO can be clearly seen to follow the double-peaked line profile shape, with a blue-shifted component centered at $-14.8$\,km\,s$^{-1}$ and a central component centered at 0.0\,km\,s$^{-1}$, relative to the systemic velocity.
 
Clear fits to three components were found across all lines from $^{29}$SiO and $^{30}$SiO. We found a possible red-shifted spike in the $^{30}$SiO $J = 26 - 25$ emission, centered at 8\,km\,s$^{-1}$ with a peak temperature of $T_{\mathrm{mb}}$ = 80\,mK, which was not seen for the other species or the low-excitation transition of $^{30}$SiO. The mean velocities $\varv_{\mathrm{m,b}}$ and $\varv_{\mathrm{m,r}}$ decrease in magnitude for increasing $E_{\mathrm{upp}}/k$ for both $^{29}$SiO and $^{30}$SiO (Figs.~\ref{fig:29sio_fits} and \ref{fig:30sio_fits}). Such a trend is less apparent for $^{28}$SiO (Fig.~\ref{fig:sio_fits}), where $\varv_{\mathrm{m,b}}$ and $\varv_{\mathrm{m,r}}$ remain roughly constant across $E_{\mathrm{upp}}/k$. This is likely a consequence of the higher optical depths in the $^{28}$SiO lines compared to the lower abundance isotopologues (see also Sect.~\ref{sect:co}). 

All three SiO isotopologues show an overall reduction in the velocity widths, $\Delta \varv$, of all components with increasing excitation temperature. The central component was found to be the dominant component in intensity for all thermal transitions of SiO. The blue-shifted component was found to be systematically higher in integrated intensity than the red-shifted component across the different transitions,though there are significant uncertainties for $^{28}$SiO (Fig.~\ref{fig:sio_fits}). For the isotopologues, the blue-shifted component is also found to be systematically stronger with only the higher $^{30}$SiO transition found to be an exception, although this line is likely impacted by the red-shifted spike discussed earlier (Figs.~\ref{fig:29sio_fits} and~\ref{fig:30sio_fits}). 

\subsubsection{H$^{12}$CN, H$^{13}$CN}\label{sect:hcn}

We analysed the emission profiles from the H$^{12}$CN $J = 1 - 0$ transition in the OSO dataset and from the $J = 13 -12$ transition observed by HIFI (Fig.~\ref{fig:hcn_lines}). The central velocities of the red- and blue-shifted components were found to remain constant with changing excitation temperature (Fig.~\ref{fig:hcn_fits}). We found strong red-shifted and blue-shifted components for H$^{12}$CN $J = 1 - 0$. The H$^{12}$CN $J = 13 - 12$ emission only showed a very weak blue-shifted component and a narrow red-shifted component (Fig.~\ref{fig:hcn_fits}). 

The emission from H$^{13}$CN $J = 1 - 0$ in the OSO observations (Fig.~\ref{fig:hc13n_oso_10}) could be fit with strong blue- and red-shifted components. The mean velocities $\varv_{\mathrm{m,b}}$ and $\varv_{\mathrm{m,r}}$ were found to be significantly different from the values measured for H$^{12}$CN.
The strengths of the blue- and red-shifted components are also not consistent between H$^{13}$CN and the behaviour found for H$^{12}$CN $J = 1 - 0$ transition, where the central component is much stronger. This may be due to optical depth effects. Measurements of additional transitions of H$^{13}$CN would be needed to rule out or prove any possible discrepancy between the isotopologues.

\subsubsection{H$_2$O}\label{sect:h2o}

We used the archival HIFI observations \citep{teyssier2012} to investigate the multi-component nature of the H$_2$O emission.
For two of the transitions observed,  o-H$_{2}$O $J_{K_{\mathrm{a}},K_{\mathrm{c}}} = 1_{1,0} - 1_{0,1}$\,$(v = 0)$ and  p-H$_{2}$O $J_{K_{\mathrm{a}},K_{\mathrm{c}}} = 1_{1,1} - 0_{0,0}$\,$(v = 0)$, we applied masking to interstellar absorption features over the range $-8$ to $13$\,km\,s$^{-1}$ (Fig.~\ref{fig:h2o_lines}). We found that the mean velocities for the different components were consistent for o-H$_{2}$O and p-H$_{2}$O emission at similar excitation temperatures, in line with expectations (Fig.~\ref{fig:h2o_fits}). The relative contribution of the central component increased with excitation temperature and the red-shifted component was consistently weaker than the blue-shifted component, for both o-H$_{2}$O and p-H$_{2}$O.

The o-H$_{2}^{16}$O transitions $J_{K_{\mathrm{a}},K_{\mathrm{c}}} = 1_{1,0} - 1_{0,1}\,(v_{2} = 1)$ at 658.006\,GHz and $J_{K_{\mathrm{a}},K_{\mathrm{c}}} = 1_{1,1} - 0_{0,0}\,(v_{2} = 1)$ at 1205.789\,GHz are expected to be affected by masering \citep{Chen2000}.  Water masers have previously been observed around NML Cyg at 22\,GHz as reported by, e.g., \cite{Zhang2012}, where velocity peaks were measured at $\varv_{\mathrm{LSR}}$ $\sim -22$\,km\,s$^{-1}$ and 5\,km\,s$^{-1}$. Due to the higher excitation energy of the HIFI lines, we expect them to trace inner regions much closer to the star and so result in smaller magnitudes for the mean velocities of the blue/red-shifted components. This is reflected in our measurements for the mean velocity of the blue-shifted component, where the values for v$_{m,b}$ for the two emission lines from HIFI were found to have a much smaller magnitude than the blue-shifted 22\,GHz water maser.

For the thermal emission lines of H$_2^{16}$O, $\varv_{\mathrm{m,b}}$ was found to range from $-13$ to $-7$\,km\,s$^{-1}$ and $\varv_{\mathrm{m,r}}$ was found to range from $5$ to $23$\,km\,s$^{-1}$. These mean velocities are different from those traced by the maser emission, implying that the thermal and maser lines are dominated by emission from different parts of the outflow. 

The emission from p-H$_{2}^{18}$O $J_{K_{\mathrm{a}},K_{\mathrm{c}}} = 1_{1,1} - 0_{0,0}\,(v_{2} = 1)$ was found to be dominated by the central component and to have a red-shifted component weaker than the blue-shifted component (Fig.~\ref{fig:h218O_hifi}).

\subsubsection{SO, SO$_{2}$}\label{sect:so_so2}

We reproduced all SO emission lines well with three components. Only in the case of the SO $J_{K} = 15_{16} - 14_{15}$ transition at 645.879\,GHz ($E_{\mathrm{upp}}/k = 253$\,K; Fig.~\ref{fig:so_lines}) was there a difficulty in disentangling the blue and central components. The narrow spike in the centre of the line profile led to a final fit with a narrow central component, a significantly lower $\varv_{\mathrm{m,b}}$ and wider $\Delta\varv_{\mathrm{b}}$ than for the other measured transitions; see Fig.~\ref{fig:so_fits}.

All three detected transitions of SO$_{2}$ contain strong blue-shifted components, in line with the previous detection of a blue-shifted spike for two SO$_{2}$ transitions \citep{pulliam2011}. The emission from the $J_{K_{\mathrm{a}},K_{\mathrm{c}}} = 16_{2,14} - 15_{3,13}$ transition, at the highest $E_{\mathrm{upp}}/k$ of the three transitions detected here, showed only a very weak red-shifted component. \footnote{We note that the SO$_{2}$ $J_{K_{\mathrm{a}},K_{\mathrm{c}}} = 3_{1,3} - 2_{0,2}$ transition at 104.029\,GHz, is located near SO$_{2}$ $J_{K_{\mathrm{a}},K_{\mathrm{c}}} = 16_{2,14} - 15_{3,13}$ at 104.033\,GHz, but the negligible red-shifted component for the latter rules out any significant contribution.}

We found an overall decrease for $\varv_{\mathrm{m,b}}$ for both SO and SO$_2$ (Figs.~\ref{fig:so_fits} and \ref{fig:so2_fits}). 
The $\varv_{\mathrm{m,r}}$ values for SO also decrease, whereas for SO$_{2}$, $\varv_{\mathrm{m,r}}$ remains fairly constant.However, we note that the 7 SO detections probe a larger range of $E_{\mathrm{upp}}/k$ than the 3 SO$_2$ detections.

The widths of the central components, $\Delta \varv_{\mathrm{c}}$, decrease with increasing excitation temperature for both SO and SO$_{2}$, though $\Delta \varv_{\mathrm{r}}$ is found to increase, as is also found to be the case for several other molecular species (discussed in Sect. \ref{sect:structurefromfits}). We see no statistically significant trends for $\Delta \varv_{\mathrm{b}}$ for either SO$_{2}$ or SO.

We found the blue-shifted component to be systematically stronger than the red-shifted component also for SO and SO$_2$ (Figs. \ref{fig:so_fits} and \ref{fig:so2_fits}). The blue-shifted component is found to have a similar strength to the central component.

\subsubsection{SiS, NH$_3$, H$_2$S, OH}\label{sect:single_lines}

We find that the observed emission profiles of SiS, NH$_{3}$, H$_{2}$S and OH\footnote{In the case of the OH $N_J = 2_{3/2} - 1_{1/2}$ transition we ignored the fine structure of the transition in the multi-component fitting.} are well reproduced with multi-component fits as shown in Figs.~\ref{fig:SiS_oso} -- \ref{fig:oh_hifi}. For all four transitions, we find a strong central component. In the case of SiS, NH$_{3}$ and H$_{2}$S, the blue-shifted component is stronger than the red-shifted component, but for the OH line we find the red-shifted component to be slightly stronger. As there is only one line observed for each of these molecules, this limits the conclusions that can be drawn.

\subsubsection{Abundances}\label{sect:rot_diagrams}

We applied rotational diagram analysis \citep[e.g.][]{goldsmithlanger} to derive rotational excitation temperatures and column densities for those molecular species with multiple detected transitions:  $^{12}$CO, $^{13}$CO, $^{28}$SiO, $^{29}$SiO, $^{30}$SiO, SO, SO$_{2}$, H$^{12}$CN, p-H$_{2}$O and o-H$_{2}$O. The fluxes measured for each of the transitions were separated into the fluxes found for the three Gaussian components for each of the lines. Unfortunately, the results were inconclusive with non-physical $T_{\mathrm{rot}}$ values derived, and large uncertainties on the final derived column density values. This was likely due in part to the impact of optical depth effects, which affect the validity of using an approach that requires the approximation of LTE conditions and assumes optically thin lines. The analysis was also limited by the simplified approximation of the multi-component Gaussian fits applied to the line profiles, and uncertainties in the exact regions traced by the various molecular transitions. Radiative-transfer modelling would be necessary to derive meaningful results, but was outside the scope of this report.

\subsection{Isotopic ratios}\label{sect:isotopes}
We have used frequency-corrected line intensity ratios to derive first-order estimates of isotopologue abundance ratios \citep[see e.g.,][]{debeck2010_comdot}. For this, we used the total integrated intensity of the measured line. A comparison of derived isotopic ratios for the red, central, and blue components separately, rather than the total integrated intensities did not reveal any significant differences, considering the uncertainties on the fits. The ratios determined suffered from a clear limitation due to the influence of the  opacity of the measured transitions, since they were derived for highly abundant molecules such as CO, SiO, HCN, and H$_2$O.

We found $^{12}$CO/$^{13}$CO$\,\approx\,3.7 - 8.1$ and H$^{12}$CN/H$^{13}$CN$\,\approx\,9.3\pm1.5$. These ratios imply a low $^{12}$C/$^{13}$C ratio (below 10), in line with those reported by \citet{milam2009} and \citet{debeck2010_comdot} for evolved stars. We note the possibility for differences between derived $^{12}$CO/$^{13}$CO and H$^{12}$CN/H$^{13}$CN considering the difference in photodissociation of CO (via lines) and HCN \citep[via the continuum;][]{saberi2020}. 

We found $^{28}$SiO/$^{29}$SiO\,$\approx$\,$10.9\pm0.9$, $^{28}$SiO/$^{30}$SiO\,$\approx$\,$9.5\pm0.7$, and $^{29}$SiO/$^{30}$SiO\,$\approx$\,$0.9\pm0.4$, whereas the solar ratios are, respectively, 20, 30, and 1.5 \citep{asplund2009}. 

We derived H$_2^{16}$O/H$_2^{18}$O\,$\approx$\,$8.7\pm0.5$, whereas the solar isotopic ratio $^{16}$O/$^{18}$O is 499 \citep{asplund2009}. We can compare this to an estimate of H$_2^{16}$O/H$_2^{18}$O for VY CMa, found to be $\approx$\,37 taken from the integrated intensity values indicated in \cite{alcolea2013_vycma_hifistars} for the p-H$_{2}$O and p-H$_{2}^{18}$O $J = 1_{1,1} - 0_{0,0}$\,$(v_{2} = 1)$ transitions. Whether there is indeed a strong (relative) enrichment in $^{18}$O compared to the solar composition would need to be confirmed by analysing emission from e.g. C$^{18}$O and Si$^{18}$O, which were not detected in the OSO and HIFI spectra, despite the expectation of detectable line strengths, following the derived line intensity ratio from the water lines. 

More transitions and more isotopologues need to be measured to improve these first estimates. Furthermore, radiative-transfer modelling of the observed lines will be needed to account for opacity when deriving isotopic ratios for C, O, and Si. With spatially resolved observations, one could also trace the possibility for spatial variations of isotopic ratios across the outflow.

\subsection{Envelope Structure}\label{sect:structurefromfits}

We derive several possible trends in the results of the applied multi-component fitting procedure. We note that their interpretation has to be taken with caution due to a few key factors; we have access to a limited number of observed molecular species and transitions, the method does not account for radiative-transfer effects, and we have no explicit geometric information on the measured emission.

\subsubsection{Mean Velocities}\label{sect:mean_velocities}

We can use the mean velocities of the different components in the line profiles to consider the spatial regions that the emission originates from (see Sect.~\ref{sect:geometry} and Fig.~\ref{fig:angles}). The mean velocities of the blue and red-shifted components, $\varv_{\mathrm{m,b}}$ and $\varv_{\mathrm{m,r}}$, provide an indication of the angle of inclination between the blue-shifted or red-shifted components and the line of sight, if one assumes the the material in the directed outflow components has the same acceleration profile and reaches the same terminal velocity as the underlying spherical wind. 

For many of the molecular species, including $^{12}$CO, $^{13}$CO, $^{29}$SiO, $^{30}$SiO, SO, and SO$_{2}$, we can see a trend where the absolute value of the offset velocities, $\varv_{\mathrm{m,b}}$ and $\varv_{\mathrm{m,r}}$, decreases for increasing  excitation temperature $E_{\mathrm{upp}}/k$ (Figs.~\ref{fig:co_fits}, \ref{fig:13co_fits}, \ref{fig:29sio_fits}, \ref{fig:30sio_fits}, \ref{fig:so_fits} and \ref{fig:so2_fits}). However, this is not seen consistently across the different isotopologues of the same molecular species. The degree of correlation between the decrease in absolute mean velocities and increase in excitation temperatures is found to be stronger for $^{13}$CO than for $^{12}$CO. This is also the case when comparing $^{28}$SiO to its isotopologues, $^{29}$SiO and $^{30}$SiO. This is likely due to the higher optical depths in the main isotopologues.

Not all molecular species show a decrease in the magnitude of the mean velocity with increasing excitation temperature. SO and SO$_{2}$ show a decrease in $\varv_{\mathrm{m,b}}$, but a roughly constant $\varv_{\mathrm{m,r}}$. In addition, as $|\varv_{\mathrm{m,r}}| < |\varv_{\mathrm{m,b}}|$ for these species, this suggests that SO and SO$_2$ possibly trace different parts of the outflow.

\subsubsection{Velocity Widths}\label{sect:vel_widths}

We can assume a terminal outflow velocity of $\varv_{\infty} = 30\,$km$\,$s$^{-1}$. Previous measurements of the terminal velocity for NML Cyg were found to reach 33$\,$km$\,$s$^{-1}$ \citep{debeck2010_comdot}. The terminal velocity limit applied here is found to represent the upper limit traced by the width of the central component of the lowest excitation transition for $^{13}$CO, which we expect to trace most of the extended envelope. Our velocity widths for all components across all molecular species are found to fall below this limit.

At higher excitation temperatures there will typically be higher optical depths. As discussed in \ref{sect:co}, this can result in higher levels of absorption for the blue-shifted emission within the line profiles. This will cause a general red-shift for the bulk of the emission with respect to the central peak of the line profiles. It is, therefore, possible that due to the assumed symmetry of the Gaussian components, some of the emission from the central component is erroneously attributed to the red-shifted component. This may lead to a widening of the red-shifted component measured, and a narrower central component.

We find this in the case of $^{12}$CO and SO$_{2}$, where the velocity width of the red-shifted component increases for some of the line transitions at higher excitation temperature (Figs.~\ref{fig:co_fits} and \ref{fig:so2_fits}), as well as a decrease for $\Delta\varv_{\mathrm{m,c}}$ with increasing excitation temperature for all species (Figs.~\ref{fig:co_fits}, \ref{fig:13co_fits}, \ref{fig:29sio_fits}, \ref{fig:30sio_fits}, \ref{fig:so_fits} and \ref{fig:so2_fits}). 

It is difficult to disentangle the effect of this from the line profiles, with radiative transfer analysis of the higher excitation lines required to better quantify this effect. 

\subsubsection{Relative Line Intensities}\label{sect:line_intensities}

We can also consider the relative line intensities measured for the outflow components across the different molecular species. This was done using the normalised line strengths $I_{\mathrm{b}}$ and $I_{\mathrm{r}}$, the ratios between the integrated intensity of the blue- and red-shifted components and the integrated intensity of the central component. In the case of line emission originating primarily from a singular spherical component, we would expect to find the relative strengths $I_{\mathrm{b}}$ and $I_{\mathrm{r}}$ significantly below 1. In the case of line emission arising predominantly from the directed outflow(s), we would, on the other hand, expect to find $I_{\mathrm{b}}$ and $I_{\mathrm{r}}$ consistently above 1.

For the majority of molecular species investigated, we found that $I_{\mathrm{b}}$ and $I_{\mathrm{r}}$ lie between 0.1 and 1, indicating that the central component is the dominant source of emission. This is not the case for most of the observed H$_{2}$O lines, where the bulk of the emission is seen to stem from one or both of the blue- and red-shifted components (Fig.~\ref{fig:h2o_fits}). For the red-shifted component, this is possibly partially due to the optical depth effect as discussed in Sect.~\ref{sect:mean_velocities}, and so may include emission that could be attributed to the central component. The directed components are also seen to dominate for a number of SO and $^{29}$SiO transitions (Figs.~\ref{fig:so_fits} and \ref{fig:29sio_fits}). We also find for many of the molecular species investigated that $I_{\mathrm{b}}\,>\,I_{\mathrm{r}}$, implying that the blue-shifted component possibly contains more mass than the red-shifted component and that the two directional outflows are hence not identical in nature, reducing the likelihood that they form a symmetric, bipolar outflow. 

For several of the molecular species, including H$^{12}$CN, $^{13}$CO and H$_{2}$O, we see that the relative strengths, $I_{\mathrm{b}}$ and $I_{\mathrm{r}}$, of the directed outflow components decrease at higher excitation temperatures (Figs.~\ref{fig:hcn_fits}, \ref{fig:13co_fits}, and \ref{fig:h2o_fits}).
The presence of weaker and narrower emission from directional outflows might indicate that the emission from higher excitation lines of these species mainly traces the inner regions of the outflows, where the material is still accelerating.

\subsubsection{Geometrical constraints on the outflows}\label{sect:geometry}

\begin{figure*}
\centering
\includegraphics[width=0.4\linewidth]{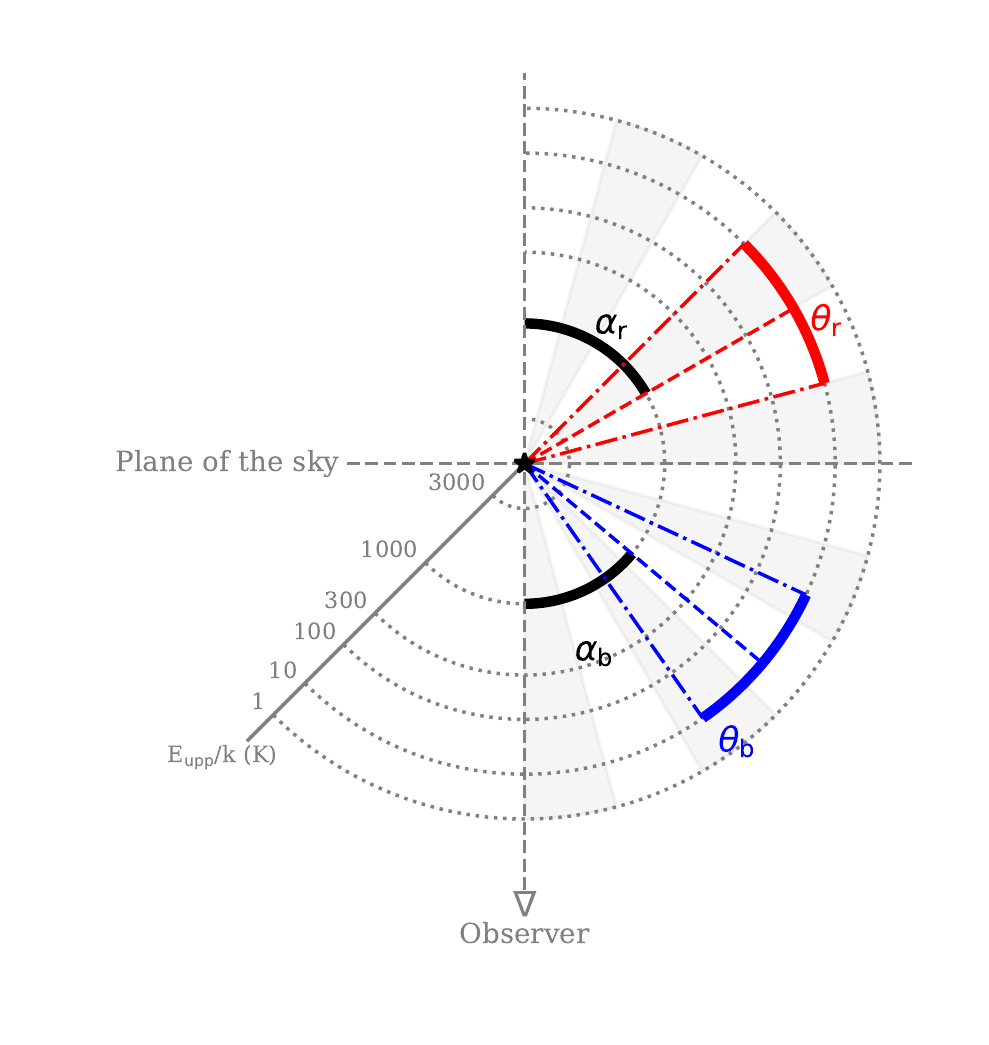}
\includegraphics[width=0.4\linewidth]{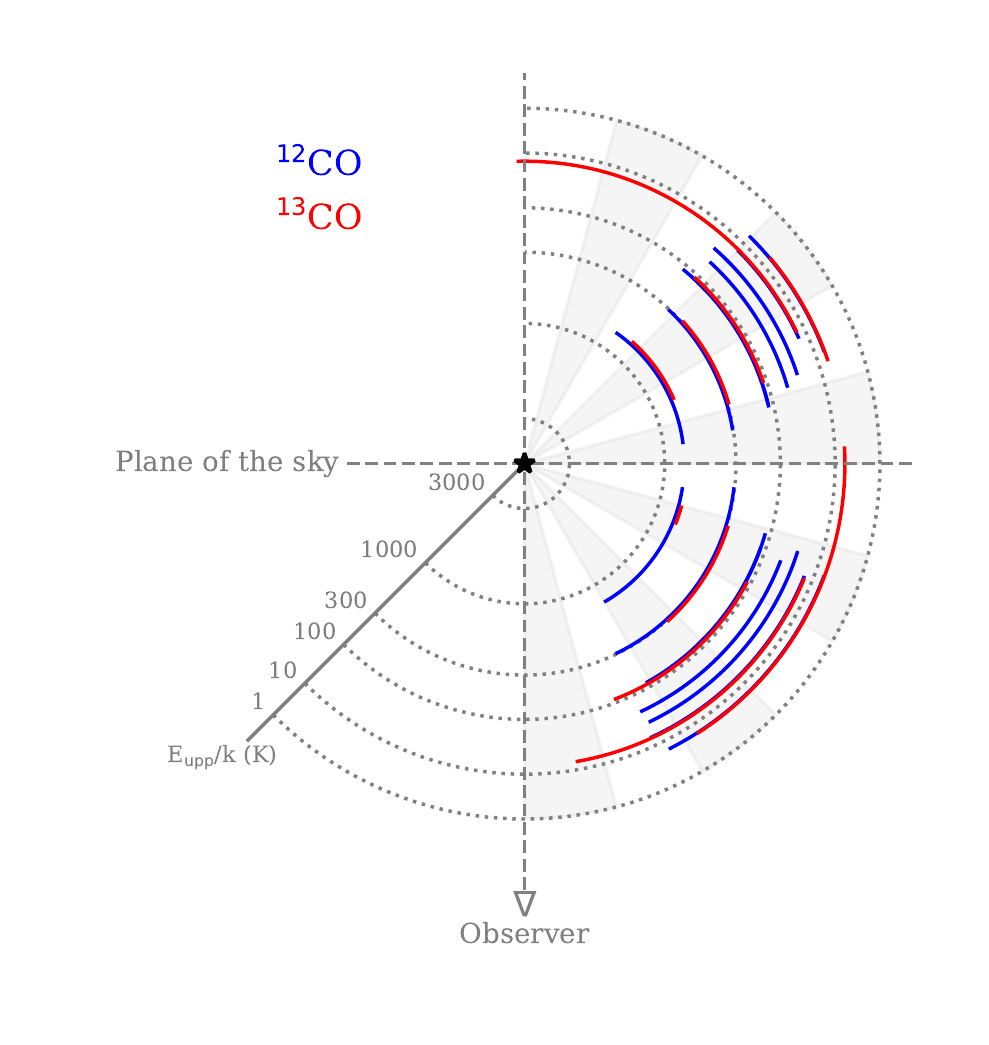}\\
\vspace{-0.75cm}
\includegraphics[width=0.4\linewidth]{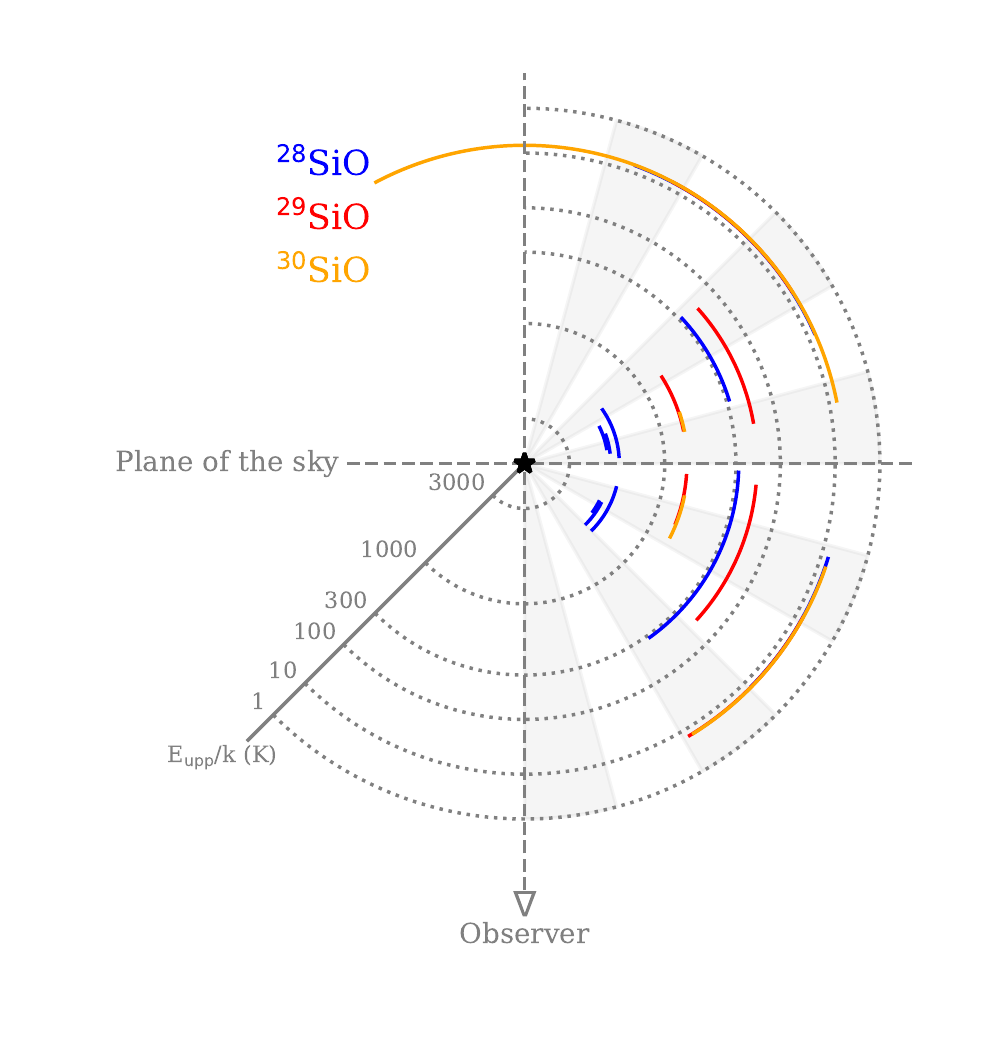}
\includegraphics[width=0.4\linewidth]{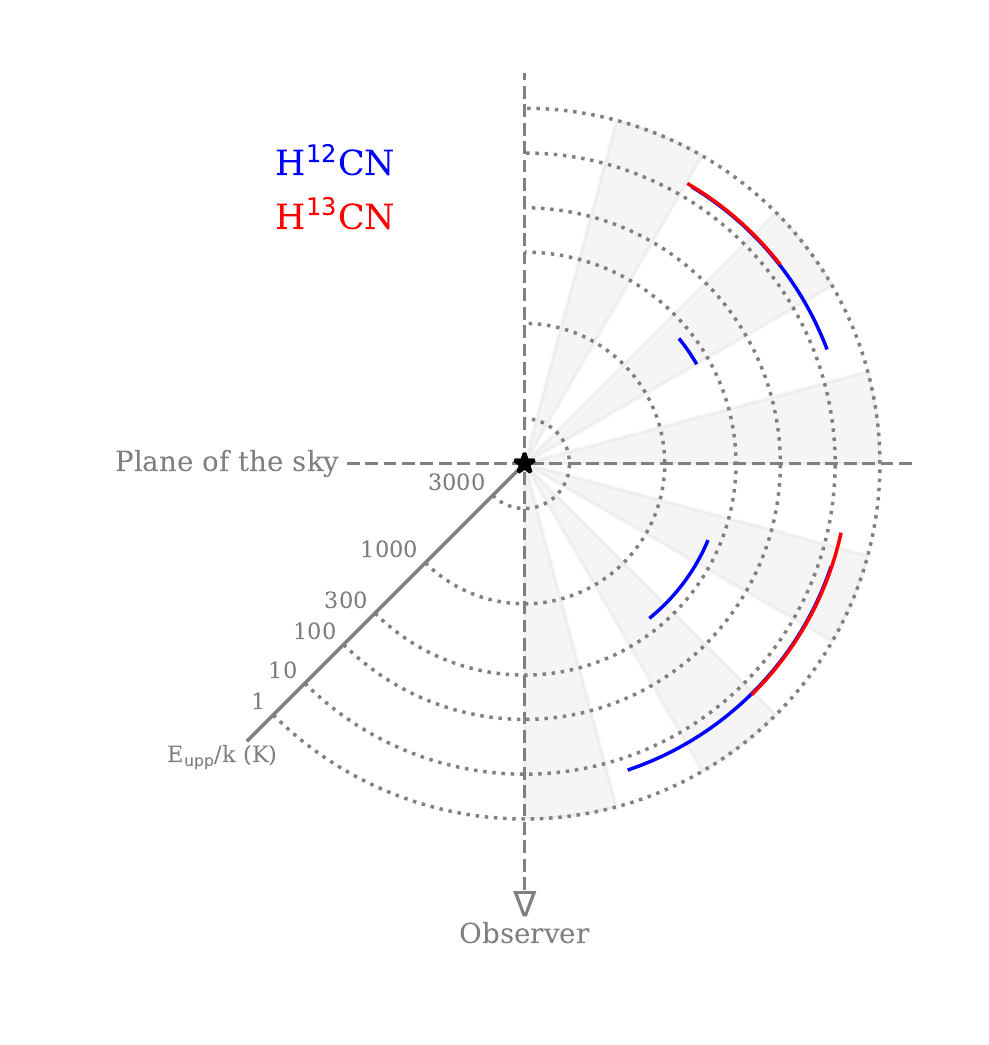}\\
\vspace{-0.75cm}
\includegraphics[width=0.4\linewidth]{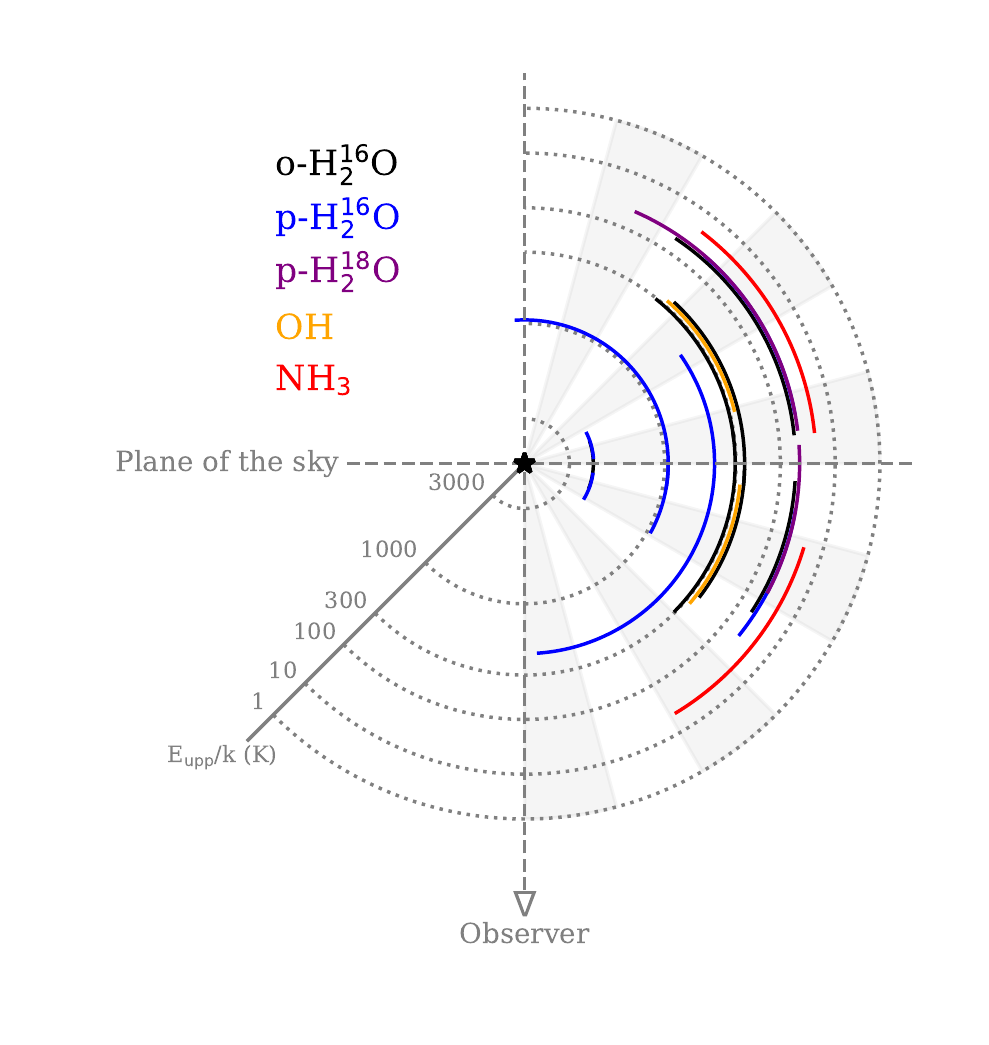}
\includegraphics[width=0.4\linewidth]{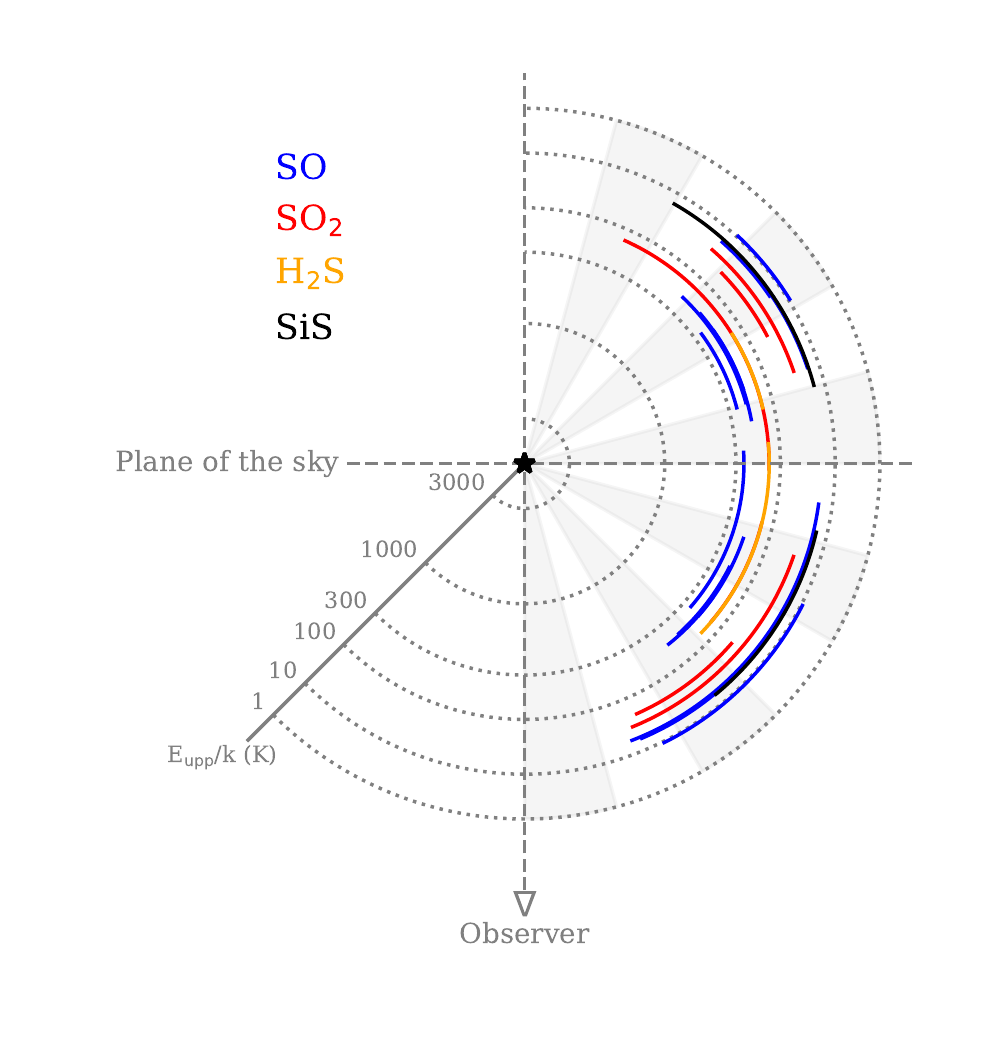}
\caption{Estimated inclination and opening angles of directed red-shifted and blue-shifted outflows based on multi-component fits for all detected transitions, including maser lines. The top left panel shows a schematic of the average inclination angles ($\alpha_{\mathrm{b}}$, $\alpha_{\mathrm{r}}$) and the opening angles ($\theta_{\mathrm{b}}$, $\theta_{\mathrm{r}}$) of the directed components -- see text for explanation. The remaining panels show the component angles derived for the molecular species labelled in the top left quadrant of each panel. Arcs are drawn at a radius corresponding to the transitions' $E_{\mathrm{upp}}/k$, as shown by the extra axis in the lower left quadrant of each panel and the corresponding dotted grid lines. \label{fig:angles}}
\end{figure*}

As discussed above, our observations reveal multiple components in the spectra that cannot be explained with a simple spherical outflow geometry. The trends we see in $\varv$ and $\Delta\varv$, across molecules and excitation energy, suggest that the measured thermal emission possibly traces red-shifted and blue-shifted components that contain gas in an accelerating outflow. 

Assuming, for simplicity, that all material around NML~Cyg moves at the same terminal velocity $\varv_{\infty}$ (as discussed earlier in Sect.~\ref{sect:vel_widths}), we can derive the average orientation with respect to the line of sight, $\alpha$, of the outflows, from the mean velocities $\varv_{\mathrm{m}}$. The opening angles they subtend, $\theta$, can then be derived by also considering the velocity widths, $\Delta\varv$: 
\begin{equation}
\alpha = \arccos\left(\frac{|\varv_{\mathrm{m}}|}{\varv_{\infty}}\right)
\end{equation}
\begin{equation}
\theta = 2\arcsin\left(\frac{\Delta\varv}{2\varv_{\infty}\sin(\alpha)}\right)
\end{equation}
The constraints we can set on the outflow geometry are limited as $\varv_{\mathrm{m}}$ only reflects the inclination $\alpha$ with respect to the line of sight. 
We are unable to constrain the rotational angle around the line of sight, $\phi$, ranging over a full circle ($0^{\circ}-360^{\circ}$). This means that, even where similar $\alpha$ and $\theta$ are derived for two spectral lines, they do not necessarily trace the same spatial region. In case of a high degree of similarity across many lines, one can suggest the presence of coherent outflows, since the fitting procedure made no assumptions on the geometry. In the top left panel of Fig.~\ref{fig:angles}, we present a schematic overview of the parameters we can determine, $\alpha$ and $\theta$. 
 
Additionally, similar $\alpha_{\mathrm{b}}$ and $\alpha_{\mathrm{r}}$ for a given spectral line might suggest a bipolar-like structure. However, given the rather random distribution of clumps and outflows reported for RSGs in the literature, we do not see similar $\alpha$ values as strong enough evidence of such symmetric structures without spatially resolved images to confirm them. 

Considering the assumption of a constant velocity of 30\,km\,s$^{-1}$ throughout the outflow and the approximation of the line profiles with a sum of three Gaussians, our results shown in Fig.~\ref{fig:angles} are first estimates to explore the geometry of NML~Cyg’s outflow. To avoid a visual suggestion of bipolar outflow structures based on first-order estimates, we have chosen to plot the derived angles $\alpha$ such that they are not in opposing quadrants. The plots have been split into different molecular species to avoid overcrowding a singular plot of the outflow geometries. 

We find that the blue-shifted and red-shifted outflows have angles with respect to the line of sight of roughly $50^{\circ}$ and $60^{\circ}$, respectively. We find opening angles of about $30^{\circ}$ for both. We note that not all transitions lead to the same results and that only a further application of a three-dimensional radiative-transfer model will be able to fully account for optical depth and geometrical effects for the different molecular species.

We note that we have also carried out the procedure assuming that the wind velocity follows a typical acceleration profile for the winds of asymptotic giant branch stars \citep[a so-called $\beta$-profile with $\beta$ in the range $0.5-3$, see e.g. ][]{decin2010,decin_2020}. The effect of assuming a different value for $\beta$ is limited to the transitions with the highest $E_{\mathrm{upp}}/k$. Furthermore, using such an approach, one also needs to make assumptions on the radial temperature profile, in addition to the acceleration profile. Since this introduces too many highly uncertain assumptions on the physical properties throughout the outflow, we refrained from including this here. 

Considering the large range of $E_{\mathrm{upp}}/k$ traced by both the red-shifted and blue-shifted components of thermal emission lines and the trends discussed earlier, it is possible that the suggested outflows extend over large radial ranges and their ejection was hence sustained for a significant amount of time, with dynamical ages  possibly reaching up to several hundreds of years.

In contrast to the consistent measured velocities for the outflow components, randomly distributed clumps would likely lead to an absence of clear trends across different excitation conditions for the derived angles $\alpha$ and $\theta$. The observations cannot rule out clumpy emission, but do suggest that the bulk of the emission originating from random locations in the outflow is not the most plausible explanation. We therefore suggest that the measured spectral lines are indicative of large-scale coherent red-shifted and blue-shifted outflows, superposed on a spherical outflow. These outflow components are then characterised by different densities and/or abundances compared to the underlying spherical wind.

The presence of directed outflows that are sustained for a long time is, however, very hard to explain based on any of the currently accepted possible wind-launching mechanisms. Mass-loss from RSGs has been linked to Alfv\'{e}n waves \citep{hartmann_avrett_alfven_1984}, which cause material to eject from the star due to the influence of magnetic fields, \citep[][as seen for Betelgeuse, CE Tau and $\mu$ Cep]{auriere_2010_bfields, tessore_2017_bfields}, though these are expected to vary over timescales of months. Large convective cells on the surfaces of RSGs are another possible underlying mechanism for multiple discrete mass-loss ejection events \citep{josselin_plez_2007}, but the variations in these cells are predicted to occur over short timescales of days and weeks, with survival periods of the convective cells believed to last over timescales of only a few years  \citep{lopezariste_2018_convect_betelgeuse}. Additional links between large convective cells on the surface of RSGs to mass-loss ejection events were drawn in the case of the recent `Great Dimming' event of Betelgeuse \citep{Montarges_betelgeuse_2021}, which occured over a period of about 150\,days. In \citet{humphreys2021}, the `Great Dimming' event of Betelgeuse was then compared to possible episodic outflows in different directions lasting a few decades around VY CMa, determined from the presence of large dust clumps around VY CMa and coincident temporal variability in the light curve of the star, though the authors point out a clear difference in the timescales and energies involved. 

In conclusion, our observations suggest the presence of longer episodes of localised ejection in the circumstellar environment of NML~Cyg. High-angular resolution observations will be essential to trace the location and extent of the emission and to discern between large-scale outflows and clumpy emission.

\subsection{Radiative transfer modelling of $^{12}$CO}\label{sect:shape}

Motivated by the result that all observed emission lines consist of multiple components, we calculated multi-dimensional radiative transfer models of $^{12}$CO emission to constrain the large-scale geometry of the circumstellar material. We used \texttt{SHAPE} \citep{SHAPE} to create a three-dimensional structure of the envelope and the integrated module \texttt{shapemol} \citep{SHAPEMOL} to calculate the radiative transfer and simulate the line emission profiles of $^{12}$CO.

\begin{figure} 
    \centering
    \tdplotsetmaincoords{70}{40}
    \begin{tikzpicture}
    [ tdplot_main_coords,
    scale=5,
    my fill/.style={fill=black, fill opacity=.1},
    my fill2/.style={fill=black, fill opacity=.5}]
  \coordinate (O) at (0,0,0);
    \draw[thick,-stealth] (-1,0,0) -- (1,0,0)coordinate(y) node[anchor=north east]{$x$}; 
    \draw[thick,-stealth] (0,-1,0) -- (0,1,0)coordinate(x) node[anchor=north west]{$z$}; 
    \draw[thick,-stealth] (0,0,-0.6) -- (0,0,1)coordinate(z) node[anchor=south]{$y$}; 
    \tdplotsetcoord{R}{1}{30}{-130}
    \tdplotsetcoord{Z}{1}{90}{-90}
    \tdplotsetcoord{R2}{1.01}{30}{-130} 
    \draw[thick, dashed] (O) -- (Rxy);
    \draw[thick, dashed] (R) -- (Rxy);
    \draw[very thick,red, -stealth] (O) -- (R2); 
    \draw [canvas is xz plane at y = 0] pic["$\Phi_y$", draw = black, text = black, angle eccentricity=1.2, angle radius=2cm, stealth-]  {angle=Rxy--O--Z};
    \draw [canvas is xy plane at z = 0] pic["$\Phi_x$", draw = black, text = black, angle eccentricity=1.2, angle radius=1.5cm, stealth-]  {angle=R--O--Rxy};
    \end{tikzpicture}
    \caption{The coordinate system used by SHAPE, including the geometrical meaning of the positioning angles $\Phi_x$ and $\Phi_y$. }
    \label{fig:shape_geometry}
\end{figure}
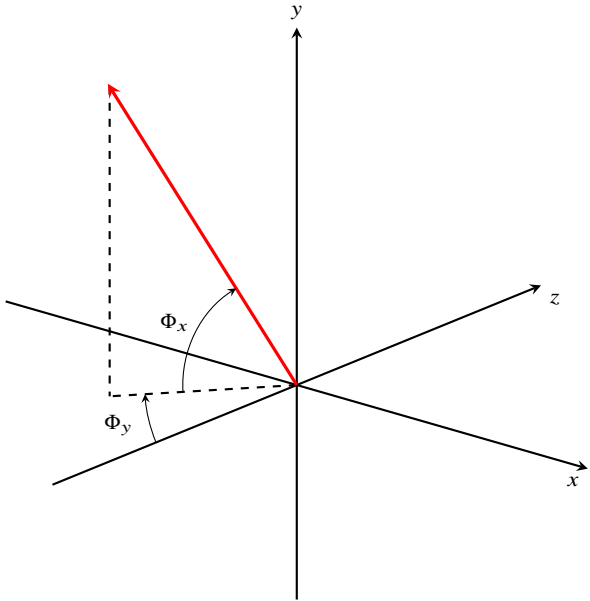

\texttt{SHAPE} can be used to model three-dimensional geometric objects, with each of the geometric properties defined by free input parameters, including the mass-loss rate, $\dot{M}$, inner and outer radii, $r_{\mathrm{inner}}$ and $r_{\mathrm{outer}}$, as well as, where necessary, associated positioning angles $\Phi_{x}$ and $\Phi_{y}$.
Figure~\ref{fig:shape_geometry} shows the coordinate system used by \texttt{SHAPE}. The $z$-axis coincides with the line of sight and the $xy$-plane corresponds to the plane of the sky. The positioning angle $\Phi_x$ corresponds to rotation about the $x$-axis in the $yz$-plane, and $\Phi_y$ coincides to a rotation about the $y$-axis in the $xz$-plane. Within \texttt{SHAPE}, the positional angles $\Phi_x$ and $\Phi_y$ can take the values $-180\,-\,180$, with $0$ along the line of sight to the observer. Physical properties, such as velocity, density, and temperature profiles for the individual objects are described with analytical expressions. 

The density, velocity and temperature profiles are also used as input for \texttt{shapemol}, alongside the logarithmic velocity gradient, $\epsilon = \frac{d\varv}{dr}\frac{r}{\varv}$, 
and the microturbulence, $\delta \varv$. \texttt{shapemol} takes the  absorption and emission coefficients, $k_{\nu}$ and $j_{\nu}$, as input, calculated from pre-generated opacity tables and the pre-calculated product $X\frac{r}{\varv}$, where $X$ is the abundance. For details on the method we refer to \citet{SHAPEMOL}.

We modelled the emission from the $^{12}$CO $J=1-0,\,2-1,\,3-2,\,4-3$, and $6-5$ transitions, and compared the results to the observations from OSO, JCMT and HIFI.
Higher excitation transitions that were available from the HIFI observations ($J=10-9$ and $16-15$) were not included owing to the limited range in density and temperature covered by the opacity tables employed by \texttt{shapemol}.

The analytical expressions were defined as $\rho(r) = \dot{M}(4 \pi \varv_{\infty}r^{2})^{-1}$ for the density profile, and $T(r) = T_{\mathrm{eff}}(r/R_{\star})^{-0.7}$ for the temperature profile. In the case of this work, the terminal expansion velocity is assumed to be constant for all components. With the inclusion of a microturbulent velocity $\delta{\varv}$ = 2.5\,km\,s$^{-1}$ \citep[in line with expectations from][]{taniguchi_2020_microturbulence_Rsgs}, 
we assumed $\varv_{\infty}$ = 30.5\,km\,s$^{-1}$ in the \texttt{SHAPE} models. This leads to a total width of the emission lines that is in line with terminal velocity of 33\,km\,s$^{-1}$ reported by \citet{debeck2010_comdot}. Due to modelling constraints, a non-zero value for the logarithmic velocity gradient, $\epsilon$, was required, and was set to the lowest possible value of 0.2. In addition, the limited range of values that could be used for the pre-calculated product $X\frac{r}{\varv}$ led to a lower limit enforced on the $^{12}$CO abundance at $X = 3\times10^{-4}$, a factor of 6 higher than previously assumed for NML~Cyg \citep{debeck2010_comdot}. This, in turn, led to a lower mass-loss rate when determining an appropriate fit to the line profiles. This does not affect the overall investigation into whether multiple components provide a satisfactory fit to the profiles, but detailed modelling of the CSE would require an extension on the parameter space currently exploitable through \texttt{shapemol} so that appropriate abundances could be assumed and higher-excitation transitions could be included in the modelling efforts.

\begin{figure*}
	\begin{subfigure}{0.49\textwidth}
		\centering
		\includegraphics[width = \textwidth]{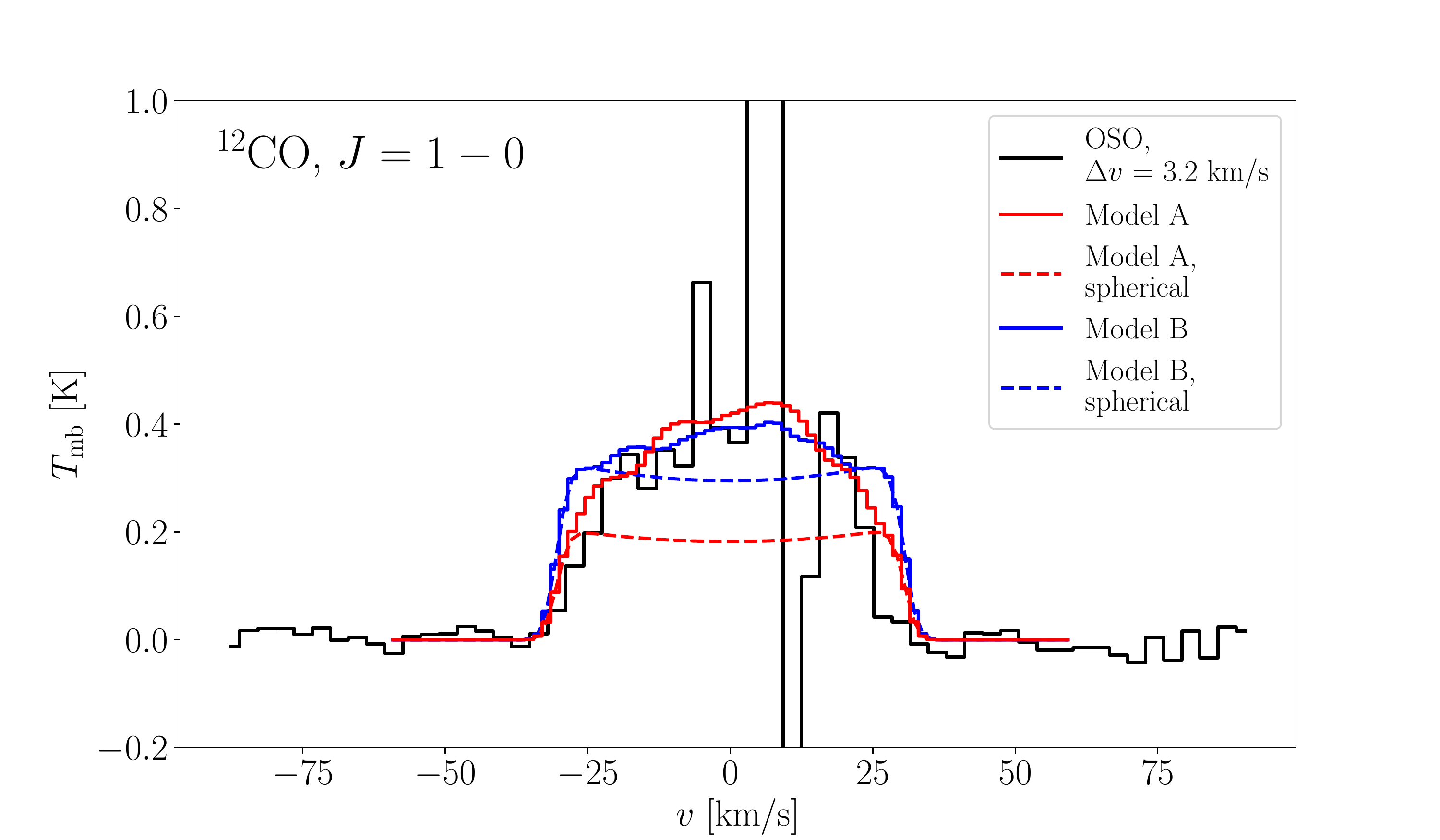}
	\end{subfigure}
	\begin{subfigure}{0.49\textwidth}
		\centering
		\includegraphics[width = \textwidth]{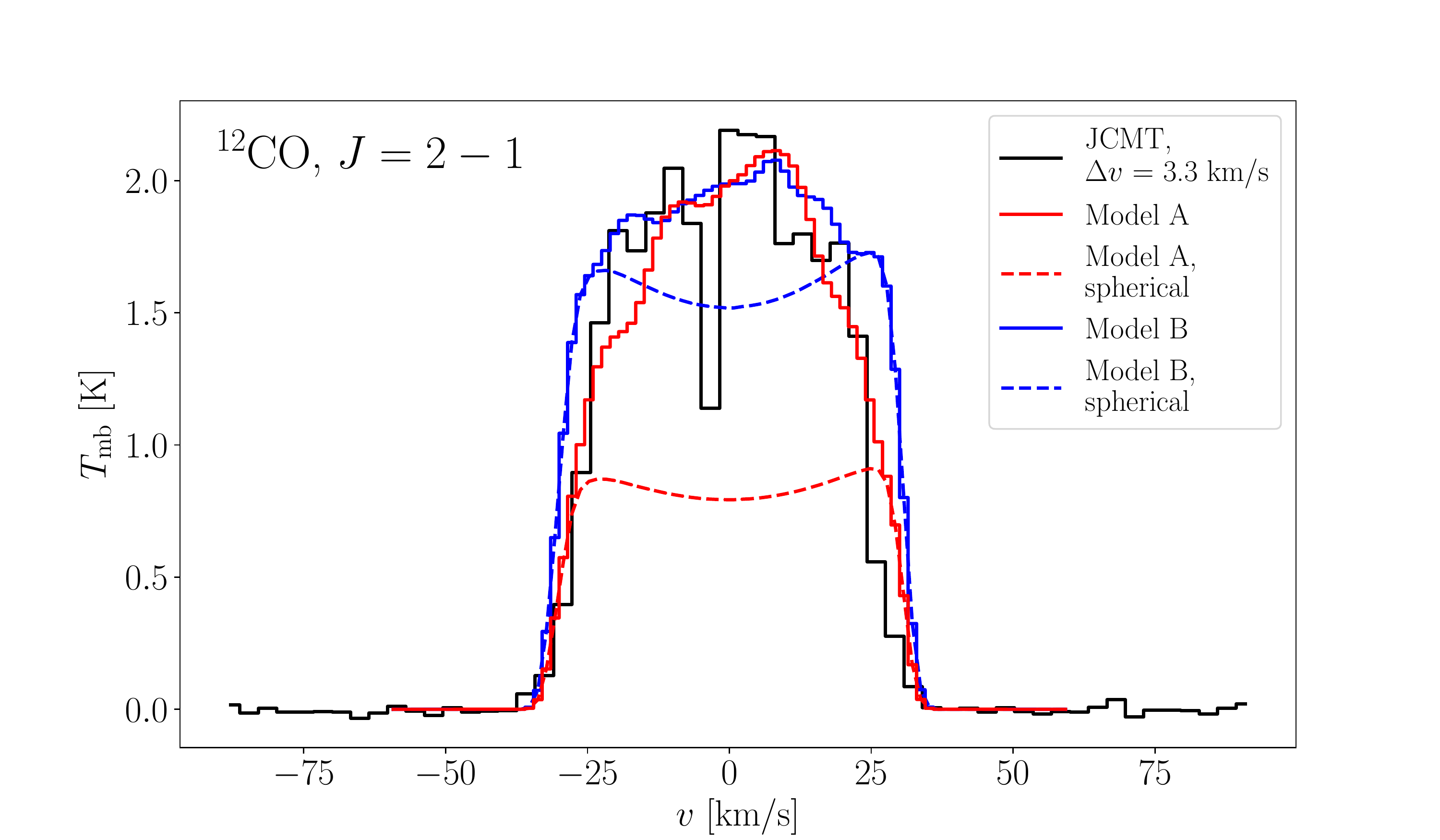}
	\end{subfigure}
	\begin{subfigure}{0.49\textwidth}
		\centering
		\includegraphics[width = \textwidth]{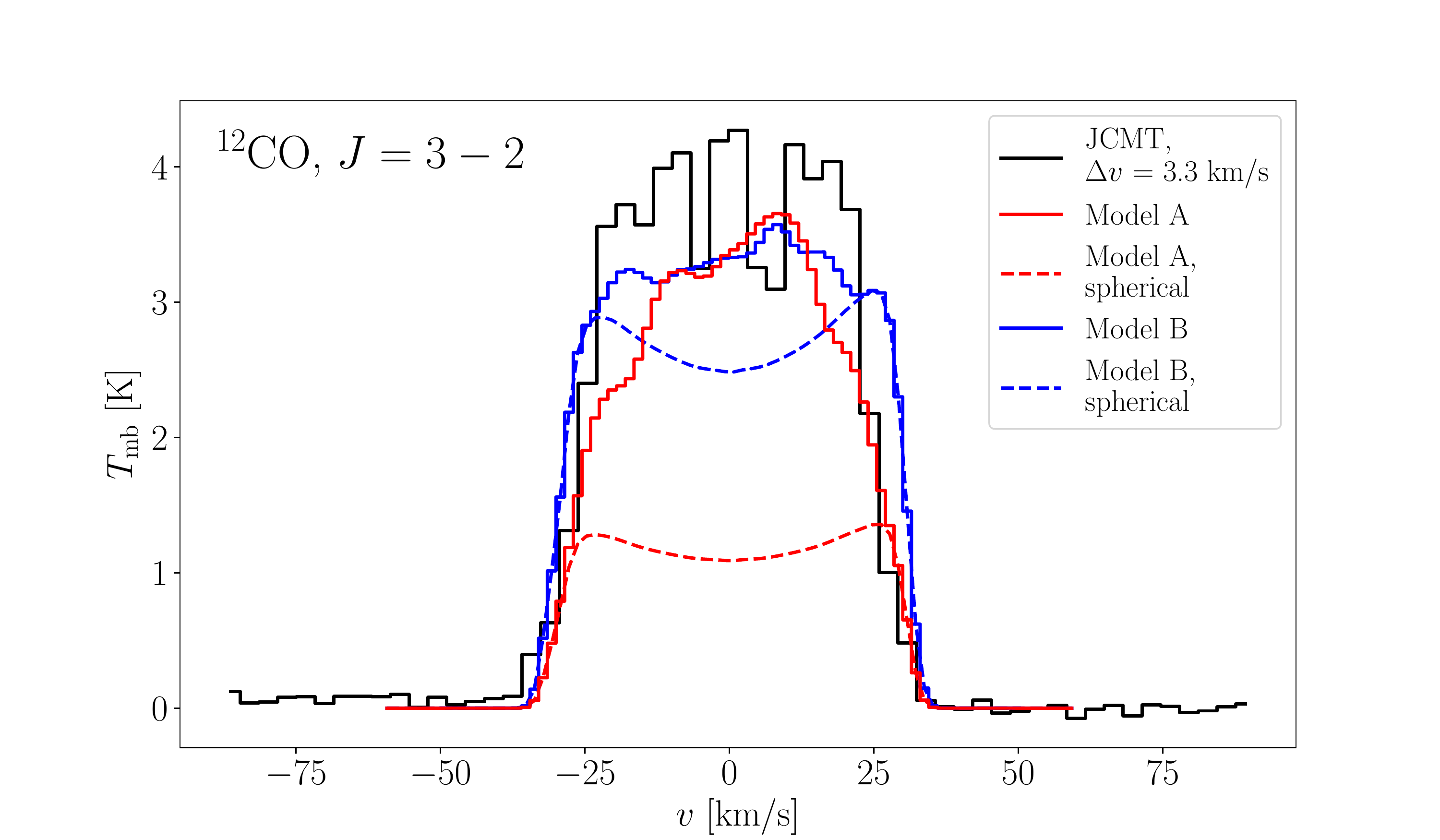}
	\end{subfigure}
	\begin{subfigure}{0.49\textwidth}
		\centering
		\includegraphics[width = \textwidth]{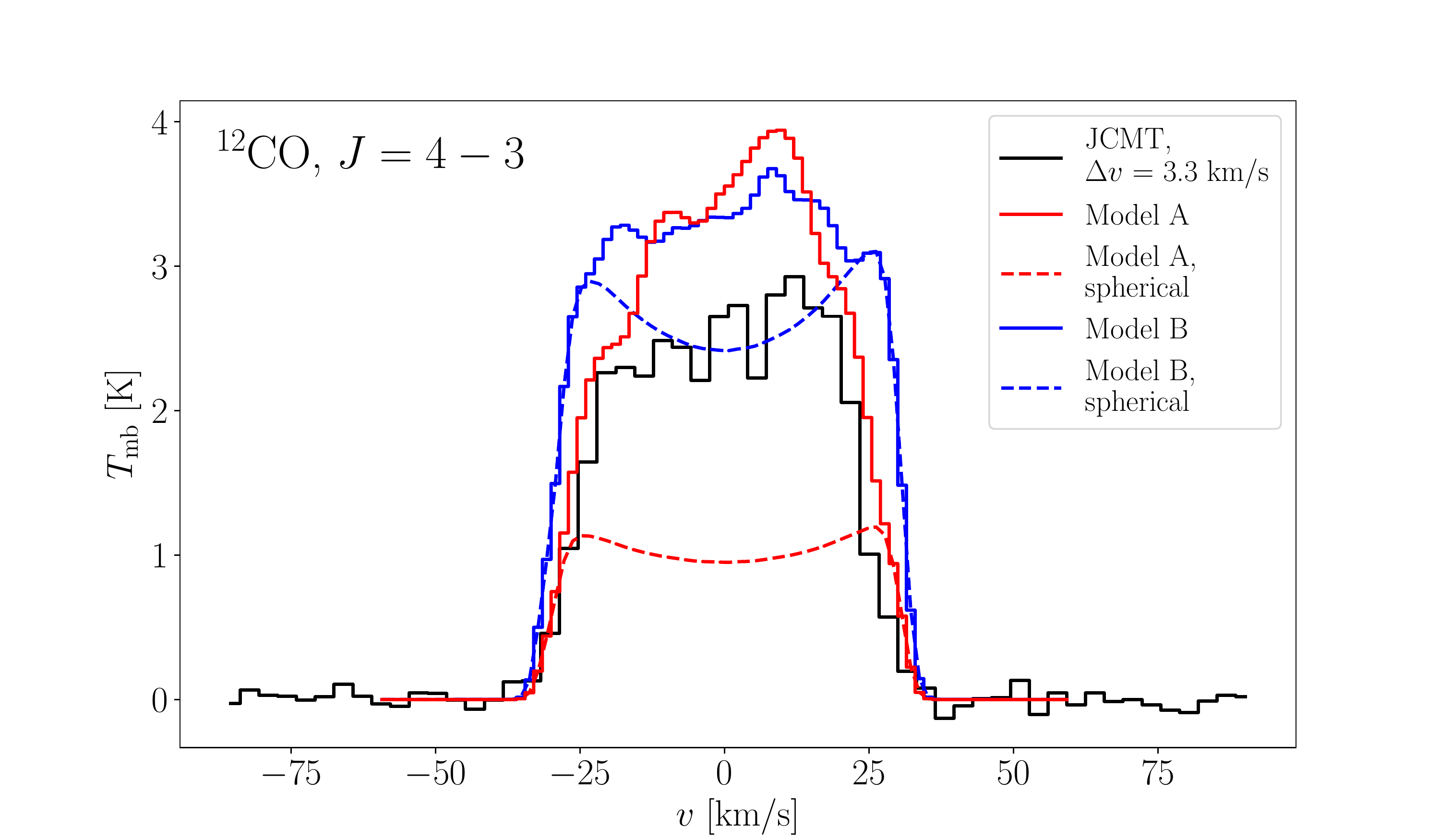}
	\end{subfigure}
	\begin{subfigure}{0.49\textwidth}
		\centering
		\includegraphics[width = \textwidth]{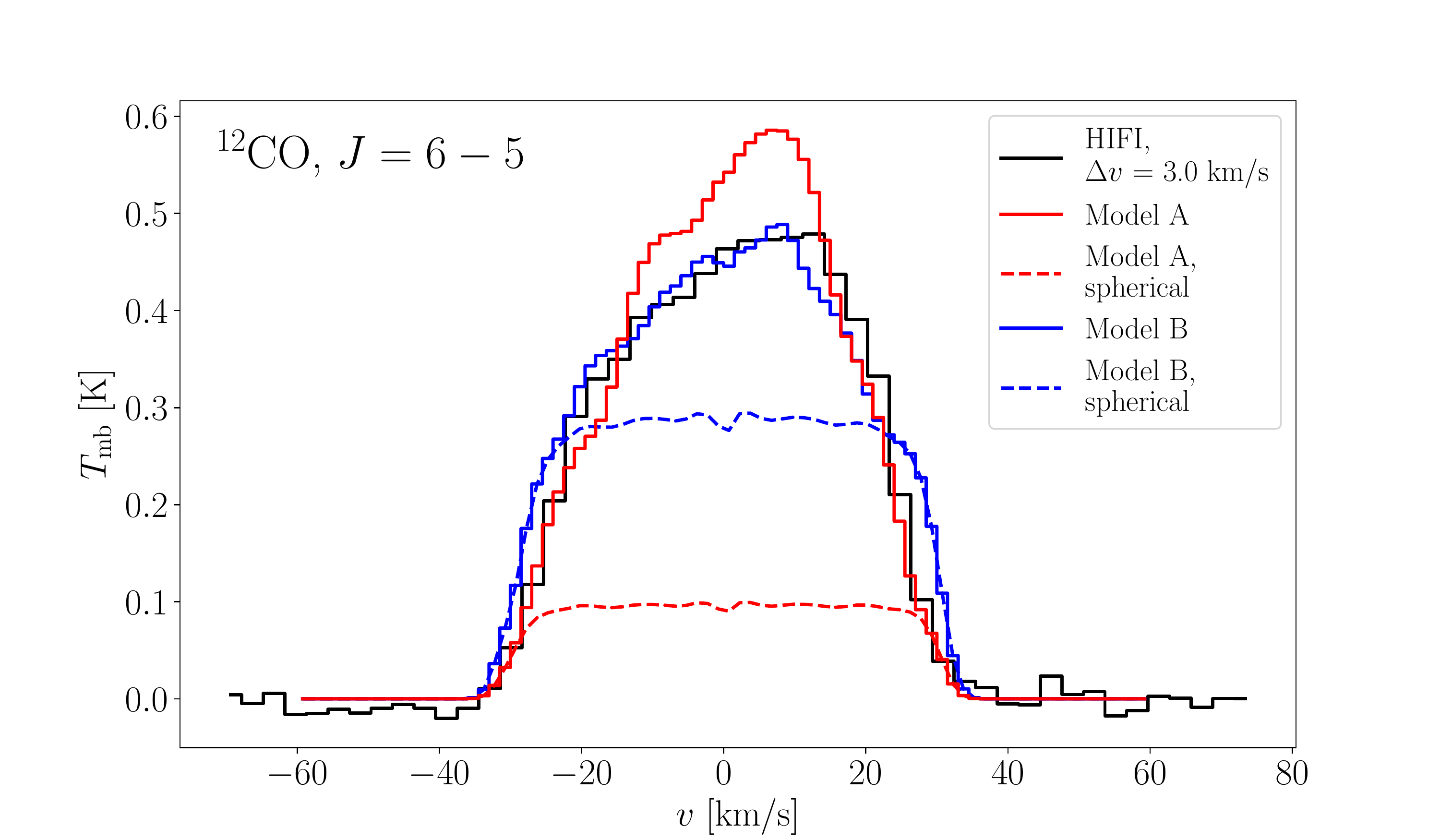}
	\end{subfigure}
	\caption{$^{12}$CO emission from NML~Cyg \emph{(black)} compared to radiative transfer models produced with \texttt{SHAPE}: Model A \emph{(red)} and Model B \emph{(blue)}. The solid lines represent the full models, while the dashed lines correspond to the spherical components.
    }
	\label{fig:co_nml_shapefits}
\end{figure*}

\begin{figure}
    \centering
	\begin{subfigure}{\linewidth}
		\centering
		\includegraphics[height=1.85in]{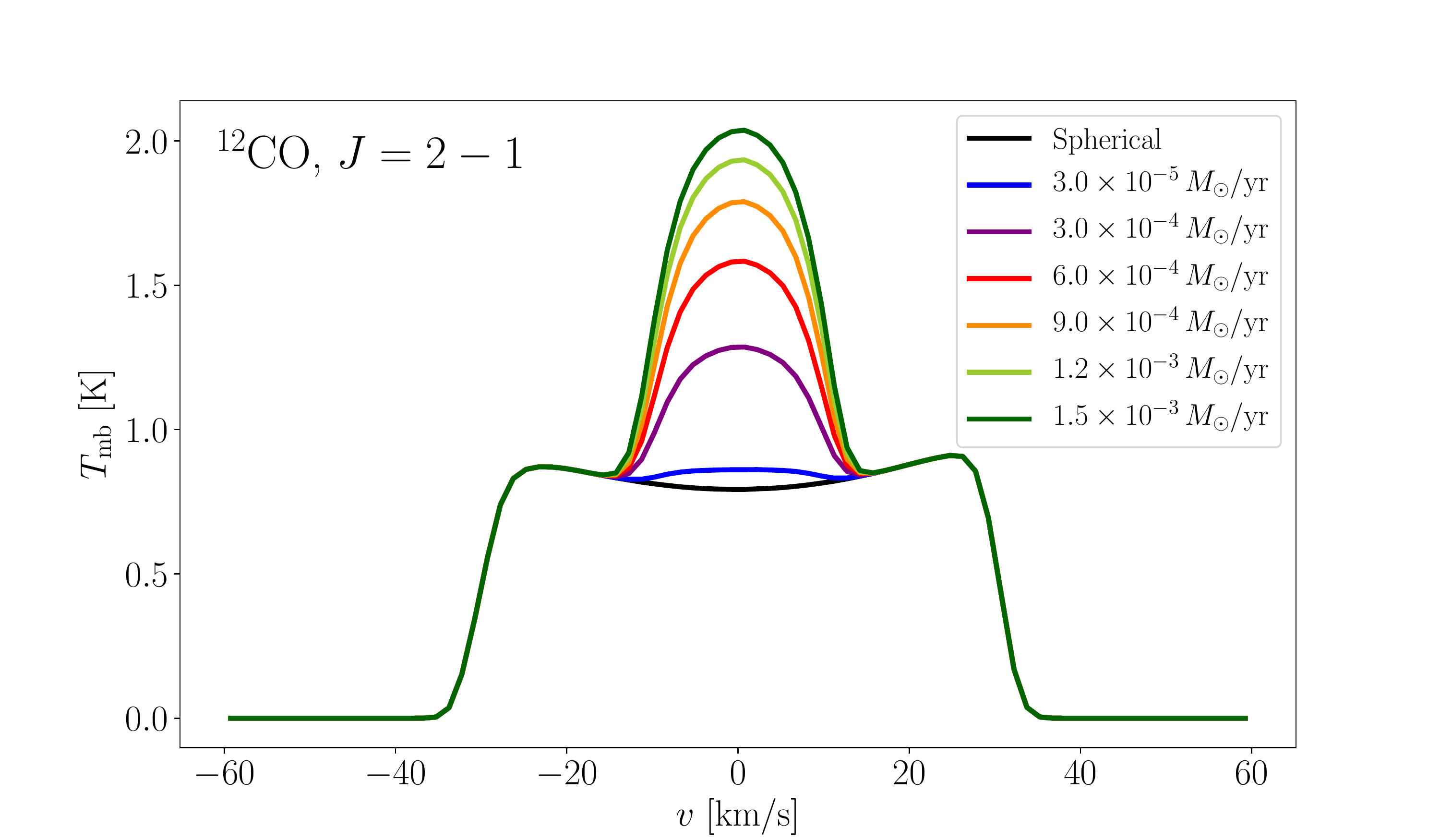}
	\end{subfigure}\hfill
	\begin{subfigure}{\linewidth}
		\centering
		\includegraphics[height=1.85in]{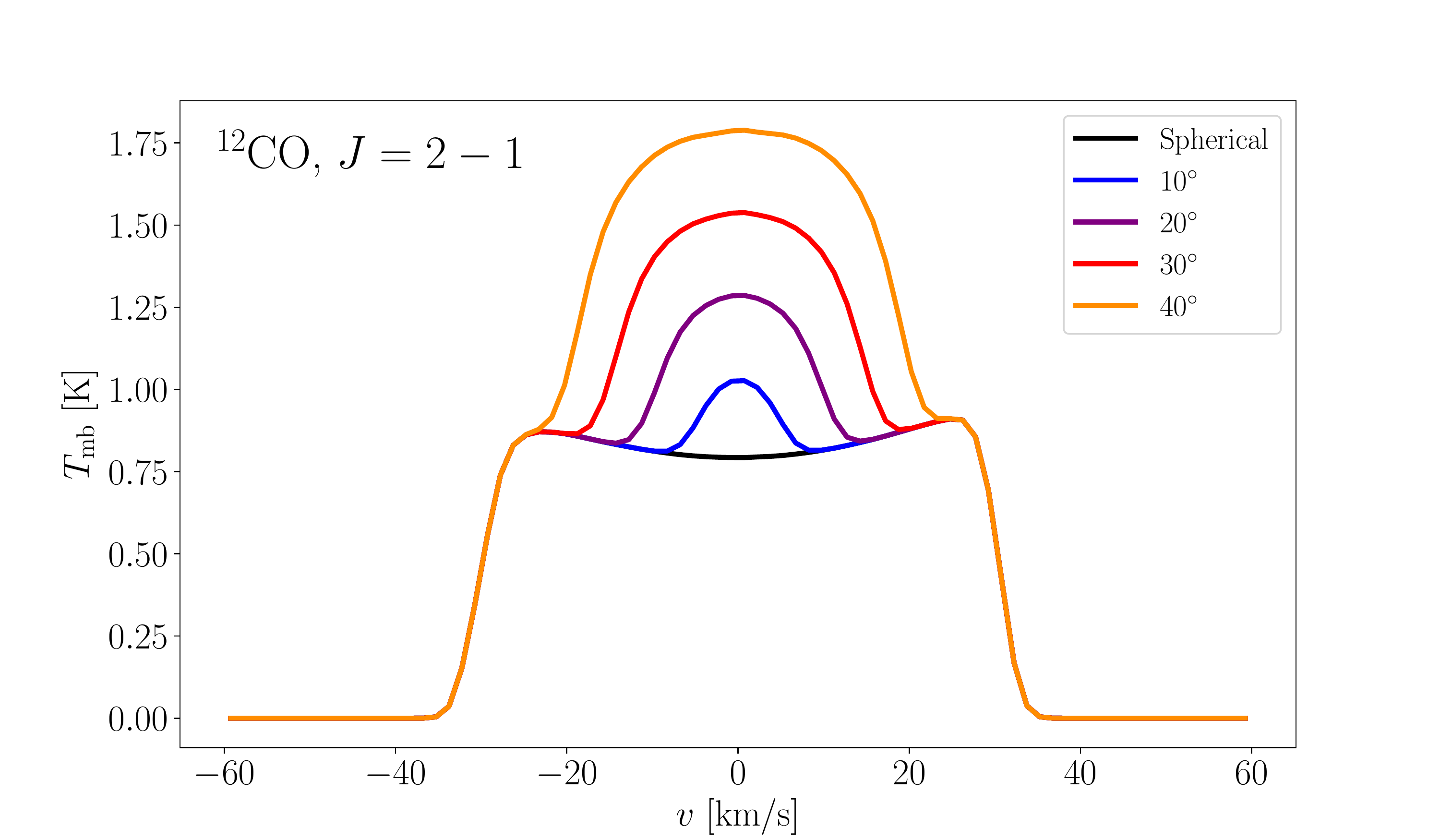}
	\end{subfigure}\hfill
    \caption{Impact of variations in mass-loss rate and opening angle on the $^{12}$CO $J=2-1$ emission. The model consists of a directed outflow perpendicular to the line of sight, with $r_{\mathrm{inner}} = 3 \times 10^{16}$\,cm, $r_{\mathrm{outer}} = 9 \times 10^{16}$\,cm, and $\Phi_x=90^{\circ}$, and a spherical component with $r_{\mathrm{inner}} =4 \times 10^{15}$\,cm, $r_{\mathrm{outer}} = 9 \times 10^{16}$\,cm and mass-loss rate $\dot{M} = 3 \times 10^{-5}\,M_{\odot}\,\mathrm{yr}^{-1}$. The solid black line is a reference model which includes only the spherical component.
    \emph{Top:} Line profiles produced for different mass-loss rates for a directed outflow of opening angle $\phi=20^{\circ}$.
    \emph{Bottom:} Line profiles produced for different opening angles for a directed outflow with mass-loss rate $\dot{M} = 3 \times 10^{-4}\,M_{\odot}\,\mathrm{yr}^{-1}$.}
    \label{fig:shape_mdot_phi}
\end{figure}

A spherical wind is reproduced in \texttt{SHAPE} with the use of three parameters: $\dot{M}_{\mathrm{sph}}$, $r_{\mathrm{inner, sph}}$, and $r_{\mathrm{outer, sph}}$. Attempts to model the emission lines with only a spherical wind did not result in acceptable fits (Fig.~\ref{fig:co_nml_shapefits}), most clearly reflected in the lack of complexity in the simulated line profiles. Based on the assumed abundance and the mass-loss rate reported by \citet{debeck2010_comdot}, we explored a grid of mass-loss rates in the range $1\,-\, 8\times10^{-5}\,M_{\mathrm{\odot}}\,\mathrm{yr}^{-1}$ to determine $\dot{M}_{\mathrm{sph}}$, over a log grid  increasing by a factor of 3 each step. Once the mass-loss rate led to an over-estimate for the line intensities, as high a mass-loss rate as possible was specified from within the interval range between two grid steps. The same  abundance values were used throughout. None of the $^{12}$CO line intensities were overestimated by this model. Model A used a final spherical wind mass-loss rate of $3 \times 10^{-5} M_{\odot} \mathrm{yr}^{-1}$, and Model B used a higher final spherical wind mass-loss rate of $6 \times 10^{-5} M_{\odot} \mathrm{yr}^{-1}$. We consider two models (Model A and Model B) because of the degeneracy faced for $\phi$ and $\dot{M}$ (Fig.~\ref{fig:shape_mdot_phi}). This degeneracy can only be concretely resolved by constraints on the geometry of the outflows from spatially resolved observations.

In order to better replicate the line profiles, the addition of directed outflows were considered. The parameters describing the directed outflows were initially constrained using the $^{12}$CO $J=2-1$ emission line, as it traces the full extent of the envelope \citep{debeck2010_comdot,ramstedt2020_deathstar}. Emission from $^{12}$CO $J=1-0$ also traces the full extent of the envelope, but showed clear signs of significant interstellar contamination along the line of sight, attributed to an interstellar clump \citep{kemper_2003_nmlcyg}. 
We modelled directional components as localised conical outflows that radially expand, with no assumptions made for the underlying ejection mechanisms. The geometry of each directional outflow component was described within \texttt{SHAPE} as a cone with five free parameters: $\dot{M}$, $r_{\mathrm{inner}}$,  $r_{\mathrm{outer}}$, the opening angle $\phi$, and the positioning angles $\Phi_{x}$ and $\Phi_{y}$. 
The initial value of $r_{\mathrm{inner}}$ for the directed outflows was set equal to $r_{\mathrm{inner,\,sph}}$; the opening angles were set to $\phi_{\mathrm{initial}} = 10^{\circ}$ and the initial mass-loss rates were set equal to that of the spherical component. The value of $\phi_{\mathrm{initial}}$ was chosen to produce clearly separated features while being large enough to avoid aliasing and artefacts in the rendered image. Initial values for the positioning angles $\Phi_x$ and $\Phi_y$ for the directed components were set by comparing the normalised line profiles of the observations and the model outputs. This was done in order to remove the need to set the mass-loss rates at this stage to any particular value. The positioning angles affect the Doppler shift of the emission, while the mass-loss rates affect the line intensity without causing any additional shift. As long as the mass-loss rates are high enough for the features in the spectra to be distinguishable the initial values for the positioning angles can be set. After the initial positioning angles were set, the mass-loss rates and opening angles were adjusted.

An increase in either the mass-loss rates or the opening angles increases the intensity of the component and alter the shape of the line profile (Figs.~\ref{fig:shape_mdot_phi}). 
The increase in the integrated intensity and peak temperature in the emission line profiles for increasing mass-loss rate becomes less significant as optical depths increase.
Narrow cones with high mass-loss rates are easy to fit to specific features, but can lead to deep notches in between the features that might not be present in the observed line profiles. They also necessitate very high mass-loss rates; a model with opening angles $\phi\approx10^{\circ}$ and $\dot{M}_{\mathrm{sphere}} = 3 \times 10^{-5}\,M_{\odot}\,\mathrm{yr}^{-1}$ might require $\dot{M}_{\mathrm{directed}}> 10^{-3}\,\mathrm{yr}^{-1}$, which is on the higher end of mass-loss rates reported in the literature for extreme sources such as IRC~+10420, a yellow hypergiant \citep[e.g.][]{QL2016}. 

\begin{figure} 
    \begin{subfigure}{0.22\textwidth}
    \includegraphics[height=1.5in, trim={12cm 0cm 10cm 0.5cm},clip] {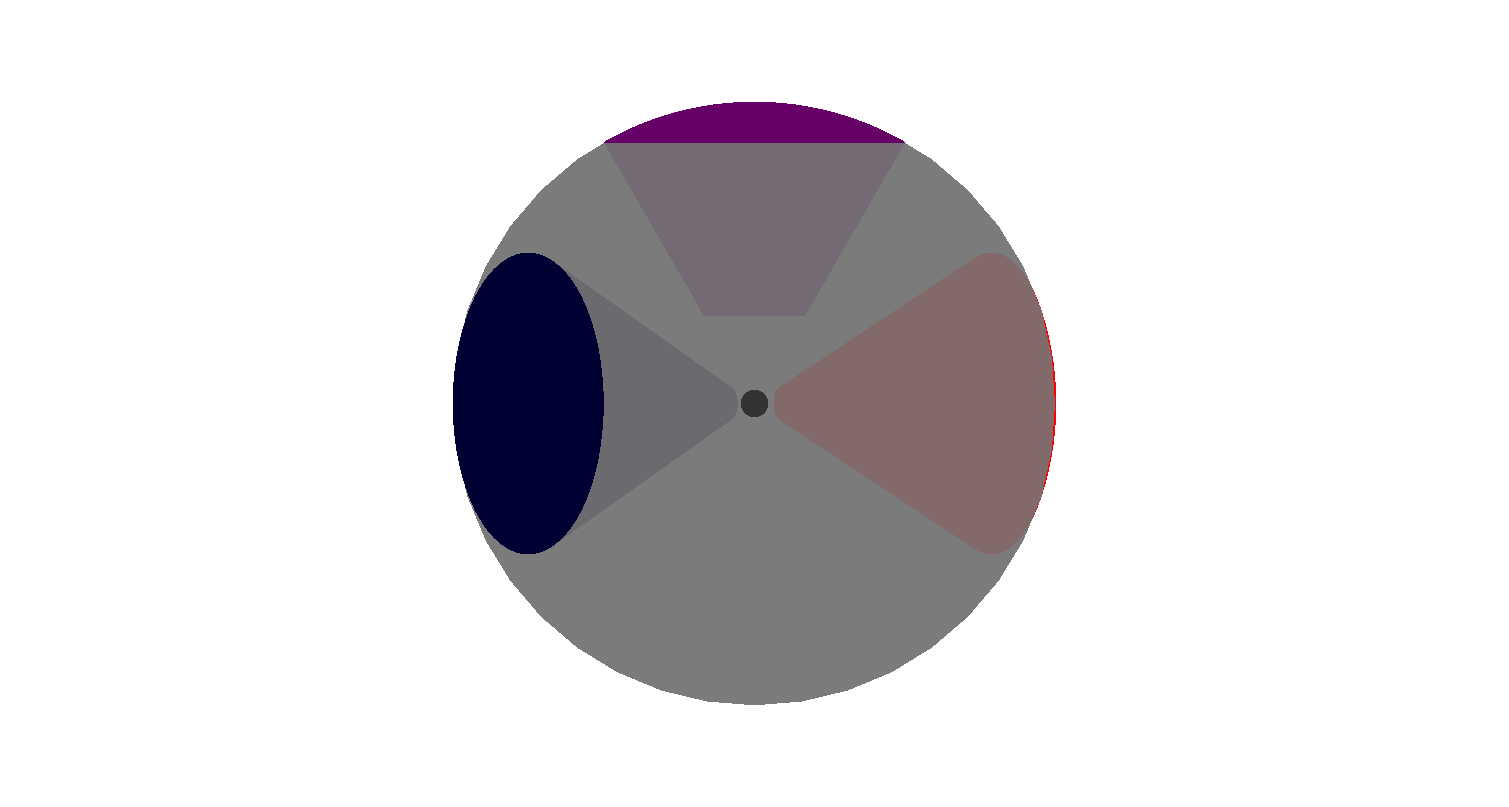}
   \caption{Head-on view \label{fig:shape_headon}}
   \end{subfigure}
   \begin{subfigure}{0.22\textwidth}
    \includegraphics[height=1.5in, trim={10cm 0cm 10cm 0.5cm},clip]{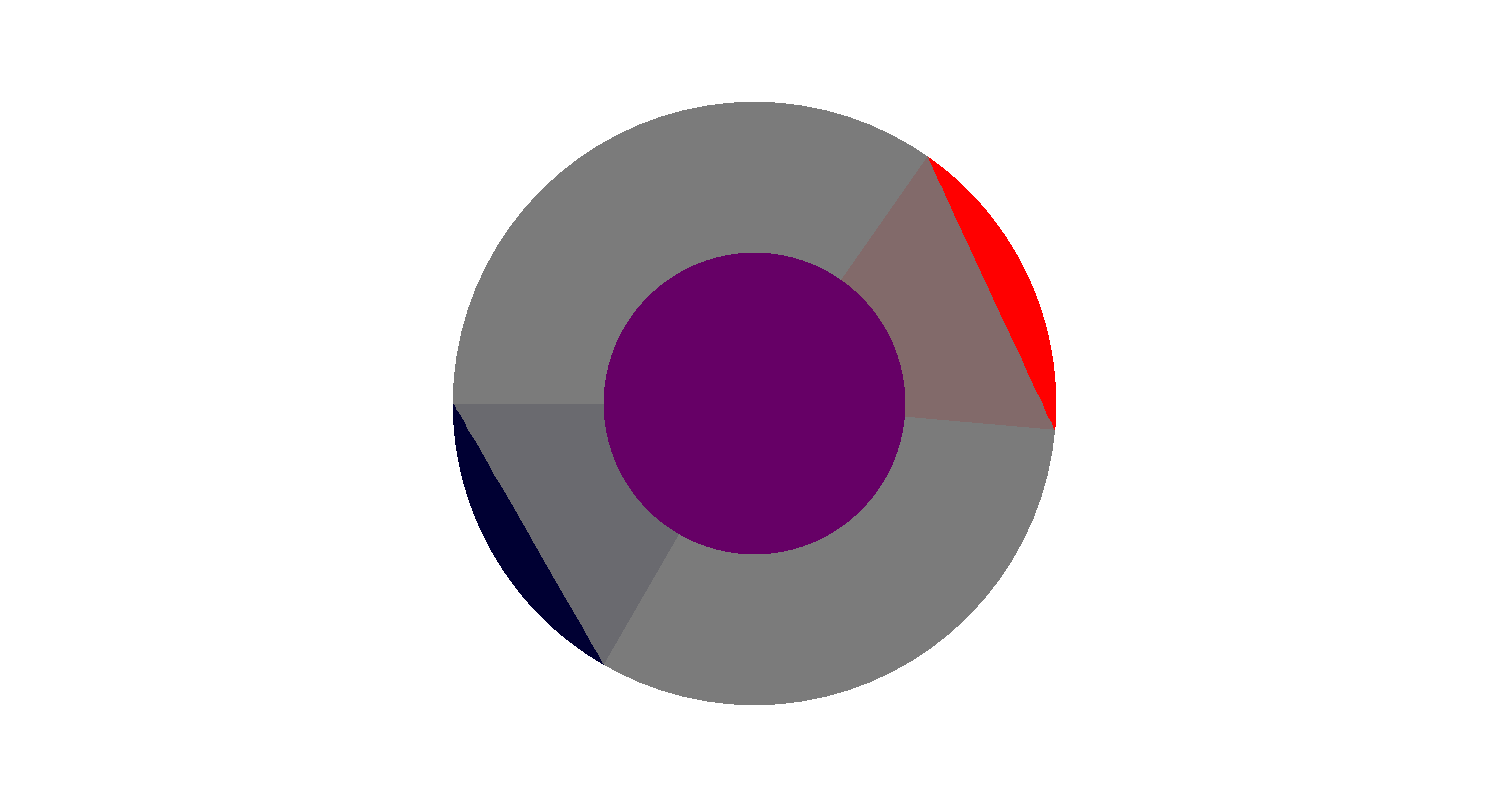}
    \caption{Bird's eye view \label{fig:shape_birdseye}}
    \end{subfigure}
    \caption{\texttt{SHAPE} model of NML~Cyg's circumstellar environment. \emph{Left:} head-on view and \emph{right:} bird’s-eye view of Model A as seen in \texttt{SHAPE}. The grey circle corresponds to the spherical component, the blue, red and purple segments correspond to the blue-shifted, red-shifted and central directed outflows. The small dark spot visible in the centre of panel \ref{fig:shape_headon} corresponds to the innermost region that is not modelled here owing to limitations in \texttt{shapemol}.}  
    \label{fig:shape_co}
\end{figure}

\begin{figure}
    \centering
    \includegraphics[width=\linewidth]{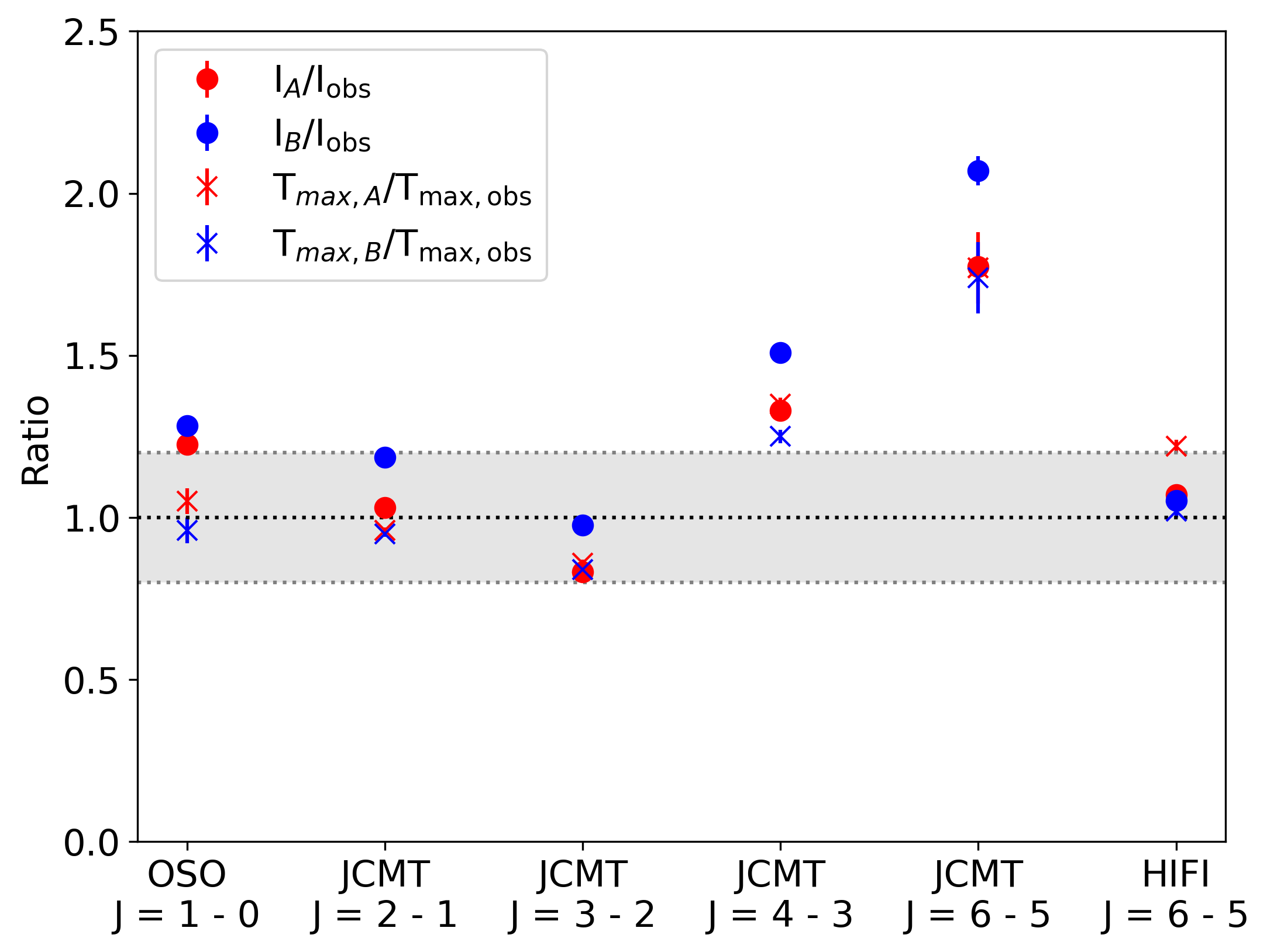}
    \caption{The ratios between the observed and modeled lines’ peak and integrated intensities, $\frac{T_{\textrm{max, model}}}{T_{\textrm{max, obs}}}$, and $\frac{I_{\textrm{model}}}{I_{\textrm{obs}}}$, respectively. The region shaded in grey indicates -20\% and +20\% thresholds for the ratios.}
    \label{fig:model_ratios}
\end{figure}

We found that it was necessary to include 3 directional components in addition to the underlying spherical wind in order to adequately replicate the observed line profiles (Figs.~\ref{fig:co_nml_shapefits} and \ref{fig:shape_co}). The best fit parameters are listed in Table~\ref{tab:shape_properties}. A measure of the fits is presented in Fig.~\ref{fig:model_ratios}, where the peak, $T_{\mathrm{max}}$, and the integrated intensities, $I$, from the models are compared to the observations. 

Models A and B were both able to reproduce the integrated intensities within 20\% of the observed values for the $J=2-1,\,J=3-2\,$ and the HIFI $J=6-5$ emission lines. For the HIFI $J=6-5$ line, only Model B recreated the peak intensity within 20 \% of the observed data. Attempts to fit the $J=6-5$ emission line observed with JCMT led to significant underestimations ($>30\%$) of all other emission lines, including the $J=6-5$ transition measured with HIFI. Owing to this and the lower calibration uncertainties on the HIFI observations, the JCMT measurement of $J=6-5$ was not used to constrain the model. As mentioned, the inclusion of higher-$J$ $^{12}$CO emission would require more complex radiative transfer models that can adequately describe the denser, inner regions of the outflow.

We found that the directed components needed to be higher in density than the spherical outflow by up to an order of magnitude. Assuming a constant expansion velocity, the enhanced mass-loss events that formed the directed outflows must have lasted on the order of up to a few hundreds of years, in contrast to typical timescales of variation seen for RSGs (as discussed in Sect.~\ref{sect:geometry}).
 
\begin{table}
\centering
\caption{The best fit parameters for the two radiative transfer models of NML~Cyg consisting of  a spherical component and three directed outflows. The parameters include the inner radii  $r_\text{inner}$, outer radii $r_\text{outer}$, mass-loss rates $\Dot{M}$, opening angles $\phi$, and the two positioning angles  $\Phi_x$ and $\Phi_y$.}
\label{tab:shape_properties}
\begin{tabular}{l c c c c c c c c} \hline
Component & $r_\text{inner}$ & $r_\text{outer}$& $\Dot{M}$ & $\phi$ & $\Phi_x$ &$\Phi_y$ \\ 
 & \multicolumn{2}{c}{($10^{16}$\,cm)} & ($M_{\odot}\,yr^{-1}$) & ($^{\circ}$) & ($^{\circ}$) & ($^{\circ}$) \\\hline
\textit{Model A} \\
Spherical   & $0.4$ &$9$ & $3.0\times10^{-5}$ &   &   &   \\
Blue        & $1$ & $9$ & $1.5\times10^{-4}$ & 30 &  0 & 60 \\
Red         & $1$ & $9$ & $2.1\times10^{-4}$ & 30 & 0  & $-115$ \\
Central     & $3$ & $9$ & $3.0\times10^{-4}$ & 30 & 90 & 0 \\
\hline
\textit{Model B}  \\
Spherical   & $0.4$ & $9$ & $6.0\times10^{-5}$ &   &   &   \\
Blue        & $1$ & $9$ & $1.5\times10^{-4}$ & 15 & 0  & 60 \\
Red         & $1$ & $9$ & $2.1\times10^{-4}$ & 15 & 0  & $-115$ \\
Central     & $3$ & $9$ & $3.0\times10^{-4}$ & 20 & 90 & 0 \\
\hline
\end{tabular}
\end{table}

\section{Discussion}\label{sect:discussion}

\subsection{Comparison between multi-component fitting procedures}\label{sect:comp_fitting}

We find clear evidence for multi-component outflows from both the simplified method of fitting multiple Gaussians to the observed line profiles, as well as the application of radiative transfer modelling of the low-excitation $^{12}$CO transitions using \texttt{SHAPE}. The use of multiple Gaussians led to accurate fits able to reproduce the line profiles through the inclusion of three components, assumed to represent a directional red-shifted and blue-shifted component alongside a central wind. Our radiative transfer models required one additional component in the plane of the sky. 
We found the intensity of the blue-shifted components consistently similar to or larger than that of the red-shifted components in our three-component fitting results, whereas the \texttt{SHAPE} models required a slightly higher $^{12}$CO mass in the red-shifted component than in the blue-shifted component. This could be, in part, due to the fact that $\varv$ and $\Delta\varv$ were derived independently for the blue- and red-shifted components, whereas the \texttt{SHAPE} model assumed the same opening angle to govern all of the model components. In addition to this, the \texttt{SHAPE} models also included an extra component that would produce some of the emission picked up by either of the three components in the simple fitting procedure.

Although neither method is conclusive on the geometry or the detailed nature of the outflow, we conclude that the results from both lend strong support to the presence of large-scale outflow components, built up over timescales of up to hundreds of years, in addition to an underlying wind. 
 
\subsection{Comparison to maser and dust emission around NML Cyg}\label{sect:comp_nmlcygliterature}

Previous studies have highlighted asymmetries in the  circumstellar environment of NML~Cyg. Maser emission from OH and H$_2$O have revealed the presence of a large-scale spherical shell alongside a more recently developed bipolar outflow oriented along a SE-NW axis \citep{richards_1996_masers_h2o,etoka_diamond_2004}.
\citet{schuster2009} found that the mid-infrared dust emission aligns with these directed maser outflows. In contrast, near-infrared observations of the circumstellar dust by \citet{monnier2004} revealed an equatorial enhancement oriented NE-SW, that is, orthogonal to the maser emission and the mid-infrared dust emission.

Our results clearly provide further support for deviations from spherical symmetry in the circumstellar material. We consistently find that the measured lines consist of multiple components, suggesting that the observed molecules are present in multiple sub-structures. We cannot currently constrain the orientation in the plane of the sky of the directed components, nor have we attempted to study the effect of differences in e.g. density, molecular abundances, or radiation field between the different components on the emission. 
However, although we refrained from assuming a bipolar structure in our fitting and modelling methods, we do find that the molecular emission likely originates from a spherical wind and from directed components opposed in velocity with roughly similar opening angles, lending support to the bipolar scenario proposed by \citet{etoka_diamond_2004}.

We generally found, from our multi-component fitting procedure, that the central component dominates the emission in the measured spectral lines towards NML~Cyg (see Sect.~\ref{sect:line_intensities}). At the same time, there is a significant contribution from both the blue- and red-shifted components, with the emission from the blue-shifted component consistently similar or somewhat stronger than from the red-shifted component. Similar spectral asymmetries were also found in the emission of OH and H$_2$O masers, although the spatial locations of their brightest (blue-shifted) components are located at opposite ends of the NE-SW axis \citep{richards_1996_masers_h2o,etoka_diamond_2004}. In addition, \citet{schuster2009} noted a marked excess in mid-infrared dust emission to the NW of the star and explained this as a consequence of the UV irradiation of the circumstellar dust by the nearby Cyg~OB2 association. Spatially resolved observations of the circumstellar molecular emission, in combination with three-dimensional radiative-transfer models, are needed to confirm or disprove similar (a)symmetries as seen in the dust and maser emission.

\subsection{Comparison to other RSGs}\label{sect:comp_rsgs}

NML~Cyg is one of only a few RSGs for which the circumstellar environment has been measured in thermal emission from multiple molecular species and isotopologues. Spatially resolved measurements of this class of objects are currently still extremely rare. For RSGs that have been studied in detail, there has been evidence for structure on a wide range of scales.
Some show predominantly small-scale clumping \citep[e.g. $\mu$~Cep;][]{Montarges2019} whereas the circumstellar environments of others are dominated by large-scale ejections visible as arcs and plumes \citep[e.g. VY~CMa;][]{SMAKaminski2013,humphreys2021} or structures suggestive of bipolar outflows \citep[e.g. S~Per, VX~Sgr, NML~Cyg;][]{richards_1999_sper,Murakawa_2003,etoka_diamond_2004}. 
Since NML~Cyg, VX~Sgr, S Per, and VY~CMa are high-$\dot{M}$ sources, whereas $\mu$~Cep loses mass at a rate one to two orders of magnitude lower \citep[as seen in the range of mass-loss rates for these sources given by][]{debeck2010_comdot, mauron_josselin_2011_masslossrsgs, Shenoy2016, gordon2018_nmlcyg_dust, gail_2020_rsgs}, this could be suggestive of a bifurcation in mass-loss behaviour (clumpy versus coherent outflows) connected to the underlying mass-loss rate, which is in turn connected to the stellar parameters. This may also be linked to differences in the underlying mechanisms involved in the mass ejection, such as Alfv\'{e}n waves or convective cells. Further investigation of the circumstellar environments of RSGs is crucial to understand whether there is indeed such a relation and how initial masses and stellar properties throughout a massive star's lifetime can influence the final fate of RSGs. 

Systematic studies of galactic RSGs are needed to test whether the above is general or individual behaviour and whether it can be correlated with stellar properties also across a larger sample. Investigations into the relation between the gas-mass loss and the dust-mass loss from galactic RSGs will, furthermore, be crucial to help interpret the observations of extragalactic populations of RSGs, which are generally used to measure the overall impact of RSGs on galactic evolution \citep[e.g.,][]{Dell_agli_2018_rsgs_galacticevol,wang2021}. We underscore the need for a systematic understanding of the molecular and dust content of the outflows of this class of star, in order to understand the mass-loss mechanism, the mass-loss history, and ultimately, the impact of mass loss on stellar and galactic evolution. 

\section{Conclusions}\label{sect:conclusions}

This paper has presented the results of a spectral survey carried out over $68 - 116$\,GHz towards the red supergiant NML~Cyg with the 3\,mm and 4\,mm receivers on the OSO 20\,m radio telescope. We detected emission from 15 lines pertaining to 10 molecular species around NML~Cyg. We complemented our observations with archival HIFI and JCMT observations, extending our coverage to a total of 49 lines from 15 different molecular species. 

The observed emission line profiles were well-represented by the sum of three Gaussian profiles. The quality of those fits and the retrieved fit parameters strongly suggest the presence of three distinctive spectral outflow components: one red-shifted, one blue-shifted, and one central component. 
Radiative transfer modelling of CO emission, using \texttt{SHAPE} and \texttt{shapemol}, led to a similar geometry comprised of a spherical wind with superposed directional outflows which could have been sustained over timescales up to a few hundred years. Our proposed component decomposition is reminiscent of the bipolar structure previously seen in maser and dust emission towards NML~Cyg. Although we found that emission from each of the observed molecules likely originates from all of the components in the outflow, there likely are differences in the relative contributions from the various components. Whether these differences stem from differences in physical properties, molecular abundances, or the radiation field will have to be tested with further radiative transfer models, necessarily constrained by spatially resolved images. 

Our results indicate the possibility for a bifurcation in the nature of mass loss from RSGs, with some characterised by large-scale ejections and others by clumpiness on small spatial scales.
Whether this difference in outflow structure is correlated to the mass-loss rate, the underlying mass-loss mechanism and/or the stellar properties such as luminosity, effective temperature, age, and/or initial mass remains to be tested.
Future systematic surveys of galactic RSGs should trace both the gaseous and  dusty components in their outflows to critically improve our understanding of this phase of massive-star evolution. A deeper comprehension of the galactic RSG population will help to determine how RSGs may contribute to the overall gas and dust budgets in other galaxies, especially for young galaxies at high redshift where the dust production is heavily reliant on the evolved stages of shorter-lived massive stars. 

\section*{Acknowledgements}

H. Andrews and E. De Beck acknowledge support from Onsala Space Observatory for the provisioning of its facilities/observational support. The Onsala Space Observatory national research infrastructure is funded through Swedish Research Council grant No 2017-00648. We thank Henrik Olofsson for his work in obtaining part of the observations and for his advice on the post-processing and reduction of the data. The authors acknowledge the reviewer, Anita Richards, for helpful comments to improve the quality of the manuscript. E. De Beck acknowledges funding from the Swedish National Space Agency. 

\section*{Data availability}

The observations were taken under the programme code O2015c-02 at the OSO 20m radio telescope. The data underlying this article will be shared on reasonable request to the corresponding author.

%
\bibliographystyle{mnras} 
\bibliography{./ref.bib} 
%

\appendix
\onecolumn
\pagebreak
\newpage
\newpage
\twocolumn
\section{OSO 20m telescope 3mm and 4mm receiver validation }\label{sect:verification}

The data were initially gathered in 2016 as science verification data for the 3\,mm and 4\,mm receivers installed in 2015 and 2016 on the OSO 20m telescope \citep{oso_3mmrec_instrumentpaper, 4mm_oso_rec_instrument2016}. Additional data were taken of IRC~+10420 on 6 November 2020 to account for artefacts in the original observations. We first show the consistency between the two receivers by comparing the data gathered in an overlapping frequency range. We then further validate the receiver performance by comparing our observations of IRC~+10420 to published IRAM 30m telescope observations \citep[][henceforth \citetalias{QL2016}]{QL2016}.

\subsection{Receiver comparison} 

\begin{figure}
	\centering
	\begin{subfigure}{0.48\textwidth}
	\centering
	\includegraphics[height=3.6cm]{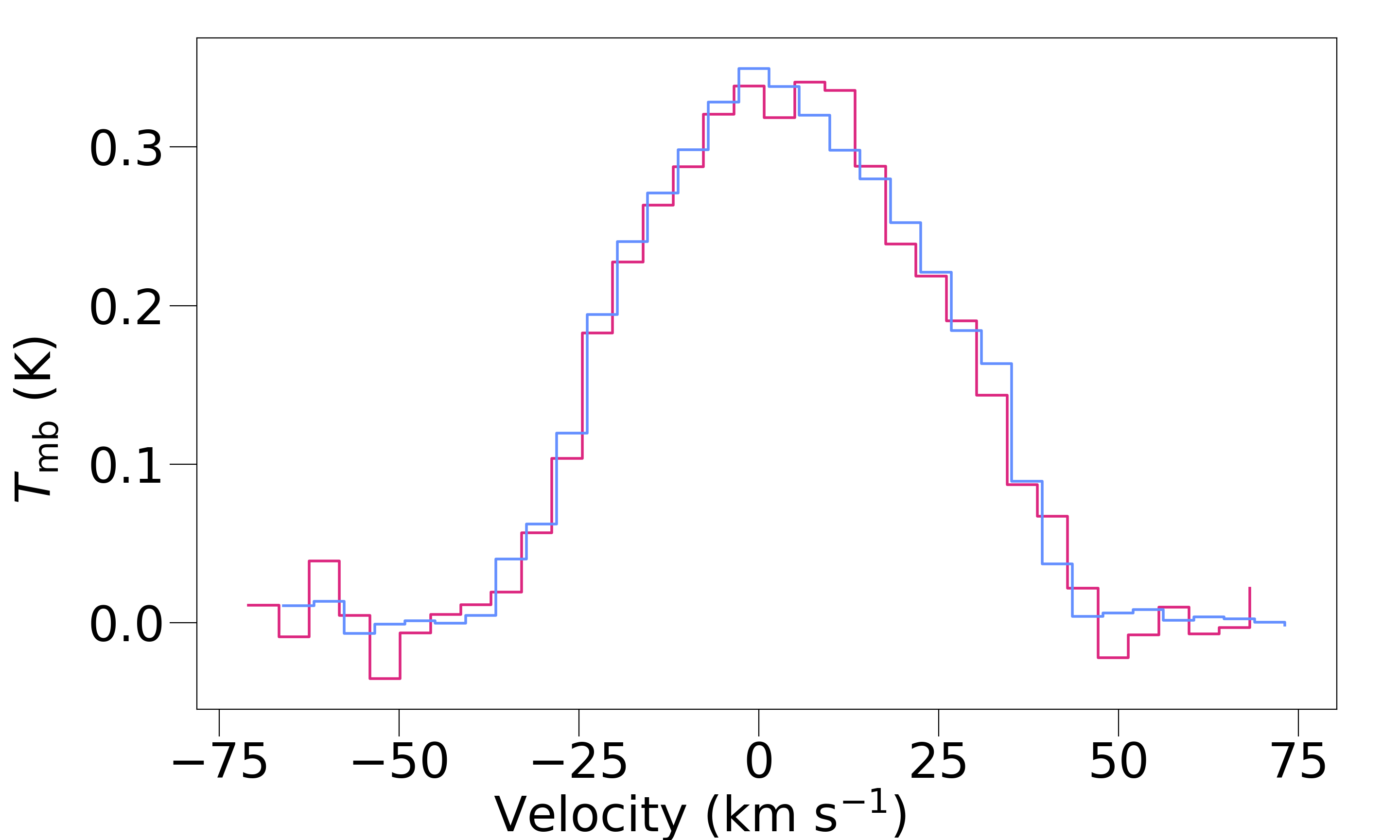}
	\caption{IRC~+10420}
	\label{fig:sv_irc}
	\end{subfigure}
	\begin{subfigure}{0.48\textwidth}
	\centering
	\includegraphics[height=3.6cm]{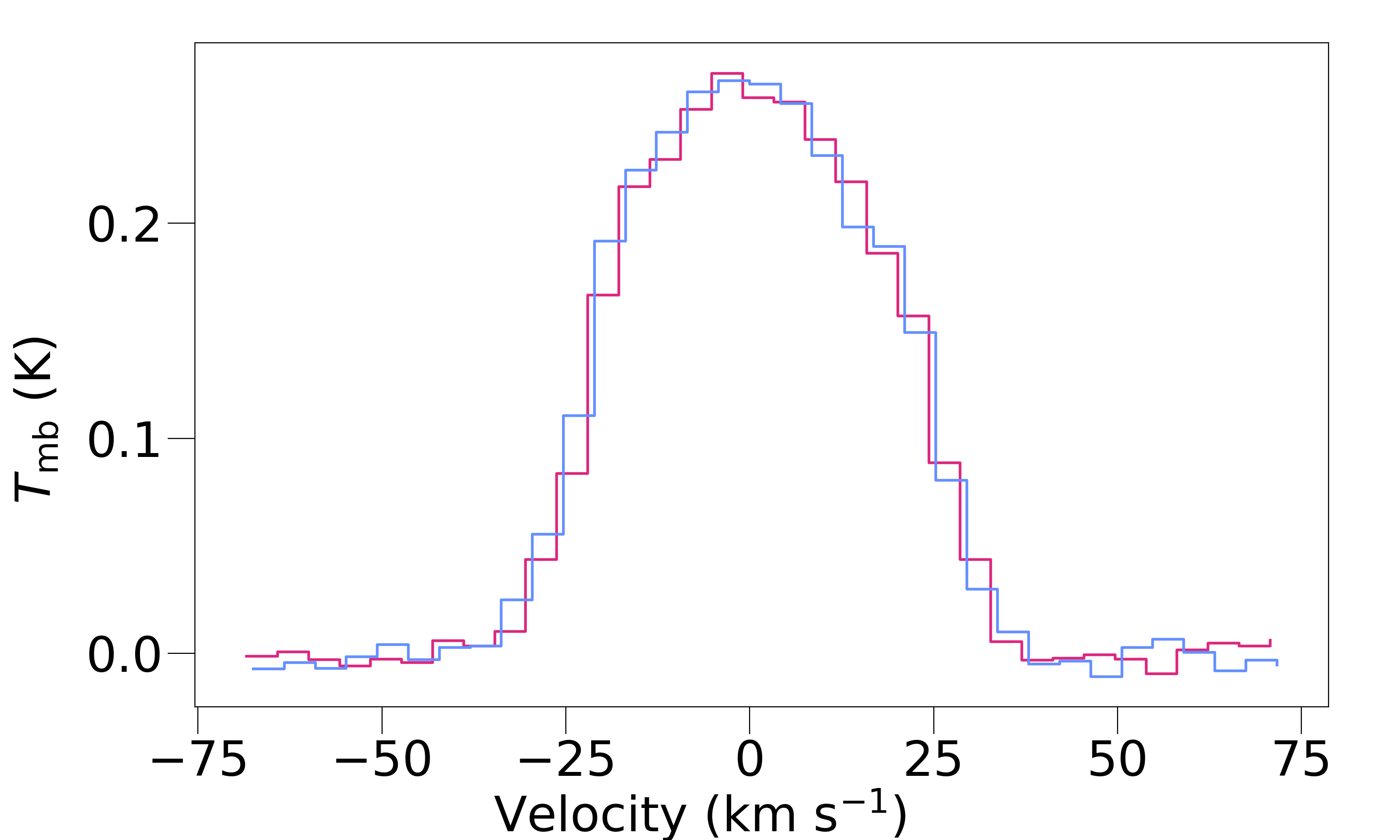}
	\caption{NML~Cyg}
	\label{fig:sv_nml}
\end{subfigure}
\caption{
$^{28}$SiO  $J = 2 - 1\,(v = 0)$ emission at 86.847\,GHz measured towards IRC~+10420 \emph{(top)} and NML~Cyg \emph{(bottom)} measured with the 3\,mm receiver tuned at 87\,GHz \emph{(blue)} and the 4\,mm receiver tuned at 85\,GHz \emph{(red)}.
}
\label{fig:rec_comp}
\end{figure}

The quality of the data and the reliability of both the 3\,mm and 4\,mm receivers can be quantitatively checked by comparing the detections of strong lines that appear in overlapping frequency ranges. The $^{28}$SiO  $J = 2 - 1\,(v = 0)$ transition at 86.847\,GHz was observed with the 4\,mm receiver at a frequency band centered on 85\,GHz and with the 3\,mm receiver at a frequency band centered at 87\,GHz. Emission from this transition was detected for both IRC~+10420 and NML~Cyg. The detections were found to be consistent across the receivers (Fig.~\ref{fig:rec_comp}), with differences between the line profiles found to lie well within the expected calibration uncertainties of 10\% \citep{Priv_Comm_Henrik}. The properties of the line profiles are listed in Table~\ref{tab:sv}. 

\begin{table}
	\caption{Properties of $^{28}$SiO  $J  = 2 - 1\,(v = 0)$  emission at 86.847\,GHz. 
	Listed in each row are the peak temperature, the integrated intensity, and the full width at half maximum (FWHM) of the line profile .}
	\label{tab:sv}
	\centering
	\begin{tabular}{lcccc}	
		\hline
		Source & Receiver & Peak & Intensity  & FWHM\\
		&&(mK)&(K\,MHz)& (km\,s$^{-1}$)\\
		\hline
		IRC~+10420  & 4\,mm  & 329 & 4.87 & 38.7 \\
		            & 3\,mm  & 329 & 4.95 & 38.4 \\
		\hline 
		NML~Cyg     & 4\,mm  & 267 & 3.30 & 32.5 \\
		            & 3\,mm &  269 & 3.36 & 32.0 \\
		\hline
	\end{tabular}
\end{table}

\subsection{IRC~+10420}

IRC~+10420 is a YHG located at a distance of 5\,kpc, with a high luminosity of $5 \times 10^{5}\,L_{\odot}$ and an initial mass estimate of $\sim50\,M_{\odot}$ \citep{Jones1993, Tiffany2010}. Observations indicate a very high mass-loss rate of $0.3 - 1\times10^{-3}\,M_{\odot}$\,yr$^{-1}$  \citep{Knapp1985, Oudmaijer1996, Castro-Carrizo2007, Dinh2009}. The star is surrounded by a dense circumstellar nebula that has been imaged at optical and infrared wavelengths and was found to display a high level of morphological complexity \citep{Humphreys1997,Dinh2017}.

\begin{figure*}	
	\centering
	\includegraphics[width=\linewidth, trim={2cm 0cm 5cm 0.5cm},clip]{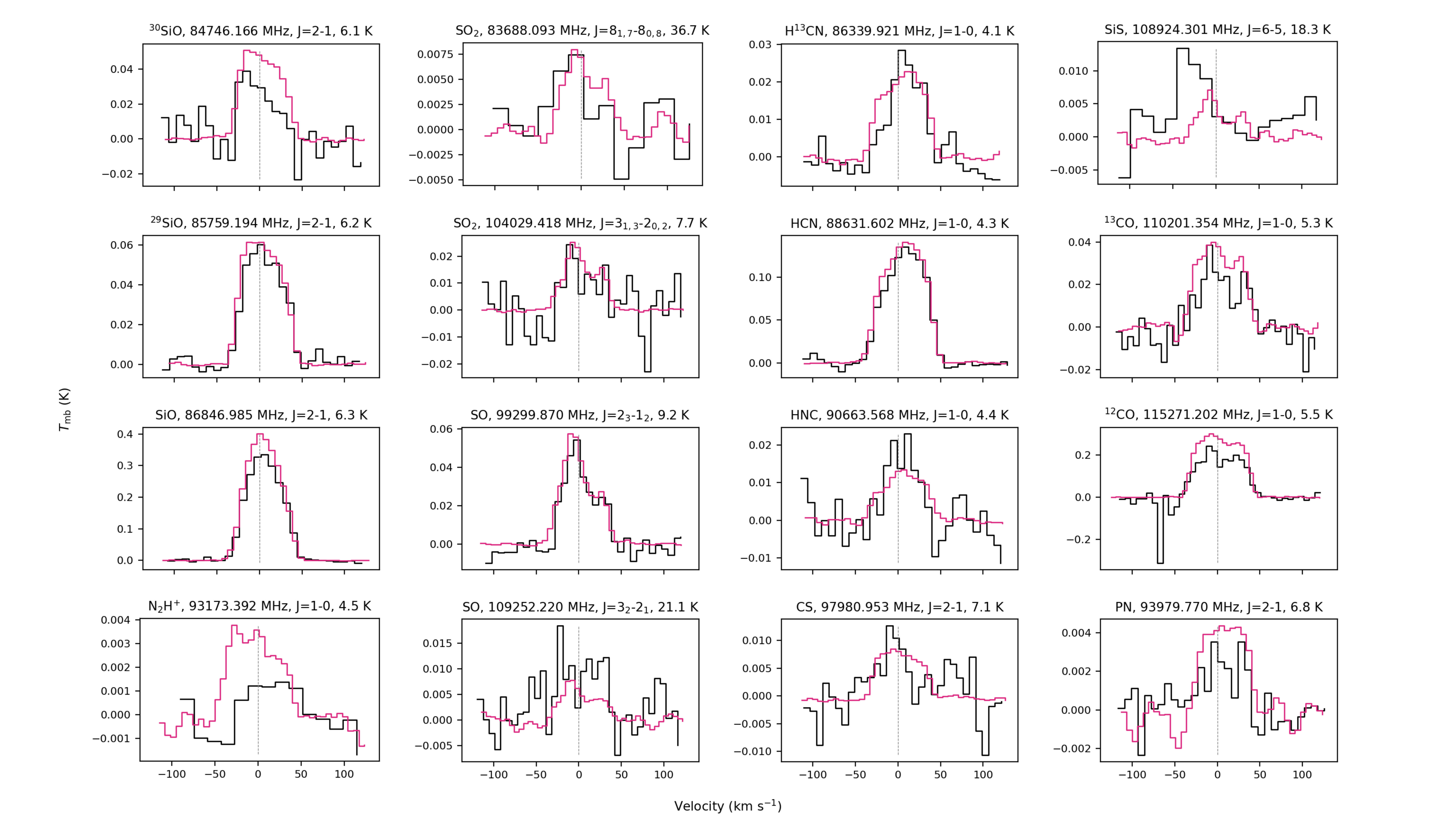}
	\caption{Molecular emission detected towards IRC~+10420 in our OSO observations \emph{(black)} and in the IRAM 30m observations from \citetalias{QL2016} with beam dilution scaling applied \emph{(red)}. The OSO observations have been smoothed to a resolution of $6 - 9$\,km\,s$^{-1}$. Tentative detections of the molecules N$_{2}$H$^{+}$, SO$_{2}$, SiS and PN in the OSO data set are also shown, with additional smoothing applied which provides a spectral resolution of  $13 - 18$\,km\,s$^{-1}$ for the OSO data.}
	\label{fig:irc_detected}
\end{figure*}

The molecular lines detected around IRC~+10420 in our observations are presented in Table~\ref{tab:linesummary}. An overview of detected (and tentatively detected) lines is shown in Fig.~\ref{fig:irc_detected}. We compare our observations to those made by \citetalias{QL2016} with the IRAM 30m telescope in an overlapping frequency range $83.087 - 115.990$\,GHz (Fig.~\ref{fig:irc_detected}) by applying scaling to the IRAM data to convert from antenna temperature to brightness temperature scale, and to account for the different beam sizes of the two telescopes. The emission from the $^{12}$CO $J = 1 - 0$ transition was expected to cover scales of up to 11\arcsec\/ \citepalias{QL2016}, filling the beam of the 30m telescope. In that case, the beam efficiency factor was not applied and the IRAM data were scaled only for the difference in beam size. The \citetalias{QL2016} observations have a spectral resolution of 6\,MHz, or $6 - 8$\,km\,s$^{-1}$. The OSO 20m observations of IRC~+10420 were smoothed to a comparable spectral resolution of $6 - 9$\,km\,s$^{-1}$, with additional smoothing applied to investigate the presence of tentative detections. The IRAM observations have a typical rms level of 1.5\,mK at 7\,km\,s$^{-1}$ resolution, which is over a factor of 3 more sensitive than our spectra which have a characteristic rms level of  5\,mK at 7\,km\,s$^{-1}$ resolution.

\begin{figure*}
	\centering
	\includegraphics[width=\textwidth]{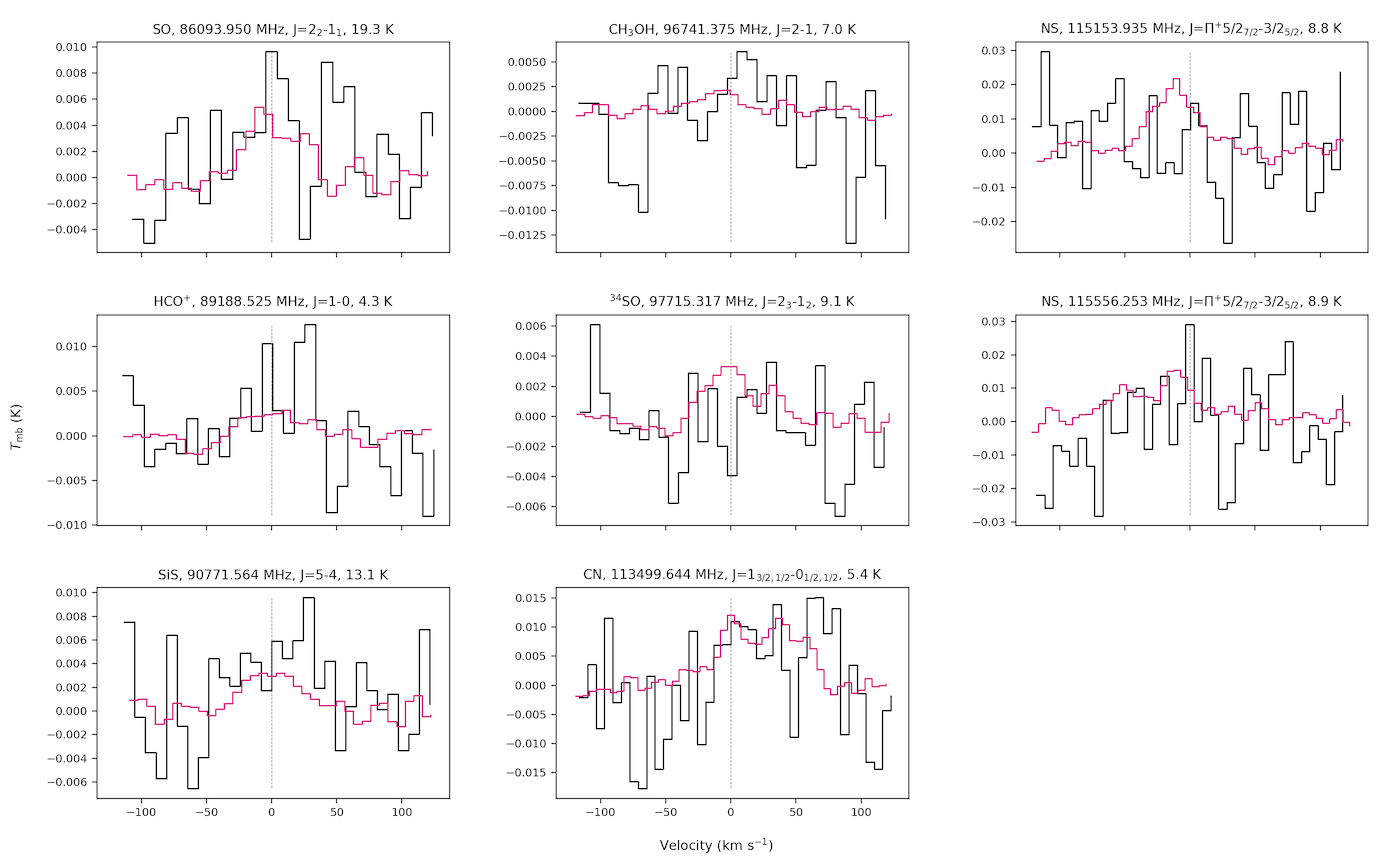} 
	\caption{Non-detections of molecular lines for IRC +10420. The corresponding detections of those molecular lines as found in \citetalias{QL2016}are overlaid in red, with beam dilution scaling applied. The OSO observations presented here are smoothed to a comparable resolution to the IRAM dataset, of $7-8$\,km\,s$^{-1}$.For the two undetected lines in \citetalias{QL2016} that exhibit hyperfine splitting, a representative line frequency has been selected for the velocity axis scaling.}
	\label{fig:irc_ql_2}
\end{figure*}

\citetalias{QL2016} detected emission from 24 transitions of 18 molecular species and isotopologues in the overlapping spectral range. In the OSO observations, 12 molecular lines were detected from 11 molecular species and isotopologues, with additional tentative detections of four lines. Lines that were only detected in the IRAM observations had line strengths in agreement with non-detection at the achieved rms noise levels in the OSO data  (Fig.~\ref{fig:irc_ql_2}). 

A visual inspection of the line profiles shows qualitatively consistent line strengths. The measured integrated intensities (Table~\ref{tab:linesummary}) are consistent within the achieved sensitivities and the 10\% calibration uncertainties of the OSO and IRAM receivers.
A particularly good agreement can be found between the IRAM and OSO measurements of the $^{29}$SiO $J = 2 - 1$, $^{28}$SiO $J = 2 - 1$, SO $N_J  = 2_{3} - 1_{2}$, and H$^{12}$CN $J = 1 - 0$ transitions. For the H$^{13}$CN $J =  1 - 0$ transition, an enhanced blue-shifted component can be seen in the IRAM data set. This may be inherent to the different observations, tracing the impact of possibly slightly different on-source pointing and/or the difference in beam sizes. It could also be a consequence of temporal variability, as evidence of temporal variability has been found in observations of thermal  emission lines of SiO, CS, and H$^{12}$CN for this source \citepalias{QL2016}.

The high level of consistency between the OSO and IRAM spectra of IRC~+10420 show that the 3\,mm receiver provides good-quality observations.
Following on from this, and the previously demonstrated consistency between the 3\,mm and 4\,mm bands in the overlapping frequency region of 85\,-\,87 GHz, we can confidently state that it has been validated that both receivers are of good quality.

\onecolumn 

\begin{table}
	\caption[]{Summary of molecular line detections for IRC~+10420. The columns list, in order: the molecular species, the transition, the energy of the upper level of the transition (in Kelvin), the rest frequency (in MHz), the measured peak temperature in Kelvin, the integrated intensity (in K\,km\,s$^{-1}$), and the FWHM of the line profile (in km\,s$^{-1}$).}
	\label{tab:linesummary}
	\centering
	\begin{tabular}{lcrrrrr} \hline
		Species & Transition & $E_{\mathrm{upp}}/k$ & Rest Freq. \footnotemark & Peak $T_{\mathrm{mb}}$ & Int. Intensity & FWHM  \\
		 & & (K)& (MHz) & (mK) & (K\,km\,s$^{-1}$) & (km\,s$^{-1}$)  \\
		\hline
		$^{30}$SiO & $J = 2 - 1$ & 6.1 & 84746.166 & 38 & 1.54 & 69.3   \\
		$^{29}$SiO & $J = 2 - 1$  & 6.2 & 85759.194 & 60 & 3.17 & 85.6  \\
		H$^{13}$CN & $J = 1 - 0$ & 4.1 & 86339.921 & 28 & 1.17 & 76.5 \\
		$^{28}$SiO & $J = 2 - 1$ & 6.3  & 86846.985 & 334 & 17.03 & 118.3  \\
		H$^{12}$CN & $J = 1 - 0$ & 4.3 & 88631.602 & 135 & 7.74 & 99.4  \\
		HNC & $J  = 1 - 0$ & 4.4 & 90663.568 & 23 & 0.78 & 72.9  \\
		CS & $J = 2 - 1$ & 7.1 &  97980.953 & 12 & 0.40 & 74.9  \\
		SO & $N_J  = 2_{3} - 1_{2}$ & 9.2 & 99299.870 & 54  & 2.08 & 73.9  \\
		SO$_{2}$ & $J_{K_{\mathrm{a}},K_{\mathrm{c}}} = 3_{1,3} - 2_{0,2}$ & 7.7 & 104029.418 & 24 & 0.72 & 63.5   \\
		SO & $N_J = 3_{2} - 2_{1}$ & 21.1 & 109252.220 & 18 & 0.65 & 73.9  \\
		$^{13}$CO &	$J = 1 - 0$ & 5.3 & 110201.354 & 39 & 1.46 & 79.9 \\
		$^{12}$CO & $J  = 1 - 0$ & 5.5  & 115271.202 & 243 & 13.17 & 108.3 \\
		\hline
	\end{tabular}
	\footnotetext{$^{5}$Transition rest frequencies are taken from the CDMS database \citep{Muller2001,Muller2005}, with the exception of those for H$_2$O and NH$_{3}$, which are taken from the JPL database \citep{JPL}.}
\end{table}

\newpage
\twocolumn
\section{Multi-component fitting results}
\onecolumn
\begin{figure}
	\begin{subfigure}[c]{0.32\textwidth}
		\centering
		\includegraphics[height=1.85in]{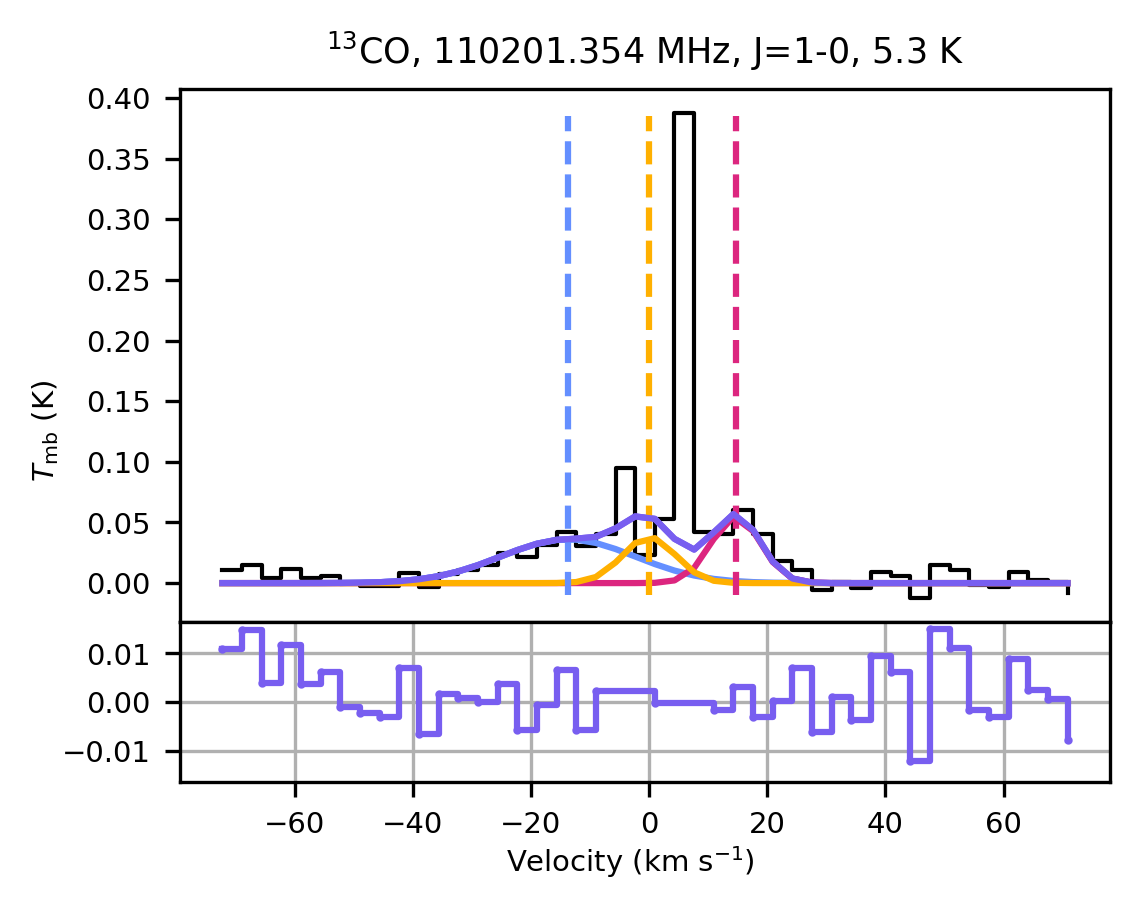}
	\end{subfigure}\hfill
	\begin{subfigure}[c]{0.32\textwidth}
		\centering
		\includegraphics[height=1.85in]{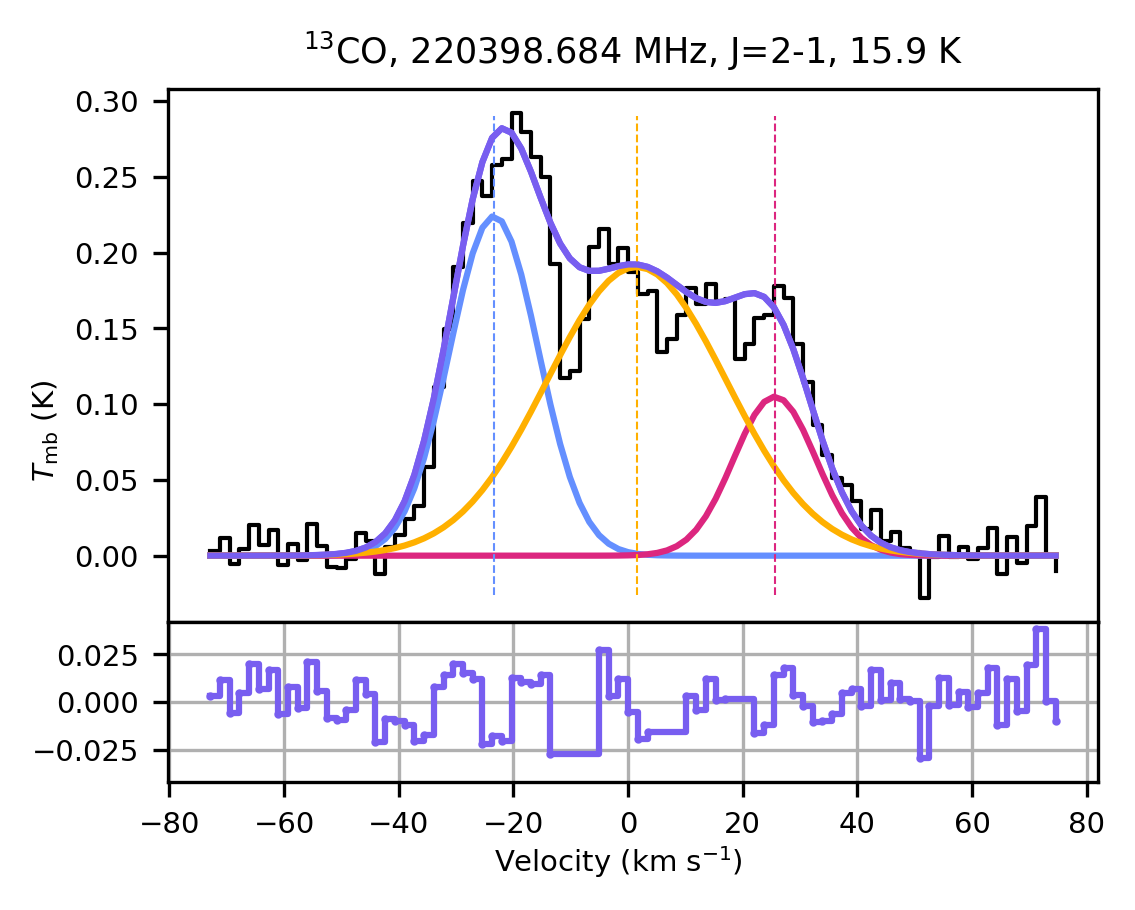}
	\end{subfigure}\hfill
	\begin{subfigure}[c]{0.32\textwidth}
		\centering
		\includegraphics[height=1.85in]{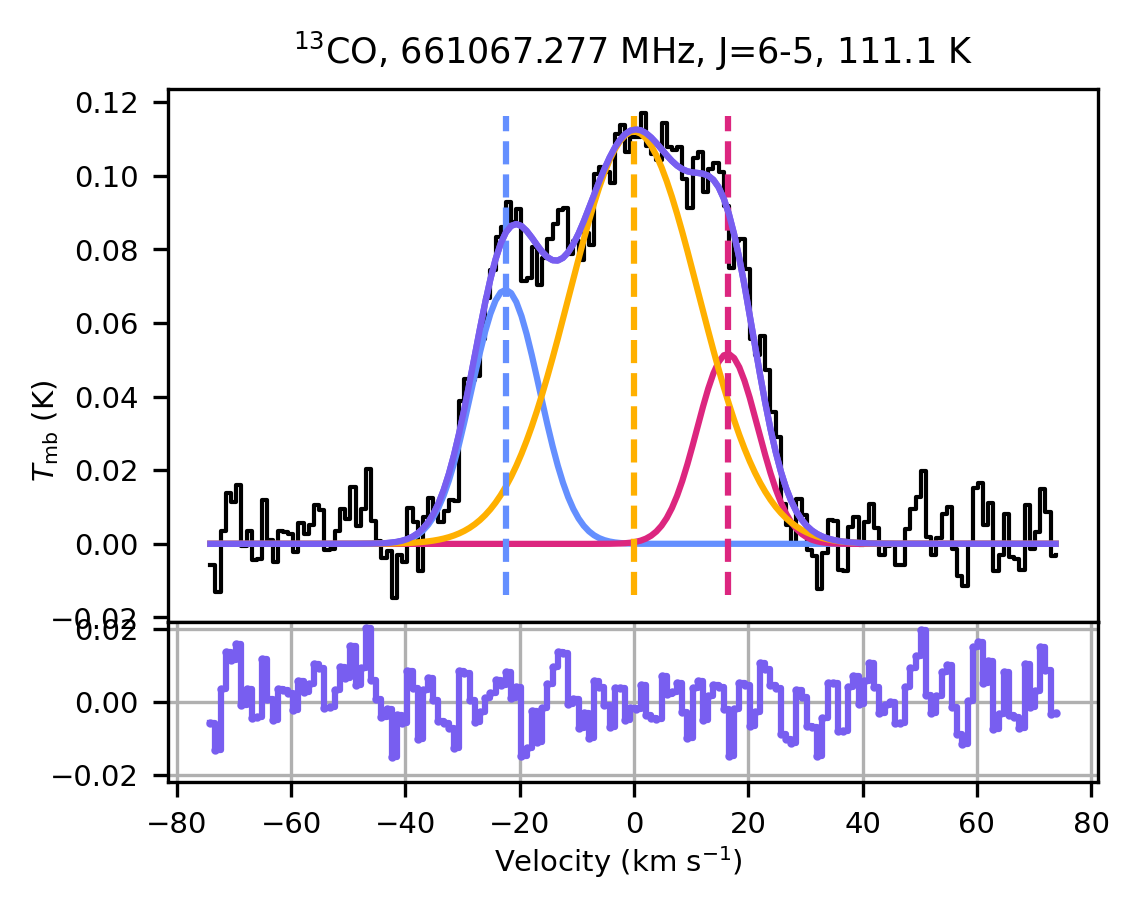}
	\end{subfigure}\hfill
	\begin{subfigure}[c]{0.32\textwidth}
		\includegraphics[height=1.85in]{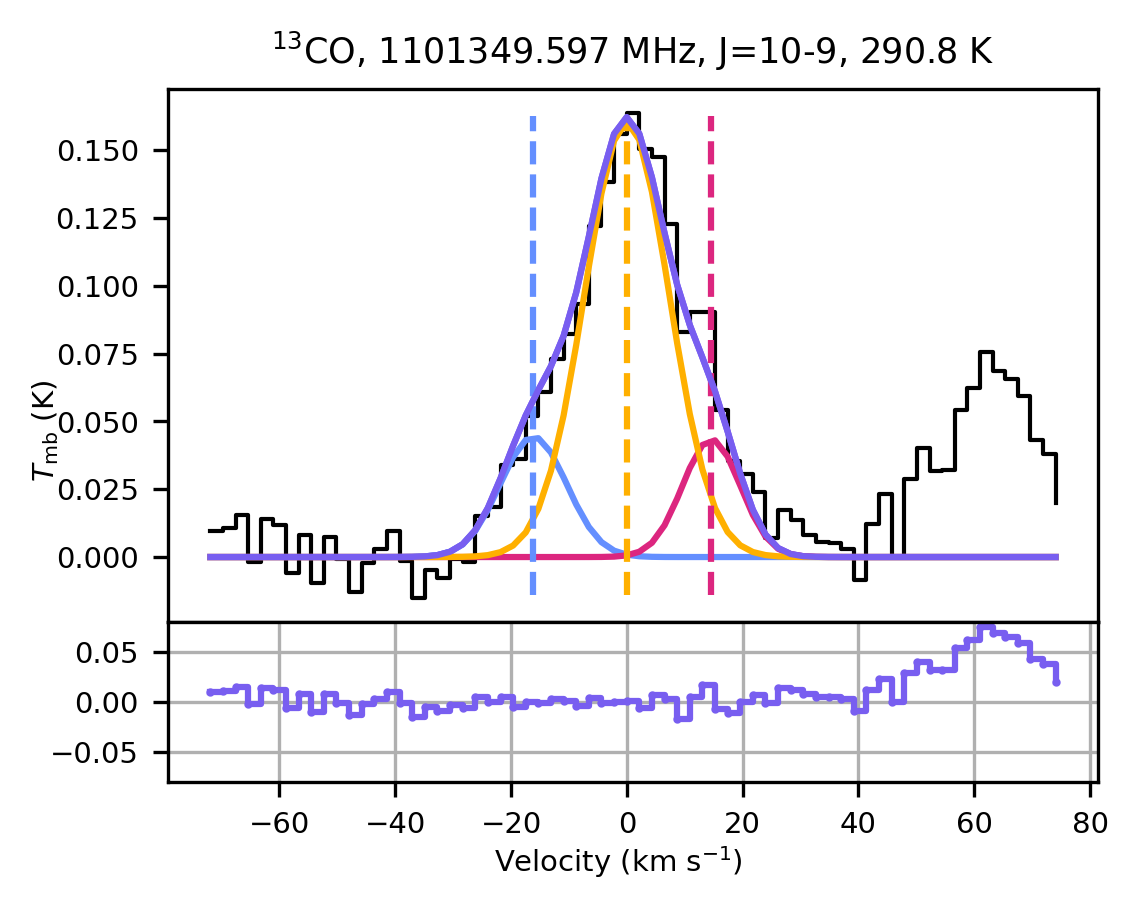}
	\end{subfigure}\hfill	
	\begin{subfigure}[c]{0.32\textwidth}
		\includegraphics[height=1.85in]{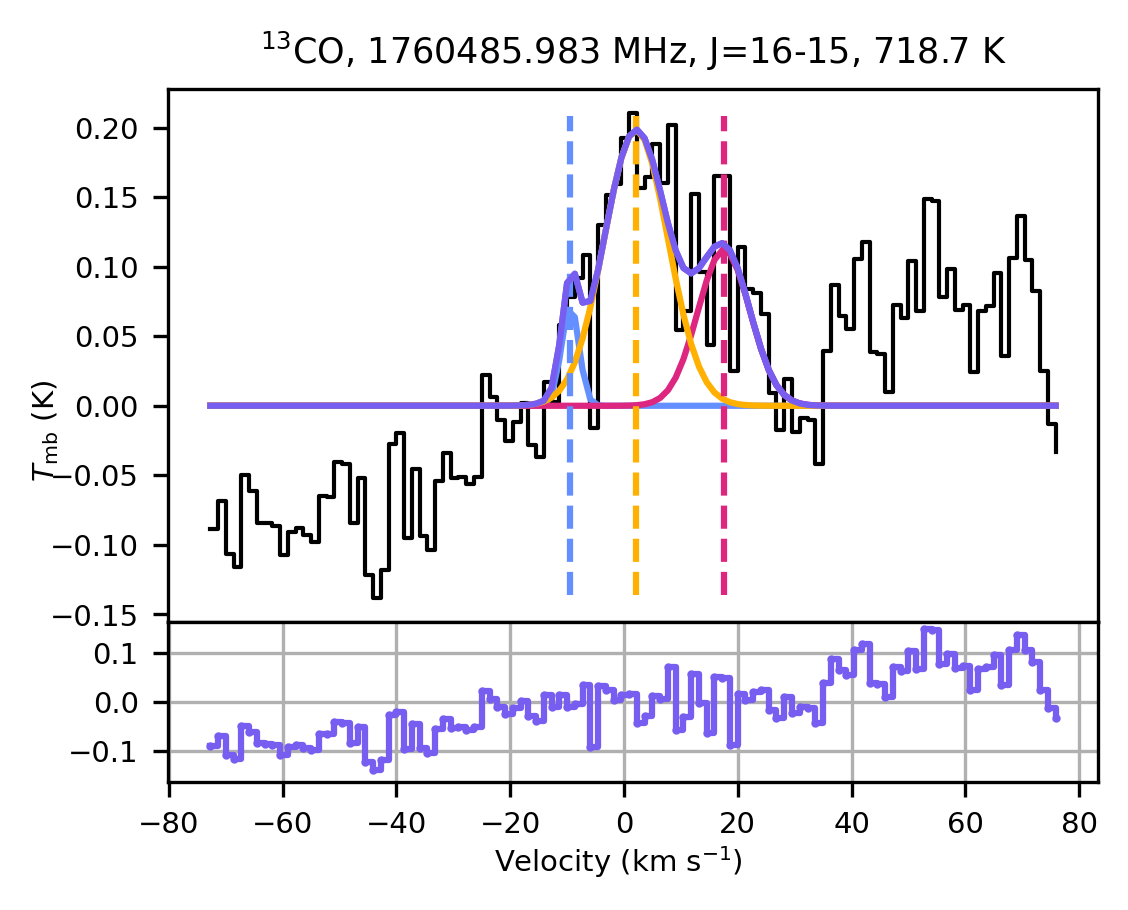}
	\end{subfigure}\hfill
	\begin{minipage}[c]{0.32\textwidth}
	\hspace{\fill}
	\end{minipage}\hfill
	\caption{Multi-component fits as described in Figure \ref{fig:co_nml_multifits}, for $^{13}$CO emission around NML~Cyg, observed with OSO, JCMT and HIFI.}
	\label{fig:13co_lines}
\end{figure}

\begin{figure*}
\begin{subfigure}[c]{0.32\textwidth}
\centering
\includegraphics[height=1.85in]{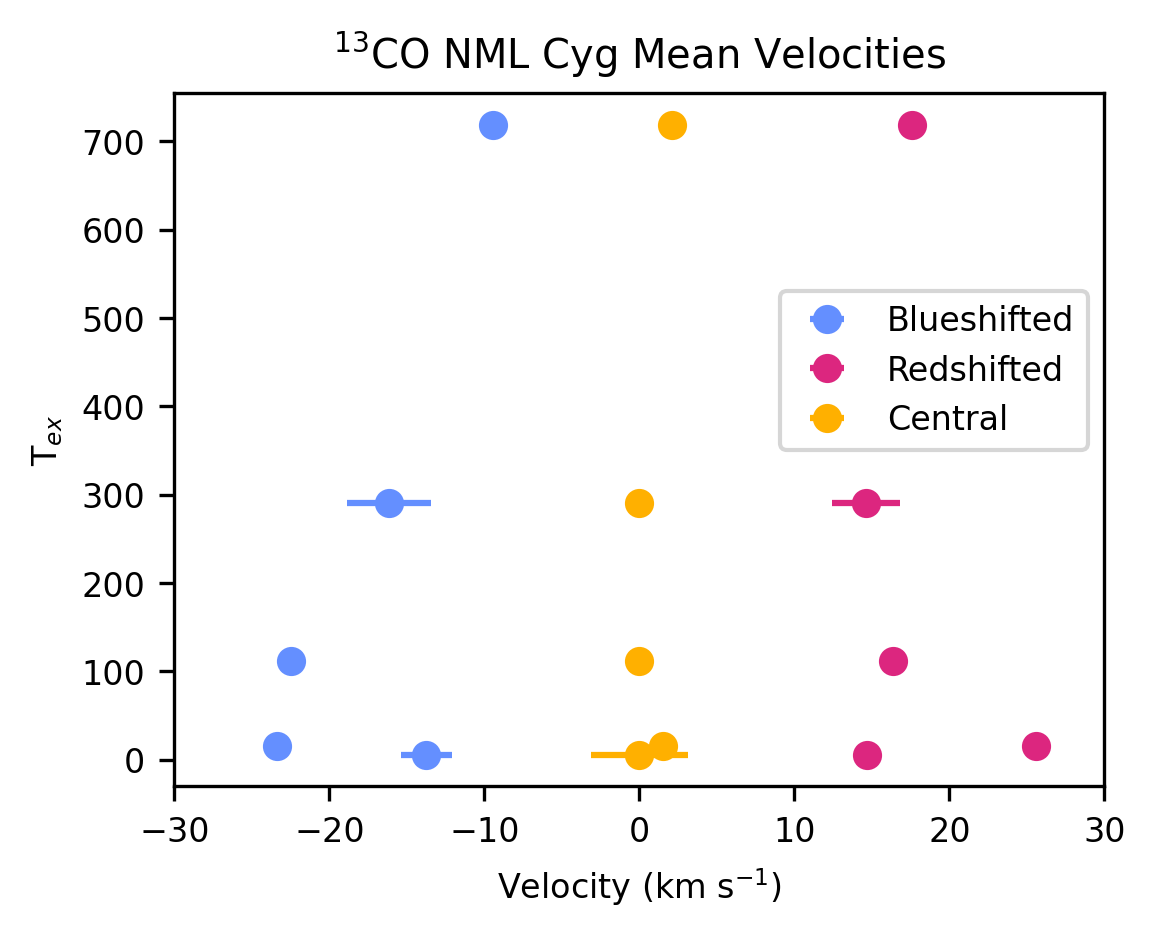}
\end{subfigure}\hfill
	\begin{subfigure}[c]{0.32\textwidth}
	\centering
	\includegraphics[height=1.85in]{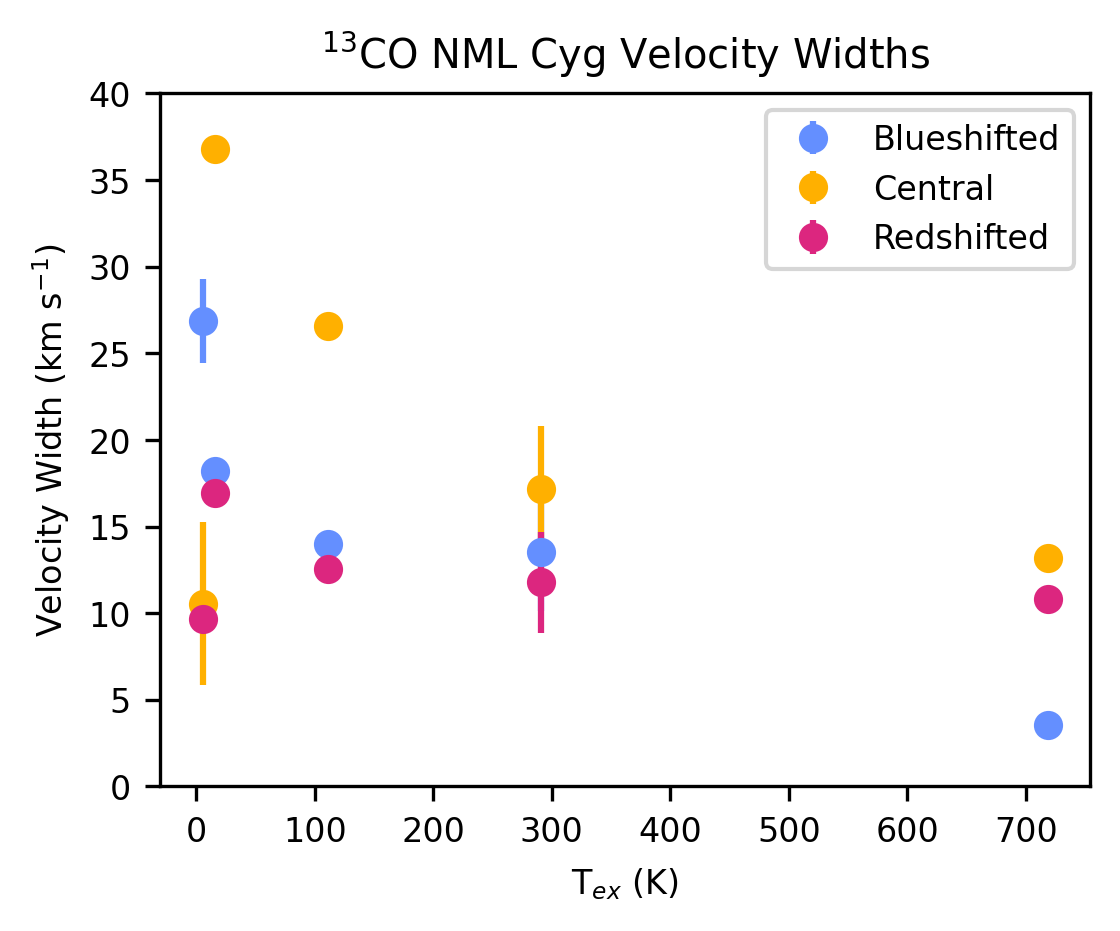}
\end{subfigure}\hfill
\begin{subfigure}[c]{0.32\textwidth}
\centering
\includegraphics[height=1.85in]{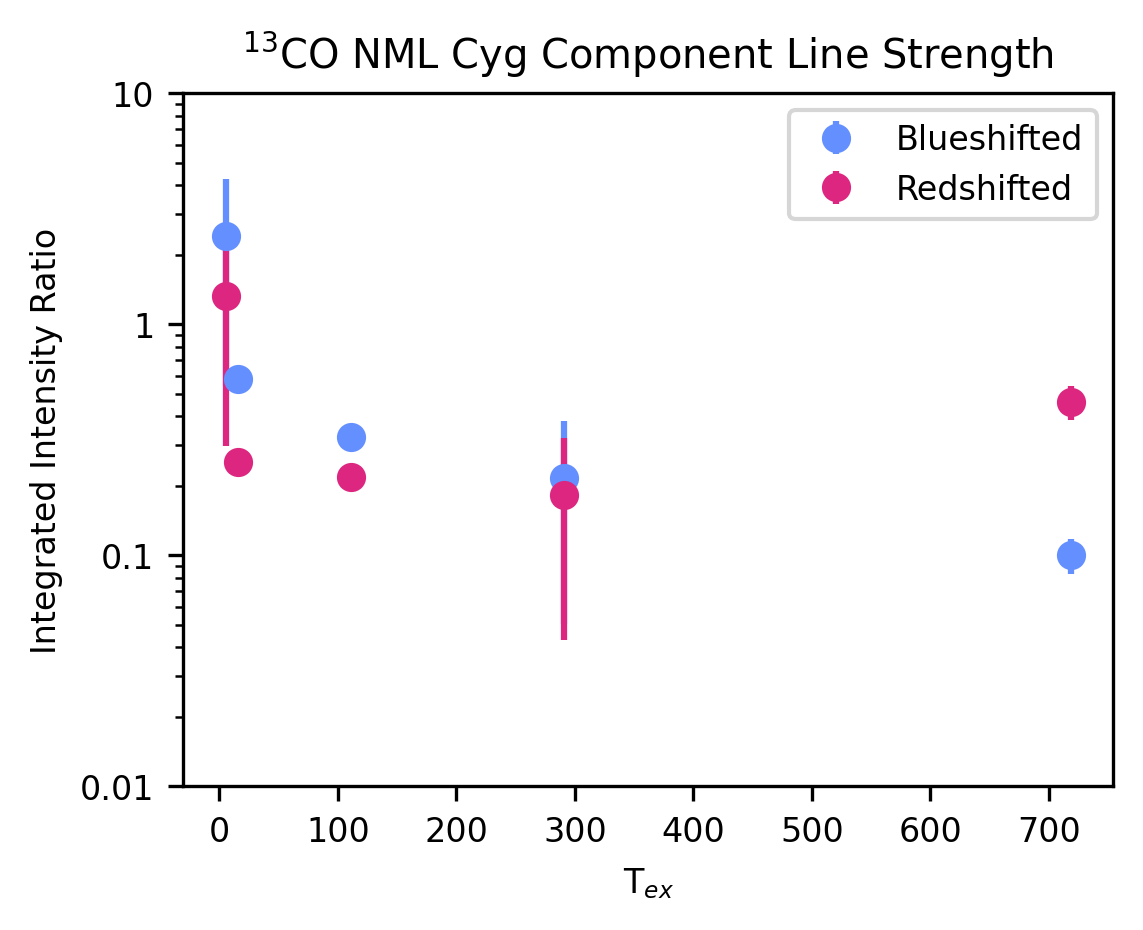}
\end{subfigure}\hfill
\caption{Same as Fig.~\ref{fig:co_fits}, for detected $^{13}$CO emission lines around NML~Cyg.}
    \label{fig:13co_fits}
\end{figure*}

\begin{figure*}
	\begin{subfigure}[c]{0.32\textwidth}
		\centering
		\includegraphics[height=1.85in]{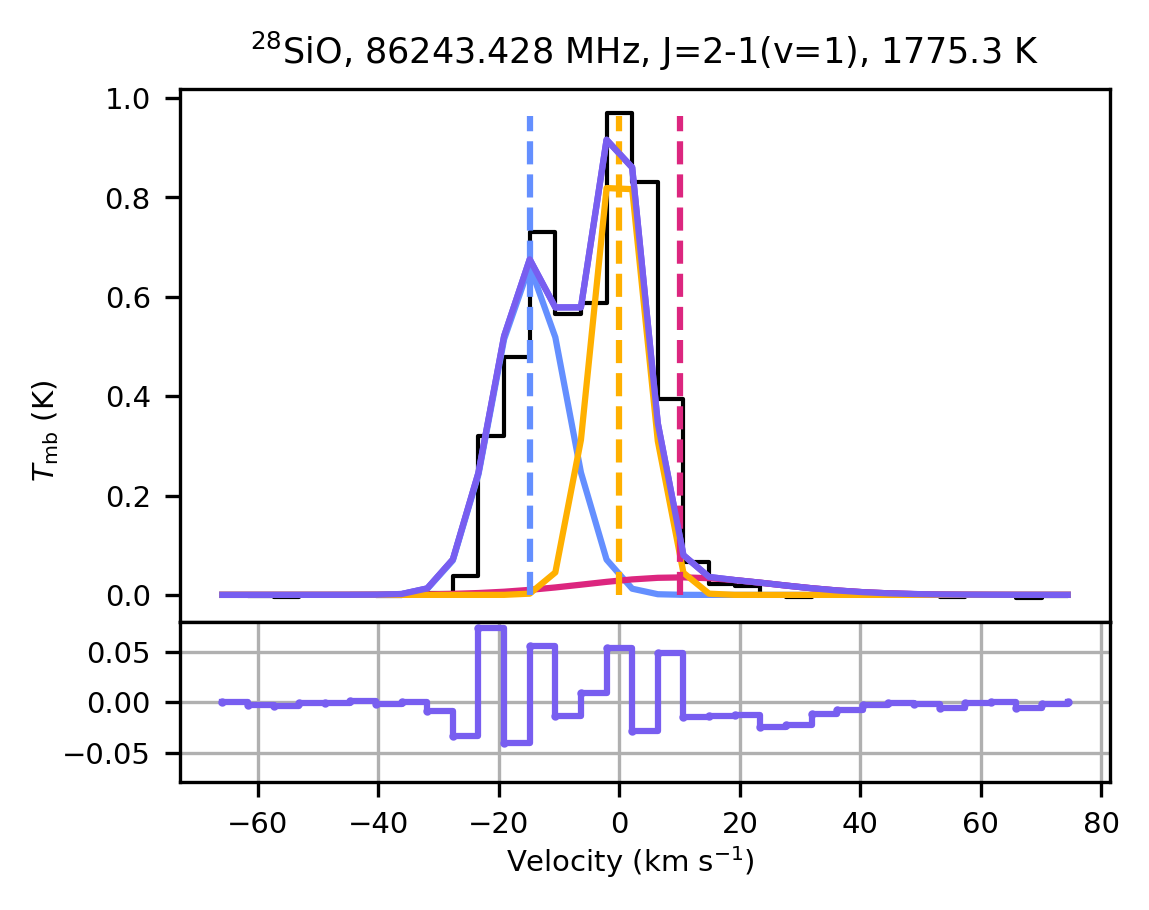}
	\end{subfigure}\hfill
	\begin{subfigure}[c]{0.32\textwidth}
		\centering
		\includegraphics[height=1.85in]{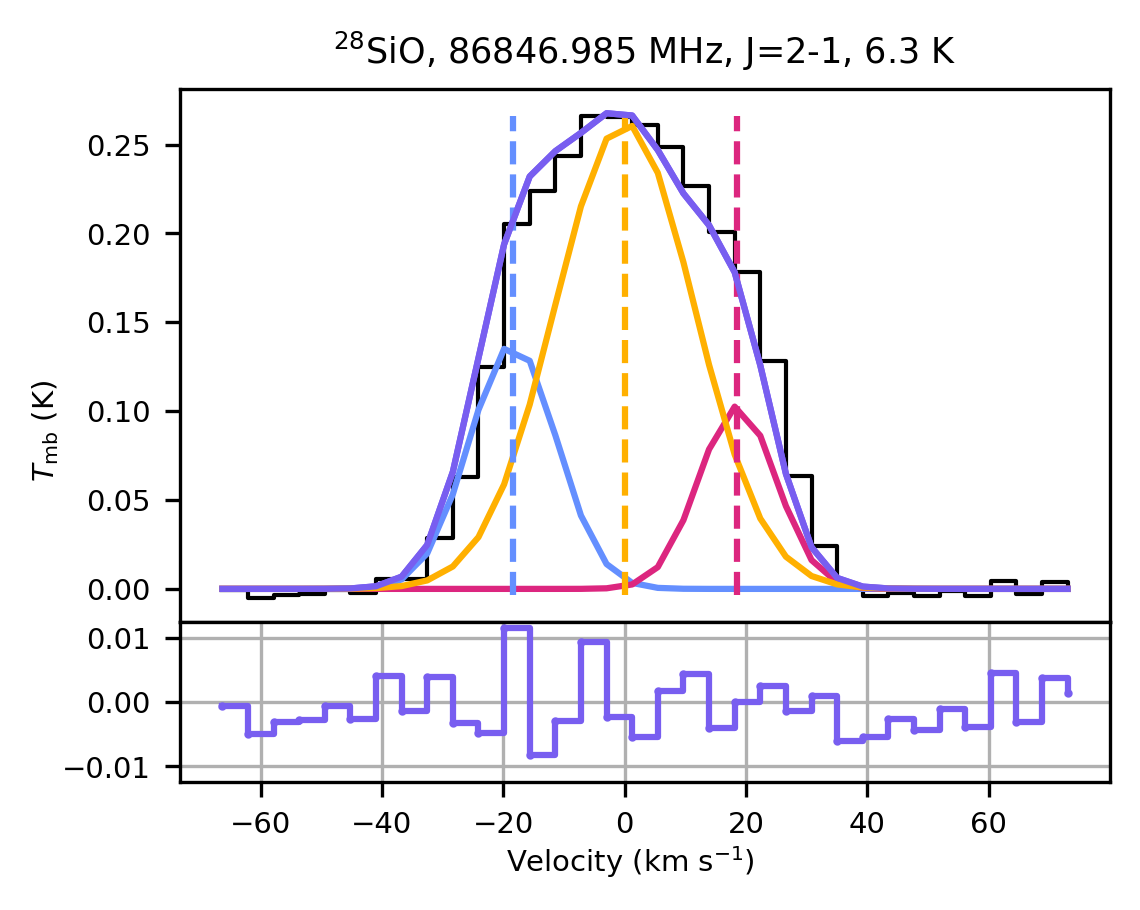}
	\end{subfigure}\hfill
	\begin{subfigure}[c]{0.32\textwidth}
		\centering
		\includegraphics[height=1.85in]{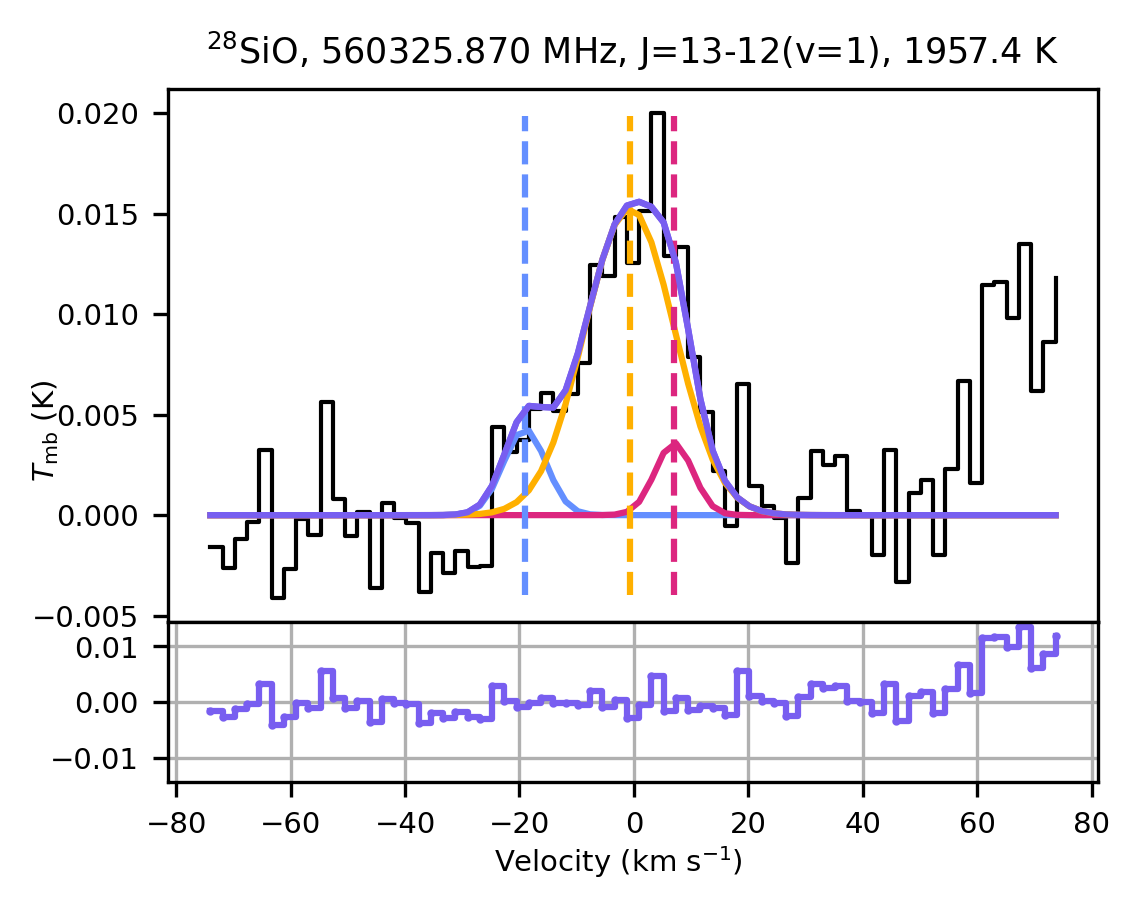}
	\end{subfigure}\hfill
	\begin{subfigure}[c]{0.32\textwidth}
		\centering
		\includegraphics[height=1.85in]{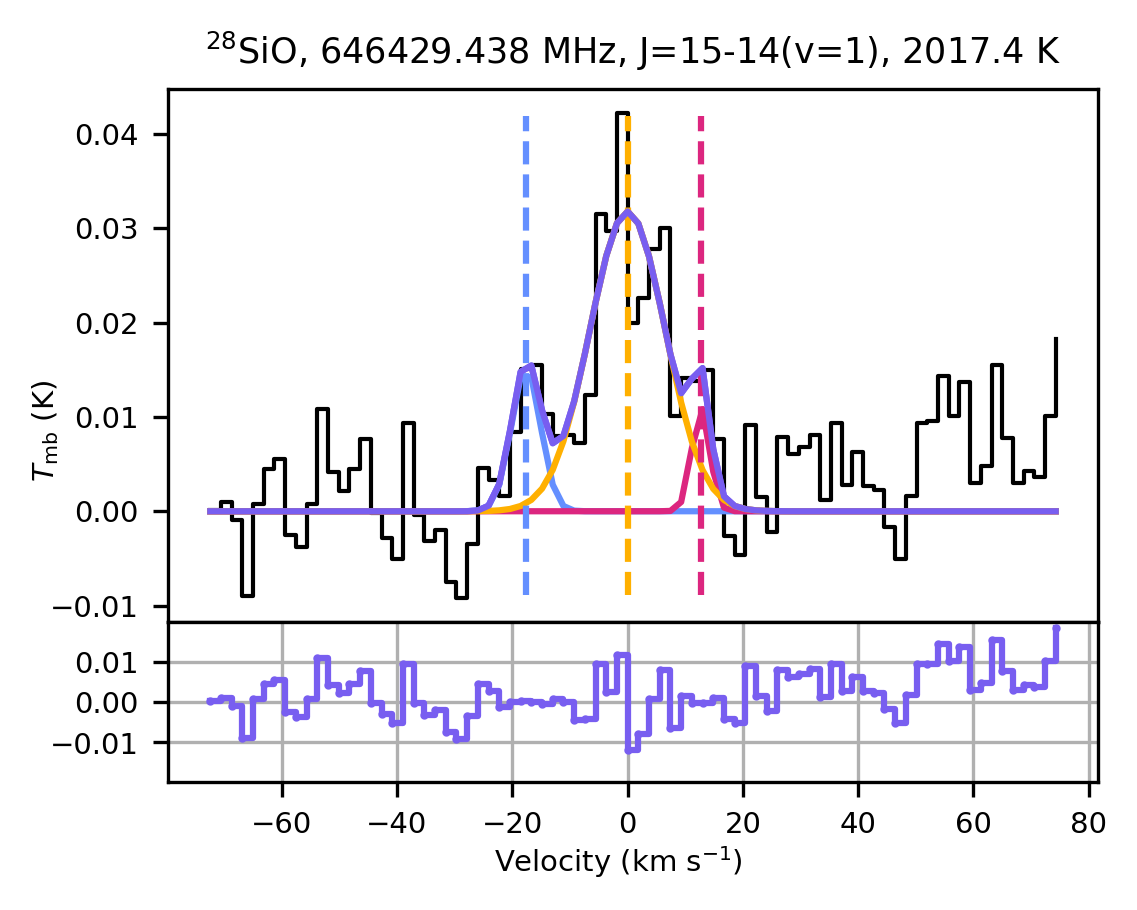}
	\end{subfigure}\hfill
	\begin{subfigure}[c]{0.32\textwidth}
		\centering
		\includegraphics[height=1.85in]{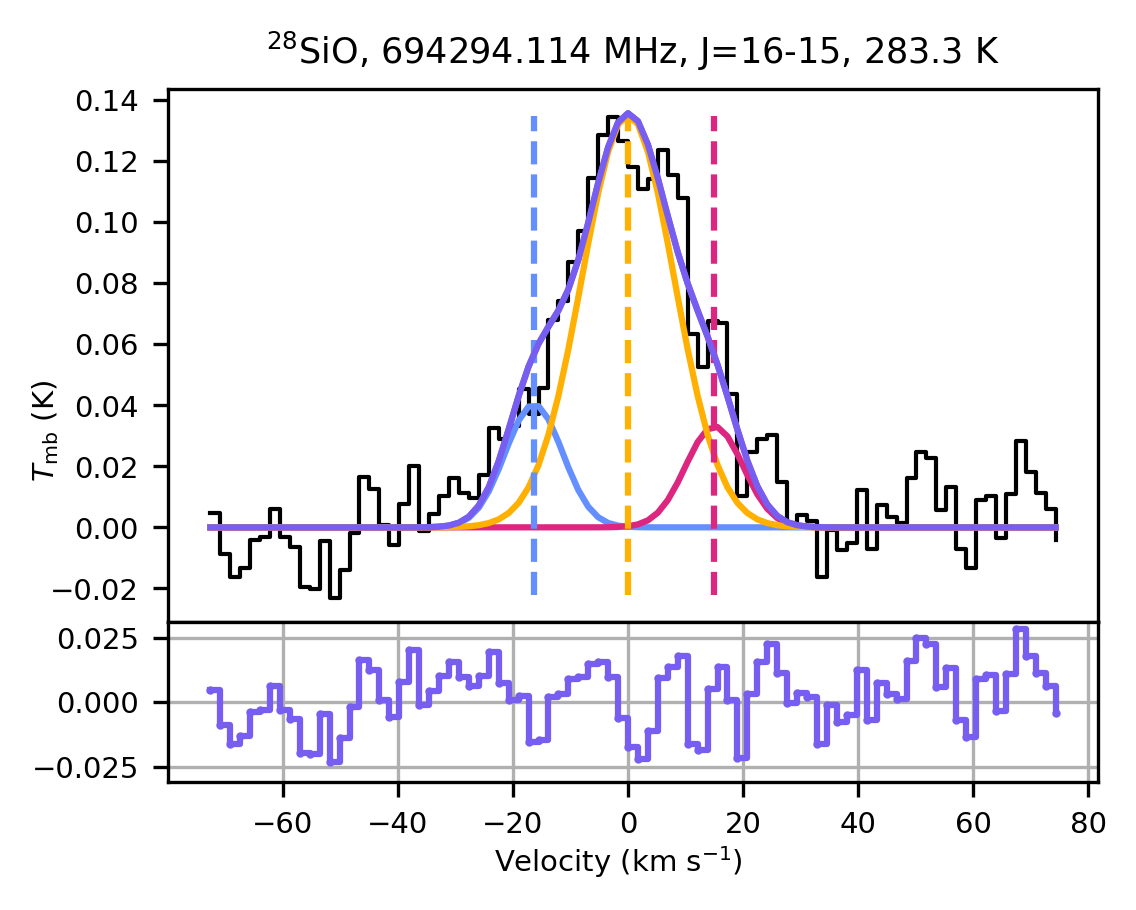}
	\end{subfigure}\hfill
	\begin{minipage}[c]{0.32\textwidth}
	\hspace{\fill}
	\end{minipage}\hfill
	\caption{Same as Fig.~\ref{fig:co_nml_multifits}, for ${^28}$SiO emission around NML~Cyg, observed with OSO and HIFI.}
	\label{fig:sio_lines}
\end{figure*}

\begin{figure*}
\begin{subfigure}[c]{0.32\textwidth}
\centering
\includegraphics[height=1.85in]{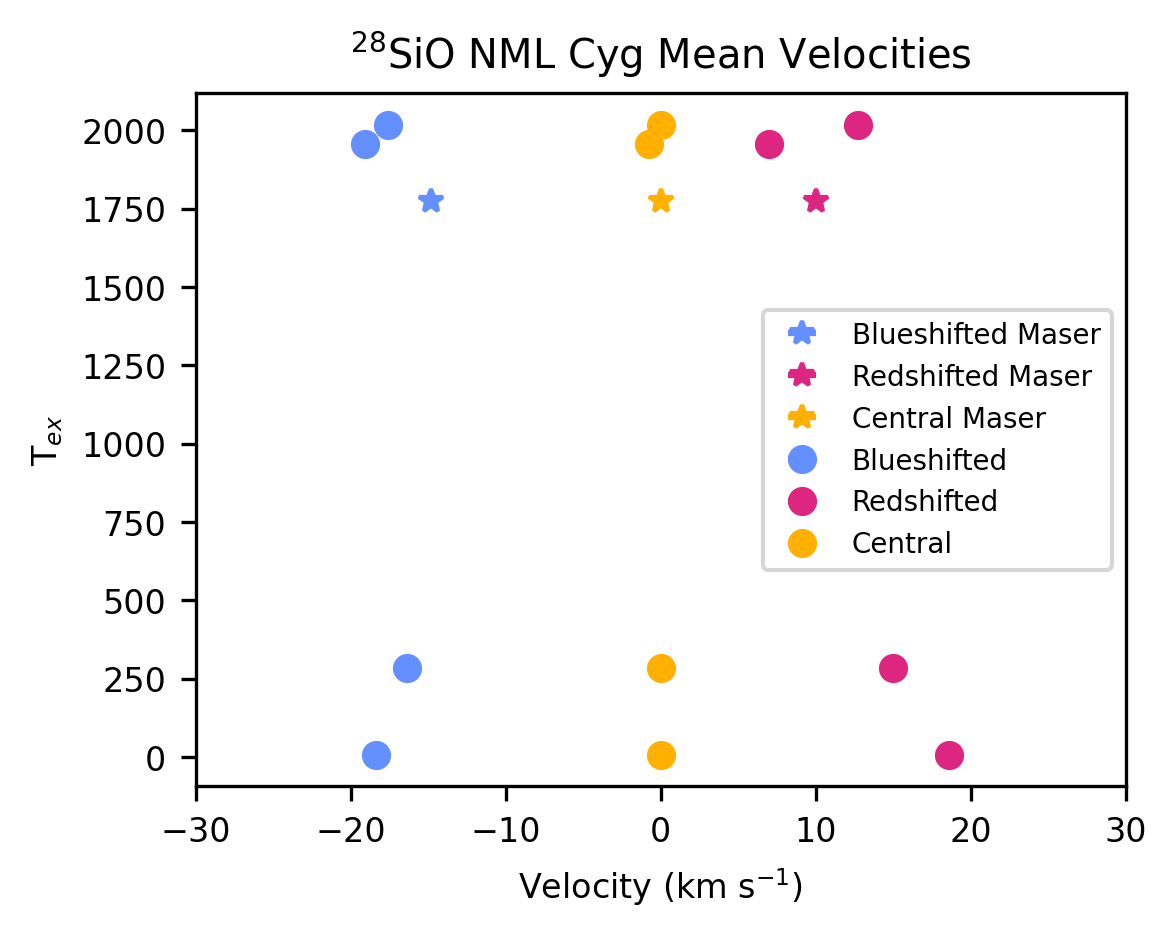}
\end{subfigure}\hfill
	\begin{subfigure}[c]{0.32\textwidth}
	\centering
	\includegraphics[height=1.85in]{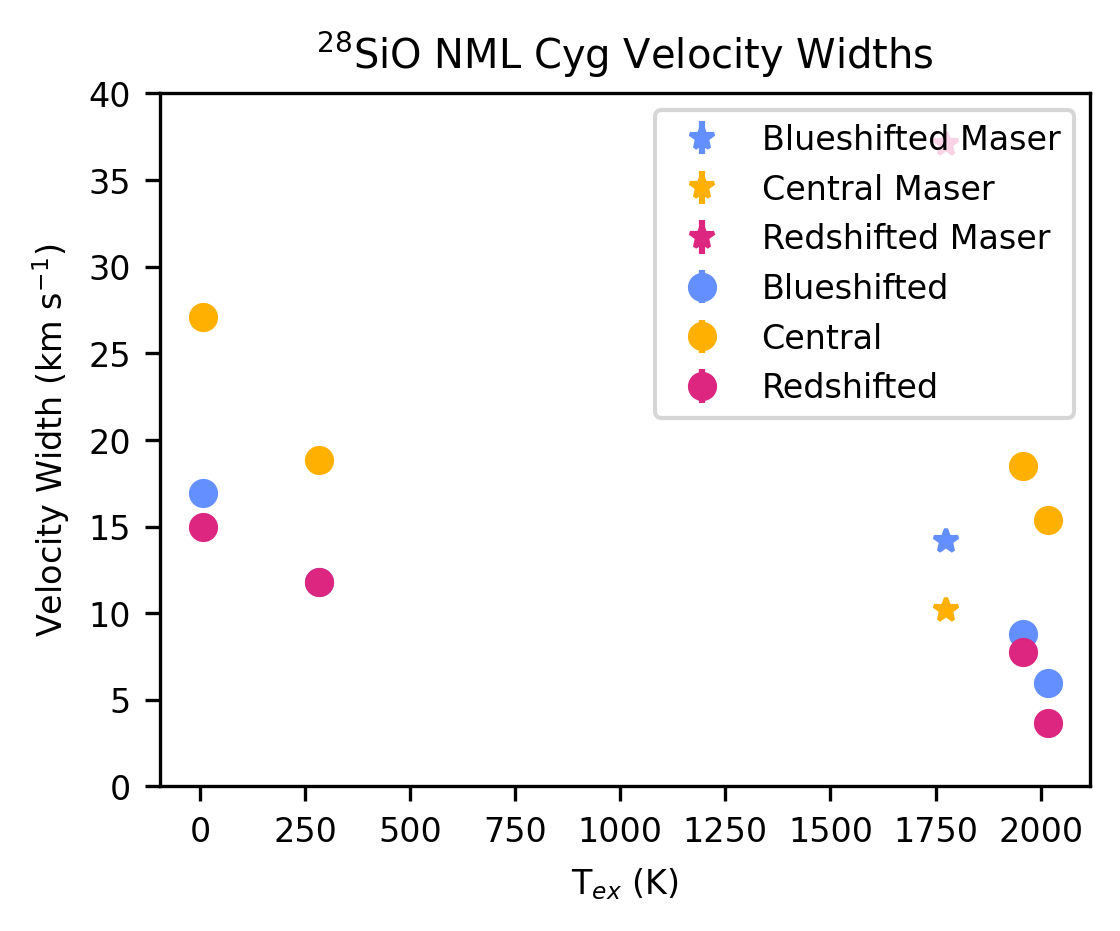}
\end{subfigure}\hfill
\begin{subfigure}[c]{0.32\textwidth}
\centering
\includegraphics[height=1.85in]{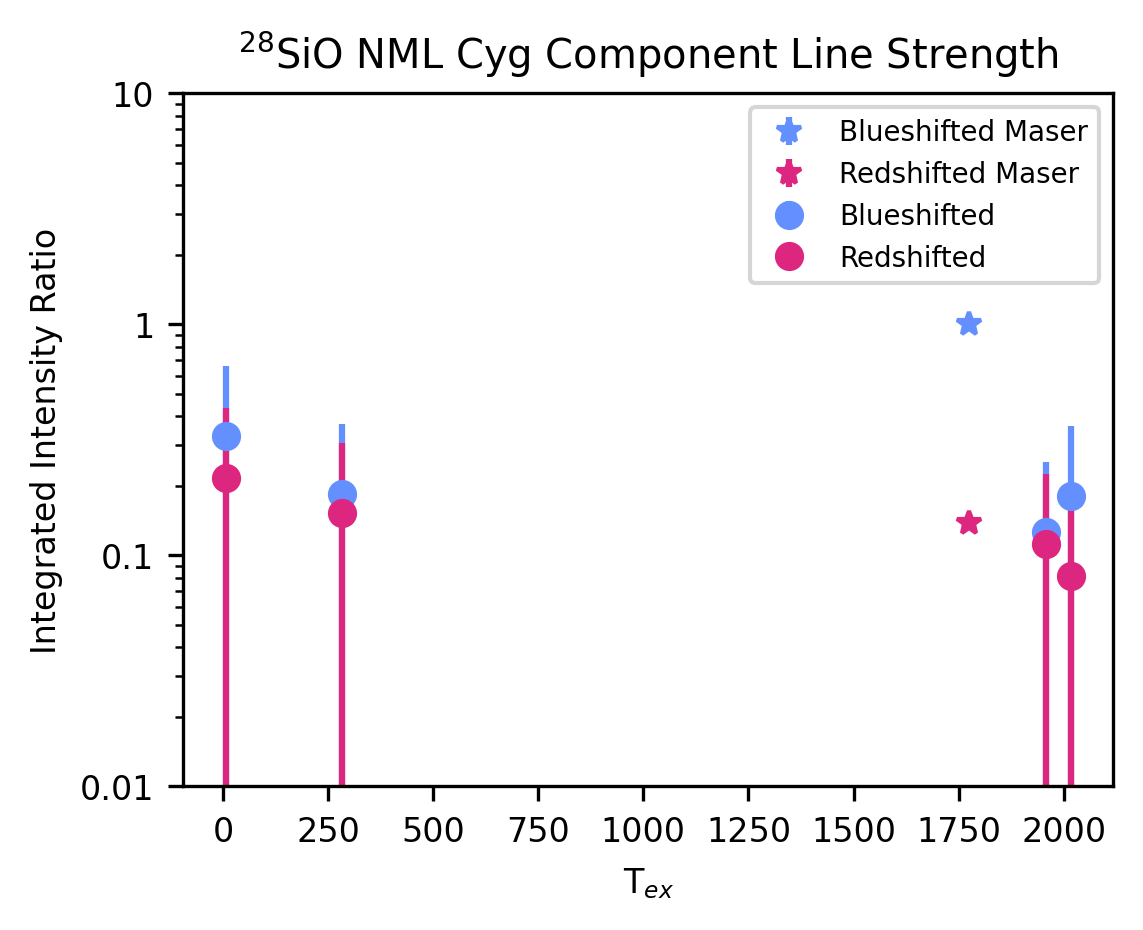}
\end{subfigure}\hfill
\caption{Same as Fig.~\ref{fig:co_fits}, for detected $^{28}$SiO emission lines around NML~Cyg.}
\label{fig:sio_fits}
\end{figure*}

\begin{figure*}
	\begin{subfigure}[c]{0.32\textwidth}
		\centering
		\includegraphics[height=1.85in]{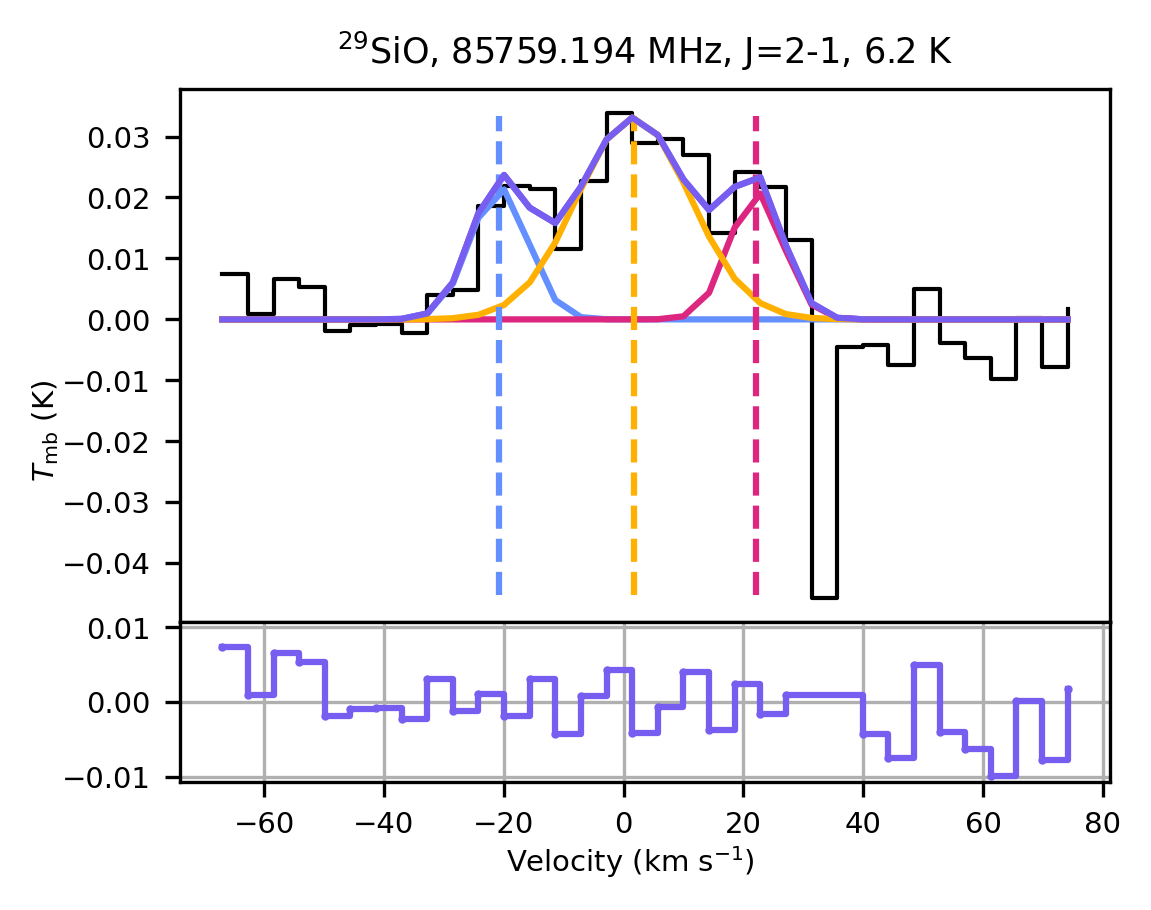}
	\end{subfigure}\hfill
	\begin{subfigure}[c]{0.32\textwidth}
		\centering
		\includegraphics[height=1.85in]{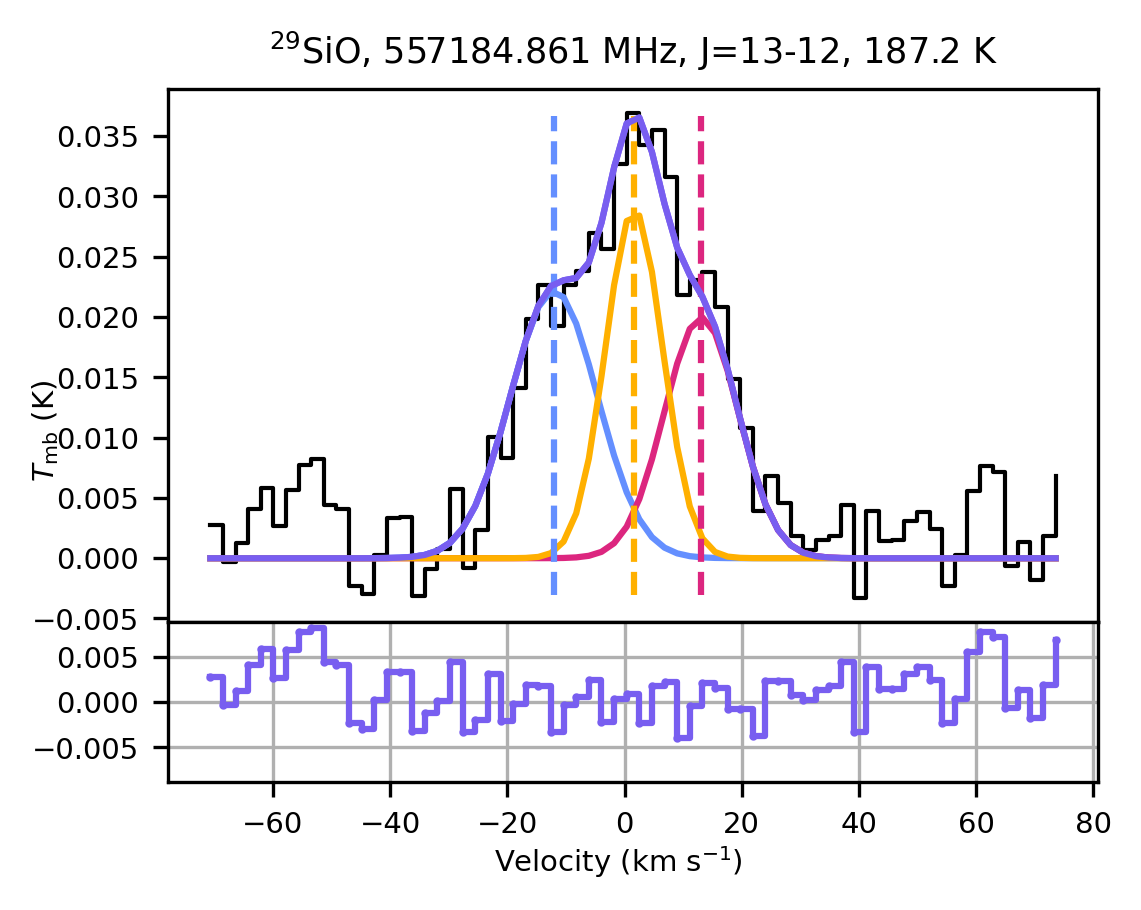}
	\end{subfigure}\hfill
	\begin{subfigure}[c]{0.32\textwidth}
		\centering
		\includegraphics[height=1.85in]{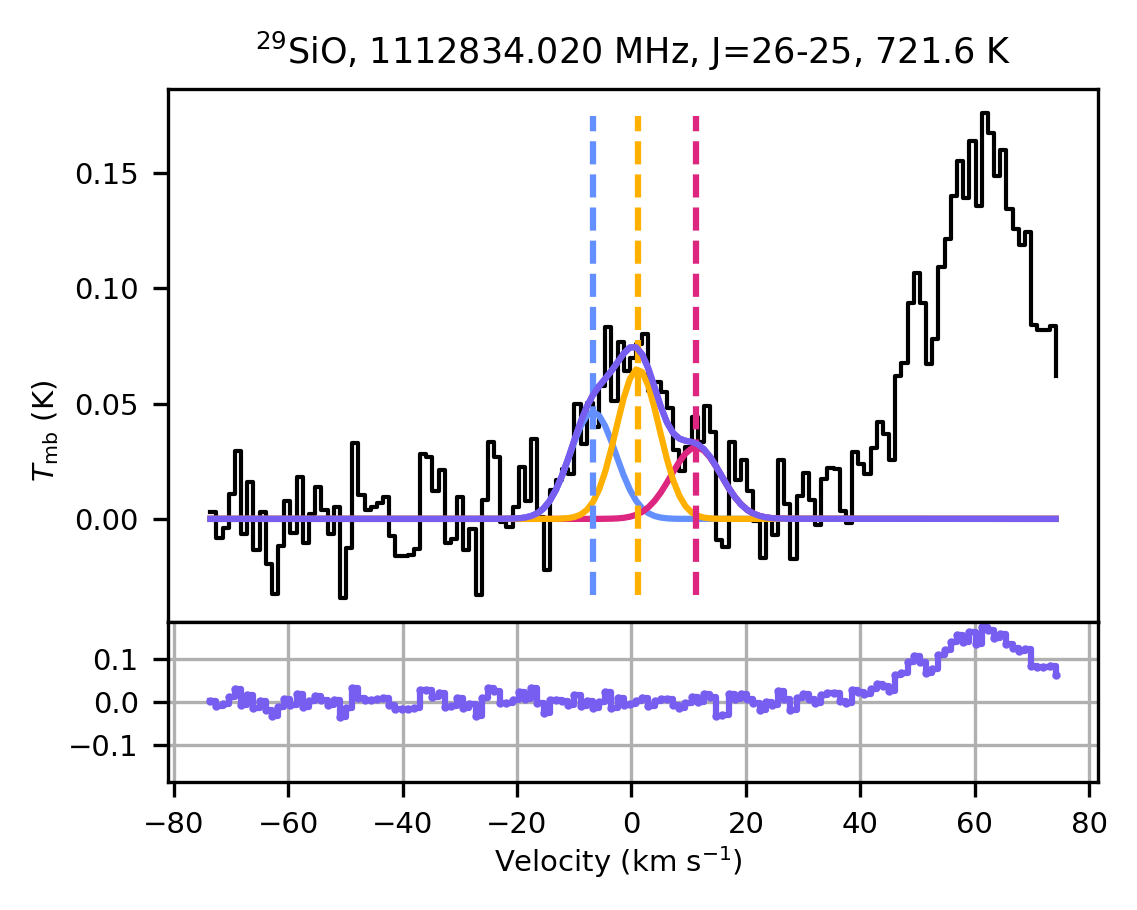}
	\end{subfigure}\hfill
	\caption{Same as Fig.~\ref{fig:co_nml_multifits}, for $^{29}$SiO emission around NML~Cyg, observed with OSO and HIFI.}
	\label{fig:29sio_lines}
\end{figure*}

\begin{figure*}
\begin{subfigure}[c]{0.32\textwidth}
\centering
\includegraphics[height=1.85in]{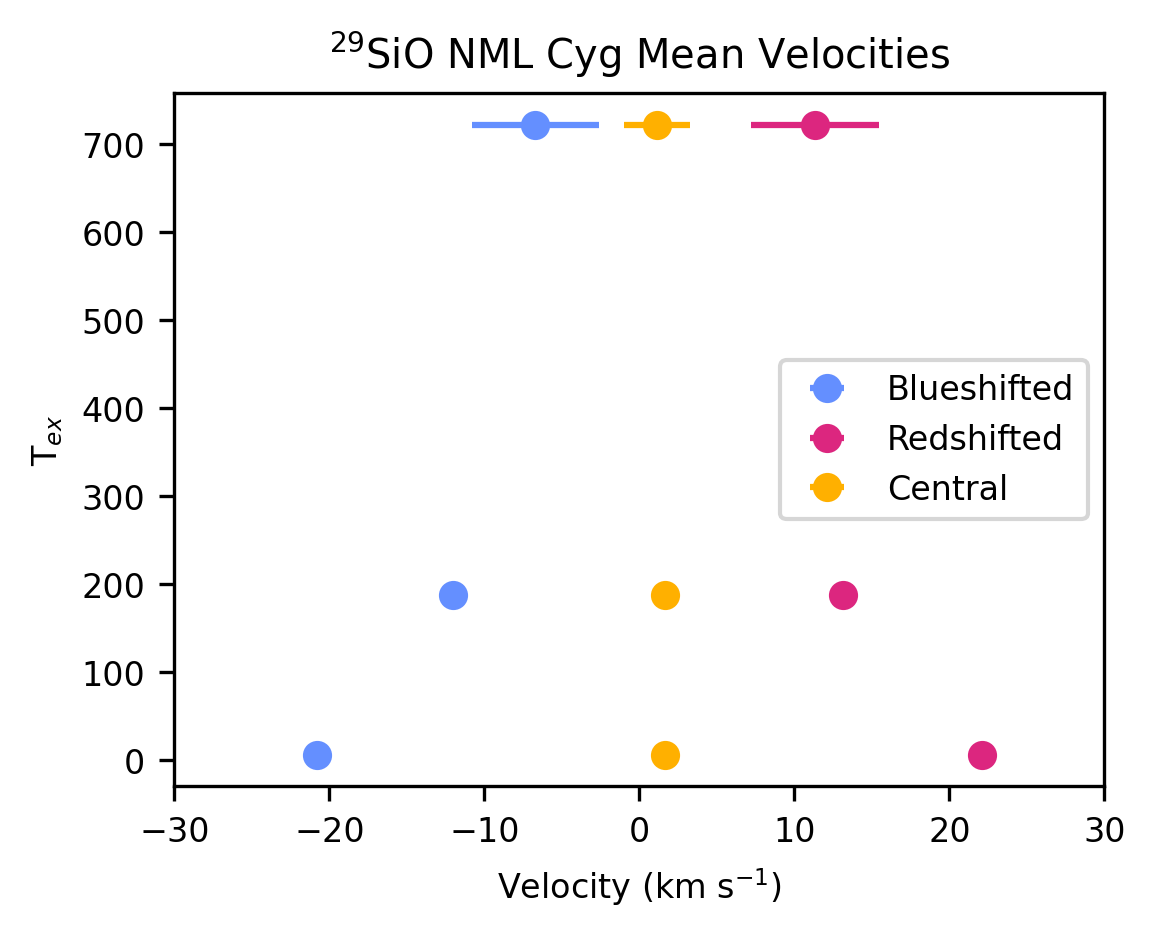}
\end{subfigure}\hfill
	\begin{subfigure}[c]{0.32\textwidth}
	\centering
	\includegraphics[height=1.85in]{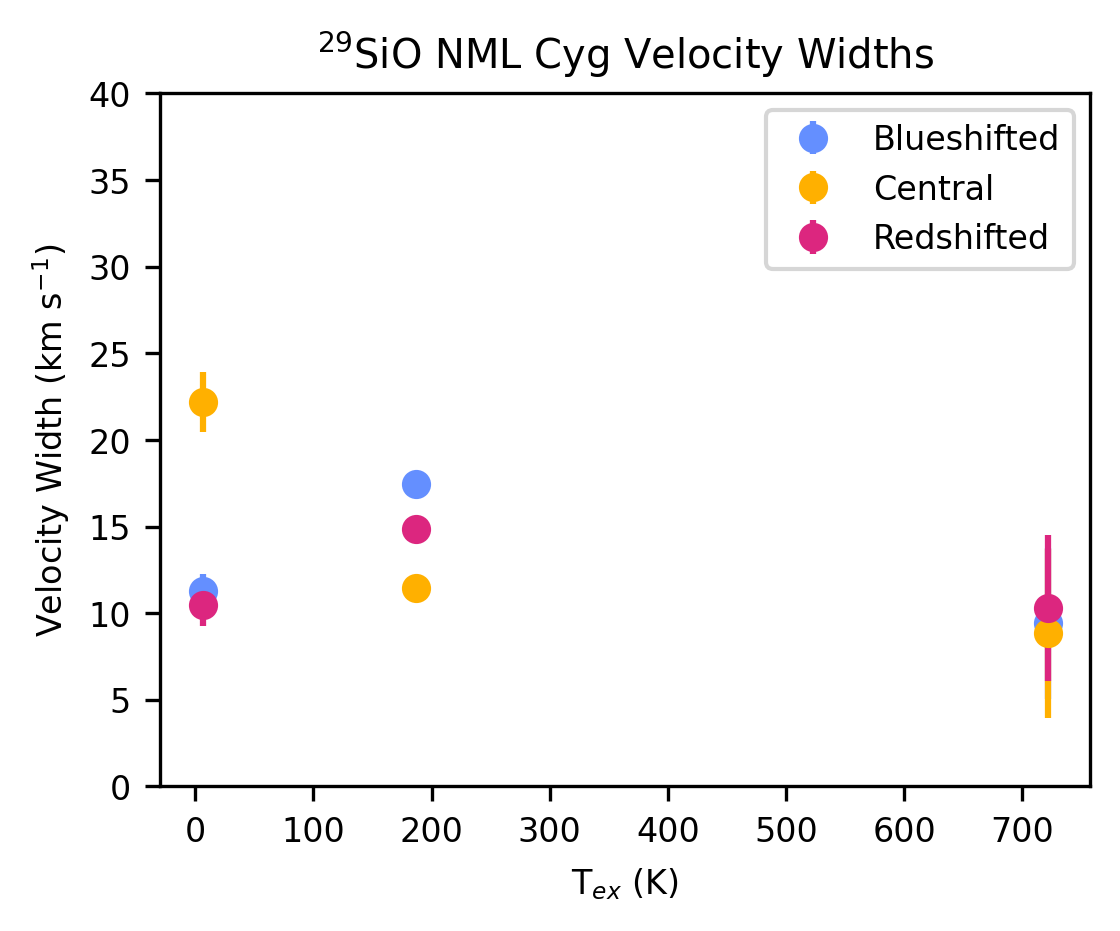}
\end{subfigure}\hfill
\begin{subfigure}[c]{0.32\textwidth}
\centering
\includegraphics[height=1.85in]{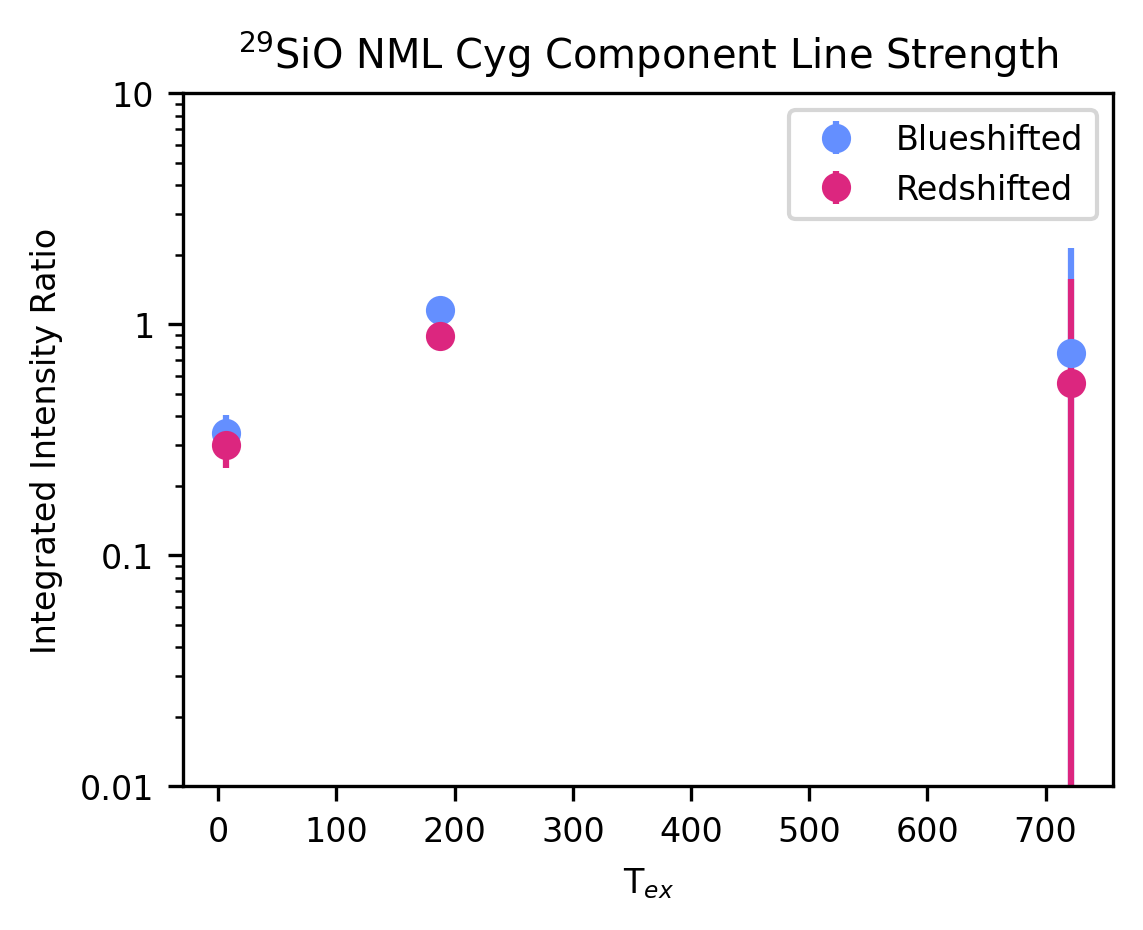}
\end{subfigure}\hfill
\caption{Same as Fig.~\ref{fig:co_fits}, for detected $^{29}$SiO emission lines around NML~Cyg.}
\label{fig:29sio_fits}
\end{figure*}

\begin{figure*}
\centering
	\begin{subfigure}[c]{0.32\textwidth}
		\centering
		\includegraphics[height=1.85in]{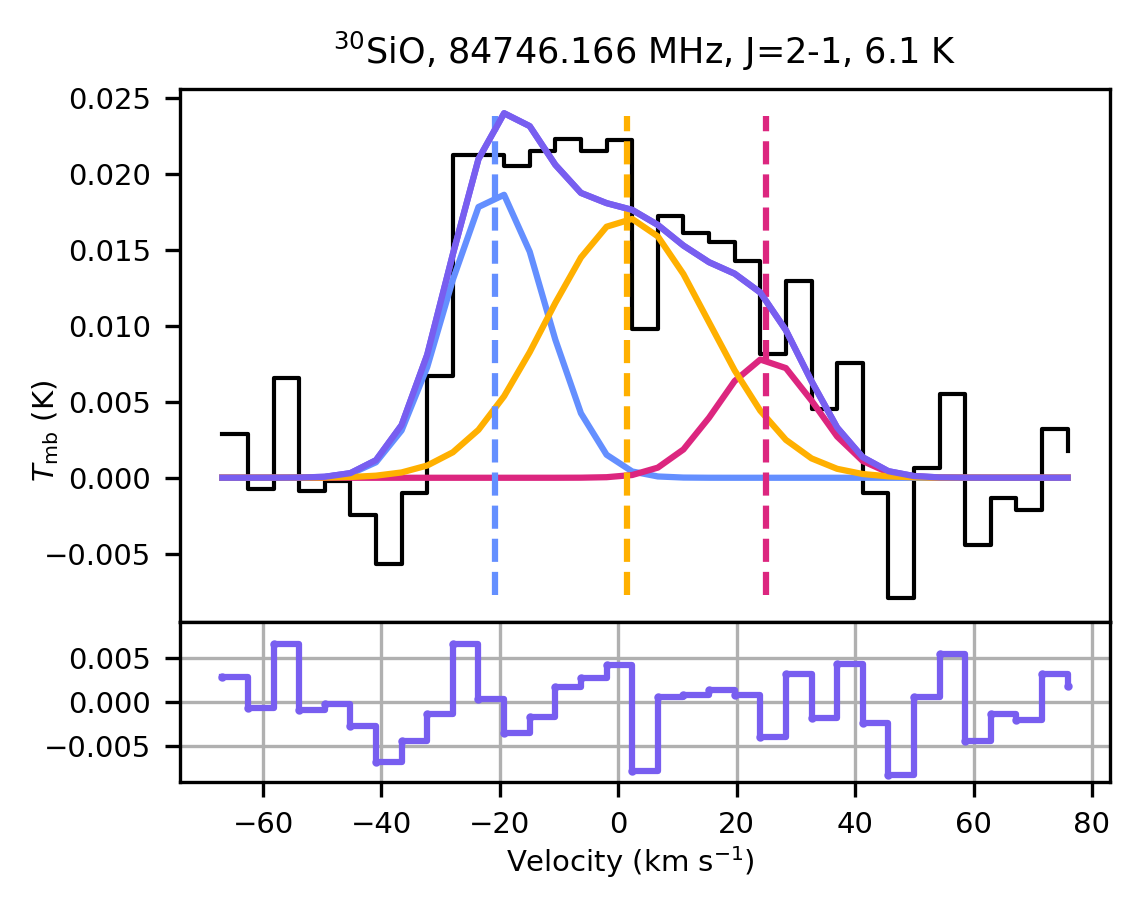}
	\end{subfigure}
	\begin{subfigure}[c]{0.32\textwidth}
		\centering
		\includegraphics[height=1.85in]{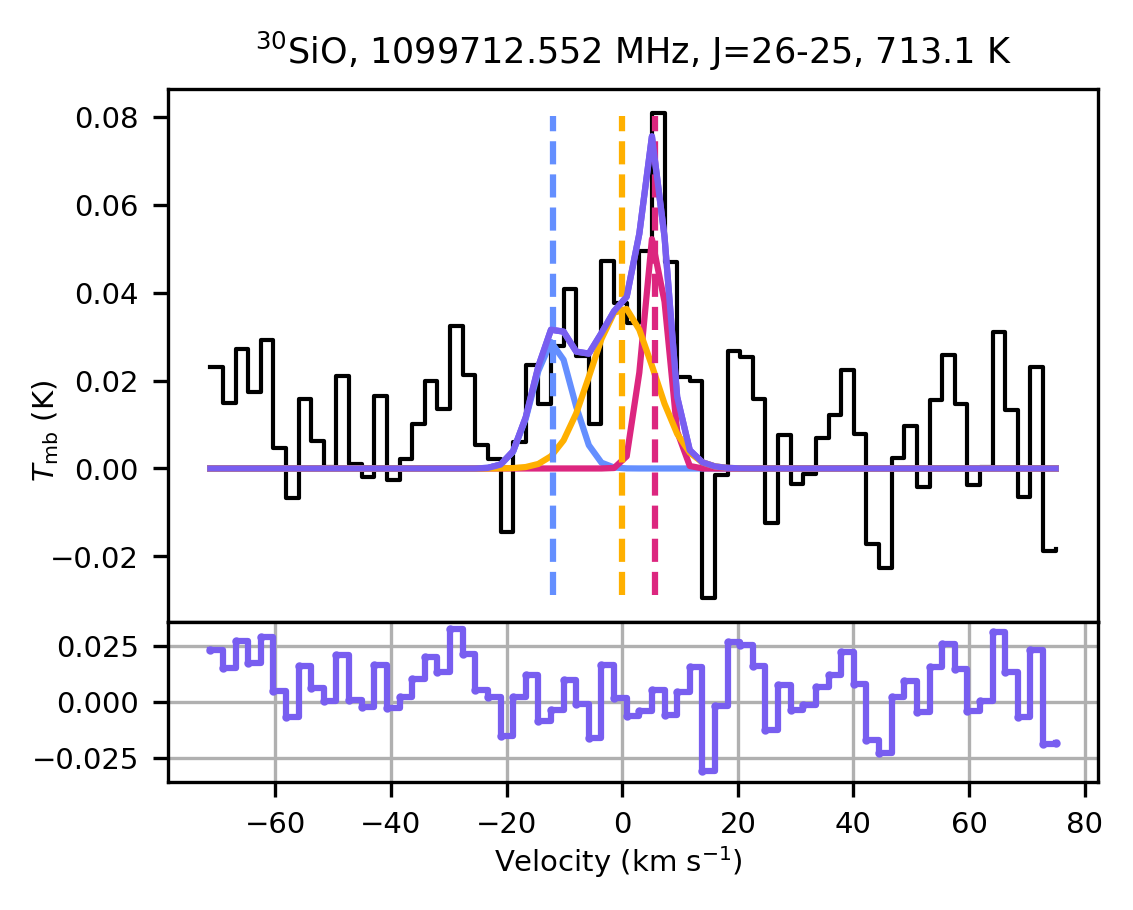}
	\end{subfigure}\hfill
	\caption{Same as Fig.~\ref{fig:co_nml_multifits}, for $^{30}$SiO emission around NML~Cyg, observed with OSO and HIFI.}
	\label{fig:30sio_lines}
\end{figure*}

\begin{figure*}
\begin{subfigure}[c]{0.32\textwidth}
\centering
\includegraphics[height=1.85in]{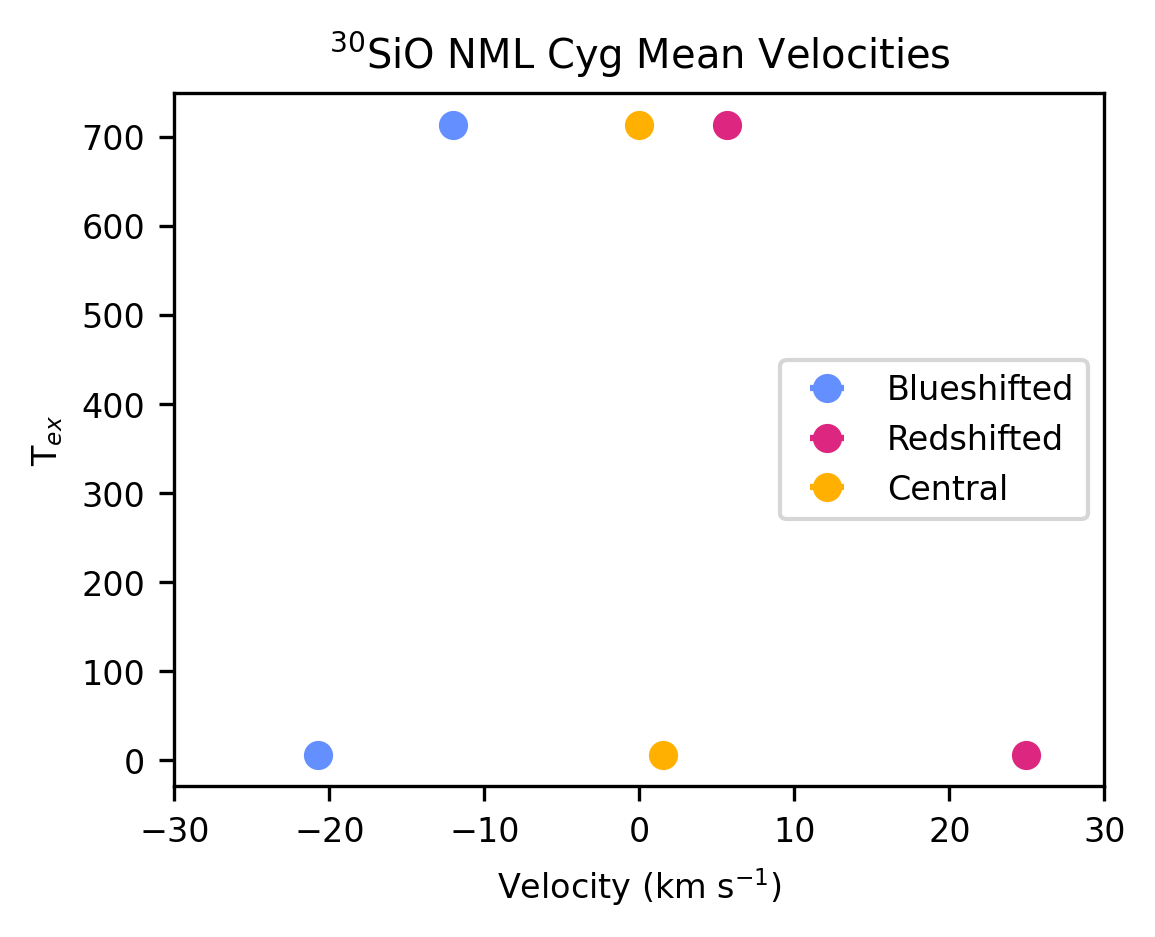}
\end{subfigure}\hfill
	\begin{subfigure}[c]{0.32\textwidth}
	\centering
	\includegraphics[height=1.85in]{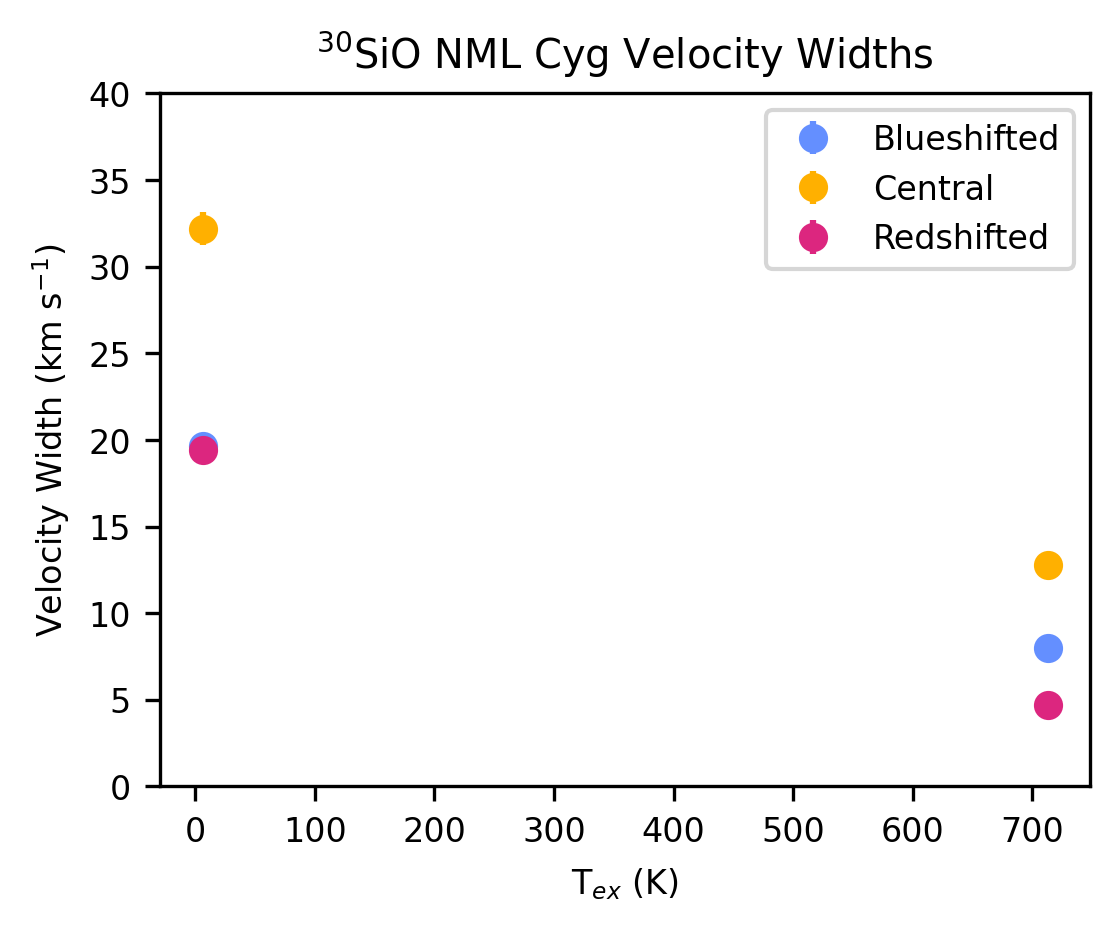}
\end{subfigure}\hfill
\begin{subfigure}[c]{0.32\textwidth}
\centering
\includegraphics[height=1.85in]{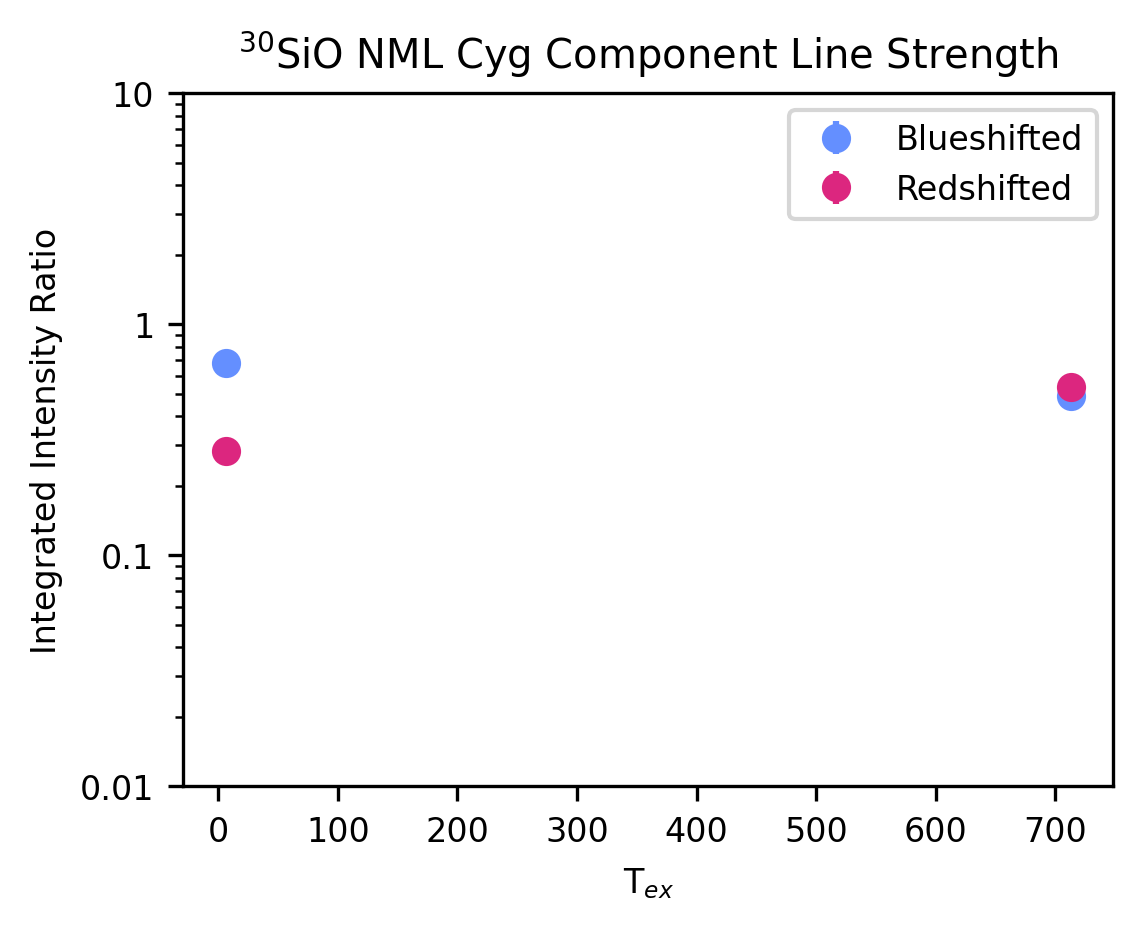}
\end{subfigure}\hfill
\caption{Plots of different fit parameters as described in Fig.\ref{fig:co_fits}, for detected $^{30}$SiO emission lines around NML~Cyg.}
\label{fig:30sio_fits}
\end{figure*}

\begin{figure*}
\centering
	\begin{subfigure}[c]{0.32\textwidth}
		\centering
		\includegraphics[height=1.85in]{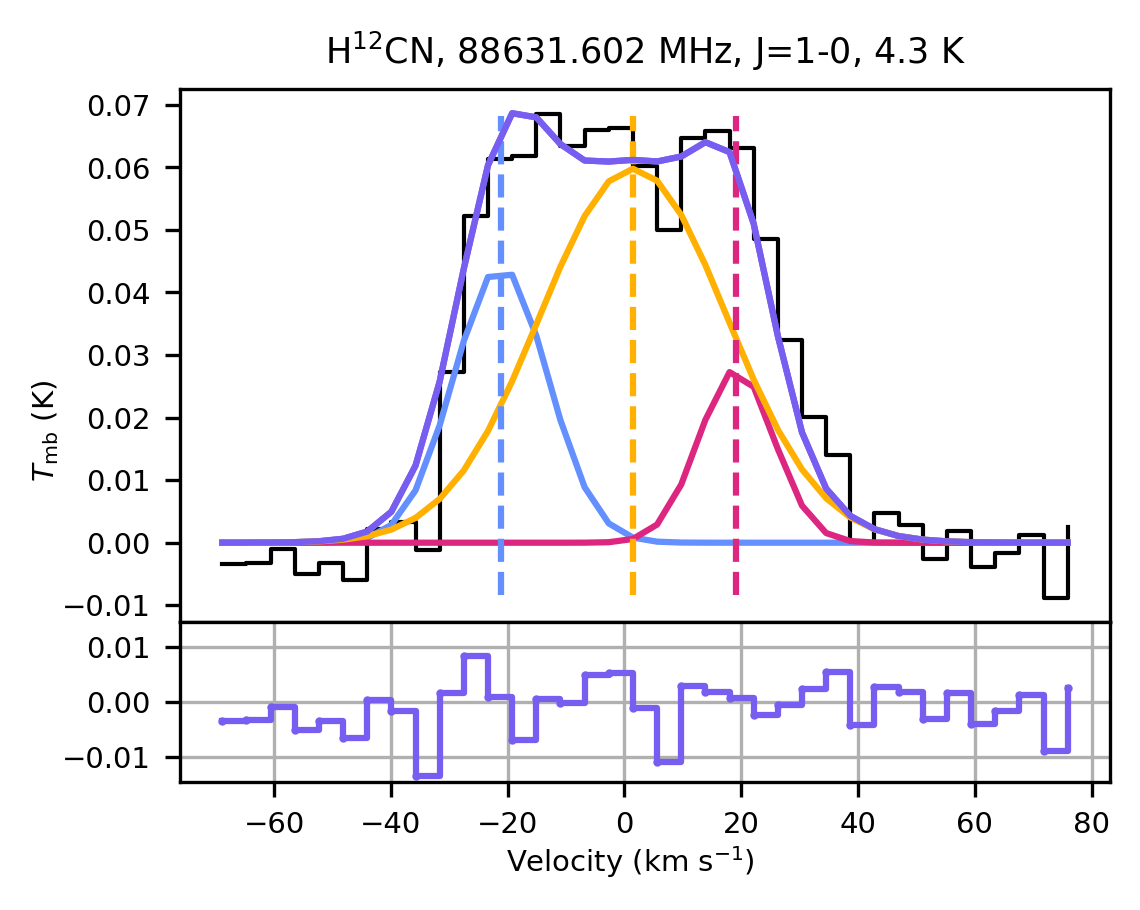}
	\end{subfigure}
	\begin{subfigure}[c]{0.32\textwidth}
		\centering
		\includegraphics[height=1.85in]{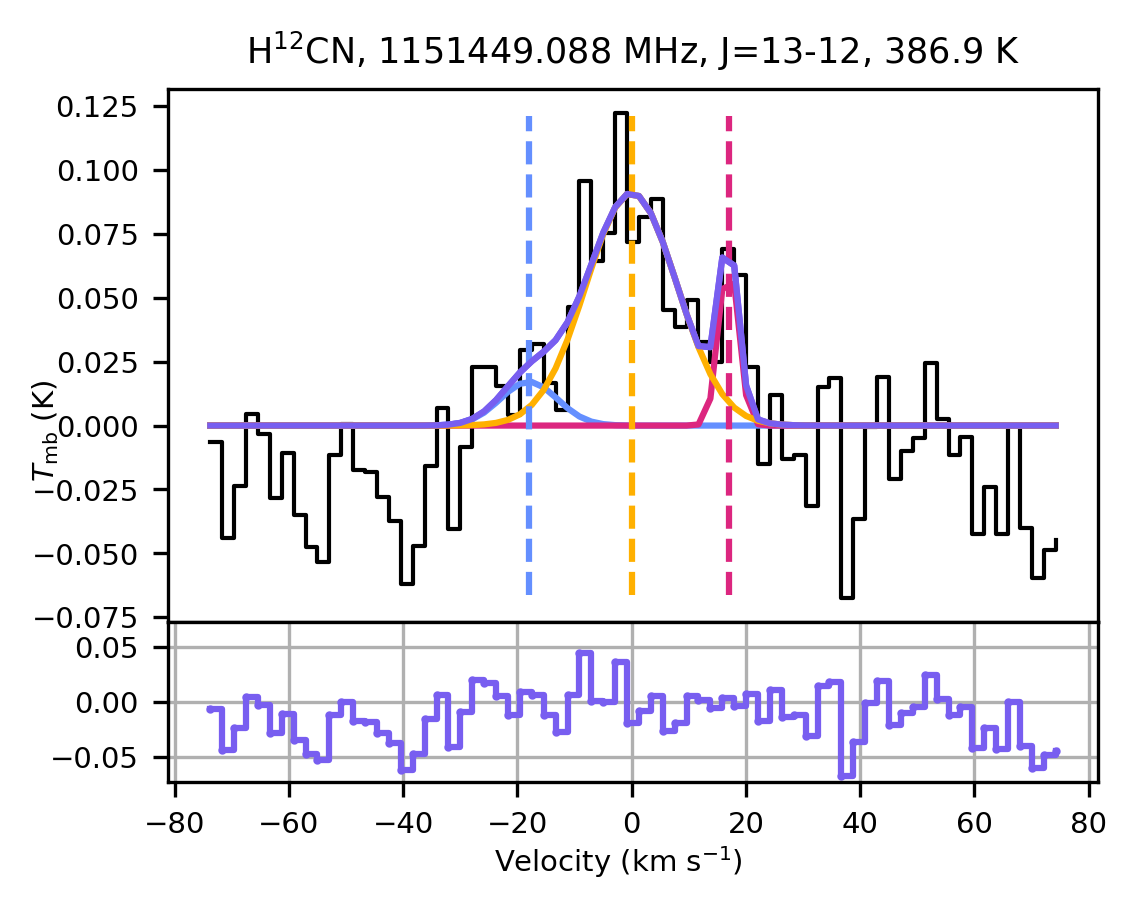}
	\end{subfigure}\hfill
	\caption{Same as Fig.~\ref{fig:co_nml_multifits}, for H$^{12}$CN emission around NML~Cyg, observed with OSO and HIFI.}
	\label{fig:hcn_lines}
\end{figure*}

\begin{figure*}
\begin{subfigure}[c]{0.32\textwidth}
\centering
\includegraphics[height=1.85in]{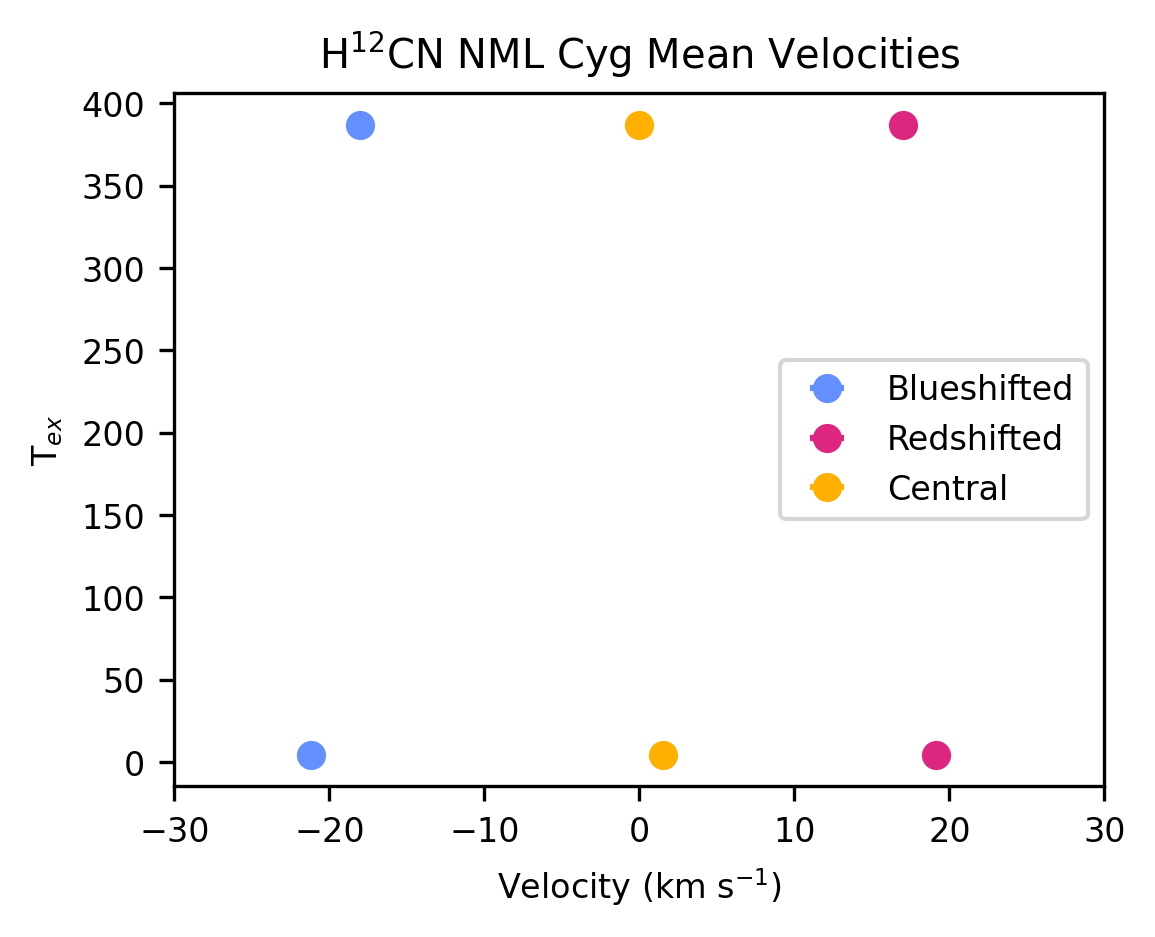}
\end{subfigure}\hfill
\begin{subfigure}[c]{0.32\textwidth}
	\centering
	\includegraphics[height=1.85in]{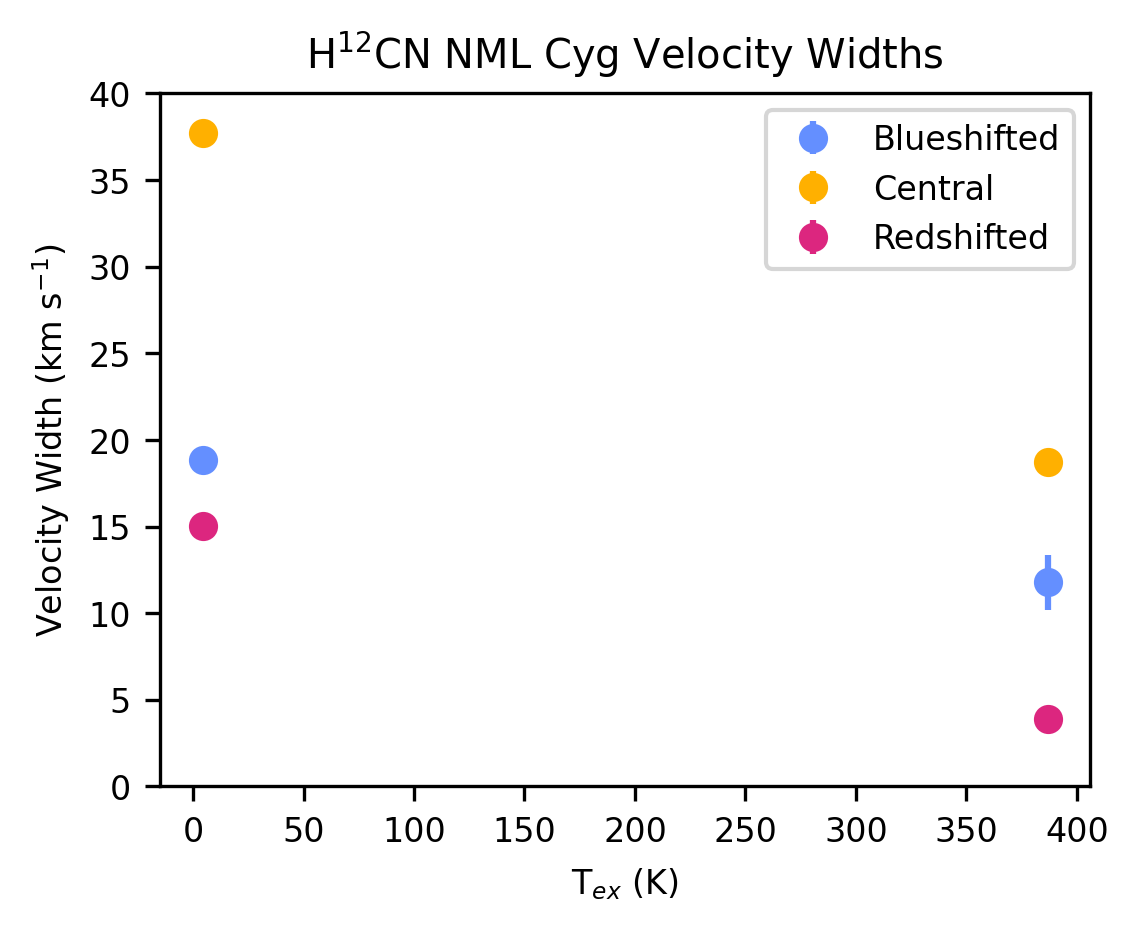}
\end{subfigure}\hfill
\begin{subfigure}[c]{0.32\textwidth}
\centering
\includegraphics[height=1.85in]{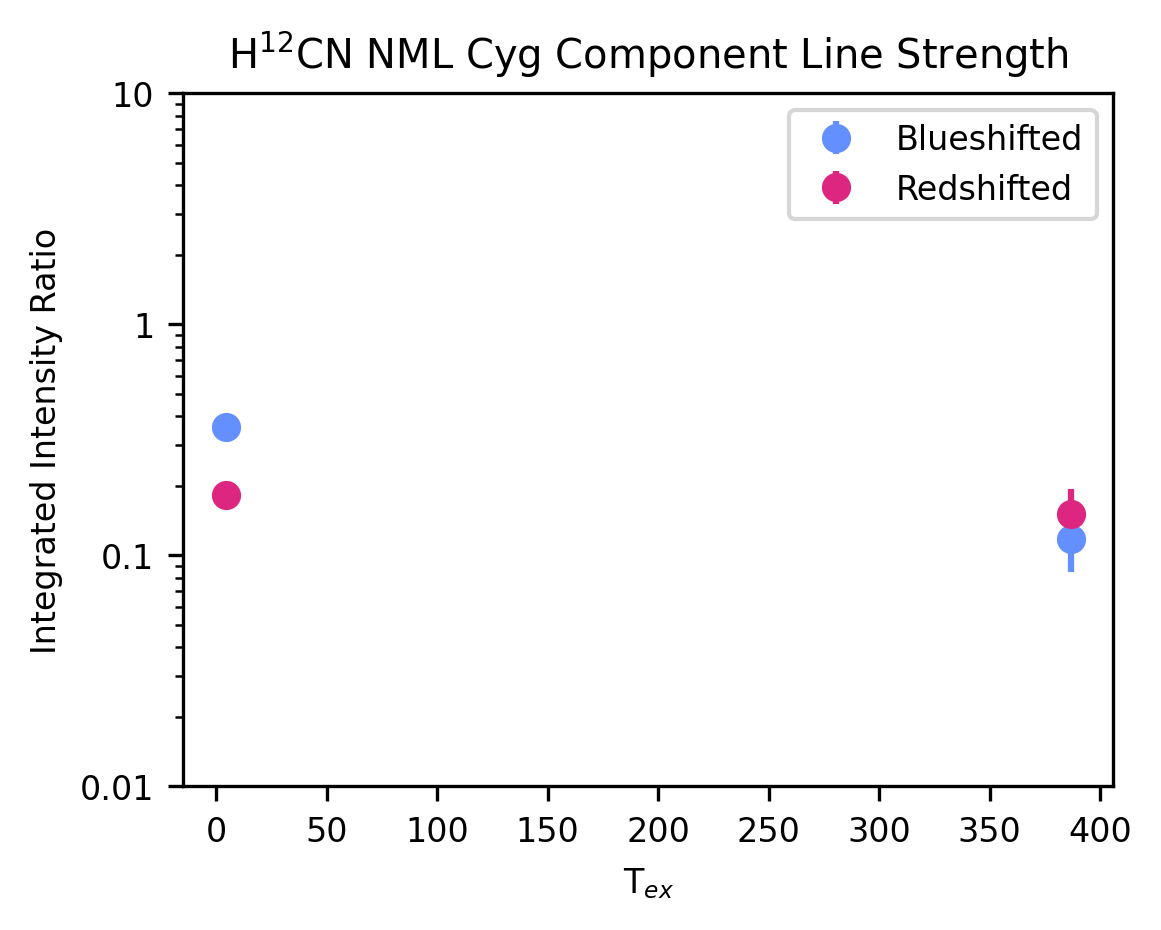}
\end{subfigure}\hfill
\caption{Same as Fig.~\ref{fig:co_fits}, for detected H$^{12}$CN emission lines around NML~Cyg.}
\label{fig:hcn_fits}
\end{figure*}

\begin{figure}
	\centering
	\includegraphics[width=0.32\textwidth]{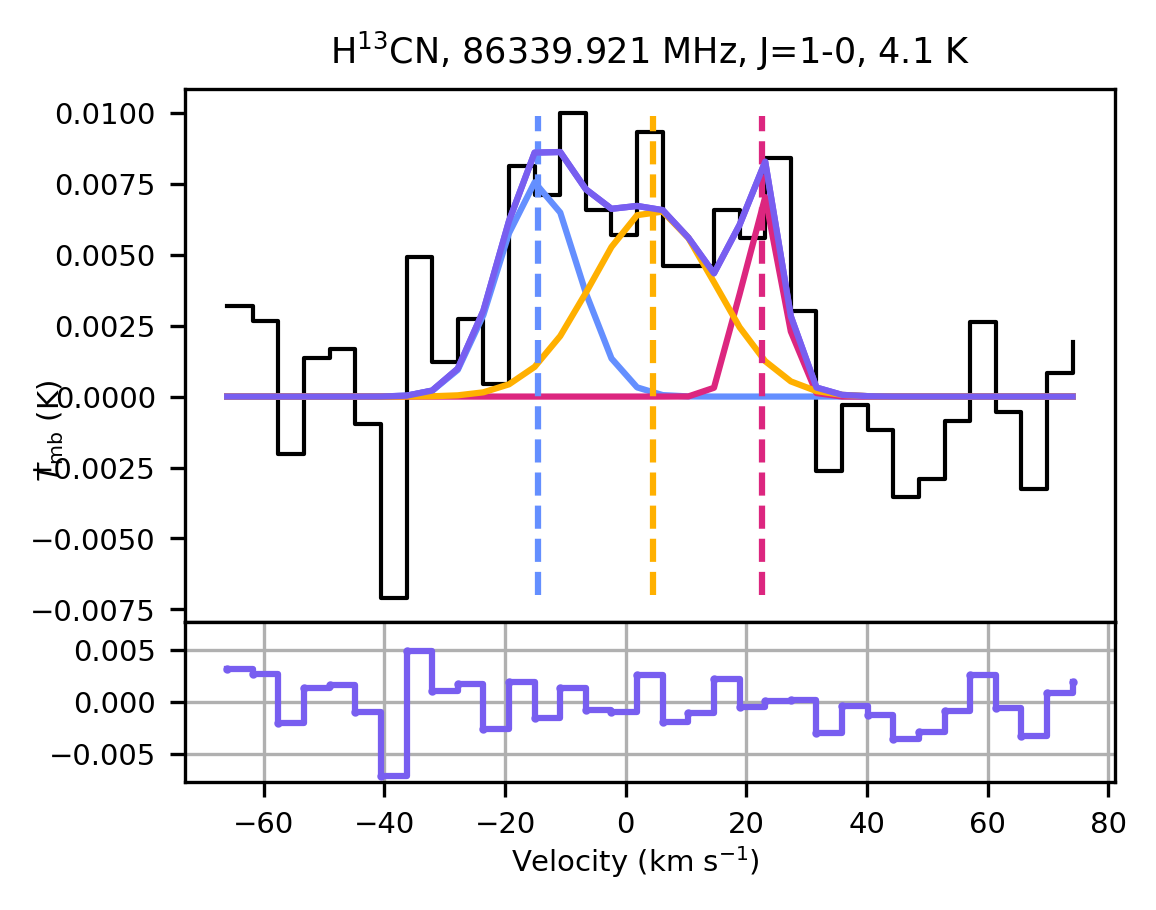}
	\caption{Same as Fig.~\ref{fig:co_nml_multifits}, for H$^{13}$CN $J = 1 - 0$ emission around NML~Cyg, observed with OSO.}
	\label{fig:hc13n_oso_10}
\end{figure} 

\begin{figure*}
	\begin{subfigure}[c]{0.32\textwidth}
		\centering
		\includegraphics[height=1.85in]{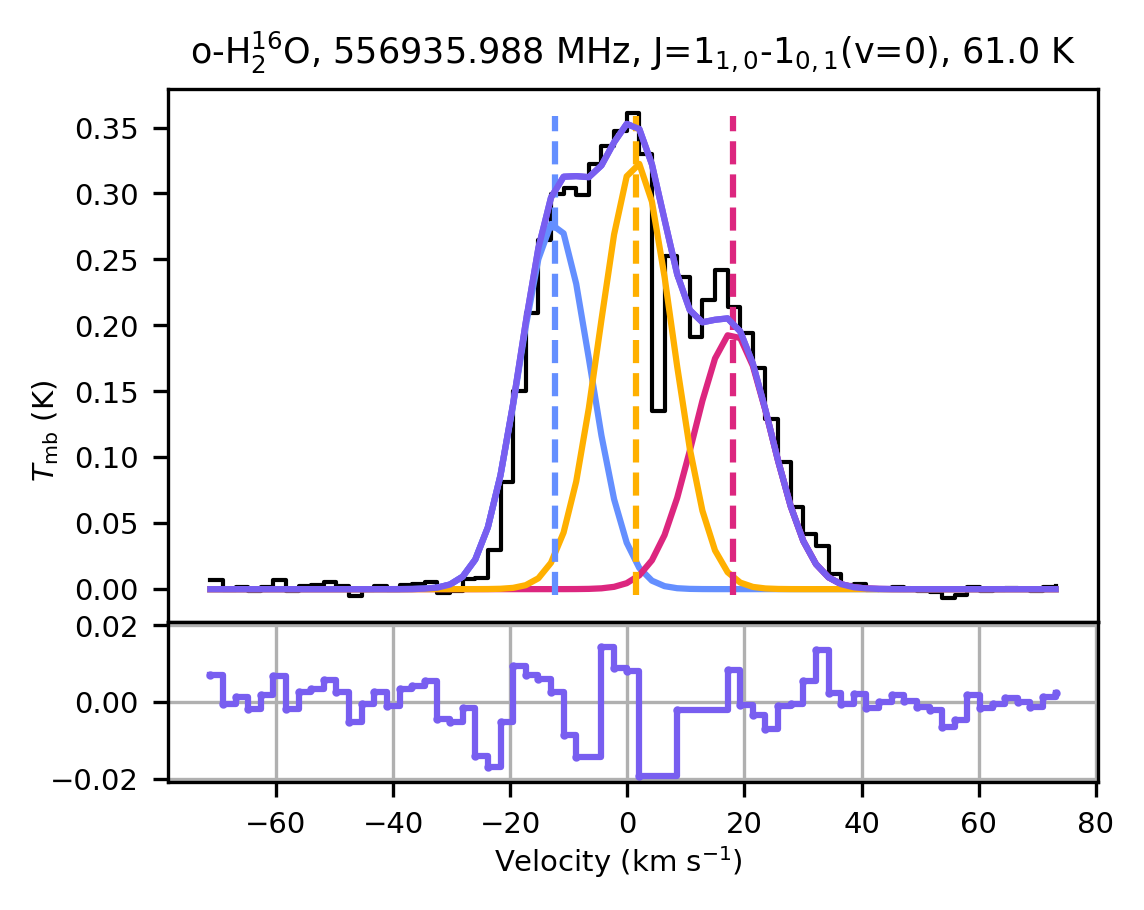}
	\end{subfigure}
	\begin{subfigure}[c]{0.32\textwidth}
		\centering
		\includegraphics[height=1.85in]{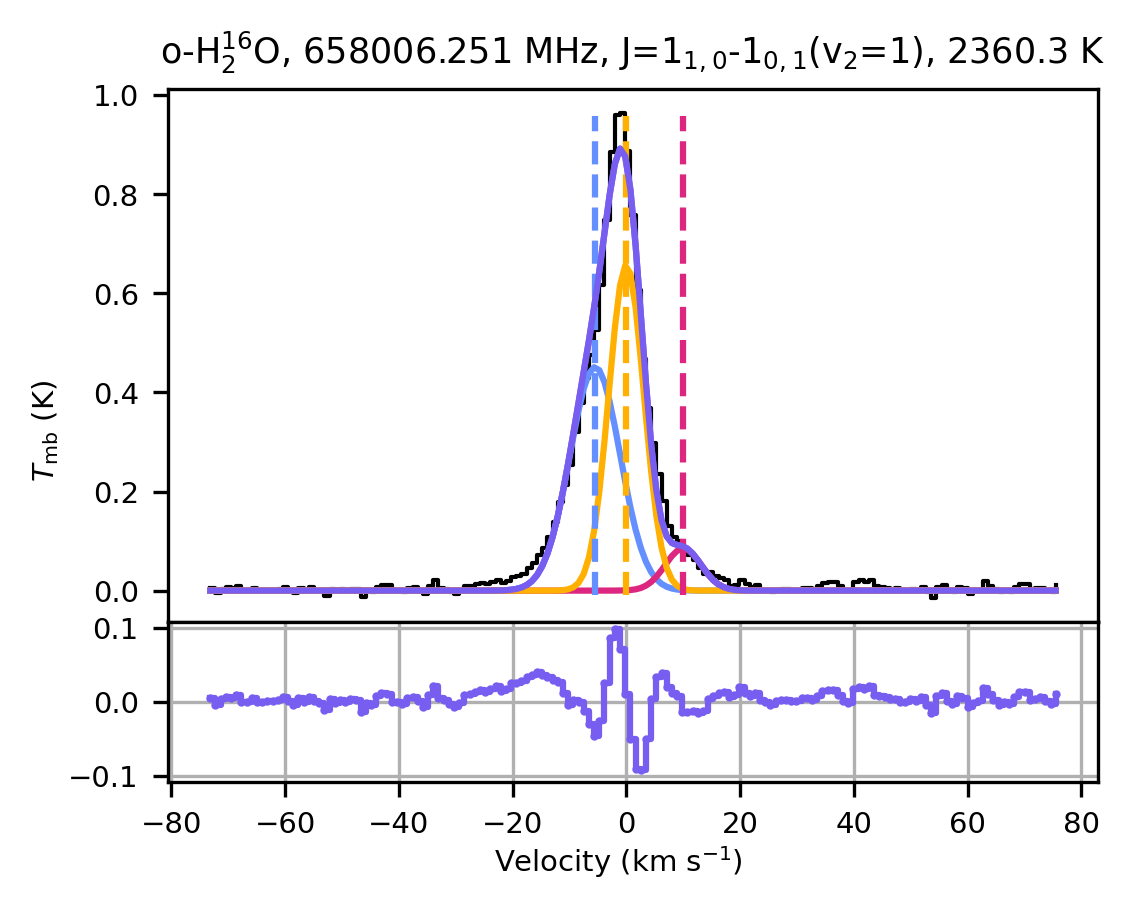}
	\end{subfigure}
	\begin{subfigure}[c]{0.32\textwidth}
		\centering
		\includegraphics[height=1.85in]{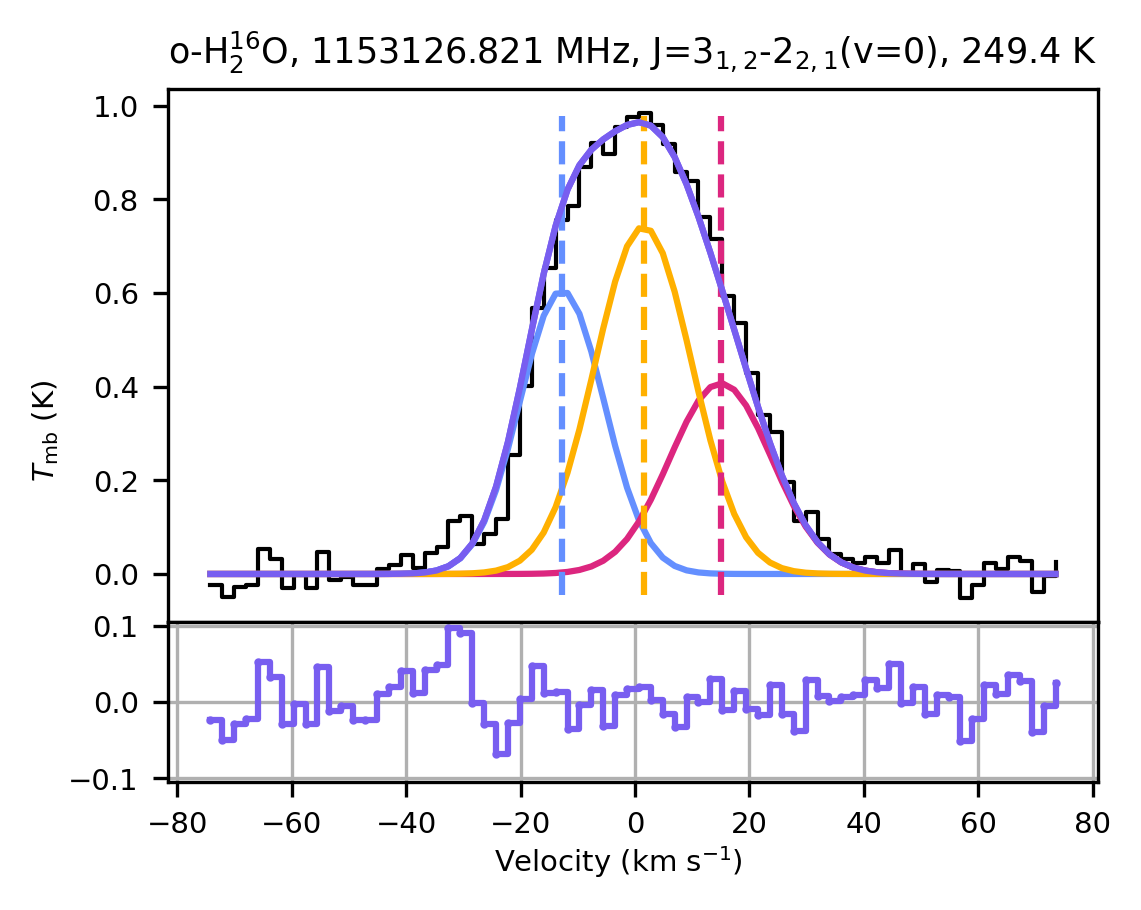}
	\end{subfigure}
	\begin{subfigure}[c]{0.32\textwidth}
		\centering
		\includegraphics[height=1.85in]{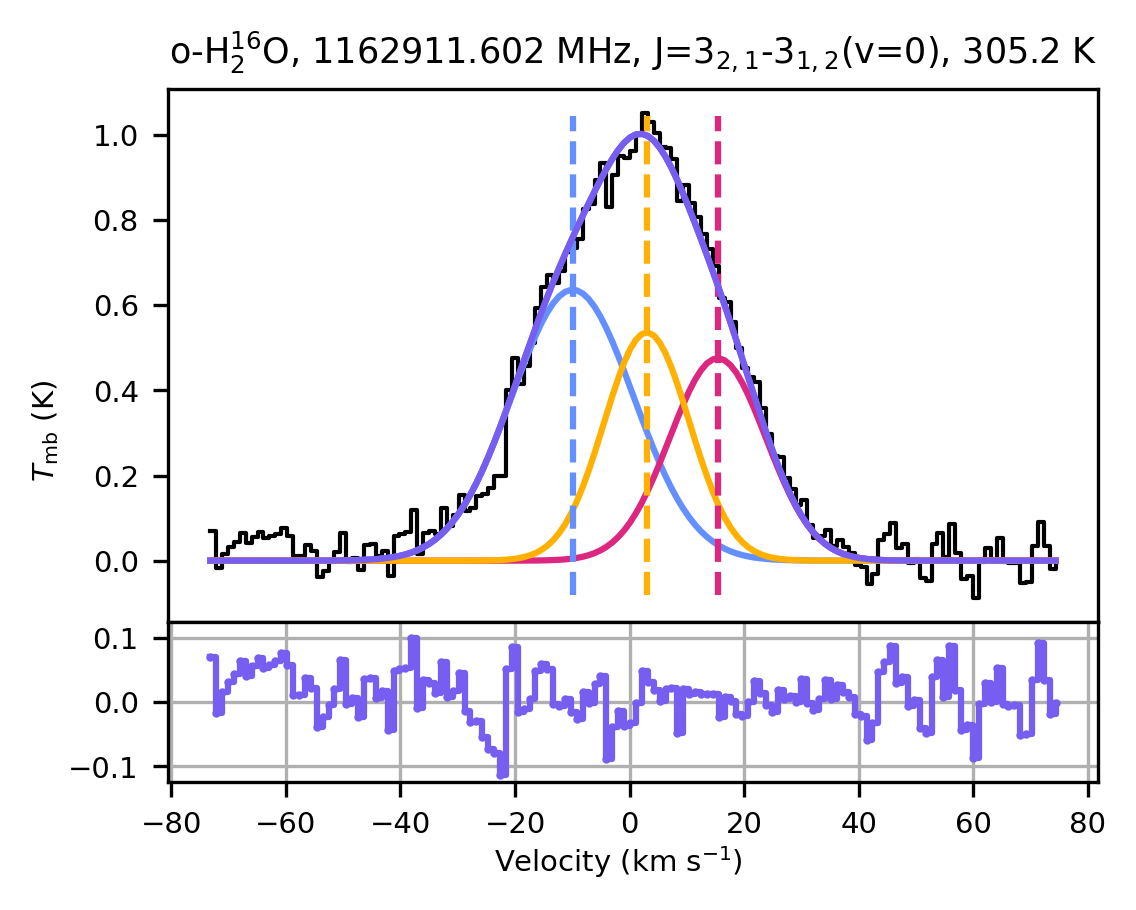}
	\end{subfigure}
	\begin{subfigure}[c]{0.32\textwidth}
		\includegraphics[height=1.85in]{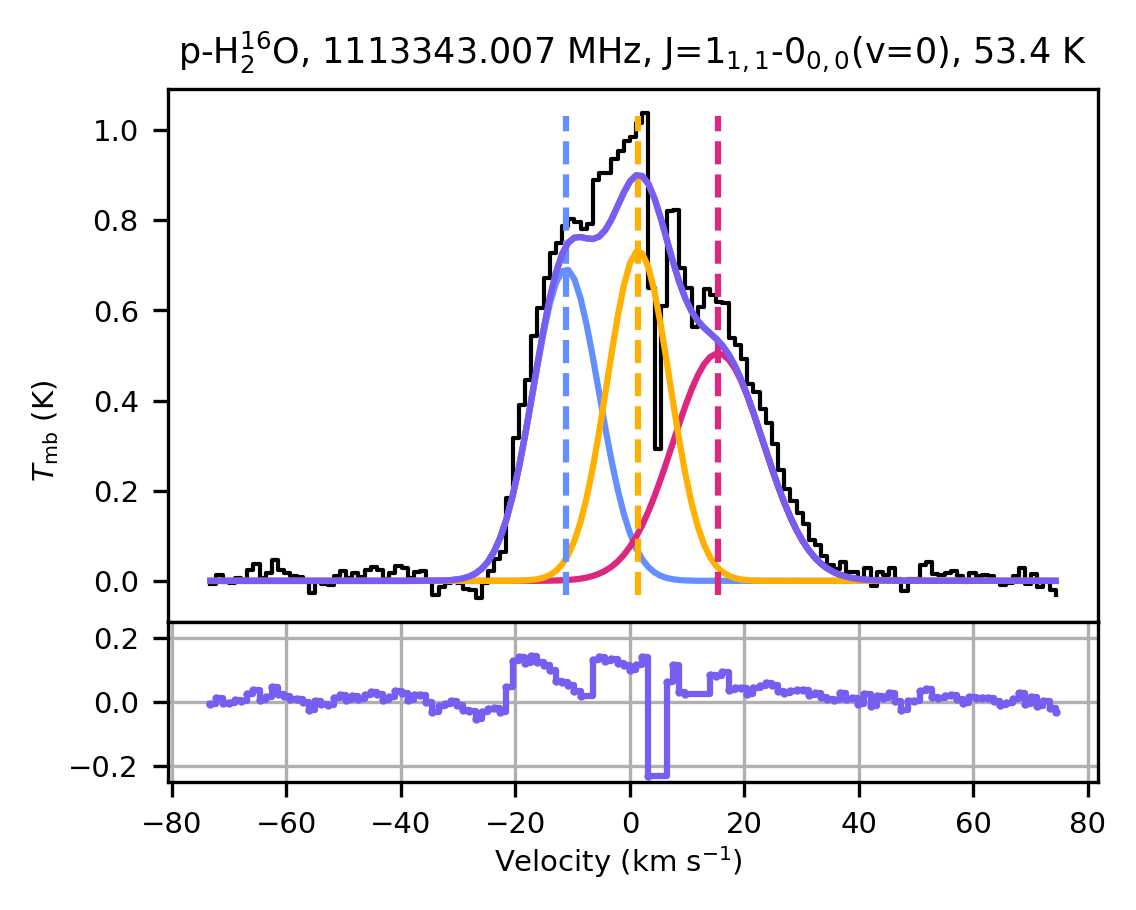}
	\end{subfigure}
	\begin{subfigure}[c]{0.32\textwidth}
		\includegraphics[height=1.85in]{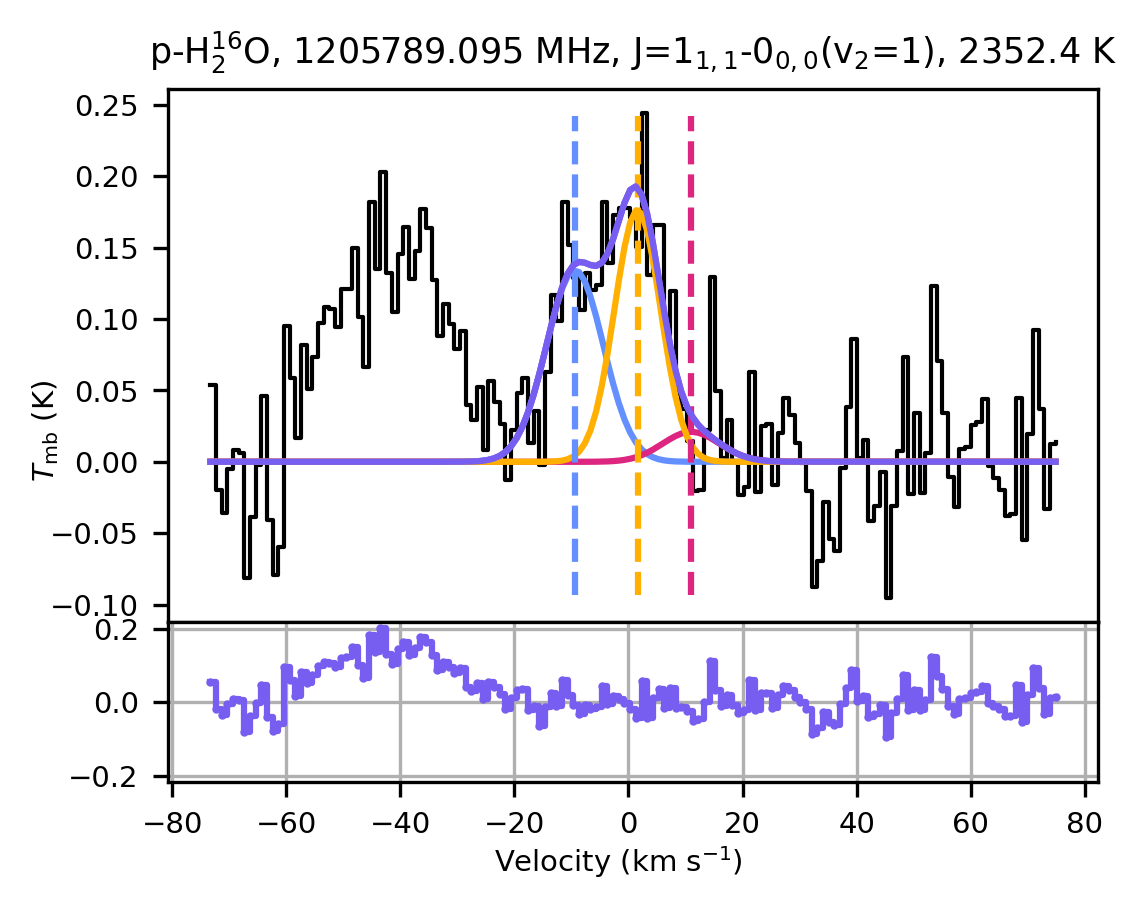}
	\end{subfigure}
	\begin{subfigure}[c]{0.32\textwidth}
		\includegraphics[height=1.85in]{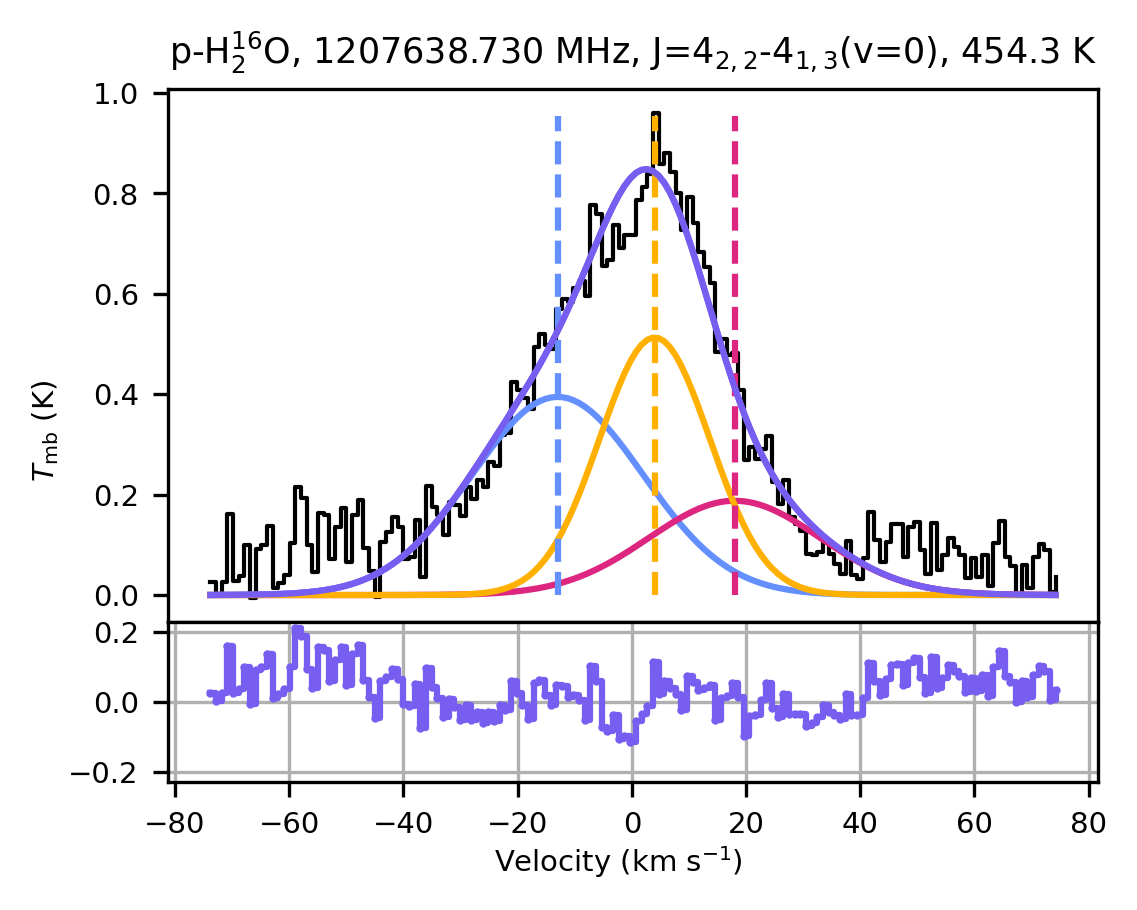}
	\end{subfigure}
	\begin{subfigure}[c]{0.32\textwidth}
		\includegraphics[height=1.85in]{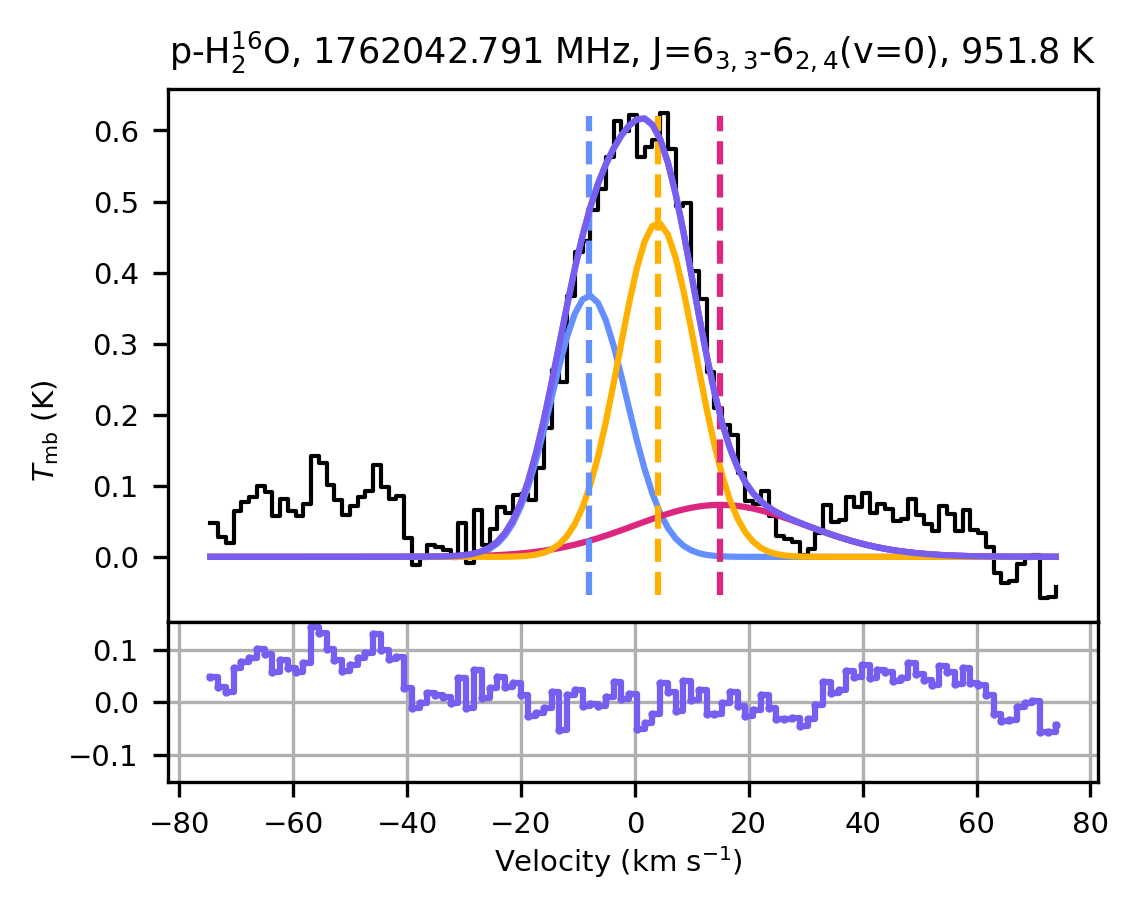}
	\end{subfigure}
	\begin{minipage}[c]{0.32\textwidth}
	\hspace{\fill}
\end{minipage}
	\caption{Same as Fig.~\ref{fig:co_nml_multifits} for H$_{2}^{16}$O emission around NML~Cyg observed with OSO and HIFI.}
	\label{fig:h2o_lines}
\end{figure*}

\begin{figure}
	\centering
	\includegraphics[width=0.32\textwidth]{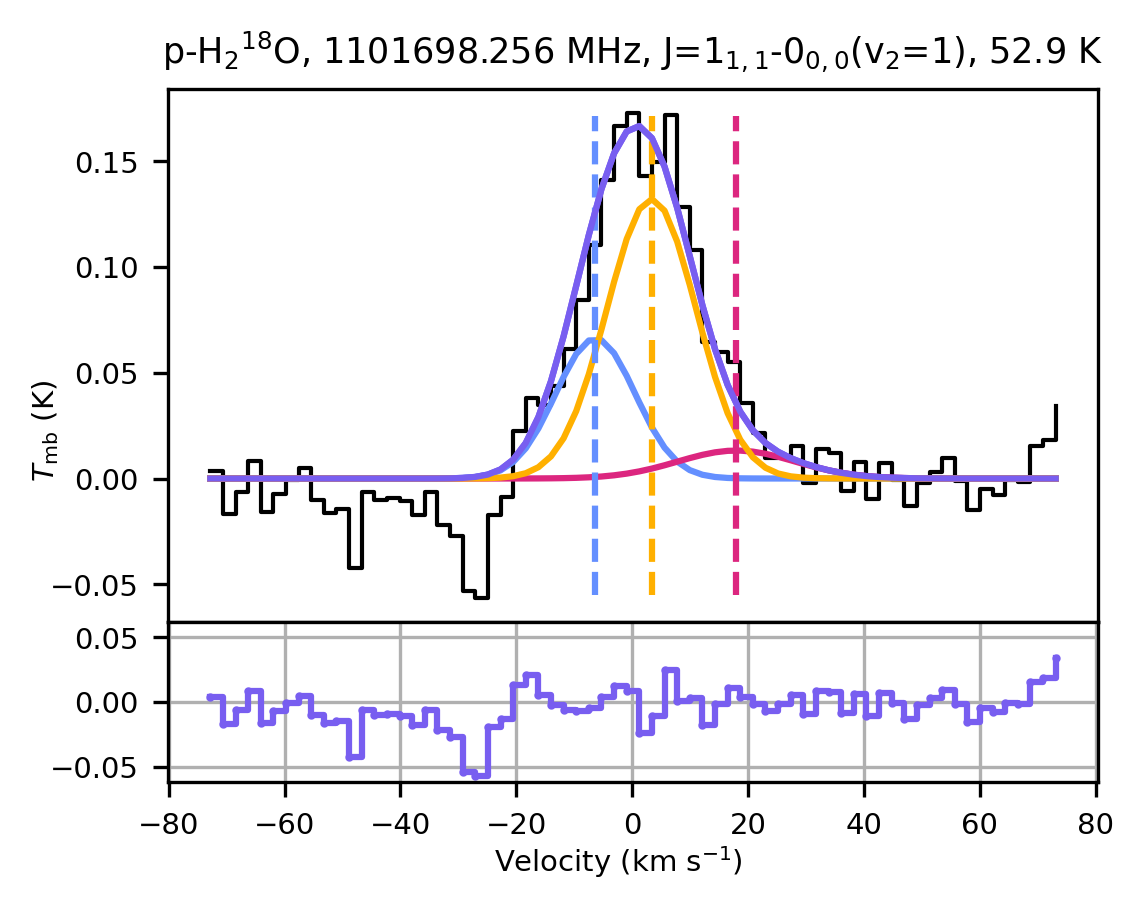}
	\caption{Same as Fig.~\ref{fig:co_nml_multifits} for p-H$_{2}^{18}$O $J_{K_{\mathrm{a}},K_{\mathrm{c}}} = 1_{1,1} - 0_{0,0}$ ($v_{2} = 1$) line observed with HIFI.}
\label{fig:h218O_hifi}
\end{figure}

\begin{figure*}
\begin{subfigure}[c]{0.32\textwidth}
\centering
\includegraphics[height=1.85in]{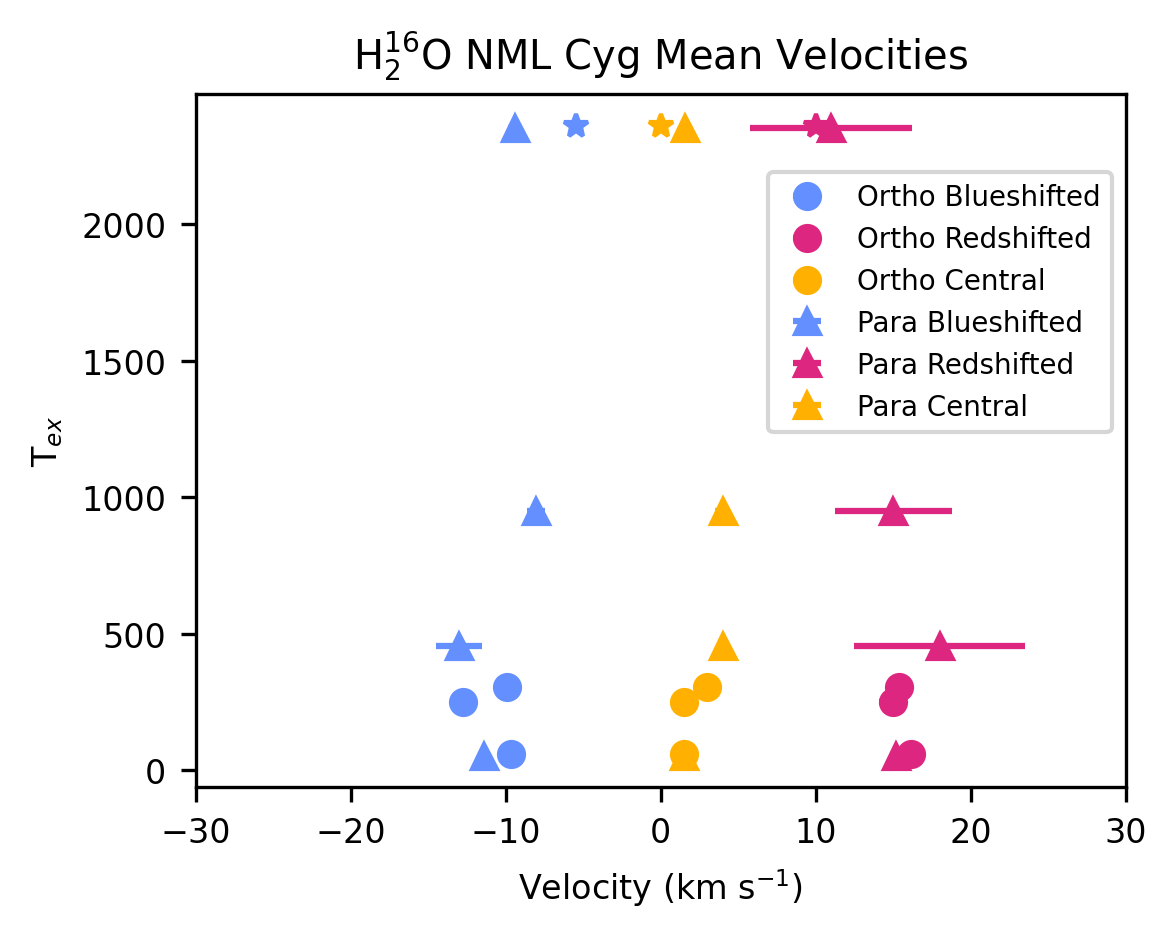}
\end{subfigure}\hfill
\begin{subfigure}[c]{0.32\textwidth}
	\centering
	\includegraphics[height=1.85in]{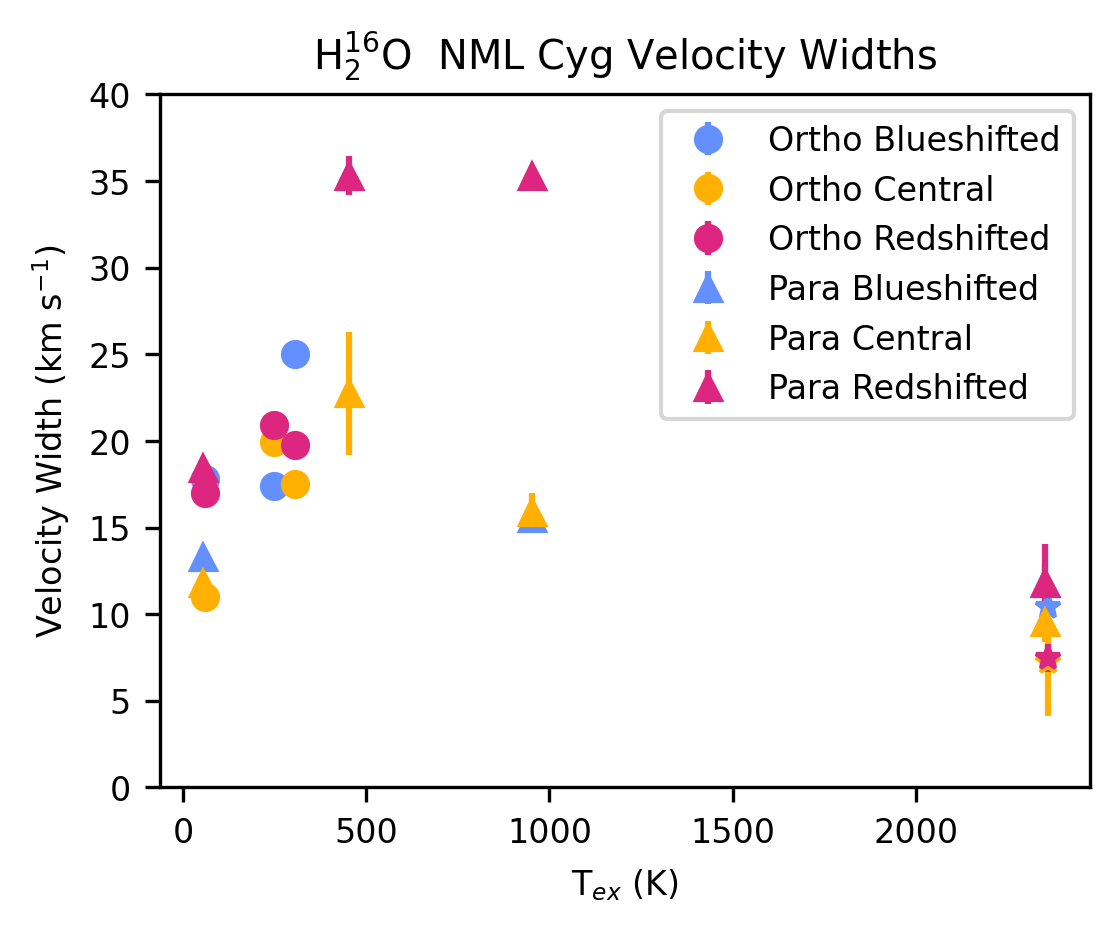}
\end{subfigure}\hfill
\begin{subfigure}[c]{0.32\textwidth}
\centering
\includegraphics[height=1.85in]{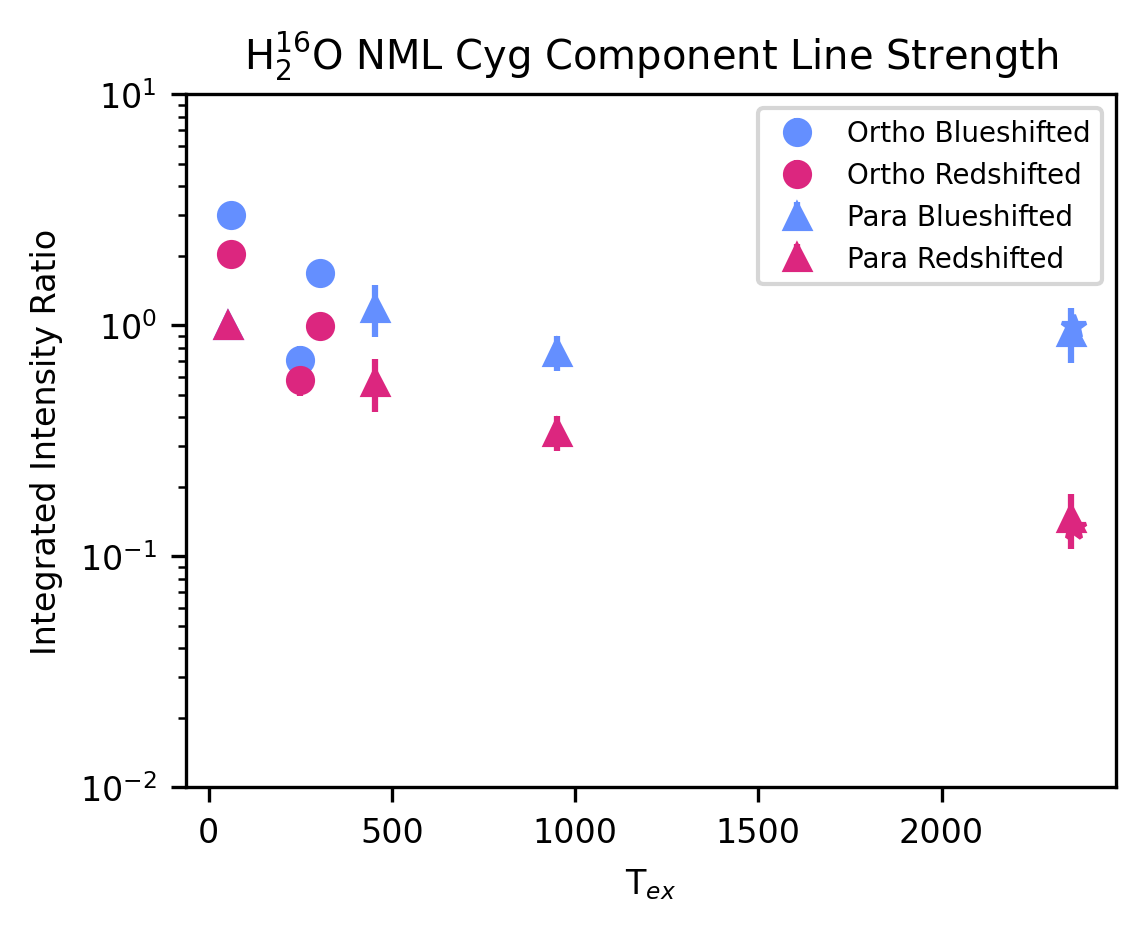}
\end{subfigure}\hfill
\caption{Same as Fig.~\ref{fig:co_fits}, for detected H$_{2}^{16}$O emission lines around NML~Cyg.}
\label{fig:h2o_fits}
\end{figure*}

\begin{figure*}
	\begin{subfigure}[c]{0.32\textwidth}
		\centering
		\includegraphics[height=1.85in]{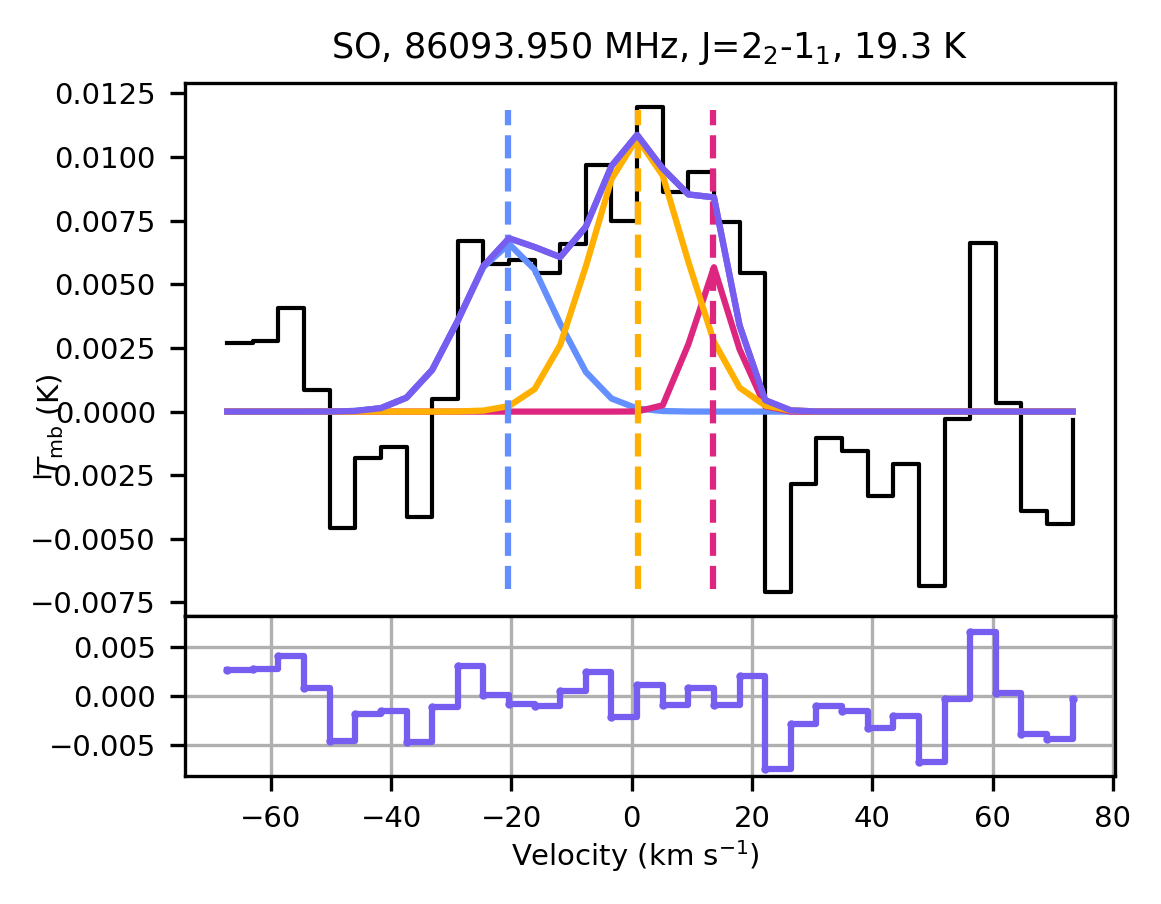}
\end{subfigure}\hfill
	\begin{subfigure}[c]{0.32\textwidth}
		\centering
		\includegraphics[height=1.85in]{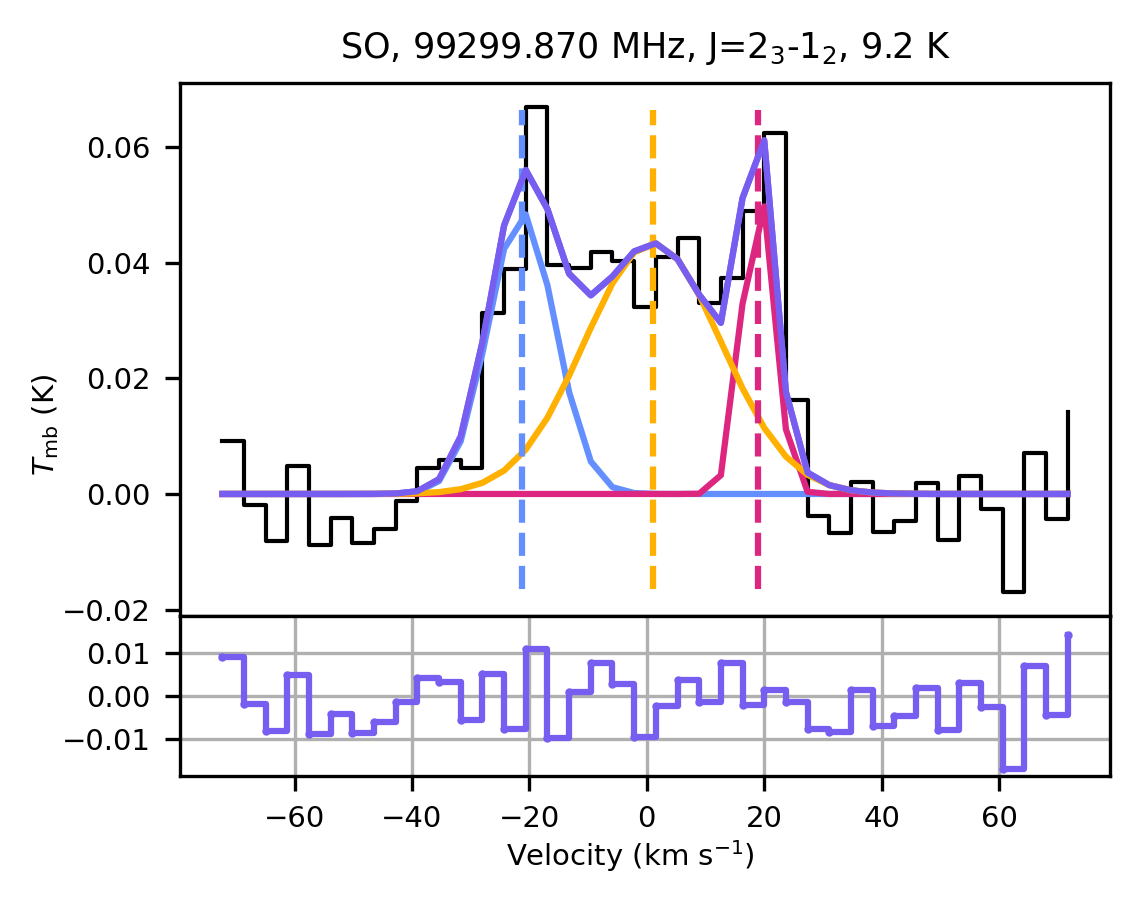}
\end{subfigure}\hfill
	\begin{subfigure}[c]{0.32\textwidth}
		\centering
		\includegraphics[height=1.85in]{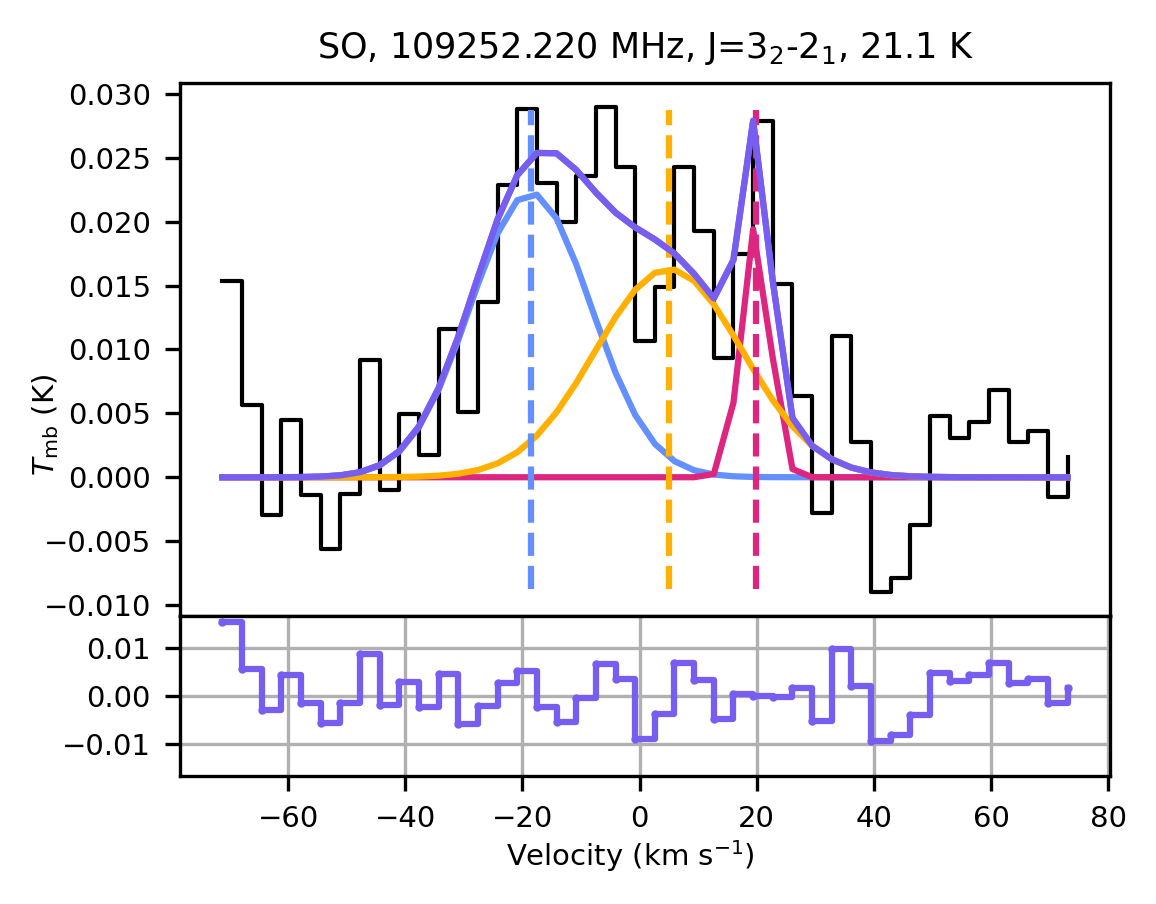}
\end{subfigure}\hfill
	\begin{subfigure}[c]{0.32\textwidth}
		\centering
		\includegraphics[height=1.85in]{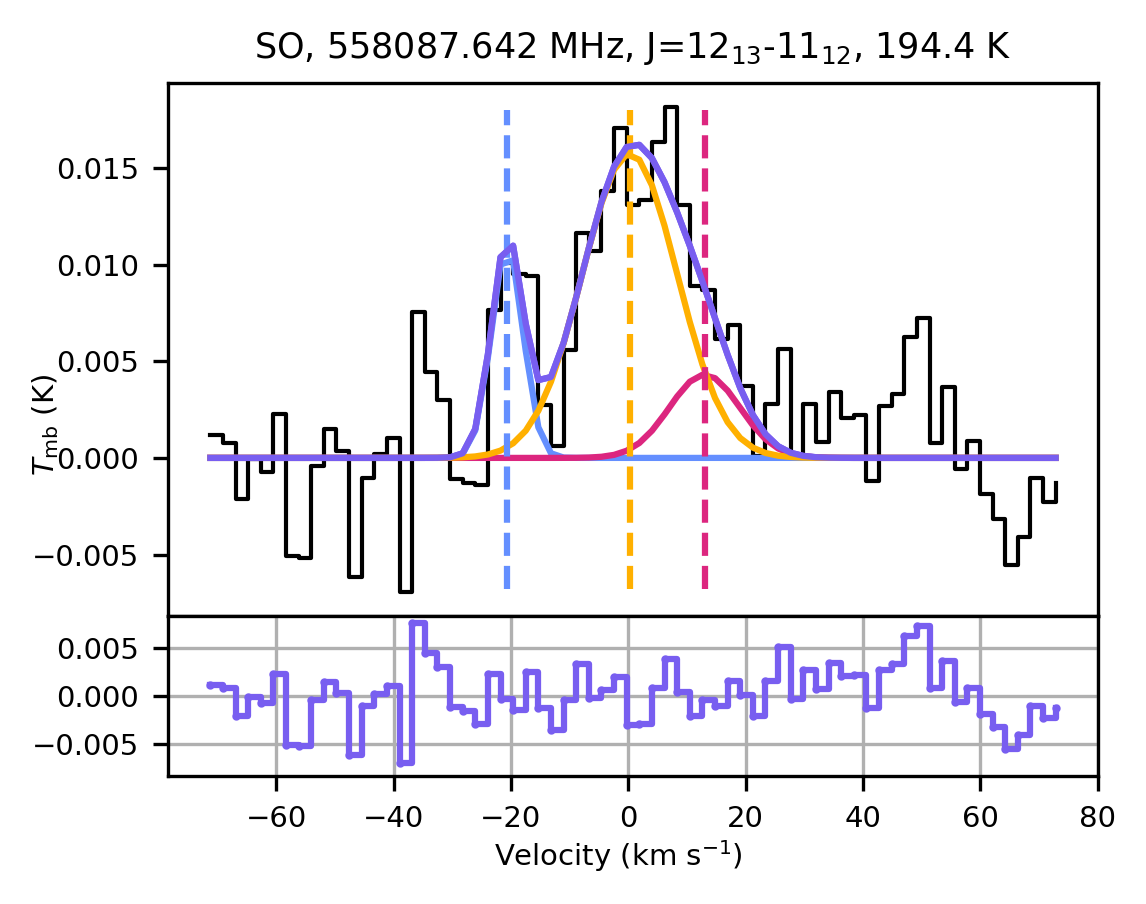}
\end{subfigure}\hfill
	\begin{subfigure}[c]{0.32\textwidth}
		\centering
		\includegraphics[height=1.85in]{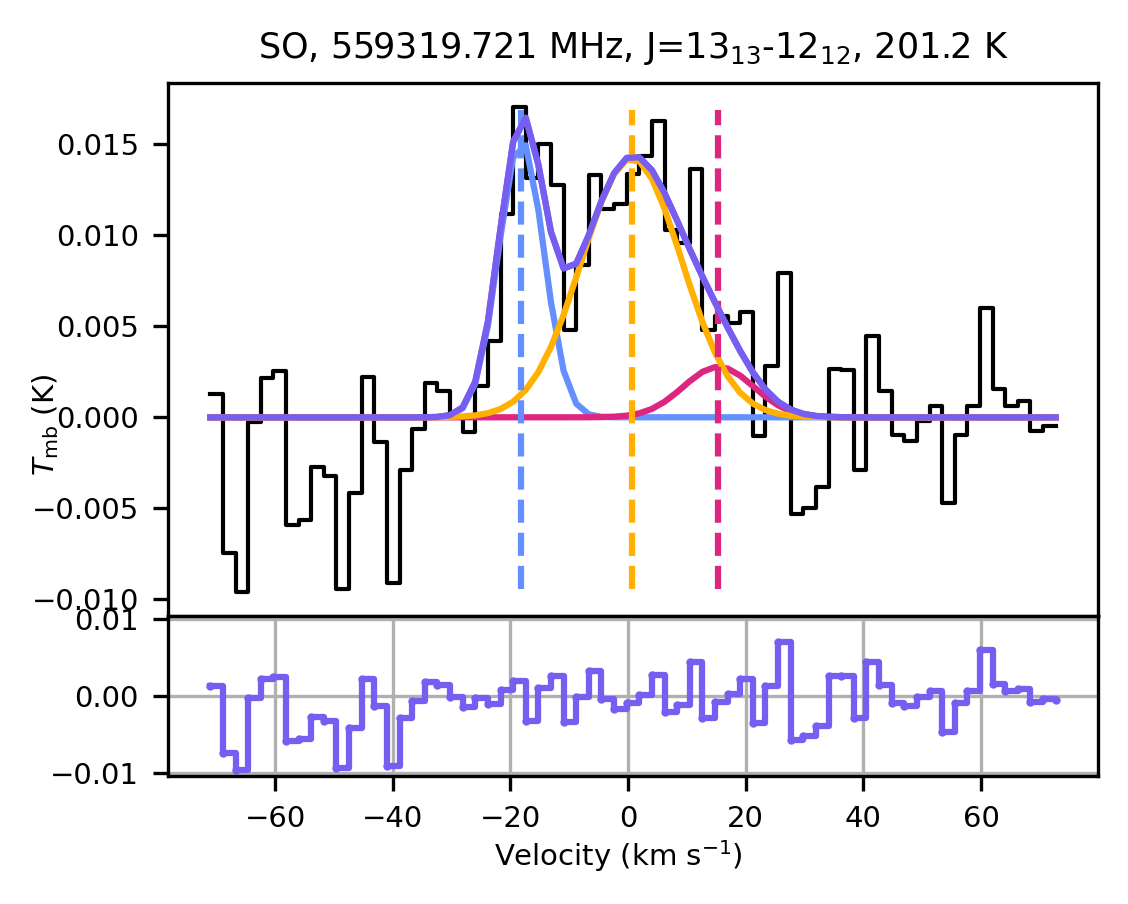}
\end{subfigure}\hfill
	\begin{subfigure}[c]{0.32\textwidth}
		\centering
		\includegraphics[height=1.85in]{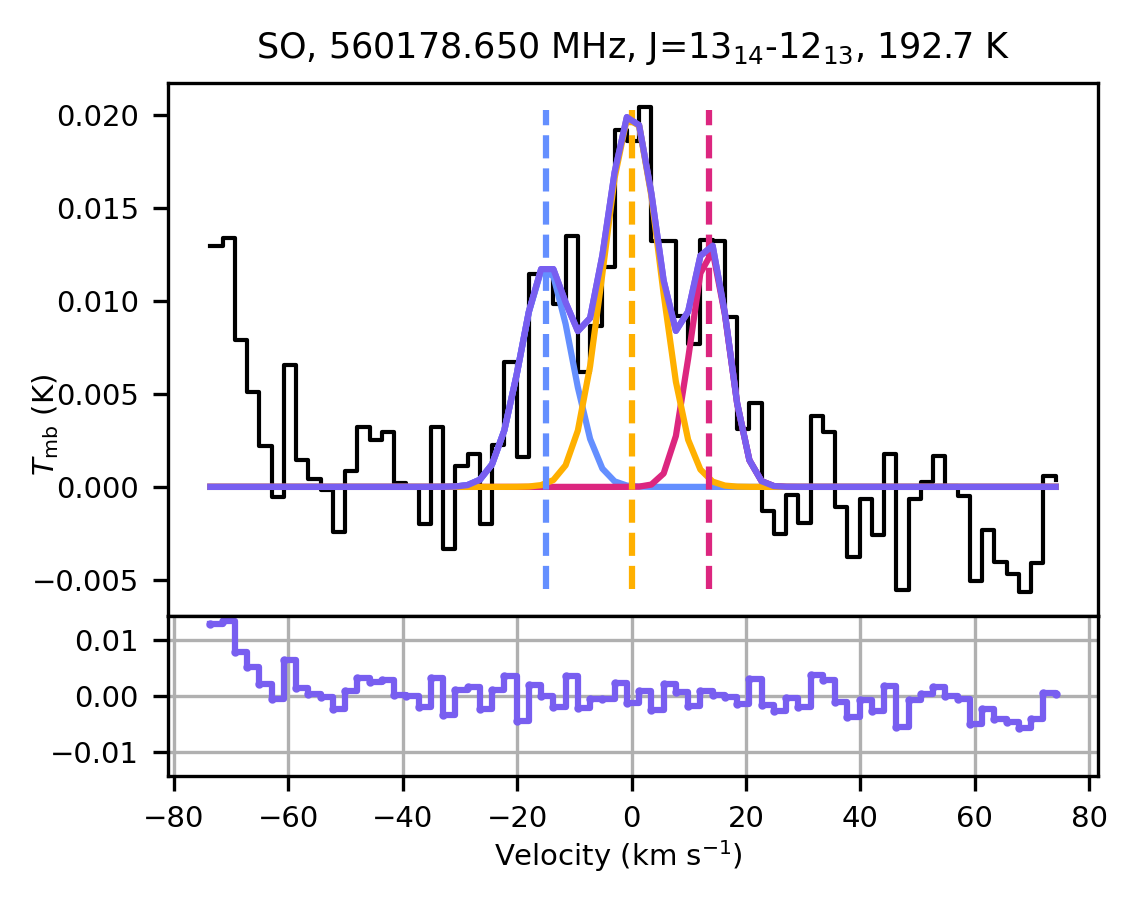}
\end{subfigure}\hfill
	\begin{subfigure}[c]{0.32\textwidth}
		\centering
		\includegraphics[height=1.85in]{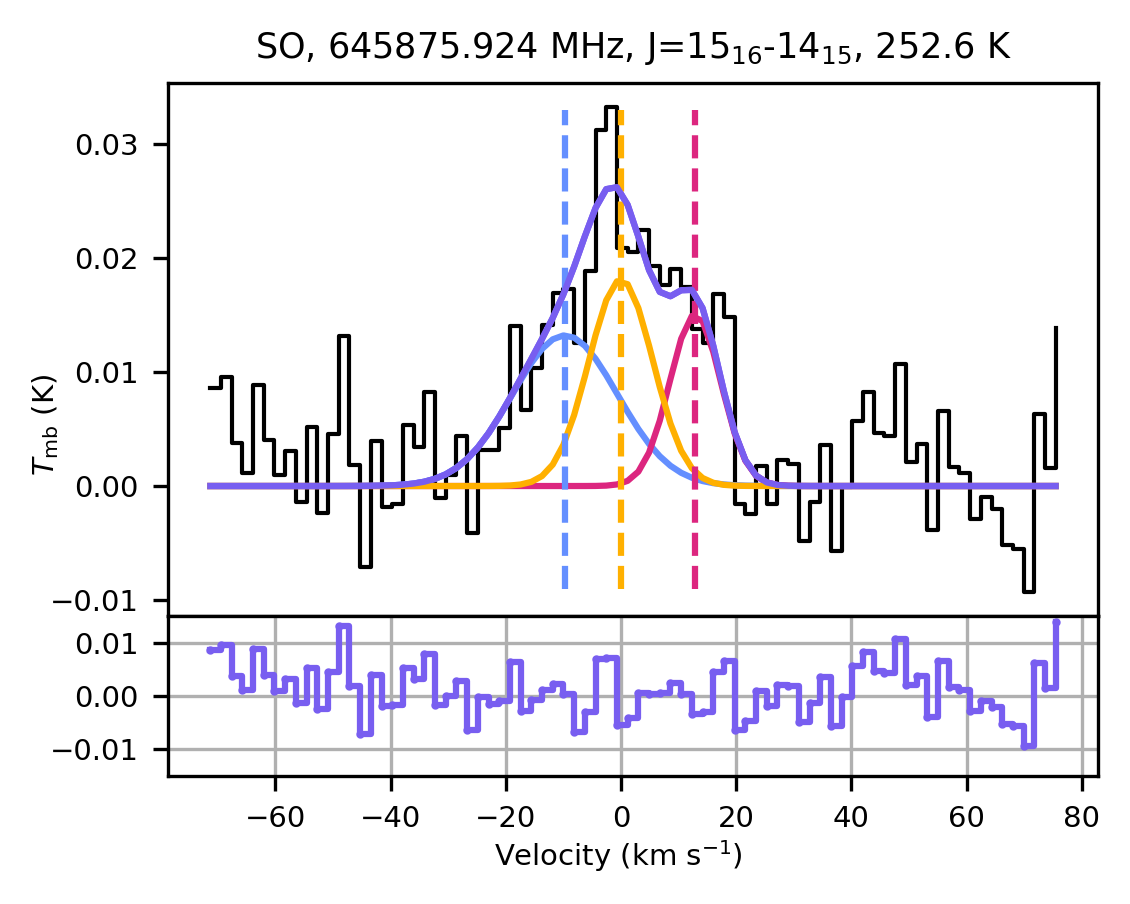}
\end{subfigure}\hfill
	\begin{minipage}[c]{0.32\textwidth}
		\hspace{\fill}
	\end{minipage}\hfill
	\begin{minipage}[c]{0.32\textwidth}
	\hspace{\fill}
\end{minipage}\hfill
	\caption{Same as Fig.~\ref{fig:co_nml_multifits} for SO emission around NML~Cyg observed with OSO and HIFI.}
	\label{fig:so_lines}
\end{figure*}

\begin{figure*}
\begin{subfigure}[c]{0.32\textwidth}
\centering
\includegraphics[height=1.85in]{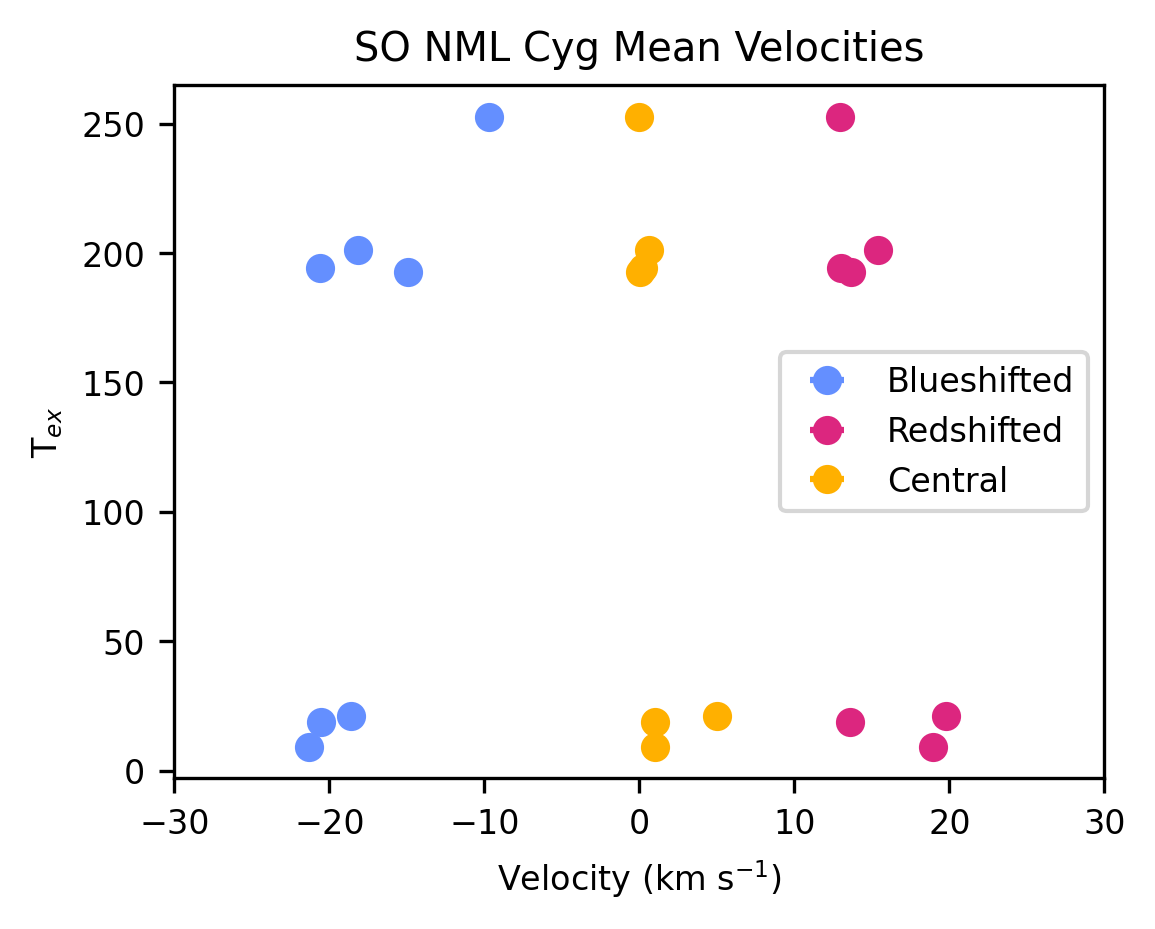}
\end{subfigure}\hfill
\begin{subfigure}[c]{0.32\textwidth}
	\centering
	\includegraphics[height=1.85in]{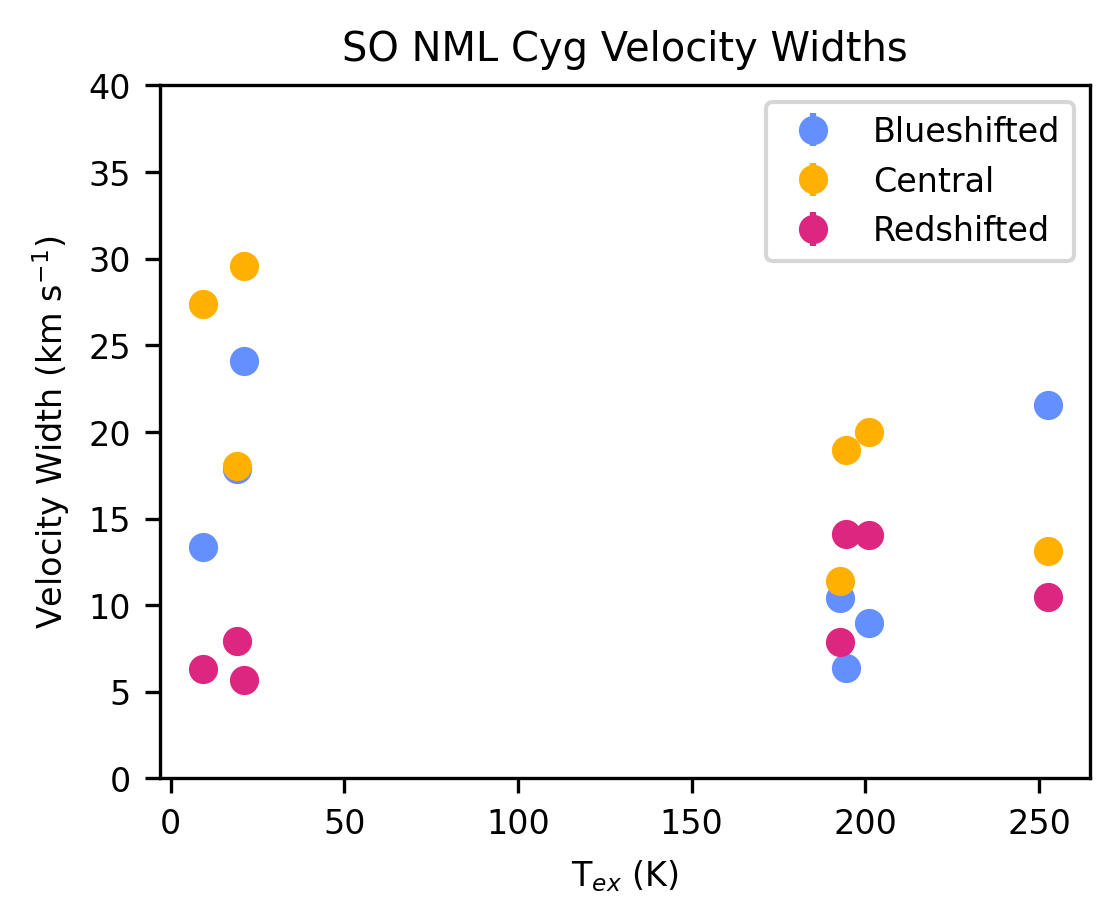}
\end{subfigure}\hfill
\begin{subfigure}[c]{0.32\textwidth}
\centering
\includegraphics[height=1.85in]{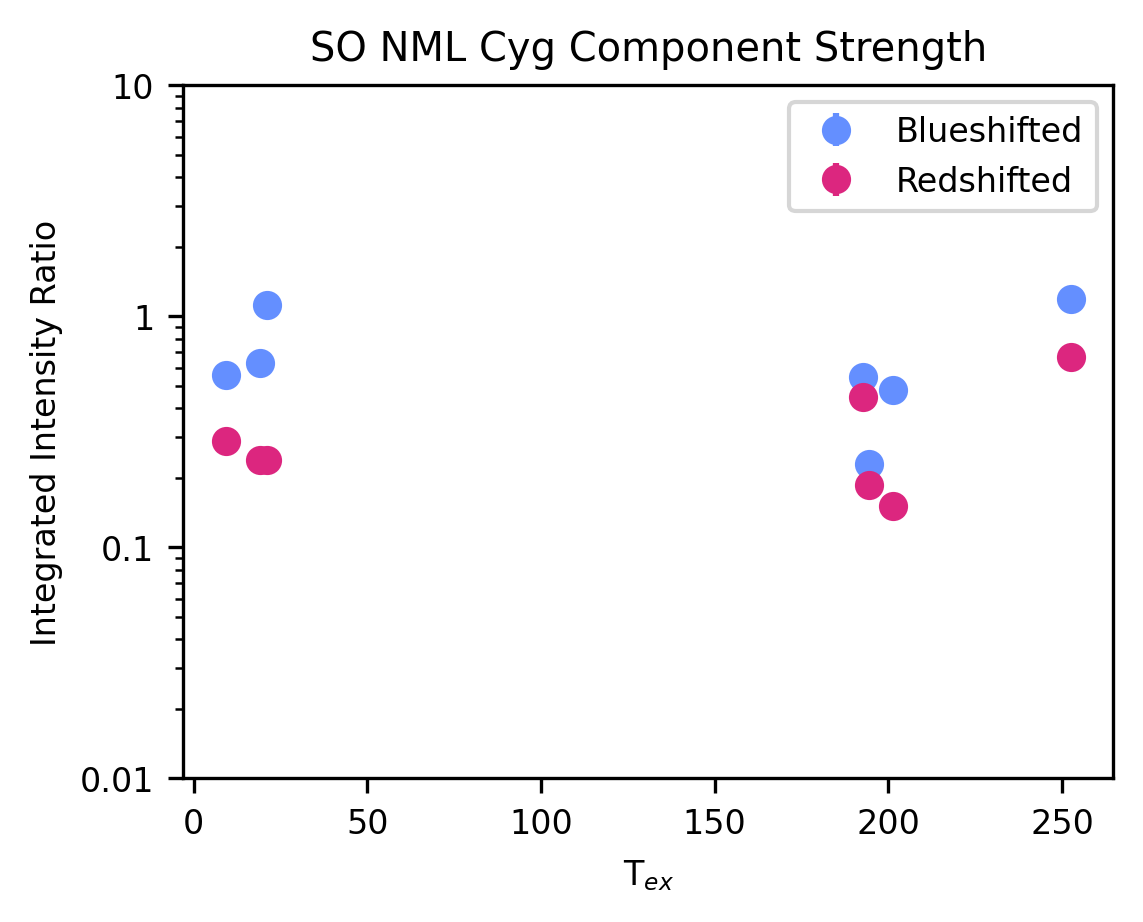}
\end{subfigure}\hfill
\caption{Same as Fig.~\ref{fig:co_fits} for detected SO emission lines around NML~Cyg.}
\label{fig:so_fits}
\end{figure*}

\begin{figure*}
\begin{subfigure}[c]{0.32\textwidth}
		\centering
		\includegraphics[height=1.85in]{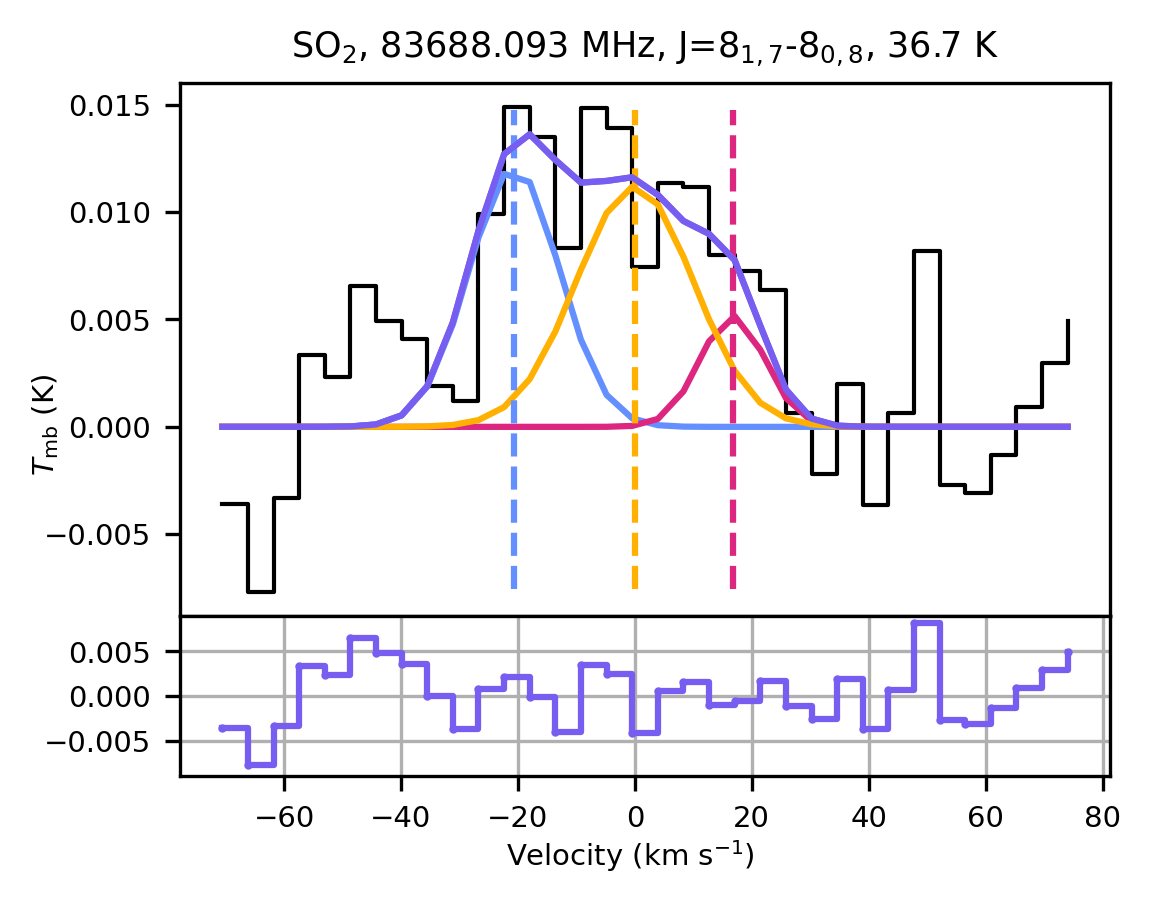}
	\end{subfigure}\hfill
\begin{subfigure}[c]{0.32\textwidth}
		\centering
		\includegraphics[height=1.85in]{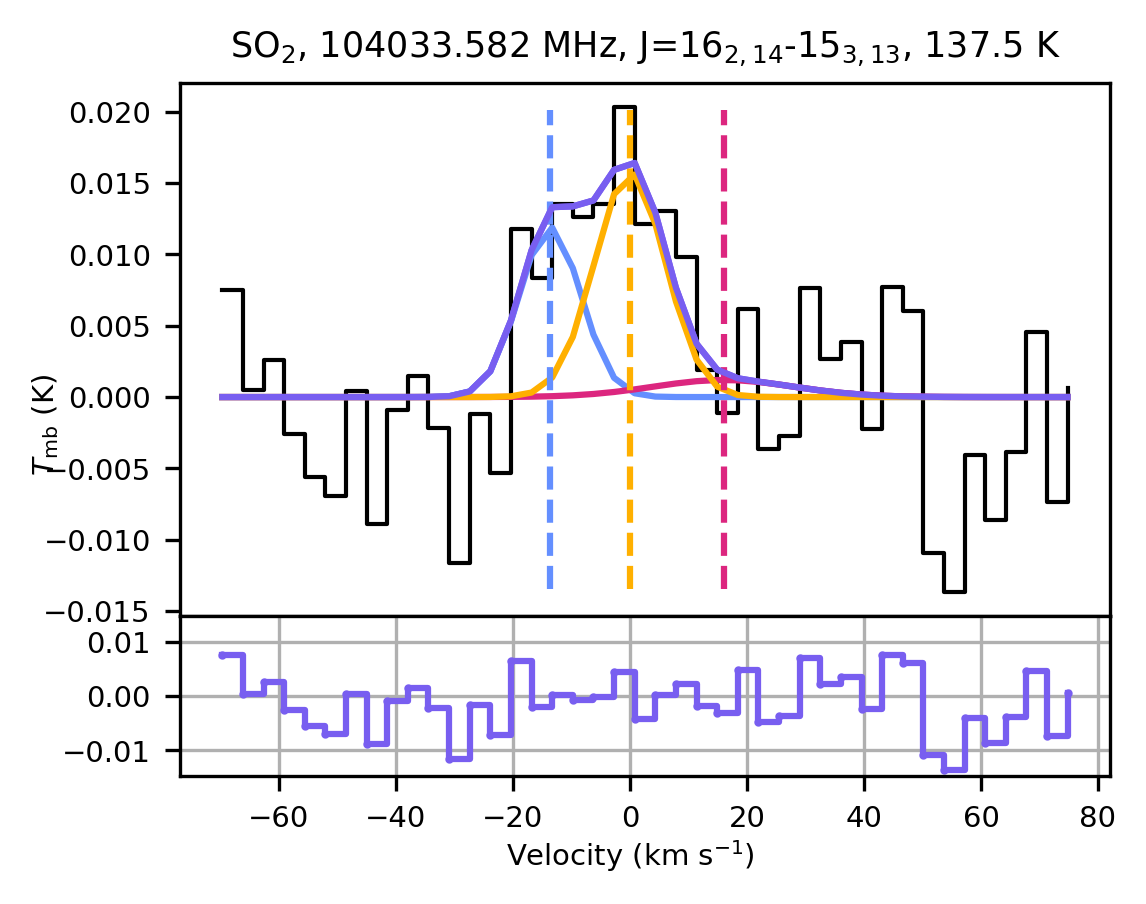}
	\end{subfigure}\hfill
\begin{subfigure}[c]{0.32\textwidth}
		\centering
		\includegraphics[height=1.85in]{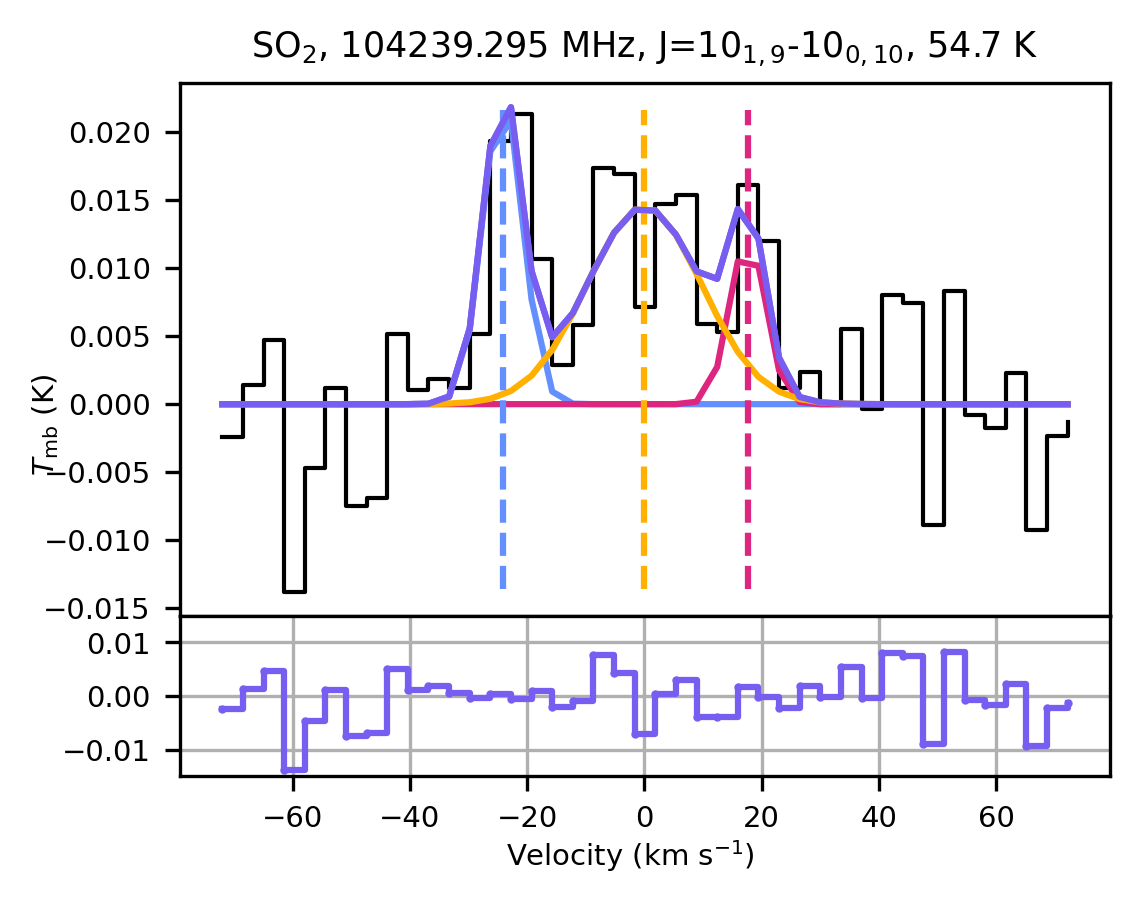}
	\end{subfigure}\hfill
	\caption{Same as Fig.~\ref{fig:co_nml_multifits} for SO$_{2}$ emission around NML~Cyg observed with OSO.}
	\label{fig:so2_lines}
\end{figure*} 

\begin{figure*}
\begin{subfigure}[c]{0.32\textwidth}
\centering
\includegraphics[height=1.85in]{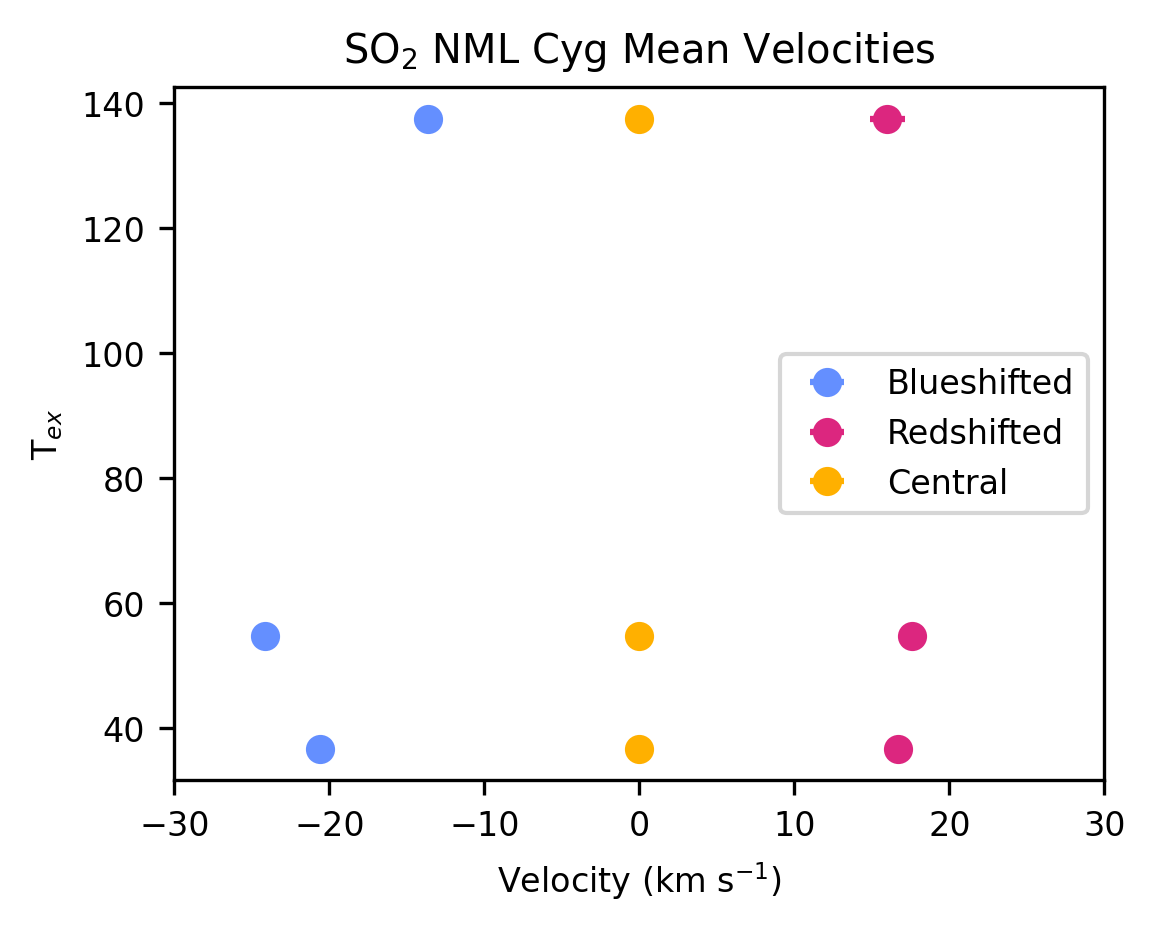}
\end{subfigure}\hfill
\begin{subfigure}[c]{0.32\textwidth}
	\centering
	\includegraphics[height=1.85in]{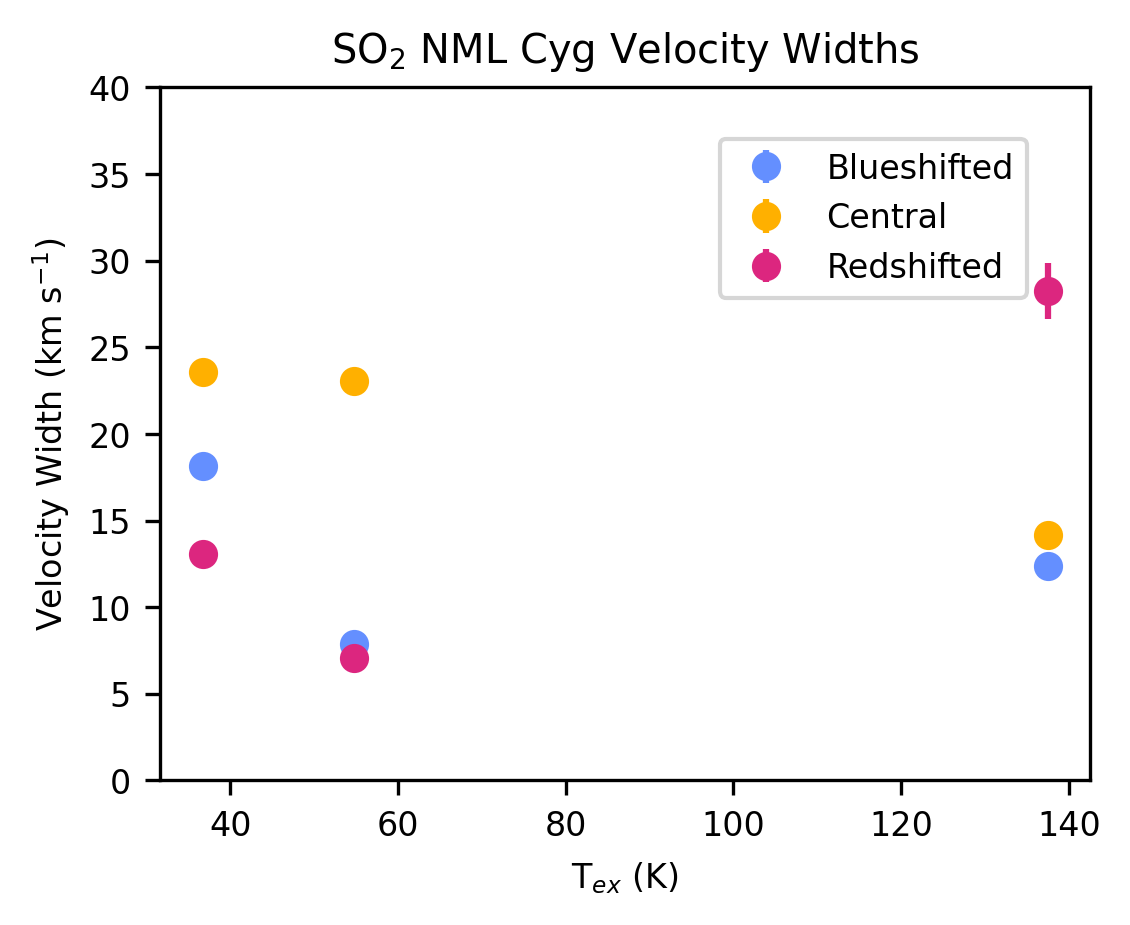}
\end{subfigure}\hfill
\begin{subfigure}[c]{0.32\textwidth}
\centering
\includegraphics[height=1.85in]{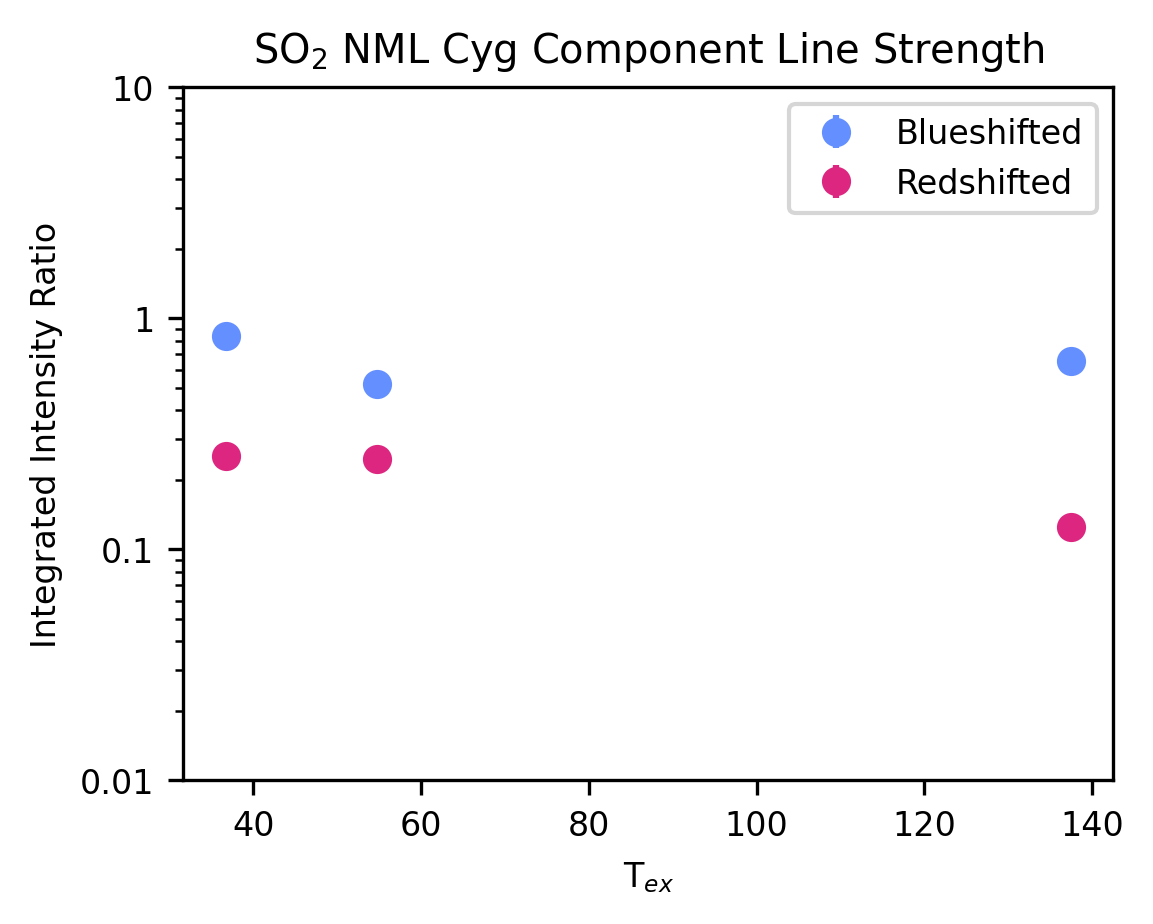}
\end{subfigure}\hfill
\caption{Same as Fig.~\ref{fig:co_fits} for detected SO$_{2}$ emission lines around NML~Cyg.}
\label{fig:so2_fits}
\end{figure*}

\begin{figure}
	\centering
	\includegraphics[width=0.32\textwidth]{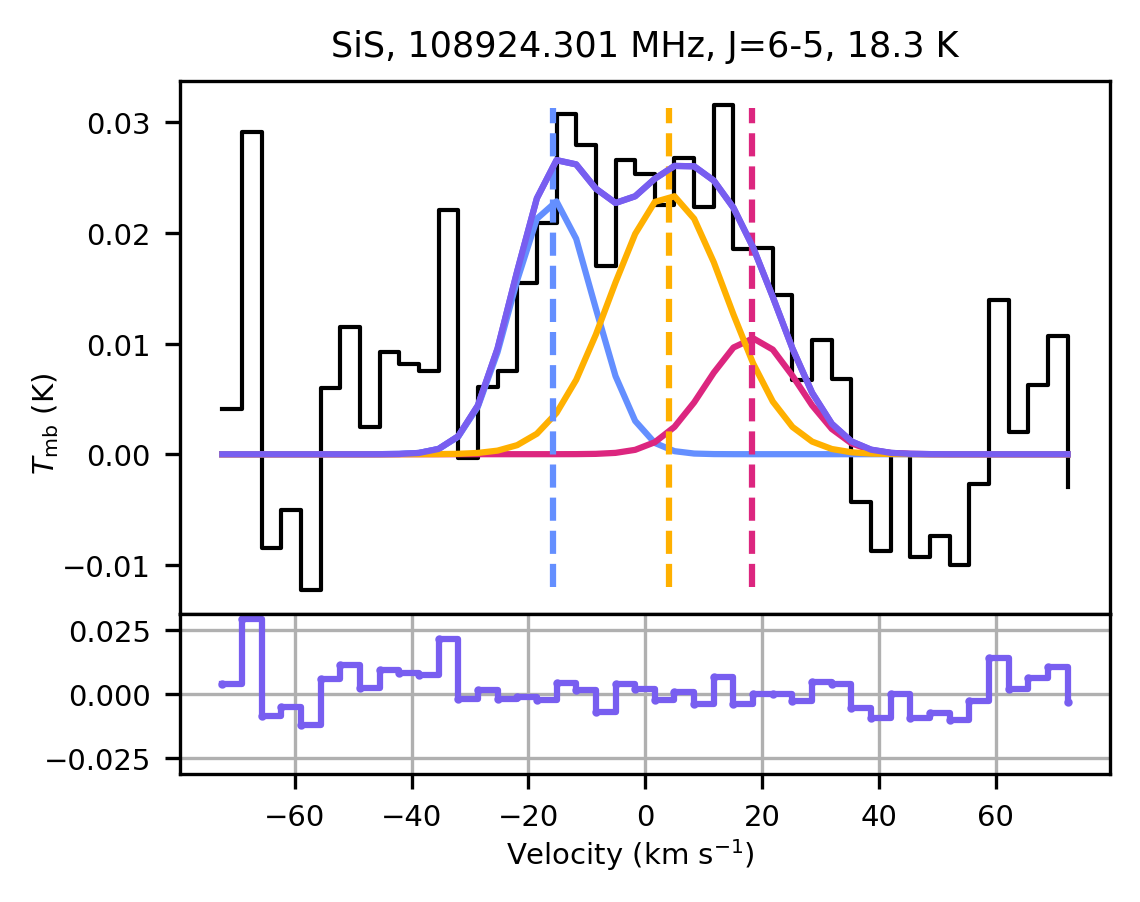}
	\caption{Multi-component fit, as described in Fig.~\ref{fig:co_nml_multifits}, for SiS $J = 6 - 5$ observed with OSO around NML~Cyg.}
	\label{fig:SiS_oso}
\end{figure}

\begin{figure}
	\centering
	\includegraphics[width=0.32\textwidth]{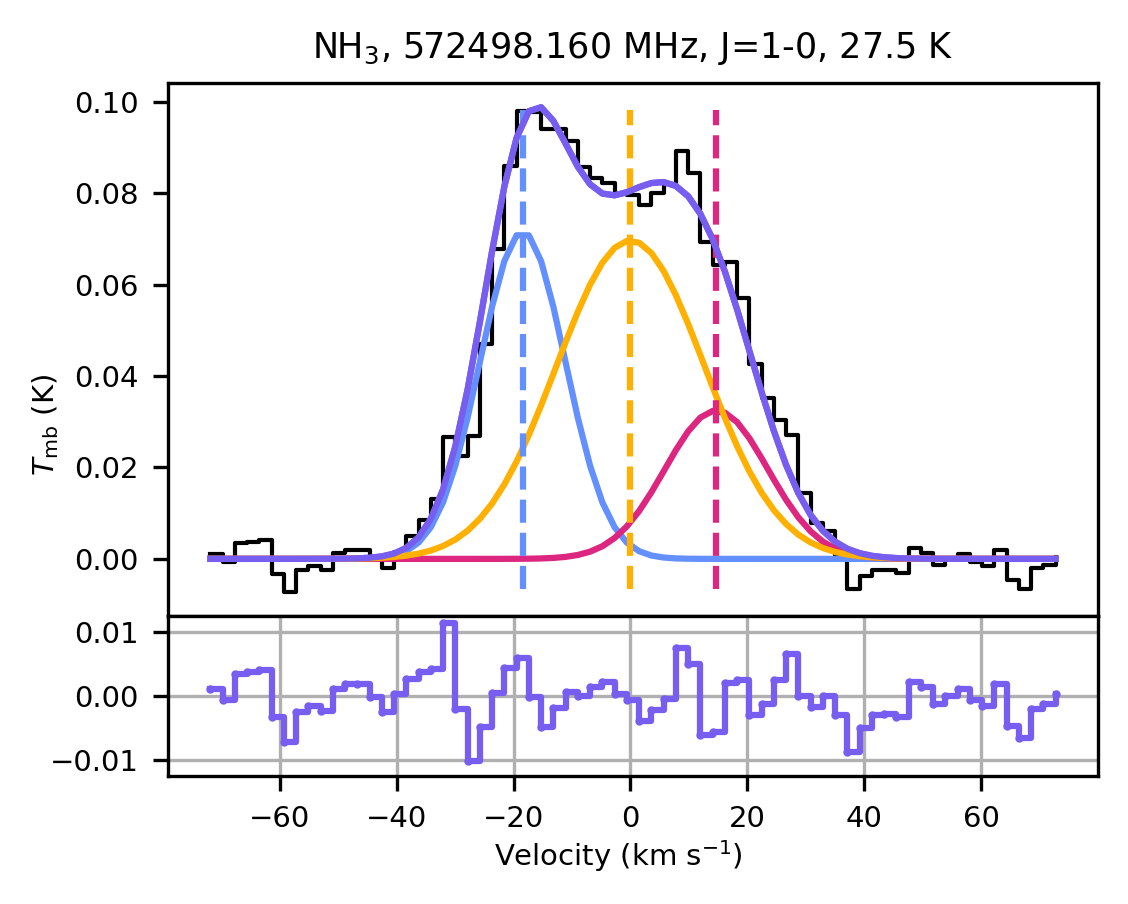}
	\caption{Multi-component fit, as described in Fig.~\ref{fig:co_nml_multifits}, for NH$_{3}$ $J = 1 - 0$ observed with HIFI around NML~Cyg.}
	\label{fig:nh3_hifi}
\end{figure}

\begin{figure}
	\centering
	\includegraphics[width=0.32\textwidth]{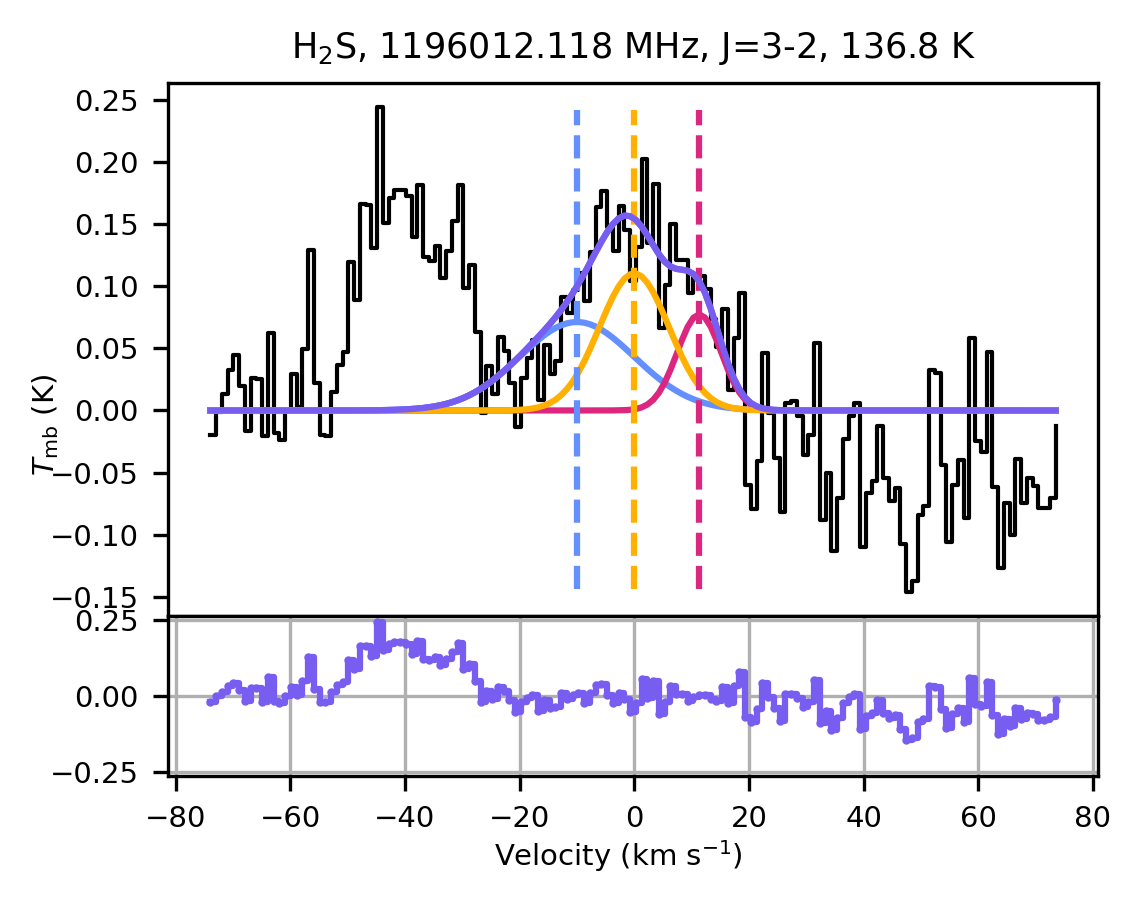}
	\caption{Multi-component fit, as described in Fig.~\ref{fig:co_nml_multifits}, for H$_{2}$S $J_{K_{\mathrm{a}},K_{\mathrm{c}}} = 3_{1,2} - 2_{2,1}$ observed with HIFI around NML~Cyg.}
	\label{fig:h2s_hifi}
\end{figure}

\begin{figure}
	\centering
	\includegraphics[width=0.32\textwidth]{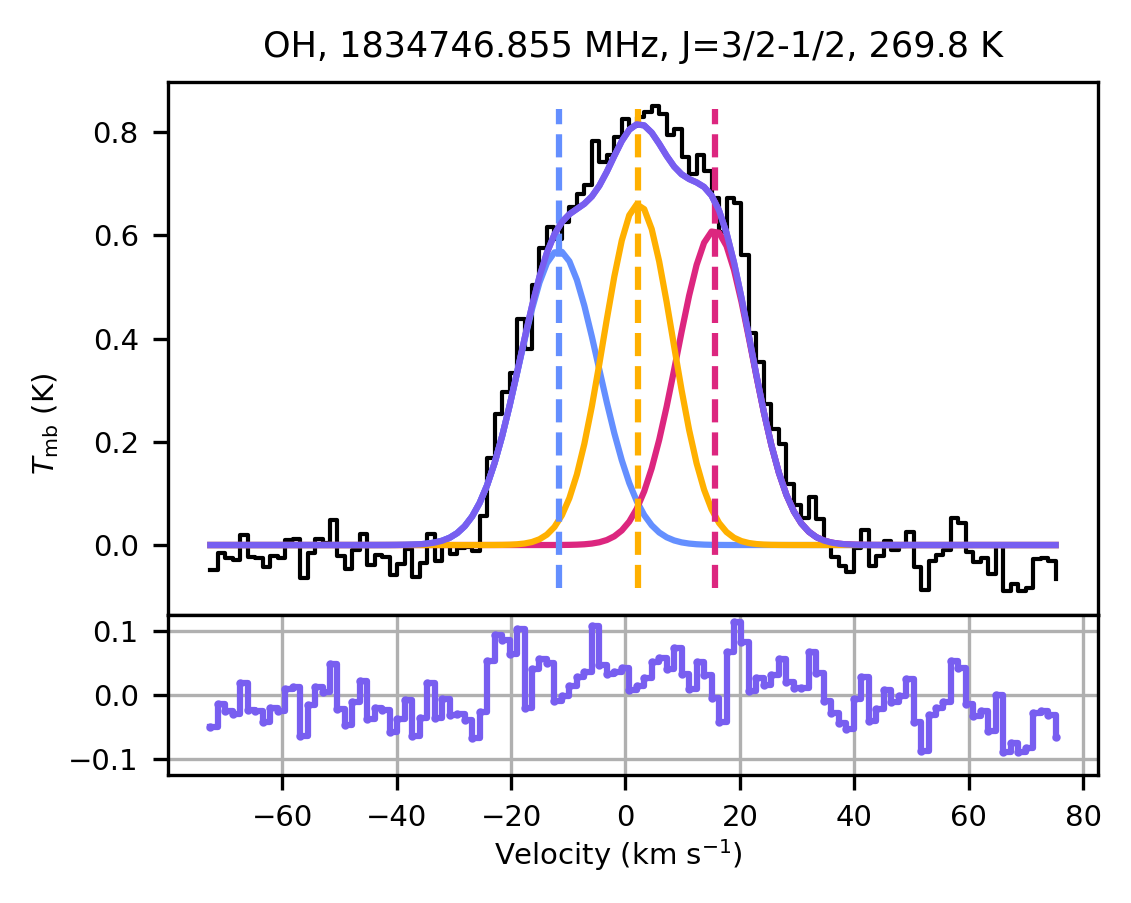}
	\caption{Multi-component fit, as described in Fig.~\ref{fig:co_nml_multifits}, for OH $N_J = 2_{3/2} - 1_{1/2}$ 
	observed with HIFI around NML~Cyg.}
	\label{fig:oh_hifi}
\end{figure}

\section{Full Spectra}

\begin{figure}
	\centering
	\includegraphics[width=0.95\textwidth]{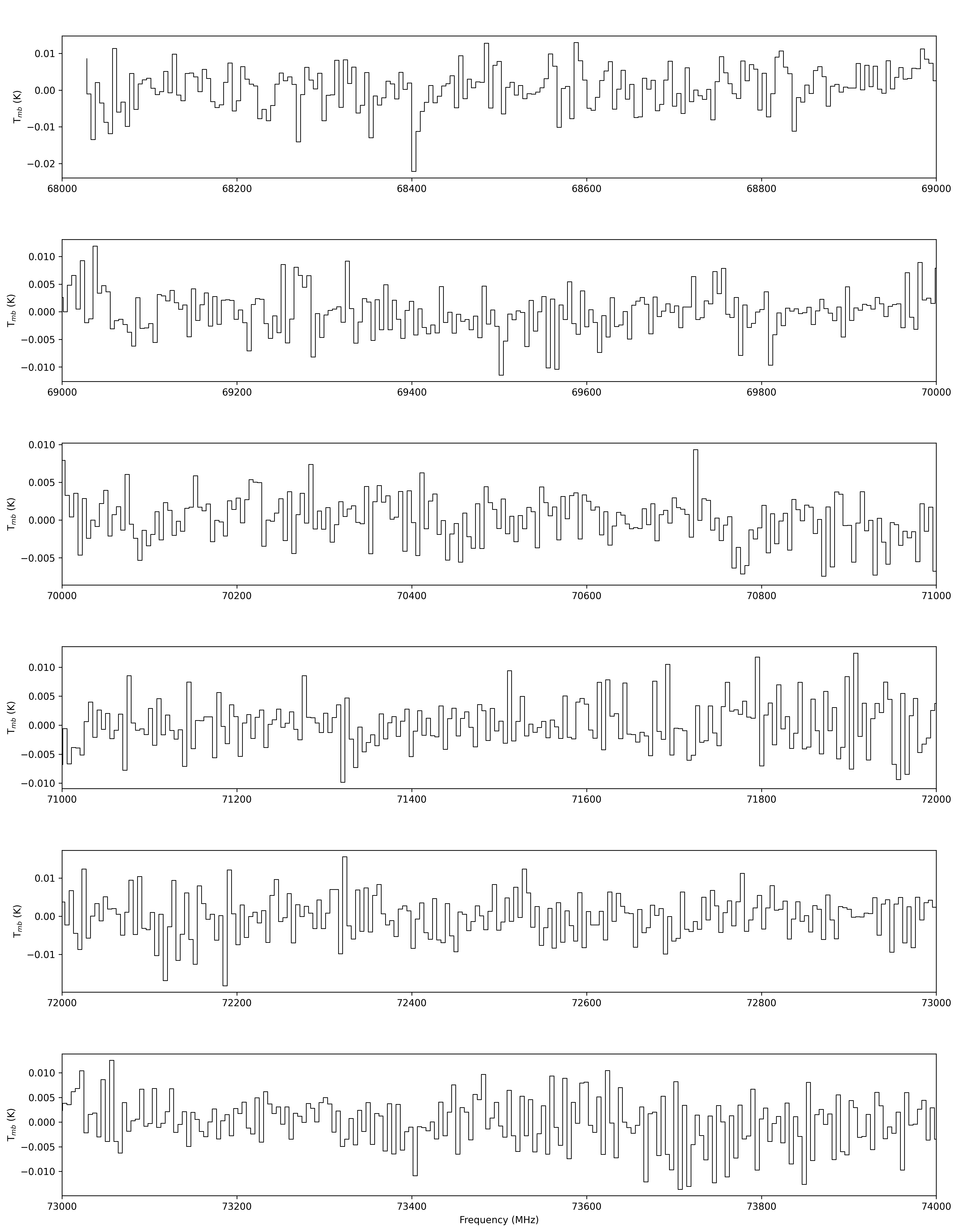}
	\caption{Full survey of NML~Cyg obtained with OSO 20m telescope over $68 - 116$\,GHz. Line detections are indicated by a red vertical line.}
	\label{fig:2index}
\end{figure}

\foreach \index in {2,...,8}
{%
	\begin{figure}[!htb] \ContinuedFloat
		\centering
		\includegraphics[width=0.95\textwidth]{figures/NML_full_spectra_\index.png}
		\caption{Continued.}
	\end{figure}
}

\newpage

\begin{figure*}
	\centering
	\includegraphics[width=0.95\textwidth]{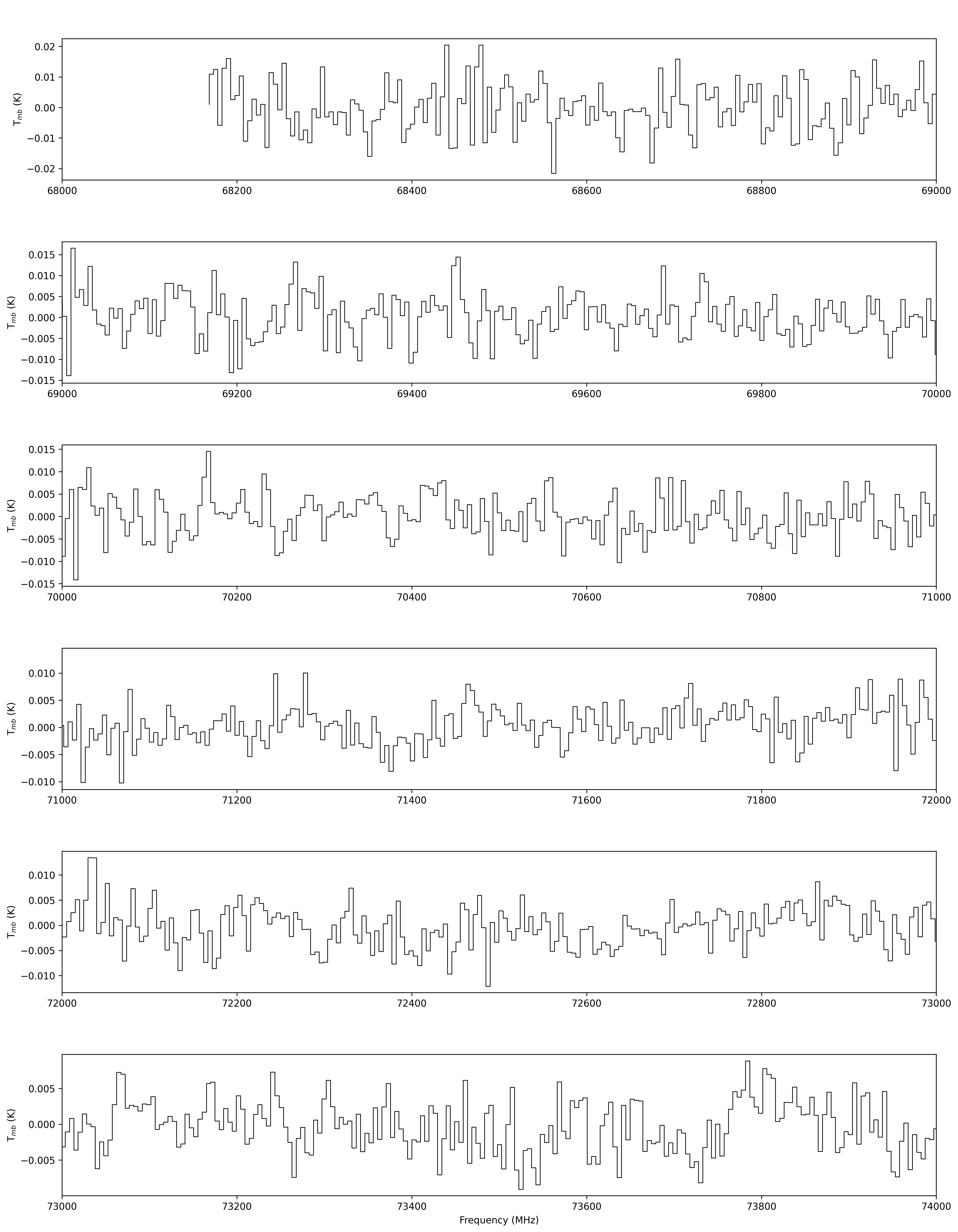}
	\caption{Full survey of IRC~+10420 obtained with OSO 20m telescope over $68 - 116$\,GHz. Line detections are indicated by a red vertical line.}\label{fig:1index}
\end{figure*}
	
\foreach \index in {2,...,8}
	 	{%
\begin{figure*}\ContinuedFloat
	\centering
	\includegraphics[width=0.95\textwidth]{figures/IRC_full_spectra_\index.png}
	\caption{Continued.} 
\end{figure*}
				}
	

\bsp	
\label{lastpage}
\end{document}